\documentclass{article}

\usepackage{fullpage}
\usepackage[margin=1in]{geometry}
\usepackage[utf8]{inputenc}
\usepackage{setspace}
\usepackage{graphicx}
\usepackage[font=normalsize]{caption}
\usepackage[font=normalsize]{subcaption}
\usepackage[round, sort, compress]{natbib}
\setcitestyle{aysep={}}
\usepackage{bibunits}
\usepackage{xspace}

\usepackage{footmisc}

\usepackage{makecell}
\usepackage{url}
\usepackage{enumitem} 
\usepackage{rotating}
\usepackage{footmisc}
\usepackage{titlesec}
\usepackage[colorlinks=true,linkcolor=black,anchorcolor=black,citecolor=black,menucolor=black,runcolor=black,urlcolor=black,bookmarks=true,breaklinks]{hyperref}
\usepackage{threeparttable}
\usepackage{pdfpages}
\usepackage{colortbl}
\usepackage{xcolor}
\definecolor{dark-red}{rgb}{0.4,0.15,0.15}
\definecolor{dark-blue}{rgb}{0.15,0.15,0.45}
\definecolor{medium-blue}{rgb}{0,0,0.5}
\hypersetup{
    colorlinks, linkcolor={dark-red},
    citecolor={dark-red}, urlcolor={dark-red}
}
\usepackage[reqno]{amsmath}
\usepackage{amssymb}
\newcolumntype{.}{D{.}{.}{-1}}
\newcolumntype{d}[1]{D{.}{.}{#1}}
\usepackage{lscape}
\usepackage{multirow}
\usepackage{booktabs}
\usepackage{latexsym}
\usepackage{verbatim}
\usepackage{array}
\usepackage{fontspec}
\usepackage{pgf, tikz}
\usepackage{amssymb}
\usetikzlibrary{arrows, automata, shapes}
\usetikzlibrary{shapes}

\titleformat*{\section}{\LARGE\bfseries}
\titleformat*{\subsection}{\Large\bfseries}
\titleformat*{\subsubsection}{\large\bfseries}
\titleformat*{\paragraph}{\large\bfseries}
\titleformat*{\subparagraph}{\large\bfseries}

\usepackage{longtable}
\usepackage{array}
\usepackage{tabularray}

\usepackage{listings}
\lstdefinestyle{prompt}{
  basicstyle=\ttfamily\footnotesize,
  breaklines=true,
  breakatwhitespace=false,
  columns=fullflexible,
  keepspaces=true,
  showstringspaces=false,
  breakindent=0pt,
  frame=single
}

\title{\huge\sc How social media creators shape mass politics: \\ A field experiment during the 2024 US elections\thanks{We thank Zihan Jin, Chenchen Li, Zaid Sahawneh, and Junyi Zhang for valuable research assistance. We thank Jennifer Allen, Jae Aron, Adam Berinsky, Milena Djourelova, Ryan Enos, Thomas Fujiwara, Andy Guess, Rafael Jiménez-Durán, Yingdan Lu, Rakeen Mabud, Sophie Mainz, Brendan Nyhan, Markus Prior, and Etienne Ollion for their valuable feedback and insights. We are also grateful to participants at the APSA Political Communication Pre-Conference, the Bocconi-Columbia Conference on Political, Social and Economic Inequality, the Center for Comparative and International Studies Colloquium at the University of Zurich and ETH Zurich, the Columbia Media Effects Empirical Workshop, the Comparative Politics and Political Economy Research Workshop at the University of Konstanz, the Duke Frontiers Workshop in Online Political Behavior, Harvard's Working Group in Political Psychology and Behavior Seminar, the Hoover Institution's Social Media and Democracy Conference, the MIT Polarization Workshop, the NBER Political Economy Summer Institute, the Northeast Political Economy Conference at Brown University, the Political Economy Colloquium and Preston Research Colloquium at the University of Wisconsin–Madison, Princeton Political Economy Seminar, Summer Institute in Computational Social Science-Paris, University College London, UCLA, and the University of Pennsylvania for their helpful comments and suggestions. We thank John Alston of Bovitz for his invaluable assistance in executing this project. The study was approved by Columbia University's Institutional Review Board (IRB-AAAV1924) and the design was pre-registered in the Social Science Registry (\href{https://www.socialscienceregistry.org/trials/13994}{www.socialscienceregistry.org/trials/13994}). The research team is grateful for funding to conduct the evaluation (but not produce any online content) from the Russell Sage Foundation and Department of Political Science and Institute for Social and Economic Research and Policy at Columbia University.}
}

\author{\hspace{10pt} 
\sc \Large Kirill Chmel\footnote{PhD student in Political Science, Columbia University} \and
\sc \Large Eunji Kim\footnote{Assistant Professor of Political Science, Columbia University} \and
\sc \Large John Marshall\footnote{Associate Professor of Political Science, Columbia University} \hspace{10pt} \and
\sc \Large Tiffany Fisher-Love\footnote{Incite Studio} \and
\sc \Large Nathaniel Lubin\footnote{Incite Studio and Berkman Klein Center, Harvard University}
}

\date{\sc \Large December 2025}

\AtBeginDocument{}

\begin{document}

\maketitle

\vspace{-48pt}

\begin{abstract} %150 word strict limit for APSR, 250 for QJE
    \hyphenpenalty=1000
    \exhyphenpenalty=1000
    \tolerance=25000
    \large \singlespacing
     \noindent Political apathy and skepticism of traditional authorities are increasingly common, but social media creators (SMCs) capture the public's attention. Yet whether these seemingly-frivolous actors shape political attitudes and behaviors remains largely unknown. Our pre-registered field experiment encouraged Americans aged 18-45 to start following five progressive-minded SMCs on Instagram, TikTok, or YouTube between August and December 2024. We varied recommendations to follow SMCs producing predominantly-political (PP), predominantly-apolitical (PA), or entirely non-political (NP) content, and cross-randomized financial incentives to follow assigned SMCs. Beyond markedly increasing consumption of assigned SMCs' content, biweekly quiz-based incentives increased overall social media use by 10\% and made participants more politically knowledgeable. These incentives to follow PP or PA SMCs led participants to adopt more liberal policy positions and grand narratives around election time, while PP SMCs more strongly shaped partisan evaluations and vote choice. PA SMCs were seen as more informative and trustworthy, generating larger effects per video concerning politics. Participants assigned to follow NP SMCs instead became more conservative, consistent with left-leaning participants using social media more when right-leaning content was ascendant. These effects exceed the impacts of traditional campaign outreach and partisan media, demonstrating the importance of SMCs as opinion leaders in the attention economy as well as trust- and volume-based mechanisms of political persuasion.

\end{abstract}

\clearpage

\addtocontents{toc}{\protect\setcounter{tocdepth}{0}}

\large

\onehalfspacing

\begin{bibunit}[apsr]

\section{Introduction}

The way Americans consume political information is undergoing a profound transformation. A decade ago, television and newspapers were the dominant sources of news; now social media is a---if not \textit{the}---leading way Americans obtain news \citep{aridor2025digital, newman2023digital}.\footnote{See also: Pew Research Center, “News Platform Fact Sheet,” September 25, 2025, \href{https://www.pewresearch.org/journalism/fact-sheet/news-platform-fact-sheet/?tabId=tab-b39b851c-e417-48ef-9b10-93ee21a0030e}{www.pewresearch.org/journalism/fact-sheet/news-platform-fact-sheet/?tabId=tab-b39b851c-e417-48ef-9b10-93ee21a0030e}.} While platforms promoting static text and images, like Facebook and Twitter, used to dominate the social media landscape while amplifing content from traditional media outlets, Gen Z and Millennials increasingly engage with user-generated news on video-based platforms \citep{aridor2024economics, aridor2025digital}. This was particularly pronounced during the 2024 US election campaign, which the \textit{New York Times} dubbed ``the TikTok election;''\footnote{\textit{New York Times}, ``The Election Has Taken Over TikTok,'' October 21, 2024, \href{https://www.nytimes.com/interactive/2024/10/21/business/media/2024-election-tiktok-trump-harris.html}{www.nytimes.com/interactive/2024/10/21/business/media/2024-election-tiktok-trump-harris.html}.} the \textit{Washington Post} further proclaimed paid-for content on Instagram, TikTok, and YouTube has become ``the new dark money.''\footnote{\textit{Washington Post}, ``The new dark money: How influencers get paid big bucks to court your vote,'' October 26, 2024, \href{https://www.washingtonpost.com/technology/2024/10/26/social-media-influencers-election-money-campaigns}{www.washingtonpost.com/technology/2024/10/26/social-media-influencers-election-money-campaigns}.} Indeed, the Democratic Party invited more than 200 digital creators to its national convention, while Donald Trump's victory speech thanked the Nelk Boys, Adin Ross, Theo Von, Bussin' With The Boys, and Joe Rogan. 

Political content on video-driven social media platforms is distinctive: instead of appearing as articles, newscasts, or talk shows from conventional news sources, much of this content is produced or shared by independent social media creators (SMCs) who develop followings by posting original content on their social media platforms. With 37\% of Americans aged 18-29 and 26\% of those aged 30-49 now regularly getting news from SMCs,\footnote{Pew Research Center, ``America's News Influencers,'' November 18, 2024, \href{https://www.pewresearch.org/journalism/2024/11/18/americas-news-influencers/}{www.pewresearch.org/journalism/2024/11/18/americas-news-influencers}.} the ever-widening array of media choices allows citizens to opt out of traditional news sources that adhere to journalistic standards of being ``accurate, fair and thorough''\footnote{Society of Professional Journalists, Code of Ethics. \href{https://www.spj.org/spj-code-of-ethics/}{www.spj.org/spj-code-of-ethics}.} and into content of their choosing \citep{arceneaux2013changing, prior2007}. Operating beyond the control of traditional media gatekeepers, SMCs often blur the lines between news and non-news by embedding policy and political ideas within their usual non-political content. 

Because traditional campaign messaging efforts and presidential debates often have limited effects on political beliefs or preferences \citep[e.g.][]{allcott2025effects, kalla2018minimal, le2023campaigns}, one might expect seemingly-frivolous SMCs without political expertise to be even less persuasive. Yet, SMCs specialize in engaging and educating their followers, which may extend their influence to politics. By reaching audiences that otherwise receive little news or political content, SMCs may provide information about salient issues, frame news and policy debates, and encourage political participation. SMCs may also be particularly effective messengers due to the trust they establish with followers by cultivating parasocial relationships---enduring but unreciprocated socio-emotional connections that audiences form with media personas \citep{hund2023influencer, lou2019influencer, schmuck2022politics}. This combination of accessibility and trust positions SMCs as modern-day opinion leaders, with the capacity to shape and interpret the political information that citizens encounter.

The political significance of SMCs may also depend on their content. At one end of the spectrum are lifestyle-oriented SMCs, whose content centers on topics like food, fashion, wellness, humor, or popular culture. On the rare---but increasingly common \citep{von2025political}---occasions they broach politics, they do so in subtle ways that can engage followers while minimizing the risk of alienation or backlash. At the other end of the spectrum are explicitly political SMCs, who focus on breaking news, policy debates, or partisan commentary. These creators tend to attract more politically-engaged and often ideologically-aligned audiences, potentially limiting their appeal or credibility among broader audiences. This heterogeneity poses a theoretical puzzle: are SMCs more politically persuasive when they limit overt signaling and engage inattentive citizens through periodic entertainment-forward political content, or when they frequently produce political content with a partisan slant? Although elites and institutional actors increasingly integrate SMCs into electoral strategies, civic campaigns, and public discourse, little is yet known about their influence over public opinion or political behavior.\footnote{Beyond election campaigns, state and local governments in Colorado, Oklahoma, and Minneapolis have enlisted micro-influencers to promote vaccines or counter misinformation. Non-profit organizations, such as the ACLU, Planned Parenthood, Turning Point USA, and World Wildlife Fund, now also create sponsored content with SMCs to educate the public on their issues or advance their policy positions and worldviews.} 

In this paper, we conduct the first field experiment of its kind to investigate whether and how sustained exposure to different types of video-based SMCs affected the political beliefs, attitudes, and behaviors of Millennial and Gen Z adults. During the 2024 US election campaign, we recruited a panel of 4,716 Americans aged 18-45 who regularly used Instagram, TikTok, or YouTube. After completing a baseline survey in early August 2024, participants were randomly assigned a treatment or control condition described below; we later measured various outcomes in midline and endline surveys fielded shortly before the 2024 elections and in January 2025. Survey responses are augmented with behavioral measures of YouTube (and some TikTok) browsing histories, donations to different types of non-profit causes, and voter file data. To maintain the study's independence, we did not partner with social media platforms and pre-registered primary analyses, outcome measurement, and hypotheses.

Our intervention encouraged participants to start following either five \textit{predominantly-apolitical} (PA) or five \textit{predominantly-political} (PP) progressive-minded SMCs between August and December 2024.\footnote{To reduce the consumption burden on participants, we focused on SMCs typically producing no more than 20 minutes of content per week. We also offered participants the option to learn more about one of two recommended SMCs producing longer-form content, but without incentives.} We distinguish these treatment conditions based on whether assigned SMCs, who are all social media natives rather than prominent due to their offline presence, principally produce explicitly political, and often partisan, videos. Our pool of 20 PA SMCs are Better Internet Initiative (BII) fellows who periodically integrate educational content on policy issues into their otherwise apolitical content. As part of the BII fellowship, these PA SMCs produced around ten non-partisan videos to inform viewers about issues relating to climate change, democracy, economic justice, or public health. In contrast, our pool of 20 PP SMCs---including 6 BII fellows---specialize in covering news and current affairs, and frequently include partisan commentary. We developed a machine learning algorithm to recommend the five SMCs within each treatment arm that best matched each participant's interests and platforms of choice. To mitigate against non-compliance in a competitive market for attention, we cross-randomized three encouragements for participants: receiving only our recommendations; an additional \$1 for each validated follow of a recommended SMC (after baseline and midline surveys); or nine biweekly quizzes each offering \$20 for correctly answering questions about recommended SMCs' non-political content.

We compare these treatment conditions with three control groups. A pure control group received no additional content. To help separate the effect of political and policy-oriented content from greater use of social media, a placebo condition similarly encouraged participants to follow five of 20 \textit{non-political} (NP) SMCs almost exclusively producing non-political content. Finally, to help distinguish the medium from its content, a text-only condition sent participants biweekly email and SMS messages summarizing the BII program's messaging. 

Our self-reported and behavioral data show that treated participants watched and internalized content from assigned SMCs, mostly due to biweekly quizzes. Quiz-incentivized participants watched an average of 11-15 videos per assigned SMC between August and December, scored highly on biweekly quizzes, and demonstrated strong recall of SMC content relative to the pure control and SMS/email groups (who very rarely consumed videos from treatment SMCs). These participants also spent around 10\% more time using social media than those in the control group, though the intervention did not significantly change the political or ideological composition of participants' broader social media consumption bundle. The latter result suggests that platform algorithms and individual search are relatively sticky at the margin. The recommendation-only encouragement also modestly increased engagement with each type of SMC, indicating limited aversion to creator-driven political content, at least during a fractious election campaign. The remaining analyses restrict attention to quiz-incentivized participants, who registered by far the strongest ``first-stage.''

Our main findings reveal that quiz incentives to start following both types of progressive-minded SMCs increased political engagement and shifted policy attitudes and systemic understandings to the left. First, participants encouraged to follow NP, PA, and especially PP SMCs became significantly more politically knowledgeable than the pure control group, at both midline and endline. Like prior studies of Facebook and Instagram's more text-based platform \citep[e.g.][]{allcott2020welfare, derksen2025instagram}, video-based SMCs also increased followers' political engagement. Second, encouragement to follow PA and PP SMCs led participants to adopt more liberal policy positions and narratives about how economic and political systems operate, whereas encouragement to follow NP SMCs instead made participants more conservative. Since the lean of their social media feeds did not perceptibly change, the latter finding most likely reflects our relatively liberal young panelists using more social media in a right-leaning online environment \citep{gauthier2025X, ibrahim2025tiktok}.\footnote{According to the Pew Center, Conservative SMCs outpaced their liberal counterparts in both engagement and content production during the 2024 election campaign% \citep{pew2024newsinfluencers, pew2025trumpharris}
, and such content could also have been more persuasive.} These attitudinal differences of about 0.1 standard deviations between the treatment and placebo SMCs conditions are most pronounced in the election-time midline survey, but persist at endline in participants' cause donation decisions. Specifically, quiz incentives to follow PA and PP SMCs increased donations to liberal over conservative causes by around 10 percentage points (or 0.15 standard deviations), relative to both the pure and placebo control groups. 

These changes in policy preferences partially translated into partisan preferences, but did not alter political participation. Particularly relative to the NP SMCs group that moved to the right, quiz incentives to follow PA and especially PP SMCs caused participants to become more favorable toward the Democratic Party and Kamala Harris and less favorable toward the Republican Party and Donald Trump. This change was again more pronounced in the election-time midline survey than the endline in January, and those encouraged to follow PP SMCs became several percentage points more likely to report voting for Harris. However, while progressive-minded SMCs led followers to favor the Democrats, they did not significantly affect political participation. Across voter registration, intended and actual voter turnout, attending protests, and various other political activities, we find no evidence that SMCs cultivated offline participation. The gap between persuasion and mobilization highlights limitations in SMCs' messaging, suggesting that the leap from online influence to real-world action remains a challenge. This said, mobilization may have been more difficult in 2024, when electoral turnout reached its second highest level since the 1960s. %A question that emerges from this finding how dedicated, longterm PA messaging related specifically to civic participation may be. Also of note, 2024 election participation was at baseline already close to a record level in American history, only slightly surpassed by 2020.

Comparing types of SMC content, the PA SMCs group experienced slightly larger effects on policy preferences and the PP SMCs group experienced slightly larger effects on partisan preferences, but these differences are largely indistinguishable. However, since PA SMCs created far less political content, they likely generated much larger effects \textit{per political video}. Our analyses of potential mechanisms show that participants encouraged to follow PA SMCs rated these SMCs as more informative and trustworthy than participants encouraged to follow PA or NP SMCs. Heterogeneous treatment effects further suggest that persuasion was greatest among respondents who felt most connected to and trusting of their assigned SMCs. These results imply that, within several months, SMCs established parasocial connections that enhanced their credibility among followers. Suggesting that SMCs' impacts could endure through the appeal of their content, consumption and internalization of assigned PA and PP SMCs' content continued after incentives were withdrawn. The more cost-efficient route to political influence thus appears to be through SMCs who rarely engage in politics, while moderators point to greater effects when reaching viewers for whom progressive content was counter-attitudinal. Since parasocial connections are likely to be stronger among individuals who self-select into following an SMC than those paid to do so, our estimates may capture a lower bound on SMCs' political influence.

Even with modest effect sizes of 0.1 to 0.2 standard deviations, SMCs appear to be at least as influential as other forms of political messaging. Within our experiment, incentives to follow progressive-minded SMCs generally produced larger effects than similar messaging via SMS/email, which required less attention from participants but likely failed to establish analogous trust in the sender. Beyond our study, we observe changes in policy attitudes and partisan favorability that exceed typical campaign outreach or partisan media in the US \citep[e.g.][]{allcott2025effects, broockman2025consuming, coppock2022does, kalla2018minimal, spenkuch2018political}, even during 2024's saturated and polarized campaign and media environment. Both PA and PP SMCs have thus become important opinion leaders.

These findings contribute to the study of political communication in the digital age. First, we provide the strongest evidence to date that SMCs shape viewers' policy and partisan positions. SMCs' influence on consumer engagement and purchases is much-studied by marketing scholars \citep[see][]{barari2025meta, leung2022influencer}, but prior studies of political persuasion on social media have focused on content generated by public figures, traditional media, and partisan ads \citep{aridor2024economics}.\footnote{Examples from social media include \cite{allcott2025effects}, \cite{bessone2022social}, \cite{ehrmann2022central}, \cite{enriquez2024mass}, and \cite{levy2021social}. Beyond social media, a large literature finds effects of slanted TV channels \citep[e.g.][]{broockman2025consuming, dellavigna2007, enikolopov2011media, martin2017}, radio stations \citep[e.g.][]{adena2015radio}, and newspapers \citep[e.g.][]{chiang2011media} as well as partisan ads \citep[e.g.][]{larreguy2018leveling, spenkuch2018political} on political preferences.} Several early studies find mixed political effects of SMCs,\footnote{\cite{Alsharawy2025TikTokInfluencers} found no attitude change in a month-long field experiment among college students. \cite{schmuck2022politics} observed greater political interest in a small German panel, while \cite{Dekoninck2022Influencers} found that political influencers can mobilize online participation.} but cannot capture the consequences of prolonged real-world engagement or the depth of parasocial connections between SMCs and followers. Our five-month field experiment establishes that sustained engagement with SMCs affects political preferences beyond advertising products, and further reveals a quantity-quality trade-off: influence can emerge through volume of political content or accumulating trust by producing less. %Also notably, neither PA nor PP SMCs operated under the one-off paid-advertising model followed by many SMCs during the 2024 election campaign. 
We thus join recent work on entertainment TV and film \citep[e.g.][]{ang2023birth, durante2019, kim2025, kim2025apprentice} in emphasizing the policy and electoral significance of popular content not produced by news outlets or political elites. 

Second, we advance the emerging literature on the political consequences of social media by focusing on content produced by SMCs.\footnote{Social media has also been linked to protest \citep[e.g.][]{enikolopov2020social, qin2024social}, hate crime \citep[e.g.][]{muller2023hashtag}, and worse mental health \citep[e.g.][]{allcott2020welfare, braghieri2022social}.} Prior studies find that platform-level interventions---such as deactivating or reconfiguring Facebook, Instagram, WhatsApp, and X---affect political knowledge, but with limited effects on political attitudes \citep{allcott2020welfare, allcott2024effects, arceneaux2024facebook, derksen2025instagram, gauthier2025X, guess2023social, liu2025short, nyhan2023like, ventura2023misinformation}.\footnote{Our findings align with studies showing that \textit{starting to use} social media \citep{bowles2025access, fujiwara2024effect, guriev20213g, melnikov2023mobile} or algorithmic feeds \citep{gauthier2025X} in slanted media environments produces larger political effects than stopping.} We instead examine the \textit{content} of social media. Studying content is challenging because SMCs' political motivations are often undisclosed, users experience individualized feeds, and consumption bundles are difficult to measure. Overcoming these challenges, we affirm that video-based platforms increase political knowledge, but also that directing individuals toward a curated set of progressive-minded SMCs shifts their political positions. We thus demonstrate the importance of \textit{which} accounts individuals follow. 

Third, we extend foundational theories of political communication---on slanted media, soft news, and selective exposure---to today's fragmented, creator-driven media environment. Ideologically-distinct television channels like Fox News \citep{dellavigna2007, martin2017} or partisan print and digital newspapers \citep{gerber2009does, king2017news, levy2021social} can shape political attitudes and behaviors. But these traditional media outlets typically operate with professional editorial standards and institutional legitimacy. In contrast, we find that independent SMCs---who can cultivate parasocial connections, but often lack expertise or formal authority---can shape political preferences by establishing trust. By experimentally varying the intensity of political content across types of SMCs, we reinforce television-based theories of soft news and ``infotainment'' in a high-choice digital media environment where boundaries between entertainment and news have dissolved \citep{baum2003, chadwick2017, kim2025}. Our results suggest avoiding political overload for both attracting attention and gaining credibility. 

Finally, our findings point to ways of engaging younger Americans disillusioned with or disconnected from politics and legacy news. Consistent with the seminal two-step flow of communication framework \citep{katz1955personal, lazarsfeld1948}, we find that SMCs have become trusted opinion leaders in the digital age. Like church leaders, union organizers, and community activists before them, SMCs serve as intermediaries filtering information from mass media or political parties to engage, inform, and sway their followers \citep{harff2025revisiting}. By building audiences with few geographic or social constraints, SMCs can complement legacy media by providing alternative pathways for engaging citizens who avoid formal political discourse. This raises welfare and regulatory questions about the accuracy of SMC content, how SMCs are compensated, and to whom they are accountable. Consequently, the rise of SMCs is not just changing where people get news, but calls for a broader reckoning over how the evolving information ecosystem shapes the foundations of democracy.

\section{Background}

A key feature of the 2024 US election campaign was the prominent role of SMCs, serving both as intermediaries between political parties and social media users and as independent sources of information and opinion \citep{harff2025revisiting}. In this section, we describe the context of our study, explain how SMCs operate, and provide a conceptual framework to understand the effects of SMCs on political preferences.

\subsection{Election campaign context}

The 2024 US election campaign was one of the most dramatic in modern history. It initially pitted President Joe Biden against former President Donald Trump in a repeat of the contentious 2020 election, which culminated in Trump supporters storming the US Capitol several weeks before Biden was inaugurated. After a quiet primary season, the first presidential debate on June 27 alarmed many in the Democratic Party about Biden’s capacity to serve and campaign effectively. An assassination attempt against Trump---the Republican Party candidate---at a Pennsylvania rally on July 13 further destabilized the race. Biden dropped out a week later, and Vice President Kamala Harris became the presumptive Democratic nominee after a majority of convention delegates pledged their support a day later.

Amid this volatility, the campaign revolved around issues of the economy, democracy, abortion rights, immigration and border security, and questions of cultural identity. Following combative presidential and vice-presidential debates in September and a second assassination attempt against Trump on September 15, Trump defeated Harris on November 5. He won the Electoral College 312–226 and the popular vote by a 1.5 point margin, with Republicans securing majorities in both chambers of Congress. In the aftermath, Democratic elites and supporters confronted internal disillusionment and political disengagement, while media attention turned to Trump’s early personnel choices---especially his creation of a Department of Government Efficiency to be led by Elon Musk, the world's richest man and a pro-Trump campaigner. 

Online content played a far greater role than in previous US election campaigns. The share of campaign and PAC media spending devoted to digital platforms---connected TV streaming services and online ads---rose from 27\% in 2020 to 36\% in 2024.\footnote{Tech for Campaigns, ``Political Digital Advertising Report,'' \href{https://www.techforcampaigns.org/results/2024-digital-ads-report}{https://www.techforcampaigns.org/results/2024-digital-ads-report}.} The Federal Election Commission's lack of disclosure requirements prevents reliable quantification of payments to SMCs, but many reports cite them as key influences.\footnote{For example, \textit{Washington Post}, ``The new dark money: How influencers get paid big bucks to court your vote,'' October 26, 2024, \href{https://www.washingtonpost.com/technology/2024/10/26/social-media-influencers-election-money-campaigns}{www.washingtonpost.com/technology/2024/10/26/social-media-influencers-election-money-campaigns}.} Both Democrat and Republican teams established and supported networks of SMCs, invited hundreds of SMCs to their national conventions, and directly sponsored SMC content. Beyond the campaigns themselves, SMCs reached large audiences with political content. For example, Trump's interview on Joe Rogan's podcast amassed more than 50 million times on YouTube in ten days before the election, underscoring how digital creators rivaled the reach of mainstream news outlets.

%Taken together, the 2024 election offered an environment in which citizens increasingly encountered political information through digital personalities rather than traditional institutions. The shifting media economy, the campaign’s unpredictability, and declining trust in legacy outlets together magnified the role of SMCs as new opinion leaders in the digital media ecosystem.

%Like many Democrats more broadly, exhausted and depressed progressive-minded SMCs scaled back their political content after the election.

\subsection{Social media creators}

In 2024, the US creator market was estimated to be worth over \$30 billion, and growing exponentially.\footnote{Exploding Topics, ``Creator Economy Market Size (2025-2030),'' November 19, 2024, \href{https://explodingtopics.com/blog/creator-economy-market-size}{https://explodingtopics.com/blog/creator-economy-market-size}.} SMCs have thus become a critical component of today's media landscape, seeking to engage, inform, advertise, and---increasingly---politically persuade. Although they come in various forms, we focus on \textit{digitally-native} SMCs---online personalities who, unlike celebrities or politicians, principally amass followings by producing online content that directly addresses their audience with minimal oversight. %Following the definition in \cite{campbell2020more}, such SMCs vary in scale from ``nano'' (1,000-10,000 followers) and ``micro'' (10,000-100,000 followers) to ``macro'' (100,000 to 1,000,000 followers) and ``mega'' (more than a million followers). 

These SMCs earn income in various ways. The most common source is sponsored content, especially from brand deals but also more flexible creator funds that establish longer-term relationships. SMCs with the largest followings can generate significant income from affiliate marketing (per click or sale) and ad revenue sharing with social media platforms.\footnote{Although the exact formulas vary, all major platforms share revenues generated from an account's content with accounts that receive substantial engagement. For example, at the time of writing, YouTube accounts that generate more than 4,000 hours of content or 10 million views per year receive 55\% of ad revenues generated by their videos and 45\% of ad revenues from their shorts.} Other SMCs develop their own brand lines or sell exclusive content or experiences. Regardless of their business model, successful SMCs combine cultivating large or well-defined audiences and persuading them to take action. This generally requires producing core content that attracts followers and, in turn, advertisers.

SMCs vary significantly in the extent to which their content is \textit{explicitly} political, which we define as covering current affairs, government, public policy, or politics. Most SMCs are exclusively non-political, gaining followers for producing lifestyle, entertainment, comedic, etc. content without explicit reference to politics. But predominantly-apolitical SMCs increasingly embed political and social education or advocacy within or alongside their normal content \citep{von2025political}, especially around elections to appeal to audiences or as election campaigns, governments, and non-profit organizations sponsor content. In contrast, predominantly-political SMCs specialize in news and politics, and often develop a consistent ideological or partisan stance.

The SMCs who venture into public affairs are distinct from traditional journalists in their expertise, autonomy, and style. First, unlike reporters who rely on rigorous training and institutional resources for background research and fact-checking, SMCs often lack formal credentials in public policy, journalism, or political commentary. Indeed, 77\% of high-profile news influencers have never been employed in the news industry.\footnote{Pew Research Center, ``America’s News Influencers,'' November 18, 2024, \href{https://www.pewresearch.org/journalism/2024/11/18/americas-news-influencers/}{www.pewresearch.org/journalism/2024/11/18/americas-news-influencers}.} Second, SMCs typically possess considerable autonomy, producing content without the oversight of institutional editors, professional guidelines, or owners that constrain traditional journalists. This enables SMCs to choose their topics, select their evidence and framing, and embed subjective perspectives. Finally, SMCs differ in their style and delivery. Rather than following more the formal style of newscast and political discussion, SMCs more often blend entertainment with commentary through informal tones, humor, personal anecdotes, and direct audience engagement.

\subsection{How SMCs might matter politically}

While SMCs regularly advertise consumer products \citep{barari2025meta, liu2024persuasive}, the extent to which they can effectively shape policy and political attitudes or behaviors is less clear. Most SMCs' lack of political expertise or institutional credibility may undermine their ability to address complex or nuanced issues \citep[e.g.][]{alt2016credible, lupia1998democratic}. SMCs may also primarily reach like-minded followers who already share their political views \citep[e.g.][]{stroud2008}. Nonetheless, SMCs could still be influential, especially among younger audiences disengaged from traditional political information sources, either by providing a gateway into political information or by leveraging distinctive forms of credibility to persuade their audience.

For SMCs to influence public opinion, audiences must first engage with their political content; as \citeauthor{prior2009improving} (\citeyear{prior2009improving}:130) notes, ``the causal chain starts with exposure.'' In today's high-choice media environment, where consumers can easily avoid politics \citep{arceneaux2013changing, de2019persuading}, SMCs may be particularly effective at reaching citizens. First, by producing content that engages many types of audience, SMCs can incidentally expose consumers to politics as a by-product of seeking their non-political content \citep{baum2003, bergstrom2018news, valeriani2016accidental} or because platform algorithms promote popular or micro-targeted content \citep{bucher2018if, cotter2019playing, suarez2022algorithmic}. Second, SMCs may increase engagement with political messages via parasocial connections with their audience \citep{giles2002parasocial, riedl2021rise}. Unlike most journalists and public figures, SMCs' informal and seemingly-unfiltered content---ranging from live streaming, content tailored to followers' comments, and Q\&A sessions---create a real-time interactive dynamic, strengthening the illusion of a two-way relationship between SMCs and their followers \citep{casalo2017antecedents, ferchaud2018parasocial, rasmussen2018parasocial, yuan2020social}. Third, exposure to political content via either mechanism could, in turn, generate interest among less politically-engaged audiences \citep{bode2016political, settle2018}.

Upon exposure, political messages can shape citizens' preferences, beliefs, and priorities through multiple pathways (\citealt{druckman2022})---by deciding which topics to emphasize \citep[e.g.][]{mccombs2018setting}, selectively providing information \citep[e.g.][]{anderson2012media, broockman2025consuming}, or framing issues and presenting perspectives \citep[e.g.][]{chong2007framing, leeper2020raising}. Several common features of SMCs position them to persuade through these channels. First, their cultivation of parasocial relationships---by appearing authentic, intimate, and relatable---makes followers more likely to trust source cues and accept agenda-setting or selective information with limited scrutiny \citep{carlson022, druckman2022, lou2022social, schmuck2022politics}. Second, to the extent that they are perceived as non-partisan, SMCs may be regarded as more credible and more likely to share their audiences' interests. Finally, their use of emotionally-resonant formats such as humorous skits, personal anecdotes, or engaging narratives also enhances the persuasive power of framing, reducing psychological resistance that can arise with overtly political communication \citep{mutz2010, settle2018, young2020irony}.

Whether predominantly-political (PP) or predominantly-apolitical (PA) SMCs are more politically persuasive, however, is theoretically ambiguous. On one hand, PP SMCs expose their followers to more political content than PA SMCs. Whether by deepening views on a few issues or shaping views on many issues, a large quantity of political content could influence followers' policy positions, issue salience, and evaluations of candidates or parties. Moreover, the explicitly partisan style of PP SMCs may help followers connect their general beliefs to concrete policy preferences and partisan evaluations. In this way, online political commentators with a clear partisan perspective---such as Philip DeFranco, Candace Owens, or Hasan Piker---could substantially influence their followers.

On the other hand, PP SMCs may reach broader audiences or more persuasively include political content. First, politically-disinterested social media users are more likely to consume the political content of PA than PP SMCs. In contrast, those who expose themselves to PP SMCs may already be well-informed or share the SMC's views. Second, by only occasionally addressing political topics in a less partisan way, PA SMCs like Alex Cooper or Joe Rogan may establish greater trust and credibility among their followers than PP SMCs.

\subsection{Implications for following progressive-minded SMCs}

The preceding discussion suggests exposure to SMC content emphasizing progressive issues and narratives could affect participants' policy and partisan preferences and behaviors. We define progressivity as a liberal-leaning ideology advocating a central role for government and active citizen engagement by democratically addressing social, economic, and political inequalities through transformative policy and systemic reform. Key planks of the current progressive policy agenda include mitigating income disparities, universal healthcare coverage, and combating climate change. Proponents often argue that, to achieve these goals, corporate power must be reduced through greater market regulation, increasing labor rights, and active and broad-based involvement in and support for democracy. This need not refer to electoral or partisan messaging. 

Our pre-analysis plan hypothesized that exposure to PA or PP SMCs' progressive-oriented content would lead participants to adopt more progressive policy positions, concerns, and narratives, while also increasing their political knowledge and participation. We expected this average effect because the average American lies to the right of progressive positions. In the 2024 US context, %in addition to increasing policy knowledge reflective of nonpartisan messages, 
such shifts may translate into greater support for the Democratic Party---including increased favorability, more positive appraisals of its candidates and performance in office, and vote choice---relative to the Republican Party, as well as increased advocacy for progressive causes. 

While these predictions may apply to progressive messaging through any medium, we further hypothesized that SMCs would be particularly impactful messengers. SMCs' reach and the trust they inspire make them more effective at engaging and persuading viewers than the same content delivered without SMCs. As explained above, it was unclear whether content produced by PP or PA SMCs would be more influential: PA SMCs may increase exposure to progressive ideas due to their greater production of such content, but their overtly political nature could reduce engagement or credibility. We proceed to test these hypotheses, and thus evaluate SMCs' influence as vehicles for political communication.

\section{Experimental design}

\begin{figure}[!h]
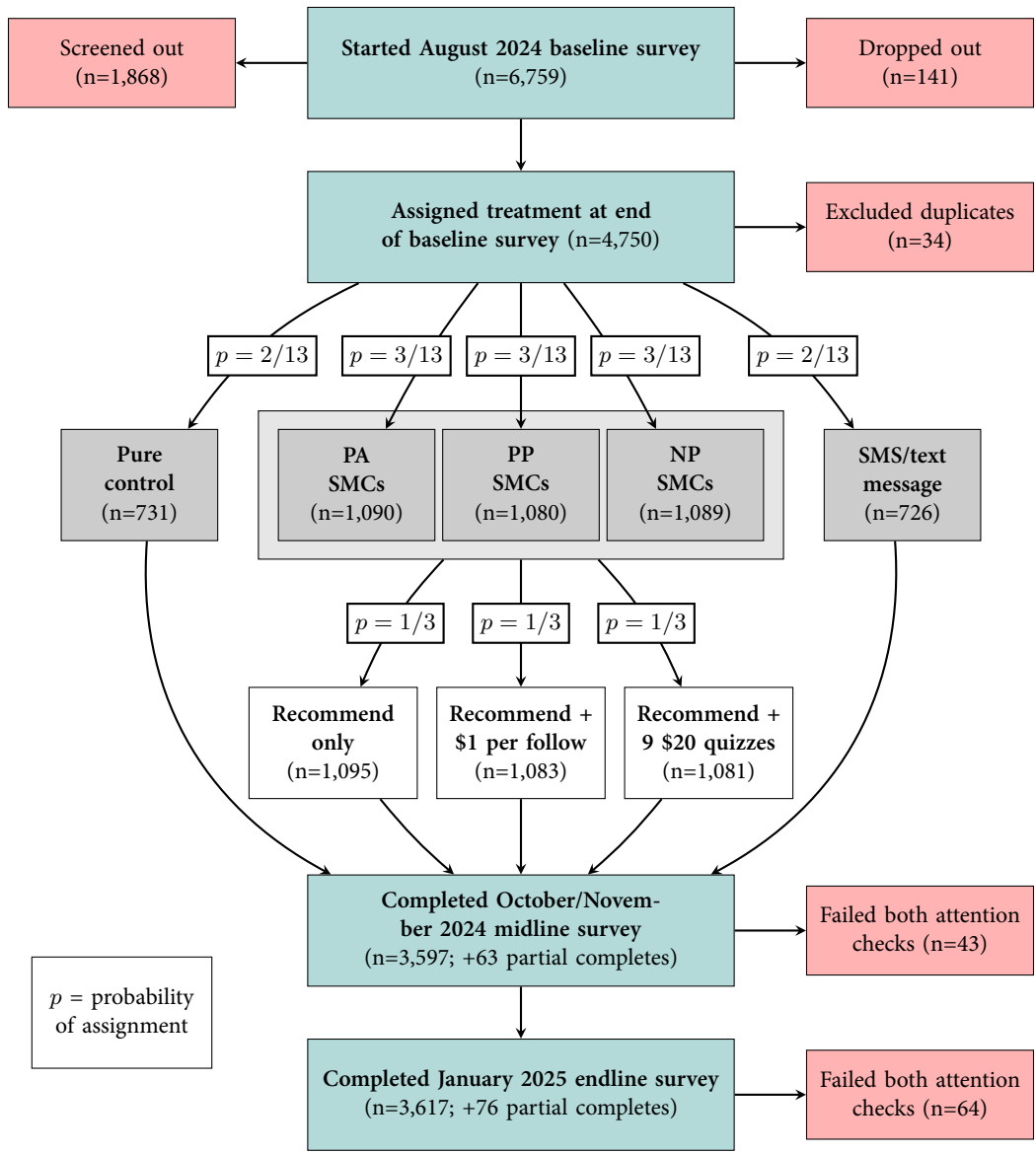


\begin{center}
\begin{small}
\tikzstyle{line} = [draw, -stealth, thick]
\tikzstyle{block} = [draw, rectangle, fill=black!20, text width=6em, text centered, minimum height=15mm, node distance=10em]
\tikzstyle{blockwide} = [draw, rectangle, fill=teal!30, text width=17.5em, text centered, minimum height=15mm, node distance=10em]
\tikzstyle{blocksmaller} = [draw, rectangle, fill=red!30, text width=9em, text centered, minimum height=12mm, node distance=10em]
\tikzstyle{blockwhite} = [draw, rectangle, fill=white, text width=6.5em, text centered, minimum height=15mm, node distance=10em]
\tikzstyle{blockwhite2} = [draw, rectangle, fill=white, text width=7em, text centered, minimum height=15mm, node distance=10em]
\tikzstyle{blocktreat} = [draw, rectangle, fill=black!10, text width=21.7em, text centered, minimum height=20mm, node distance=10em]


\end{small}
\end{center}

\vspace{-4pt}
\caption{Overview of experimental design}
\label{figure:flow_chart}
\end{figure}

\noindent During the 2024 US election season, we measure political consequences on younger adults of encouragement to start following five progressive-minded SMCs over a four and a half month period. Specifically, our field experiment evaluates the extent to which inducing sustained exposure to SMCs---who produce either predominantly-apolitical content or predominantly-political content---shapes political engagement, policy attitudes, and partisan preferences. We next describe our survey panel, treatment conditions and encouragements, randomization, data collection, and estimation strategy. Figure \ref{figure:flow_chart} summarizes the experimental design.

\subsection{Sample of panelists}

We recruited 4,716 American social media users aged 18-45 for our three-wave panel study. This sample is drawn primarily from Bovitz's Forthright pool of approximately 300,000 US-based panelists, and supplemented by panelists from Esearch (2.6\% of the sample) and SurveySavvy (19.8\% of the sample).\footnote{We had targeted 4,550 baseline survey completes, but retain participants who were assigned treatment at the end of the survey but did not fully complete the survey. We excluded 28 ineligible participants for one of four reasons: out of study age range; consumed social media irregularly and were not screened out due to a technical error; SMC recommendations (which were generated for all participants regardless of treatment status) did not pipe within Qualtrics; and multiple responses from the same participant.} These panelists usually take non-political consumer surveys and were paid in redeemable credits within their platforms for completing 20-minute baseline, midline, and endline surveys in Qualtrics, respectively commencing mid-August 2024, late October 2024, and early January 2025. Appendix Section \ref{appendix:ethics} provides detailed information on payment and research ethics.

Quota sampling ensured balance over gender and generated a broadly representative sample of young adults in terms of age, education, and partisan identification.\footnote{Appendix \ref{tab:samplecompare} shows that our sample is fairly similar to a nationally representative sample of regular social media users aged 18-45 from the 2020 American National Election Studies.} To increase the likelihood that participants consumed assigned SMC content, we screened out the 14.3\% of individuals who did not already use any of Instagram, TikTok, or YouTube more than two days a week, the 1.2\% of participants who did not provide a social media handle, and the 12.0\% of participants who failed either of our attention checks. The summary statistics in Panels A and B of Table \ref{table:summary} show that our mean participant is 32.6 years old, has a household income of almost \$70,000, followed politics somewhere between ``somewhat closely'' and ``rather closely'' during the election campaign, and was 50\% more likely to intend to vote for Harris than Trump. The sample thus skews more Democratic than the 18-44 age group in the 2024 election exit polls.\footnote{NBC News, Exit Polls, November 5; \href{https://www.nbcnews.com/politics/2024-elections/exit-polls}{https://www.nbcnews.com/politics/2024-elections/exit-polls}.} 

\begin{table}
\caption{Summary statistics at baseline for panel survey of American regular social media users aged 18-45 \label{table:summary}}
\centering
\scalebox{0.91}{
\begin{tabular}{llll}
\toprule
\multicolumn{1}{r}{} &
  \multicolumn{1}{c}{\textbf{Mean}} &
  \multicolumn{1}{c}{\textbf{Standard deviation}} &
  \multicolumn{1}{c}{\textbf{Median}} \\
\midrule
\textit{Panel A: Socioeconomics and demographics} \\
\multicolumn{1}{l}{Age (in years)} &
  \multicolumn{1}{r}{32.64} &
  \multicolumn{1}{r}{6.40} &
  \multicolumn{1}{r}{33} \\
\multicolumn{1}{l}{Male} &
  \multicolumn{1}{r}{0.44} &
  \multicolumn{1}{r}{0.50} &
  \multicolumn{1}{r}{0} \\
\multicolumn{1}{l}{Female} &
  \multicolumn{1}{r}{0.54} &
  \multicolumn{1}{r}{0.50} &
  \multicolumn{1}{r}{1} \\
\multicolumn{1}{l}{Non-binary} &
  \multicolumn{1}{r}{0.02} &
  \multicolumn{1}{r}{0.14} &
  \multicolumn{1}{r}{0} \\
\multicolumn{1}{l}{White} &
  \multicolumn{1}{r}{0.71} &
  \multicolumn{1}{r}{0.46} &
  \multicolumn{1}{r}{1} \\
\multicolumn{1}{l}{Black} &
  \multicolumn{1}{r}{0.23} &
  \multicolumn{1}{r}{0.42} &
  \multicolumn{1}{r}{0} \\
\bottomrule
\end{tabular}
}
\end{table}

Panels C shows that our sample regularly consumed SMC content at baseline, but was willing to follow new SMCs. The median participant reported spending 13 hours per week on Instagram, TikTok, and YouTube, mostly on their cell phone. They followed or subscribed to 1-10 social media accounts and watched 11-15 videos per week from these accounts, which they felt ``moderately connected'' to on average. The mean and median participant reported being ``very open'' to following new social media accounts. While they were neutral toward SMCs talking about politics, SMC content about news, public policy, or politics was less interesting to the average participant than lifestyle, entertainment, humorous, or science/tech/gaming content.

\subsection{Treatment conditions}

Our intervention encouraged regular social media users in our panel to follow five progressive-minded SMCs they did not already follow from mid-August until late December 2024. To assess the influence of different types of SMCs, our two primary treatment conditions varied whether participants were encouraged to follow only SMCs whose core content is either predominantly-apolitical or predominantly-political. We defined \textit{predominantly-political} SMCs as those who almost exclusively cover news, current affairs, and political actors or institutions and do so from a clear partisan perspective. In contrast, \textit{predominantly-apolitical} SMCs rarely or never explicitly address public policy or political issues, except as part of the BII's non-partisan programming.\footnote{Of course, almost any content could resonate politically \citep[e.g.][]{kim2025}. We thus differentiate political and non-political SMCs by the quantity and partisanship of their explicit policy or political content, rather than the underlying values or narratives it promulgates or the reactions it elicits from different audiences.} We compare these treatment conditions to a placebo group encouraged to follow \textit{non-political} SMCs whose content never explicitly addresses public policy or politics.

\begin{figure}

\begin{subfigure}{1\textwidth}
\begin{center}
\caption{Predominantly-apolitical SMCs}
\vspace{6pt}
\includegraphics[scale=0.415]{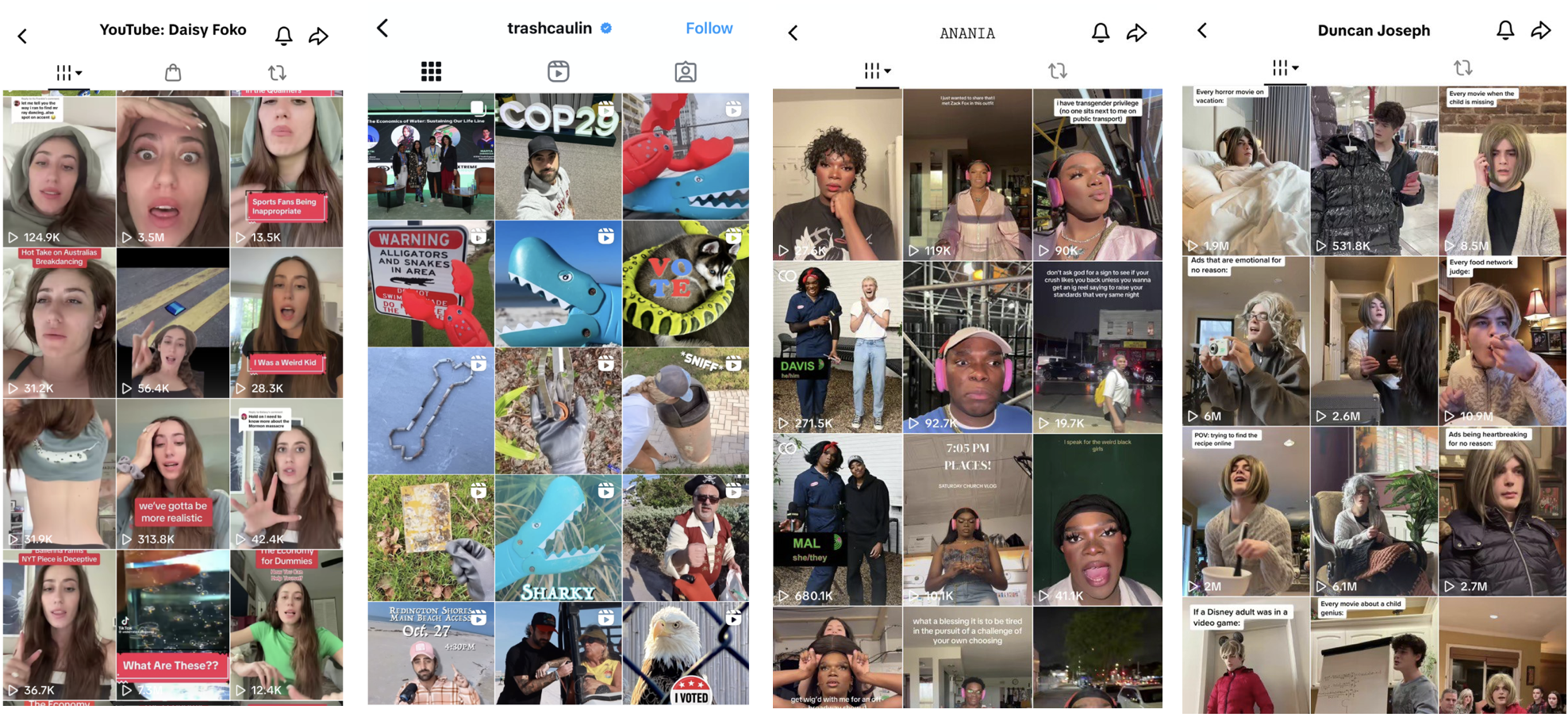}
\label{figure:apolitical_example}
\end{center}
\end{subfigure}%

\vspace{4pt}

\begin{subfigure}{1\textwidth}
\begin{center}
\caption{Predominantly-political SMCs}
\vspace{6pt}
\includegraphics[scale=0.34]{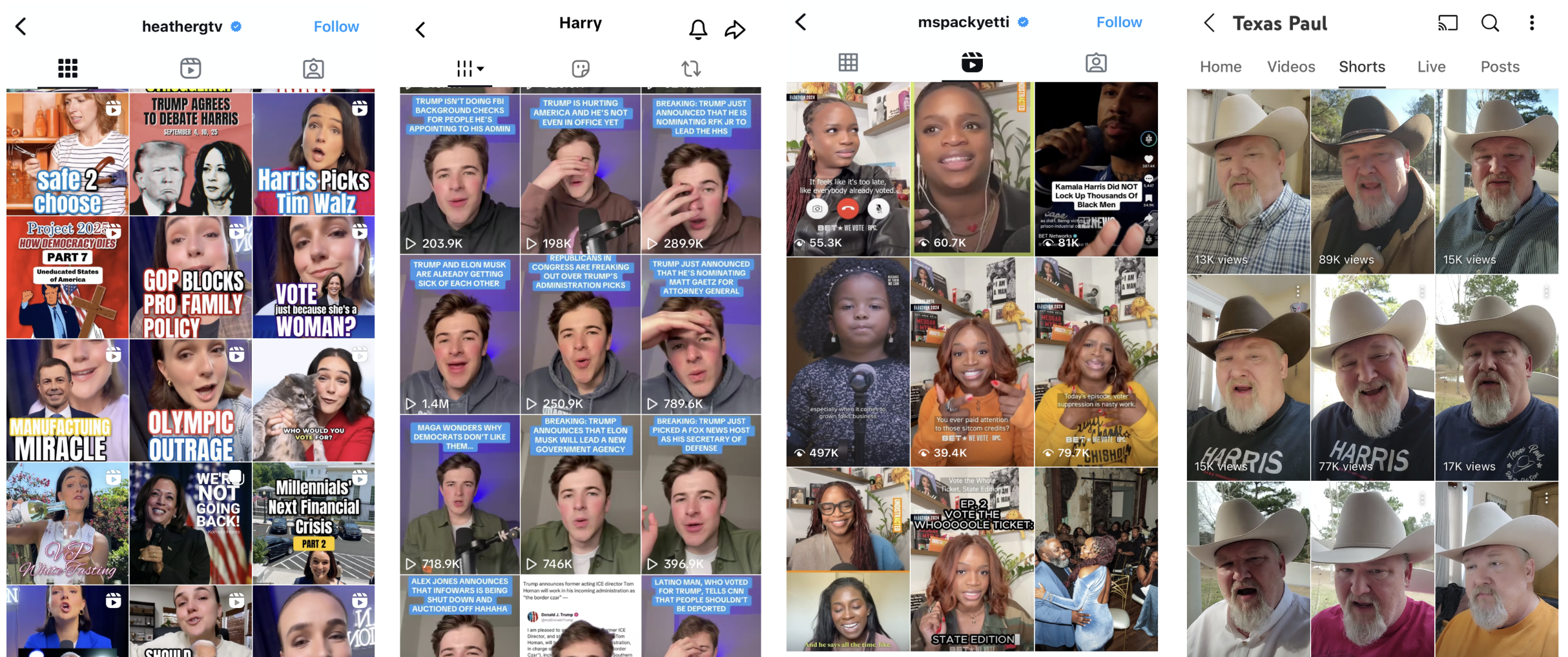}
\label{figure:political_example}
\end{center}
\end{subfigure}%

\vspace{4pt}

\begin{subfigure}{1\textwidth}
\begin{center}
\caption{Non-political/placebo SMCs}
\vspace{6pt}
\includegraphics[scale=0.34]{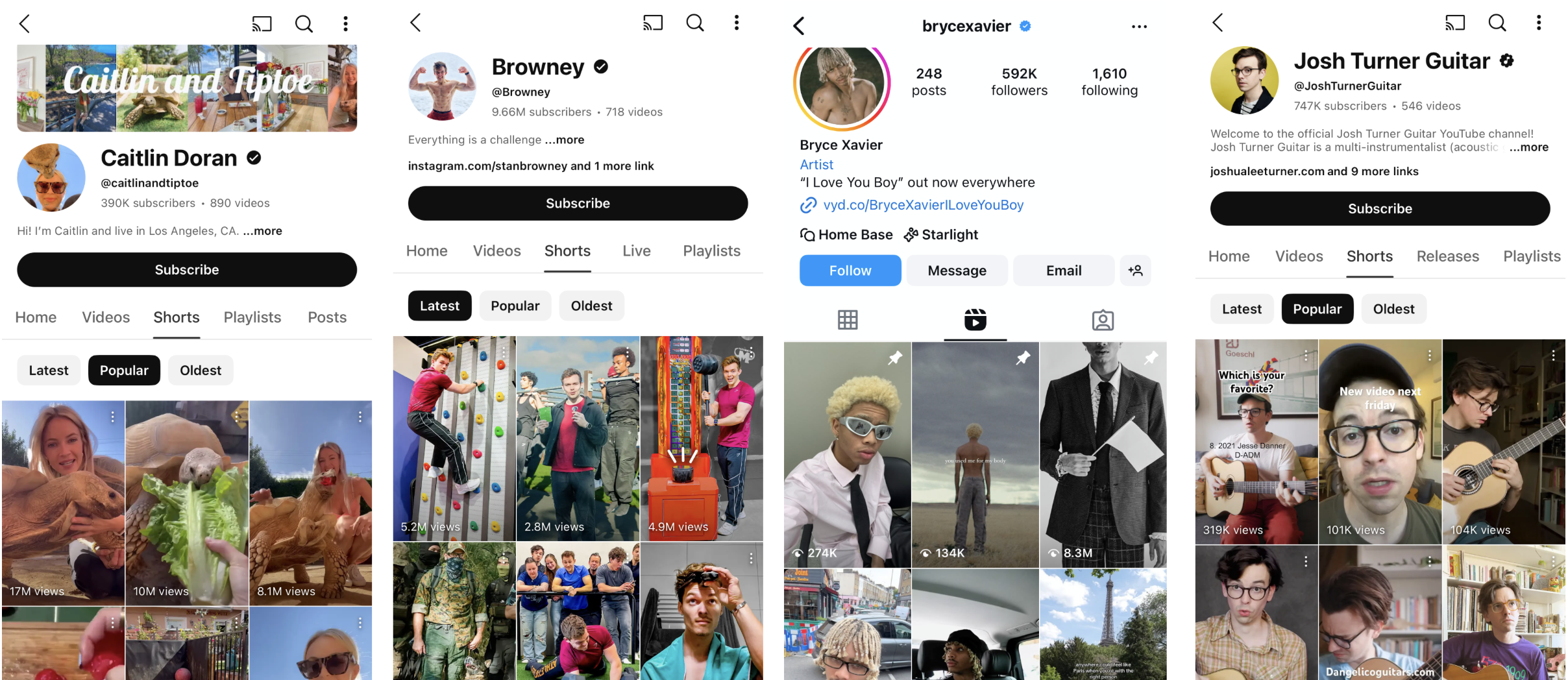}
\label{figure:nonpolitical_example}
\end{center}
\end{subfigure}

\begin{center}
\caption{Examples of assigned SMCs in this study, by treatment condition}
\label{figure:SMCs}
\end{center}

\end{figure}

Our PA, PP, and NP conditions each comprised a mutually exclusive pool of 20 ``shorter-form'' SMCs generally producing less than 20 minutes a week in the form of 1-5 minute videos. Each condition included at least ten SMCs regularly posting content on each of Instagram, TikTok and YouTube. At the end of the baseline survey, we recommended five SMCs to each treated participant from the platform (or platforms) they regularly use.\footnote{The platforms a participant regularly used are the platform they used on most days a week and any other platform they used no more than two fewer days per week than the most-used platform. The baseline survey provided links to recommended SMCs via participants' most-used platform, but reminders provided links on all regularly-used platforms.} Our recommendation algorithm used a neural network to match participants to SMCs based on participants' stated interests and feedback on SMC content from a pilot study.\footnote{In short, we first coded the characteristics and regular content of all SMCs in terms of SMC gender, age group, and ideology as well as their content's visual (e.g. fashionable, polished) and verbal (e.g. formal, assertive) styles, topics, and potential for polarization. Using baseline survey questions, our 278-person pilot study generated recommendations to minimize the distance between participants' preferences across multiple dimensions and these characterizations. Based on respondent feedback from the pilot endline survey, we trained a neural network recommendation model to predict the SMCs a participant in the full study would most enjoy consuming. Specifically, we used a hybrid collaborative and content-based filtering approach \citep{he2017neural}, where our model architecture included embedding layers for user and creator IDs to capture 200 latent factors, with embeddings concatenated and passed through multiple dense layers with Batch Normalization, LeakyReLU activation, and Dropout for regularization. The model was trained using early stopping to prevent overfitting, converging in 21 epochs to ensure stable performance on the validation set. Appendix \ref{appendix:matching_algorithm} provides further details about this matching algorithm.} All participants in these treatment conditions were informed that these recommendations derived from their baseline survey responses, and further received a several-sentence summary, image, and platform hyperlink(s) for each assigned SMC. Figure \ref{figure:SMCs} provides examples from each category of SMCs, while Appendix Section \ref{appendix:smc_descriptives} lists all SMCs.\footnote{Each condition also included five ``longer-form'' SMCs producing more than 20 minutes of content a week, usually as podcasts, live shows, or extended videos. Our recommendation algorithm identified two of these longer-form SMCs to offer participants the opportunity to receive more information about. Since only 39\% of individuals requested further information about longer-form SMCs and consuming their content is a significant time commitment, these SMCs unsurprisingly received limited engagement from participants. While we registered some significant self-reported increases, Table \ref{table:FS1_long_form} reports limited consumption of recommended longer-form SMCs: treated individuals reporting consuming these SMCs 5-10 times less frequently than the average quiz-incentivized shorter-form SMC, and ultimately consumed negligible numbers of YouTube videos (around 50 times fewer). We henceforth treat this component of treatment as excludable.}

Before describing each treatment condition's SMCs in detail, we note two features of this intervention. First, following \textit{new} accounts for four months may not establish strong connections between participants and all recommended SMCs. We are thus likely to underestimate the effect of SMC content on long-term followers, which is harder to evaluate.\footnote{This alternative estimand would require exogenous variation in the content received from an SMC among pre-existing followers. But this faces two challenges: finding a large group of SMCs willing to change the content that drives their business in a harmonized way and at an affordable price; and identifying (and then recruiting) followers for surveys, which is especially challenging on platforms like TikTok that limit researchers' capacity to scrape follower lists or to do within surveys for rarely-followed SMCs. Deactivation studies that do not also vary SMC content would underestimate the effects of SMCs if participants have already internalized the effects of prior exposure \citep{gauthier2025X}. One related study that addressed these challenges is \cite{kim2025chef}, which identified long-time frequent commenters of Instagram influencers and found that when those influencers suddenly became political after the Israel–Hamas conflict, their loyal followers disengaged from the influencers themselves.} Second, the SMC treatment conditions are internally heterogeneous due to our participant-specific recommendations, but also because we could not control platform algorithms or participant consumption choices beyond our encouragements. While treatment conditions are distinguished by a particular aspect of social media---progressive-minded SMCs, who vary in the extent of their political content---each treatment is ultimately the bundle of assigned content together with the content it led participants to consume.

\subsubsection{Selection of SMCs by treatment condition}

To capture the distinctive nature of modern SMCs, we selected SMCs who primarily gained their six- or seven-figure followings from their social media content.\footnote{This entailed excluding popular social media accounts hosted by politicians, journalists, and others with professional credibility outside social media. A couple of SMCs have appeared on TV programs, such as news shows or \textit{The Bachelorette}, and in one case had run for office, but became prominent for their online content.} %All constitute ``macro'' with 100,000 to 1,000,000 followers or ``mega'' influencers with more than 1,000,000 followers rather than celebrities \citep{campbell2020more}. 
Our treatment conditions differ by the extent to which SMCs in each group cover politics.

\textbf{\textit{Predominantly-apolitical progressive SMCs}}. Our pool of PA SMCs was drawn from the BII's fellowship program. In 2024, this program helped more than 100 online content creators with mid-sized followings---ranging from tens of thousands to millions---to produce Instagram, TikTok, and YouTube videos to educate their audiences about policy issues associated with public health, democracy, economic justice, and climate change.\footnote{The BII draws from partner organizations that specialize in these four areas to source topical information for SMCs to base their content around. The BII's project manager closely monitors and provides feedback on all BII-supported content to ensure its quality, relevance, and accuracy. In line with the BII's non-partisan status, all BII-supported content posted by fellows is compliant with section 501(c)(3) of the Internal Revenue Code. Outside of BII-coordinated content, fellows are free to create content or pursue any other opportunity in their ordinary course provided that content does not promote inaccurate information. BII fellows are largely selected on the basis of their reach, audience rapport, and compatibility with the BII's values.} We restricted selection to BII fellows contracted to produce content between August and December 2024. Most of these fellows normally produced non-political content---including celebrity and entertainment commentary, food, fashion, art production, lifestyle advice, interesting facts and history, and humorous sketches---but collaborated with the BII to produce around ten fact-checked and non-partisan social media videos during their yearlong fellowship. Fellows were paid by BII (at below-market rates) to produce these videos on BII issues of their choice in their own voice and publish them on contracted social media platforms alongside or within their regular content. Unlike one-off sponsored content, the form of social media content most frequently produced for the 2024 election, the BII program was designed to foster longer-term relationships with both SMCs and their followers.

%full summaries here: https://docs.google.com/document/d/1qbRu-Ab_WbR8HmADZQ3IU8HAOBtFXjGOzEerLFhSMPc/edit?tab=t.0

BII-supported videos during the intervention conveyed liberal messages in non-partisan ways. Between August and October 2024, content predominantly focused on the democracy, economy, and health policy domains. Common topics included the importance of fair elections, the adverse and inflationary effects of corporate power, the economic and practical consequences of climate change and government efforts to mitigate them, and the scope for government to reduce health care costs. Separately, other topics included encouragements to vote or register to vote (without advocacy to vote for any particular candidate or political party). After the elections, in November and December, content was more balanced across domains. BII-supported videos covered ballot initiative outcomes (e.g. abortion) and the importance of government responsiveness to citizens, minimum wage changes and using public power to curb the cost of living, local efforts to reach net zero emissions and tips to avoid sharing climate misinformation, and encouragements to enroll for health insurance and get COVID-19 and flu vaccinations. Despite being non-partisan, BII-supported videos highlighting salient policy issues represent a significant departure from these SMCs' typical non-political content. 

\textbf{\textit{Predominantly-political progressive SMCs}}. Our pool of PP SMCs comprise 6 BII fellows and 14 other SMCs producing news and current affairs commentary from a progressive and often partisan perspective. These BII fellows produced BII-supported videos containing educational content on policy issues and non-partisan electoral content alongside their regular, separate political content. Because most BII fellows are predominantly-apolitical, this group was augmented with non-BII SMCs similarly producing educational political content from a progressive standpoint. The content of PP SMCs varies from polished programming (somewhat similar to news broadcasts or late-night talk shows) to SMCs talking into a camera about issues of the day.

\textbf{\textit{Non-political/placebo SMCs}}. To hold constant the encouragement to follow five new SMCs on social media, we compare the preceding treatment conditions with a placebo group of NP SMCs. These SMCs produced content that was not explicitly political at the start of the study, but otherwise cover an eclectic mix of topics. This placebo group includes SMCs specializing in comedy skits, food, travel, celebrity news, community exploration, lifestyle advice, animals, and musical covers. Like the primary treatment conditions, we selected SMCs with broadly similar followings and posting cadences; selection was inspired by applicants for the BII fellowship who were not selected. 

\subsubsection{Characterization of treatment conditions}

Table \ref{influencer_stat} summarizes SMCs' attributes and content across treatment conditions. Panel A shows that SMCs in each treatment condition are fairly equally split across Instagram, TikTok, and YouTube, with many operating on multiple platforms. The median SMC in each condition has less than a million followers across platforms, with NP SMCs registering the largest followings. Although PA SMCs were less likely to be male, SMCs were similar across treatment conditions in terms of age, race, and whether they identified as or allied with LGBTQ+. 

\begin{table}
\caption{Characterization of SMCs, by treatment condition}
\vspace{-12pt}
\label{influencer_stat}
\begin{center}
\scalebox{1}{
\begin{tabular}{lccc}
\toprule
& \textbf{PA SMCs} & \textbf{PP SMCs} & \textbf{NP SMCs} \\ 
\midrule
\multicolumn{4}{l}{\textit{Panel A: Account characteristics}} \\
Produced content on Instagram & 75\%       & 45\%      & 95\% \\
Produced content on TikTok    & 75\%       & 65\%      & 80\% \\
Produced content on YouTube   & 75\%       & 30\%      & 60\% \\
%Mean followers on Instagram & 577,573 & 328,378 & 740,631 \\
%Median followers on Instagram & 212,000 & 240,000 & 590,000 \\
%Mean followers on TikTok    & 1,993,034 & 1,030,553 & 3,634,312 \\
%Median followers on TikTok & 836,500 & 770,700 & 2,650,000 \\
%Mean followers on YouTube   & 773,620 & 93,650 & 3,635,333 \\ 
%Median followers on YouTube & 219,000 & 68,900 & 1,570,000 \\ 
Median followers across Instagram, TikTok, and YouTube & 347k & 543k & 992k \\ 
\midrule
\multicolumn{4}{l}{\textit{Panel B: Personal characteristics}} \\
Male                      & 15\%       & 35\%      & 65\% \\
Aged in their 20s                    & 25\%       & 45\%      & 35\% \\
Aged in their 30s                    & 65\%       & 35\%      & 55\% \\
Aged in their 40s                    & 0\%        & 10\%      & 10\% \\
LGBTQ+/Ally                     & 25\%       & 25\%      & 20\% \\
Black                  & 20\%       & 35\%      & 10\% \\
Hispanic               & 5\%        & 0\%       & 10\% \\
White                  & 60\%       & 55\%      & 60\% \\ 
\midrule
%\multicolumn{4}{l|}{\textit{Panel C: SMC topics}}                                                                                        \\
%News                              & 20\%       & 100\%     & 5\%     & 0\%        & 100\%     & 0\%     \\
%Lifestyle                         & 80\%       & 5\%       & 80\%    & 40\%       & 0\%       & 80\%    \\
%Entertainment                     & 35\%       & 5\%       & 75\%    & 60\%       & 0\%       & 20\%    \\
%Humor                             & 15\%       & 30\%      & 40\%    & 0\%        & 40\%      & 20\%    \\
%Family                            & 15\%       & 0\%       & 5\%     & 0\%        & 0\%       & 20\%    \\
%Money                             & 0\%        & 0\%       & 0\%     & 0\%        & 0\%       & 0\%     \\
%Science                           & 0\%        & 0\%       & 0\%     & 20\%       & 0\%       & 0\%     \\
%Environment                       & 5\%        & 0\%       & 5\%     & 20\%       & 0\%       & 0\%     \\
%Culture                           & 20\%       & 0\%       & 5\%     & 40\%       & 0\%       & 0\%     \\
%Education                         & 20\%       & 35\%      & 25\%    & 40\%       & 20\%      & 0\%     \\ \bottomrule
\multicolumn{4}{l}{\textit{Panel C: Classification of content produced during the intervention}} \\
Number of Instagram image posts during intervention	&	12.9	&	28.7	&	9.8	\\
Number of Instagram videos during intervention	&	39.2	&	77.3	&	40.9	\\
Number of TikTok videos during intervention	&	49.2	&	88.7	&	43.3	\\
Number of YouTube videos during intervention (including shorts)	&	29.5	&	26.3	&	36.2	\\
Number of videos during intervention & 117.8 & 192.3 & 120.4 \\							
Median duration of videos during intervention (seconds)	&	53	&	88	&	59.3	\\
Broadly political content	&	19.7\%	&	77.0\%	&	9.7\%	\\
Covers international relations and policy issues	&	0.8\%	&	8.1\%	&	0.7\%	\\
Covers domestic public policy issues	&	9.1\%	&	47.9\%	&	1.5\%	\\
Covers government actions or performance	&	5.0\%	&	43.4\%	&	2.3\%	\\
Covers politicians or political parties	&	3.9\%	&	58.2\%	&	1.4\%	\\
Covers public opinion or collective action	&	6.5\%	&	28.2\%	&	0.7\%	\\
Covers news and current affairs	&	7.8\%	&	55.0\%	&	1.8\%	\\
Includes political content on climate topics & 3.7\% & 4.8\% & 0.6\% \\							
Includes political content on democracy topics & 3.5\% & 17.3\% & 1.6\% \\							
Includes political content on economy topics & 5.9\% & 32.9\% & 2.8\% \\							
Includes political content on health topics & 7.3\% & 25.2\% & 0.9\% \\							
References the 2024 US presidential election campaign	&	5.2\%	&	35.6\%	&	0.7\%	\\
References Democratic Party	&	3.0\%	&	38.3\%	&	0.7\%	\\
References Republican Party	&	2.3\%	&	45.9\%	&	0.6\%	\\
Political narrative embedded	&		&		&		\\
Total number of liberal issues mentioned	&	0.5	&	2.1	&	0.1	\\
Total number of conservative issues mentioned	&	0	&	0.1	&	0	\\
Conservative ideology slant (7 point scale)	&	3.82	&	2.63	&	4	\\
Conservative ideology of policies supported (7 point scale)	&	3.82	&	3.1	&	4	\\
Republican partisan slant (7 point scale)	&	3.96	&	3.05	&	4	\\
Advocates for any Democrat candidate	&	0.4\%	&	8.8\%	&	0.1\%	\\
Advocates for any Republican candidate	&	0\%	&	0.2\%	&	0\%	\\
Criticizes Democratic Party	&	0.4\%	&	4.8\%	&	0.1\%	\\
Criticizes Republican Party	&	0.9\%	&	29.9\%	&	0.2\%	\\
\bottomrule
\end{tabular}%
} 
\end{center}
\vspace{-6pt}
\begin{minipage}{0.98\textwidth} \footnotesize 
\textit{Notes:} Panel A uses public data obtained from social media accounts on November 11th, 2025. Panel B relies on author classifications of SMC characteristics a month before the beginning of the intervention. Panel C is based on classifications of the title (for YouTube), description (for Instagram and TikTok), date, and transcript of Instagram reels and clips, TikTok videos, and YouTube videos and shorts by GPT-5-mini; in each panel, summary statistics weight all 20 shorter-form or all five longer-form SMCs equally within each treatment condition. The only classified Instagram videos for the three SMCs who did not also have TikTok or YouTube accounts; data collection and prompt engineering are described in Appendix \ref{appendix:content_analysis}. 
\end{minipage}
\end{table}

As expected, the starkest differences relate to content produced during the intervention period. We downloaded all available Instagram, TikTok, and YouTube videos produced by experimental SMCs, and then used Open AI's GPT-5-mini model to classify the title, description, and transcript of each video along various dimensions; see Appendix Section \ref{appendix:content_analysis} for details. The results in Panel C show that just 10\% of NP SMCs' content was classified as broadly political---that is, including content related to government, agents of the state, politics, current affairs, or public policy. NP SMCs' content averaged exactly neutral---4.00 on a seven-point Likert scale---in terms of liberal-conservative ideology, liberal-conservative policy positions, and Democrat-Republican partisan slant. As expected, the PA SMCs' videos were more political and progressive-minded: 20\% were classified as political and they exhibited a small liberal ideological and policy---but no partisan---lean. Far more strikingly, 77\% of PP SMCs' videos were political and the average video was classified as between liberal and slightly liberal in ideology and policy and slightly pro-Democrat in stance. PP SMCs' content was thus substantially more political (four times that of PA SMCs and eight times that of NP SMCs), more clearly liberal (four to seven times more liberal \textit{per political video}), and the only group to register a pro-Democrat partisan stance.

\subsubsection{Encouragements to consume content from assigned SMCs}

SMCs rely on producing engaging content for their livelihood, but content appeals to different audiences and asking participants to follow five SMCs is time-consuming. Our recommendation algorithm mitigated this by seeking to approximate how platform algorithms and word-of-mouth expose individuals to content of interest. This basic form of encouragement allowed us to estimate effects of treatment absent material inducement to follow assigned SMCs. 

To ensure high levels of exposure to assigned SMCs' content, we also provided two increasingly-powerful financial incentives. First, we incentivized treated participants to follow or subscribe to recommended SMCs. At the end of the baseline survey, and again within the midline survey, participants were offered \$1 per SMC to upload a valid screenshot showing that they followed or subscribed to assigned SMCs.\footnote{Screenshots were deemed invalid if an image showed the participant muted an assigned account. As part of a treatment assignment reminder, we sent a follow-up message asking participants to upload screenshots if they did not do so during the baseline survey.} Ultimately, 69\% of participants complied at baseline and 67\% did so at midline.\footnote{The numbers indicate the share of participants who uploaded all five validated screenshots. The share of participants who did not upload any screenshots is 18\% at baseline and 23\% at midline, respectively.} Following \cite{levy2021social}, we expected this would lead platform algorithms to show participants more content from assigned SMCs.

Second, to approximate the rate at which individuals consume the SMCs they elect to follow, we incentivized participants to answer questions about their five assigned SMCs in nine biweekly quizzes. In each quiz, participants who correctly answered at least four of five multiple choice questions (one per assigned SMC) received \$20 in credits on their survey platform.\footnote{Each quiz was available for ten days and each question had four options. Similar financial incentives have achieved high and sustained compliance rates when encouraging study participants to watch CNN instead of Fox News \citep{broockman2025consuming}, use VPNs to access foreign news in China \citep{chen2019impact}, and consume fact-checking podcasts in South Africa \citep{bowles2023sustaining}. To reduce attrition, participants were informed that \$5 of each quiz incentive won would be withheld until completion of the following midline or endline survey.} To avoid informing participants via the quiz itself, questions focused on non-political elements throughout SMC videos. Participants obtained the quiz bonus 63\% of the time (86\% conditional on taking the quiz).

To mitigate the potential for differential attrition, the maximum potential payment was equalized across participants in our study. All participants who did not receive the quiz encouragement entered two lotteries, one with \$100 prizes conditional on completing the midline survey and the other with \$80 prizes conditional on completing the endline survey.\footnote{Both lotteries yielded \$5 less in credits for winners receiving the algorithmic feedback condition that could yield \$10 worth of credits for uploading validated screenshots.}

\subsubsection{Comparison groups}

We leverage three comparison groups to isolate effects of following progressive-minded SMCs. Relative to a pure control group that was not encouraged to consume any content for this study, the PA and PP SMCs conditions were encouraged to increase use of social media \textit{and} increase exposure to progressive-minded content. We designed two additional comparisons to disentangle these components of treatment. First, to isolate the effects of exposure to different types of SMC content while holding encouragement to use social media constant, we compare the PA and PP SMCs conditions to the placebo NP SMCs condition. Second, to hold the issues and perspectives SMCs drew from constant but vary the messenger, we compare the PA and PP SMCs conditions to a group that received similar progressive messages via SMS and email every two weeks. This newsletter consisted of five bullet points, each briefly summarizing the set of issues BII fellows selected to produce their videos from; Appendix \ref{appendix:sms_example} provides an exemplar message.

\subsection{Randomization}

Our treatment and encouragement conditions were cross-randomized within blocks of 13 participants with the same baseline political partisanship and similar levels of social media consumption (above or below than 10 hours per week). These blocks were generated sequentially within Qualtrics, which further blocked treatment assignment by participants who completed the baseline survey around the same time. As Figure \ref{figure:flow_chart} shows, one participant per block was assigned to each SMC $\times$ encouragement condition and two participants per block were assigned to both the SMS/email message and pure control conditions.

\begin{comment}

\begin{table}
\begin{center}
\caption{Distribution of panel survey participants by treatment condition and type of encouragement}
\label{table:design}
\scalebox{0.98}{
\begin{tabular}{llccccc}
 \toprule
&& \multicolumn{5}{c}{\textbf{\textit{Treatment condition:}}} \\
&& \textbf{Predominantly-} & \textbf{Predominantly-} & \textbf{Non-} & \textbf{Messaging} \\
&& \textbf{apolitical} & \textbf{political} & \textbf{political} & \textbf{via SMS} & \textbf{Pure} \\
&& \textbf{SMCs} & \textbf{SMCs} & \textbf{SMCs} & \textbf{and email} & \textbf{control} \\
  \midrule
& \textbf{Recommendation only} & 364 & 366 & 365 \\
\textbf{\textit{Encouragement}} & \textbf{\$1 per validated follow} & 361 & 360 & 362 \\
\textbf{\textit{type:}} & \textbf{Biweekly \$20 quiz} & 355 & 364 & 362 \\
& \textbf{None} & & & & 726 & 731 \\
\bottomrule
\end{tabular}
}
\end{center}
\vspace{-4pt}
\footnotesize \textit{Notes}: We intended to assign 350 participants to each SMC condition and 700 to each of the SMS/email and pure control condition, but slightly exceeded these targets. All participants were block-randomized as described in the text.
\end{table}

\end{comment}

The experimental design passes standard validation tests. First, 78\% of participants completed the midline survey and 75\% completed the endline survey, and we fail to reject the null hypothesis of equal attrition rates across all treatment conditions ($p=0.35$ at midline and $p=0.17$ at endline; see Appendix Table \ref{table:attrition}). The quiz-incentivized PP SMCs group was less likely to attrite than the pure control group at endline, but attrition rates were similar across quiz-incentivized conditions ($p=0.57$ at midline and $p=0.08$ at endline). Second, treatment conditions remained well-balanced in all survey waves (Appendix Tables \ref{table:baseline_balance}–\ref{table:endline_balance}). Across 50 predetermined covariates, we only reject the null hypothesis of equal means across treatment conditions 2, 4, and 4 times (at the 5\% level) in the baseline, midline, and endline surveys, respectively.

\subsection{Outcome measurement}

We measure outcomes using midline and endline surveys, video browsing histories from the 466 (119 usable) and 1,149 endline survey participants who respectively provided their TikTok and YouTube browsing histories,\footnote{At the end of the endline survey, participants were offered \$5 to upload their YouTube browsing history; willing participants then received detailed instructions about how to download their data. Participants were also offered \$5 to provide their TikTok browsing history, but uptake was lower because TikTok files needed to be uploaded in a separate survey because they take the file take several days to be generated. Of the 466 participants who submitted their TikTok data, only 119 provided files that contained watch history. This omission may be due to low or recent account activity, privacy configurations that limit data retention, or technical inconsistencies in the export process that cause watch history to be missing from some files.} and voter file data for 1,771 participants who agreed to allow Bovitz to match their records with Target Smart's database. Our primary outcomes include: (i) the quantity and type of content consumed on Instagram, TikTok, and YouTube; (ii) political engagement, policy preferences and salience, and narratives; and (iii) political party evaluations, voting behavior, and non-electoral political participation. We describe specific measures as we introduce the results below, following our pre-analysis plan to construct outcomes and aggregate them into standardized inverse covariance weighted (ICW) indexes \citep{anderson2008multiple}.\footnote{We pre-specified that missingness in outcome variables and covariates would be addressed as follows: (a) observations for which data are unobserved, due to survey attrition, refusal to answer questions, or unavailable administrative or audit records, are treated as missing data and dropped from our analyses; (b) ``don't know'' survey responses are assigned the median value on survey scales (except likelihood scales) and assigned zeros for binary or binarized variables (e.g. yes/no, correct/incorrect, etc. questions) and the lower end of the likelihood scales; and (c) where a question is not relevant to a respondent (e.g. because they did not vote), we assigned zeros for binary or binarized variables and the lower end of likelihood scales. Minor deviations from the pre-analysis plan are justified in Appendix Section \ref{appendix:deviations}.}

\subsection{Estimation and hypotheses}

We estimate differences between the PA and PP SMC conditions and our various control groups in any given survey wave using pre-specified OLS regressions of the following form:\footnote{As pre-specified, all analyses of midline and endline survey outcomes restrict attention to participants who did not fail both attention checks in the corresponding survey.} \begin{eqnarray} \label{equation:full}
    Y_i = \alpha_b + \beta_0 Y_i^{pre} + \boldsymbol\beta_1 [(Y_i^{pre}-\overline{Y}^{pre})\times \textbf{T}_i] + \boldsymbol\gamma \textbf{X}^{pre}_i + \boldsymbol\tau \textbf{T}_i + \varepsilon_i,
\end{eqnarray} where $Y_i$ is an outcome for respondent $i$, $\alpha_b$ are randomization block fixed effects, $Y_i^{pre}$ is the closest pre-treatment measure of the outcome (where possible) and $\overline{Y}^{pre}$ is its sample mean, $\textbf{X}^{pre}_i$ is a vector of pre-treatment covariates (and their demeaned interaction with treatment conditions) selected by the \cite{belloni2014inference} double-LASSO selection procedure, and $\textbf{T}_i$ is the vector of treatment assignments (with the pure control group serving as the omitted category).\footnote{Our pre-analysis plan also proposed to pool across encouragements. Given far greater uptake in the quiz incentives group (see below), we ultimately focus on quiz-incentivized conditions. Because encouragement groups are equally sized, the pooled effects across encouragements are approximately the mean across the three encouragements; these are reported in panels C and D of Appendix Tables \ref{table:H1_full}-\ref{table:H10_full}.} Robust standard errors are reported throughout, reflecting the individual-level randomization. 

Given the limited risk of spillovers between disparately-located panelists, $\hat{\boldsymbol\tau}$ captures two primary estimands: (i) the average treatment effect (ATE) of any treatment $\times$ encouragement cell relative to the pure control group; and (ii) differences in these ATEs between the PA, PP, and NP SMCs conditions at any given encouragement (by differencing elements of $\hat{\boldsymbol\tau}$). Following our pre-specified inference strategy, we conduct one-tailed $t$ tests for the hypotheses where we pre-specified a directional expectation, and two-tailed $t$ tests where no direction was pre-specified or we obtain an estimate opposite to our hypothesis. We later aggregate primary outcome families and apply \citeauthor{anderson2008multiple}'s \citeyearpar{anderson2008multiple} method to control the false discovery rate.

Our preregistered hypotheses---fully enumerated in Appendix Section \ref{appendix:PAP}---posited that all progressive-minded treatment conditions would increase political engagement, progressive policy attitudes, the salience of progressive policy issues, progressive worldviews, favorability toward the Democratic Party over the Republican Party, electoral participation, electoral support for the Democratic Party, progressive non-electoral political behaviors, and institutional and interpersonal trust. While we expected SMCs to be more influential than biweekly summary messages sent via SMS and email, we did not specify a directional hypothesis for whether PA or PP SMCs would be more influential due to its theoretical ambiguity. 

As a collaboration between academics and managers of the BII, we took several steps to ensure the integrity of the research. First, only the academic team were involved in the data collection, randomization, and analysis; the coauthors from the BII could not access study data, but contributed to the experimental design and interpretation of results. Second, our pre-analysis plan was unusually detailed \citep{ofosu2023pre} and closely adhered to.

\section{Changes in social media content consumption}

We begin our analysis by demonstrating that the treatment conditions---particularly when incentivized by quizzes---generated sustained engagement with assigned SMCs and greater use of social media more generally. Throughout, Panel A of each table reports the ATE relative to the pure control condition, whereas Panel B reports pairwise differences between quiz-incentivized treatment groups (and the SMS/email group). The foot of each table provides sample outcome means and standard deviations in the pure control group. Even-numbered columns include covariates selected by the doubly-robust LASSO procedure, which increases estimate precision.

\subsection{Engagement with content from assigned SMCs}

We first measure the frequency with which participants consumed assigned SMCs' content. Turning first to a standardized index of consumption, we combine two self-reported indicators: (i) the average across assigned SMCs of a five-point scale eliciting the regularity---never, monthly, biweekly, weekly, or multiple times a week---with which participants encountered each SMC's content on Instagram, TikTok, or YouTube over the past three/five months; and (ii) the average number of videos respondents recalled watching across assigned SMCs over the past three/five months. Participants in the control and SMS/email groups were asked about the SMCs they would have been assigned in a randomly-selected SMCs group. 

Columns (1)-(2) of Table \ref{table:FS1} show that the average individual who received biweekly \$20 quiz-based incentives consumed around two standard deviations more content from assigned SMCs than the control group by our midline survey; columns (7)-(8) show this reached up to 2.5 standard deviations by endline. Breaking this index into its constitutive items reveals an increase of almost two categories on the regularity of consumption scale from between ``never'' and ``once a month'' in the control group to between ``once every two weeks'' and ``once a week'' and watching 11-15 more videos per SMC over the course of the intervention than the 2.5 reported in the control group. 

Panel B shows that the PA and NP SMCs groups experienced statistically indistinguishable levels of engagement, while respondents assigned PP SMCs exhibited around 20\% greater engagement. This likely reflects the larger number of videos produced by PP SMCs, since columns (3)-(4) and (9)-(10) report similar consumption at the extensive margin and columns (5)-(6) and (11)-(12) demonstrate similar levels of accurate recall of content across these groups. In both the midline and endline surveys, quiz-incentivized participants from the PA, PP, and NP SMCs groups were all about 40 percentage points more likely to answer several (unincentivized) questions about assigned SMCs' content correctly than the 8\% rate in the control group. 

\begin{sidewaystable}
\caption{Treatment effects on consumption of assigned SMCs}
\vspace{-12pt}
\label{table:FS1}
\begin{center}
\scalebox{0.61}{
\begin{tabular}{lcccccc|cccccc|cccc}
\toprule
& \multicolumn{6}{c|}{\textbf{\textit{Midline survey outcomes}}} & \multicolumn{6}{c|}{\textbf{\textit{Endline survey outcomes}}} & \multicolumn{4}{c}{\textbf{\textit{Browsing history watch counts during intervention}}} \\
& \multicolumn{2}{c}{\textbf{Average SMC}} & \multicolumn{2}{c}{\textbf{Average SMC}} & \multicolumn{2}{c|}{\textbf{Share of}} & \multicolumn{2}{c}{\textbf{Average SMC}} & \multicolumn{2}{c}{\textbf{Average SMC}} & \multicolumn{2}{c|}{\textbf{Share of}} & \multicolumn{1}{c}{\textbf{TikTok videos}} & \multicolumn{1}{c}{\textbf{TikTok videos}} & \multicolumn{1}{c}{\textbf{YouTube videos}} & \multicolumn{1}{c}{\textbf{YouTube videos}} \\
& \multicolumn{2}{c}{\textbf{consumption}} & \multicolumn{2}{c}{\textbf{consumption}} & \multicolumn{2}{c|}{\textbf{quiz questions}} & \multicolumn{2}{c}{\textbf{consumption}} & \multicolumn{2}{c}{\textbf{consumption}} & \multicolumn{2}{c|}{\textbf{quiz questions}} & \multicolumn{1}{c}{\textbf{by assigned}} & \multicolumn{1}{c}{\textbf{by assigned}} & \multicolumn{1}{c}{\textbf{by assigned}} & \multicolumn{1}{c}{\textbf{by assigned}} \\
& \multicolumn{2}{c}{\textbf{frequency index}} & \multicolumn{2}{c}{\textbf{at all}} & \multicolumn{2}{c|}{\textbf{correct}} & \multicolumn{2}{c}{\textbf{frequency index}} & \multicolumn{2}{c}{\textbf{at all}} & \multicolumn{2}{c|}{\textbf{correct}} & \multicolumn{1}{c}{\textbf{SMCs}} & \multicolumn{1}{c}{\textbf{SMCs (log)}} & \multicolumn{1}{c}{\textbf{SMCs}} & \multicolumn{1}{c}{\textbf{SMCs (log)}} \\
& (1) & (2) & (3) & (4) & (5) & (6) & (7) & (8) & (9) & (10) & (11) & (12) & (13) & (14) & (15) & (16) \\
\midrule
\textit{Panel A: differences relative to control condition} \\
\end{tabular}
}
\end{center}
\vspace{-10pt}
\footnotesize \textit{Notes}: Each specification is estimated using OLS, and includes randomization block fixed effects (with the exception of the smaller samples for TikTok and YouTube behavioral outcomes). Except for the TikTok and YouTube behavioral outcomes, even-numbered columns additionally include predetermined covariates selected by the \cite{belloni2014inference} LASSO procedure, and their (demeaned) interaction with each treatment condition. Robust standard errors are in parentheses. Pre-specified one- and two-sided tests: * $p<0.1$, ** $p<0.05$, *** $p<0.01$. Two-sided tests of estimates in the opposite of a pre-specified one-sided test: $^{+}$ $p<0.1$, $^{++}$ $p<0.05$, $^{+++}$ $p<0.01$.
\end{sidewaystable}

Our behavioral data for the 1,149 endline respondents who shared their YouTube browsing histories validates these consumption patterns. Column (15) of Table \ref{table:FS1} show that quiz-incentivized participants watched an average of 6-8 (non-unique) videos per assigned SMC on YouTube during the intervention period, whereas the average control participant watched only one video from any of the five SMCs they would have been recommended.\footnote{The analogous average for unique videos watched per assigned SMC is 5-7. Unreported results show that consumption rates were 9.5 times greater during the week quizzes were open. Some participants were likely watching or re-watching videos just for the quiz, while others may have normalized consuming the content of SMCs they enjoy around quizzes.} For the 119 participants who shared their TikTok watch history, column (13) shows that quiz-incentivized participants watched roughly double the number of videos on TikTok as YouTube. Although respondents who shared their browsing histories may not be representative and we could not obtain browsing data from Instagram users, these behavioral measures align with the 11-15 videos per assigned SMC that quiz-incentivized respondents reported watching. This equates to 6\% of the YouTube (non-ad) videos that respondents in the PA and PP SMC treatment groups watched.

We follow \cite{marbach2020profiling} to further measure levels of compliance and types of compliers. First, Appendix Figure \ref{figure:compliance_rates} reports 50-80\% compliance rates for watching at least 1, 5, 10, 25, or 50 videos from assigned SMCs during the intervention, with similar rates across treatment conditions. The share of never-takers exceeds the share of compliers for watching 100 videos. Second, Appendix Tables \ref{table:complier_types_self_reports} and \ref{table:complier_types_youtube} show that the covariate profile of participants consuming at least 5 videos from assigned SMCs is largely similar to the full sample, suggesting that compliers are---unsurprisingly, given high compliance rates---not an unusual subgroup.

Absent biweekly quizzes, our recommendations increased engagement by far less. The recommendation-only and recommendation-plus-following/subscription encouragements both increased self-reported consumption of assigned SMCs by between 0.5 and 1 standard deviations, about 30\% of the effect of quiz incentives. For correct recall of content, the approximately 5 percentage point increase constitutes only 15\% of the treatment effect of quiz incentives. The estimates in columns (14) and (16) using the natural logarithm of TikTok and YouTube views affirm statistically significant but small increases among participants without quiz incentives.\footnote{Throughout, we add 1 before taking the natural logarithm of variables that take a value of 0. This approximates a percentage change when treatment does not affect the outcome's extensive margin \citep{chen2024logs}.} %Since these groups did not have financial incentives to regularly engage with assigned SMCs, these results show that our recommendation algorithm modestly increased engagement with previously-unfollowed SMCs and---more interestingly---that consumers were not turned off by subtler or overt political content. 
As expected, the SMS/email treatment did not alter consumption or knowledge of SMC content.

\subsection{Engagement with social media content in general}

To understand how following new SMCs affects participants' political views, we further characterize broader consumption during the 2024 election campaign. Beyond direct exposure, encouragement to follow assigned SMCs could alter the quantity or type of content participants consumed more generally.

For the \textit{quantity} of social media consumption, we asked respondents to report their use of various types of media. In the pure control group, the mean and median participant recalled spending 20 and 14 hours per week on Instagram, TikTok, and YouTube during the intervention up to midline, and a mean and median of 18 and 12 hours per week by endline. Columns (7)-(8) of Table \ref{table:FS2} show that the quiz-incentivized PA and NP SMCs groups reported spending around 13\% more hours per week on Instagram, TikTok, and YouTube by endline; the corresponding 6\% increase in the PP SMCs group is not statistically significant.\footnote{These differences do not show up in the total number of TikTok and YouTube videos watched among those providing browsing histories. However, increased self-reported use of these platforms was also less pronounced among this subgroup, suggesting a sample composition effect rather than inaccurate self-reports.} Participants receiving quiz incentives were also more likely to report consuming various famous social media accounts (Appendix Table \ref{table:FS1_other}), with greater time spent on social media coming at the expense of time spent watching television (Appendix Table \ref{table:FS3}).

\begin{sidewaystable}
\caption{Treatment effects on consumption of social media content in general}
\vspace{-12pt}
\label{table:FS2}
\begin{center}
\scalebox{0.64}{
\begin{tabular}{lcccccc|cccccc|ccccc}
\toprule
& \multicolumn{6}{c|}{\textbf{\textit{Midline survey outcomes}}} & \multicolumn{6}{c|}{\textbf{\textit{Endline survey outcomes}}} & \multicolumn{5}{c}{\textbf{\textit{Behavioral outcomes (excluding assigned SMCs)}}} \\
&&&&&& \multicolumn{1}{c|}{} &&&&&& \multicolumn{1}{c|}{} & \multicolumn{1}{c}{\textbf{Total TikTok}} & \multicolumn{1}{c}{\textbf{Total YouTube}} & \multicolumn{1}{c}{\textbf{YouTube}} & \multicolumn{1}{c}{\textbf{YouTube}} & \multicolumn{1}{c}{\textbf{YouTube}} \\
& \multicolumn{2}{c}{\textbf{Weekly hours of}} & \multicolumn{2}{c}{\textbf{Share of social}} & \multicolumn{2}{c|}{\textbf{Liberal lean of}} & \multicolumn{2}{c}{\textbf{Weekly hours of}} & \multicolumn{2}{c}{\textbf{Share of social}} & \multicolumn{2}{c|}{\textbf{Liberal lean of}} & \multicolumn{1}{c}{\textbf{videos}} & \multicolumn{1}{c}{\textbf{videos}} & \multicolumn{1}{c}{\textbf{share}} & \multicolumn{1}{c}{\textbf{liberal}} & \multicolumn{1}{c}{\textbf{Democrat}} \\
& \multicolumn{2}{c}{\textbf{IG/TT/YT (log)}} & \multicolumn{2}{c}{\textbf{media political}} & \multicolumn{2}{c|}{\textbf{social media}} & \multicolumn{2}{c}{\textbf{IG/TT/YT}} & \multicolumn{2}{c}{\textbf{media political}} & \multicolumn{2}{c|}{\textbf{social media}} & \multicolumn{1}{c}{\textbf{watched (log)}} & \multicolumn{1}{c}{\textbf{watched (log)}} & \multicolumn{1}{c}{\textbf{political}} & \multicolumn{1}{c}{\textbf{ideology}} & \multicolumn{1}{c}{\textbf{slant}} \\
& (1) & (2) & (3) & (4) & (5) & (6) & (7) & (8) & (9) & (10) & (11) & (12) & (13) & (14) & (15) & (16) & (17) \\
\midrule
\textit{Panel A: differences relative to control condition} \\ Predominantly-apolitical SMCs (quiz)&       0.086   &      0.089*   &       0.030   &       0.024   &       0.005   &       0.000   &     0.137**   &     0.121**   &       0.012   &       0.002   &       0.009   &       0.003   &      -1.560   &      -0.166   &   -0.032***   &      -0.006   &      -0.005   \\
\bottomrule
\end{tabular}
}
\end{center}
\vspace{-6pt}
\footnotesize \textit{Notes}: Each specification is estimated using OLS, and includes randomization block fixed effects (with the exception of the smaller samples for TikTok and YouTube behavioral outcomes). Except for the TikTok and YouTube behavioral outcomes, even-numbered columns additionally include predetermined covariates selected by the \cite{belloni2014inference} LASSO procedure, and their (demeaned) interaction with each treatment condition. Robust standard errors are in parentheses. Pre-specified one- and two-sided tests: * $p<0.1$, ** $p<0.05$, *** $p<0.01$. Two-sided tests of estimates in the opposite of a pre-specified one-sided test: $^{+}$ $p<0.1$, $^{++}$ $p<0.05$, $^{+++}$ $p<0.01$.
\end{sidewaystable}

We next turn to the \textit{nature} of social media content consumed. As with assigned SMCs' content, we used GPT-5-mini to classify the 1,457,016 unique YouTube videos our participants watched from non-assigned accounts during the intervention period between August 15 and December 26, 2024. The median respondent in the pure control group who provided their YouTube browsing history watched 930 non-ad videos, of which 7.4\% were classified as political on average. Appendix Table \ref{table:FS2_nonassigned} shows that these videos were evenly-balanced in terms of ideological, policy, and partisan slant as well as less explicit liberal moral values. Participants' overall YouTube consumption---the most frequently used platform in our sample---was thus mostly non-political and politically-balanced. Given our average respondent's liberal predisposition, their relatively more conservative social media bundle many participants consumed may have been particular to an election campaign where Republican content appeared ascendant. It is possible that content on Instagram and TikTok was more political than YouTube. %In particular, unprompted recommendations in late 2024 were much more likely to be political on other platforms (like TikTok) relative to YouTube, where this analysis was sourced from.

The biweekly quiz also somewhat shifted participants' social media consumption, but largely due to consuming assigned SMCs' progressive-minded content. Columns (3)-(4) and (9)-(10) of Table \ref{table:FS2} show that the share of their social media content that the PP SMCs group perceived as political rose from around half by 2-3 percentage points. Moreover, columns (5)-(6) and (11)-(12) indicate that the liberal share of content had increased by 4 percentage points from around 30\% by endline. In contrast, the PA SMCs group perceived no significant change in the political lean of their social media consumption relative to the control group, suggesting that PA SMCs imperceptibly imparted non-partisan but politically-relevant content. The NP SMCs placebo group experienced slightly more conservative content by midline, although the differences from the PA SMCs and pure control groups are not statistically significant.

We find little evidence to suggest that our intervention altered participants' social media consumption beyond assigned SMCs. Column (15) of Table \ref{table:FS2} shows that the PA and PP SMCs groups did not watch significantly more videos classified by GPT-5-mini as broadly political, suggesting that the treatment did not change the type of content participants actively sought out or were recommended by YouTube's algorithm. In line with prior evidence that X amplifies right-wing over left-wing content \citep{gauthier2025X, huszar2022algorithmic} and YouTube pushes moderately conservative content \citep{brown2022echo}, columns (16) and (17) show that participants' YouTube browsing histories were evenly split on seven-point ideology and partisanship scales---despite our sample leaning liberal on average---and were not significantly affected by treatment. Appendix Table \ref{table:FS2_nonassigned} reports similar results for indexes covering other classifications, including content promoting liberal moral values, liberal policies, and explicit or implicit support for the Democrats. Given that the median participant consumed almost 1,000 YouTube videos during the intervention, it is perhaps unsurprising that consuming around 70 videos produced by progressive-minded SMCs did not substantially change their broader consumption bundle.

\subsection{Summary}

\begin{figure}[!h]

\begin{subfigure}{0.5\textwidth}
\begin{center}
\caption{Midline consumption of assigned SMCs (ICW index)}
\vspace{6pt}
\includegraphics[scale=0.42]{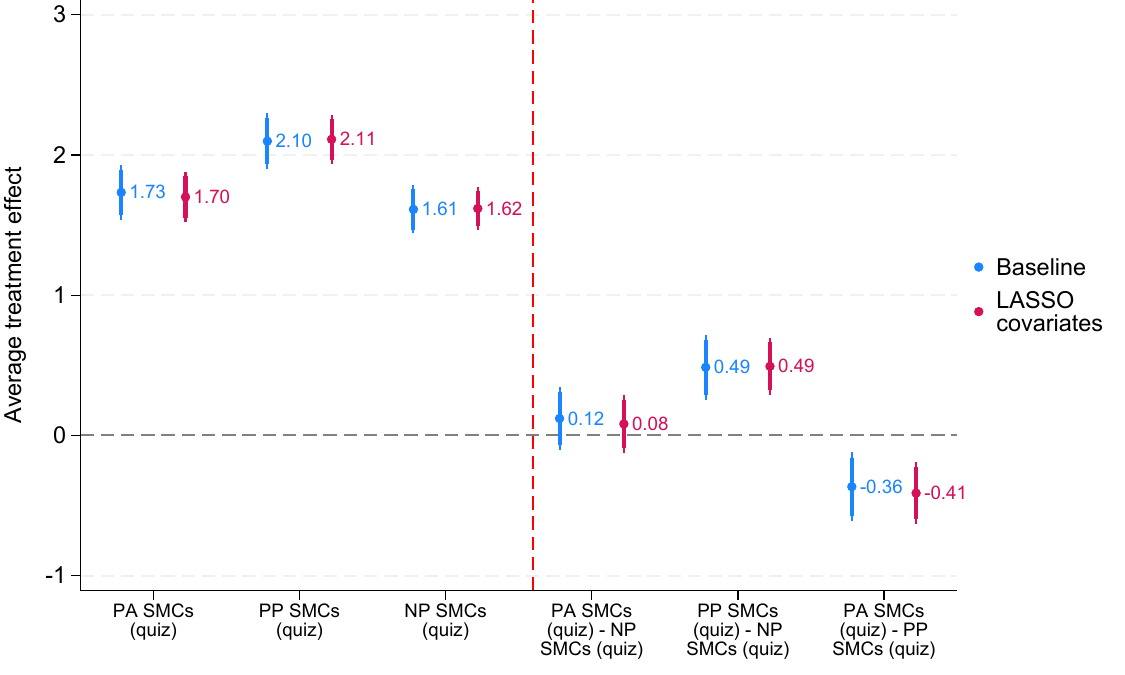}
\label{figure:midline_consumption}
\end{center}
\end{subfigure}%
\begin{subfigure}{0.5\textwidth}
\begin{center}
\caption{Endline consumption of assigned SMCs (ICW index)}
\vspace{6pt}
\includegraphics[scale=0.42]{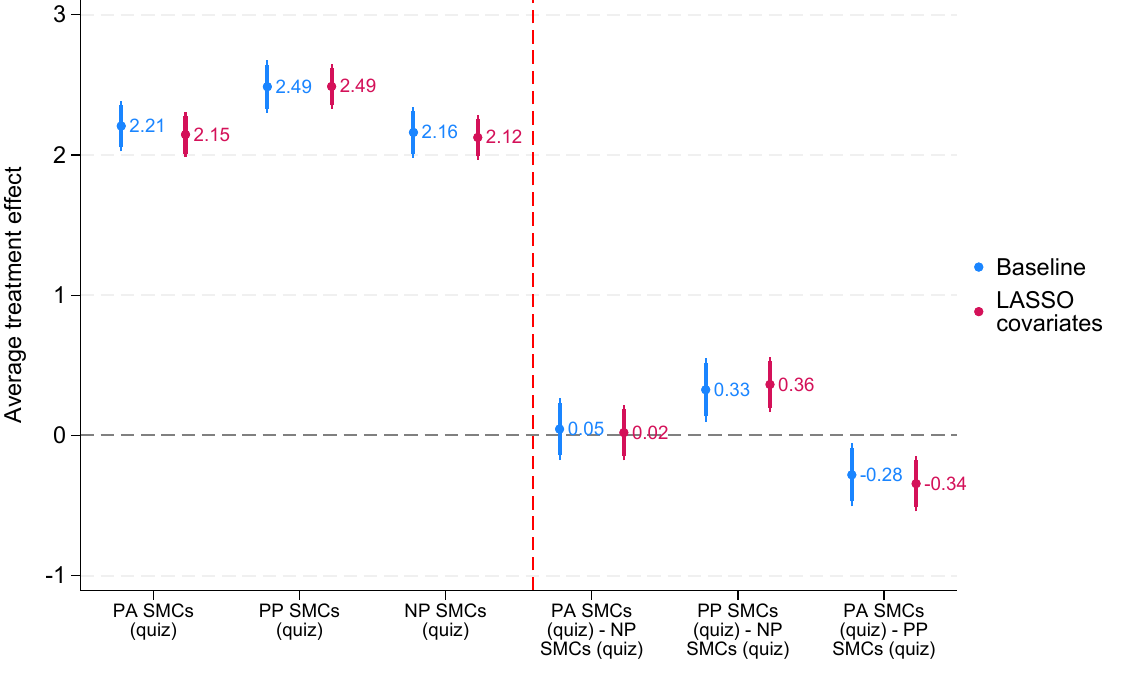}
\label{figure:endline_consumption}
\end{center}
\end{subfigure}%

\vspace{-4pt}

\begin{subfigure}{0.5\textwidth}
\begin{center}
\caption{Midline share of correct answers about assigned SMCs}
\vspace{6pt}
\includegraphics[scale=0.42]{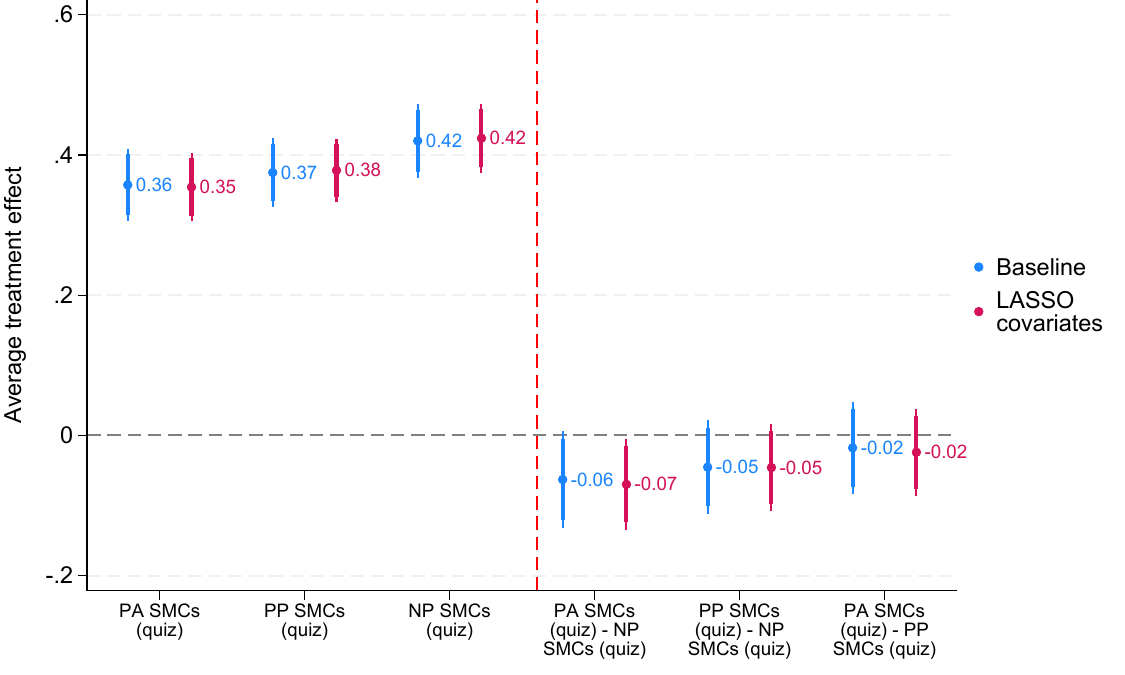}
\label{figure:midline_smc_knowledge}
\end{center}
\end{subfigure}%
\begin{subfigure}{0.5\textwidth}
\begin{center}
\caption{Endline share of correct answers about assigned SMCs}
\vspace{6pt}
\includegraphics[scale=0.42]{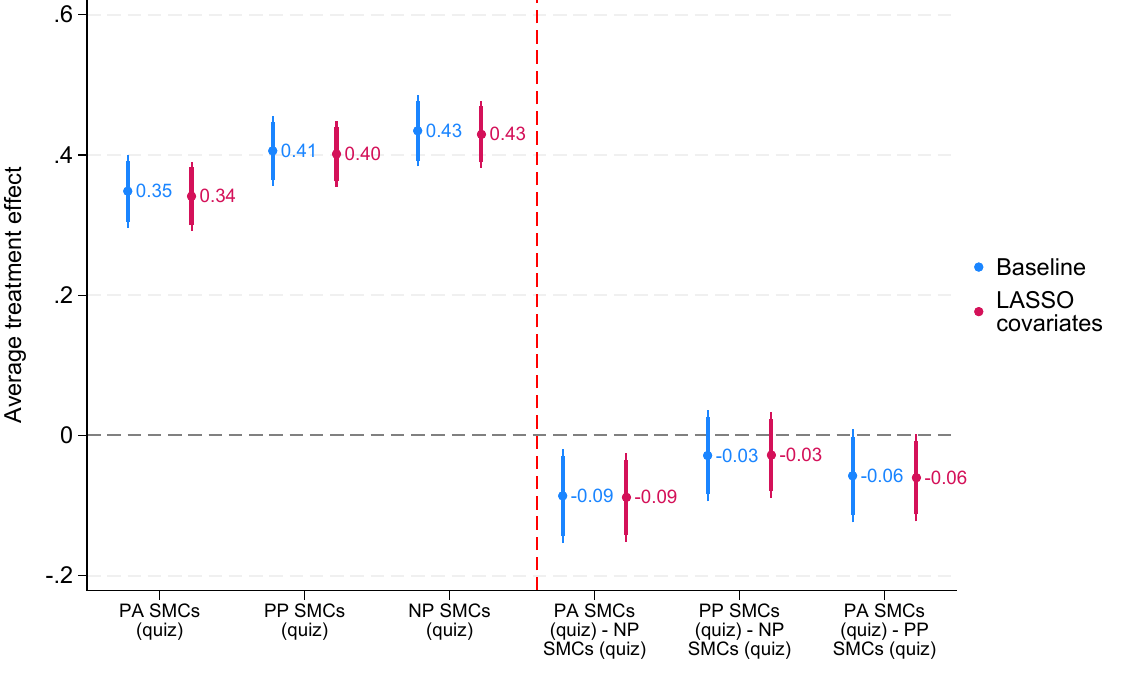}
\label{figure:endline_smc_knowledge}
\end{center}
\end{subfigure}%

\vspace{-4pt}

\begin{subfigure}{0.5\textwidth}
\begin{center}
\caption{Midline total weekly hours of social media (log)}
\vspace{6pt}
\includegraphics[scale=0.42]{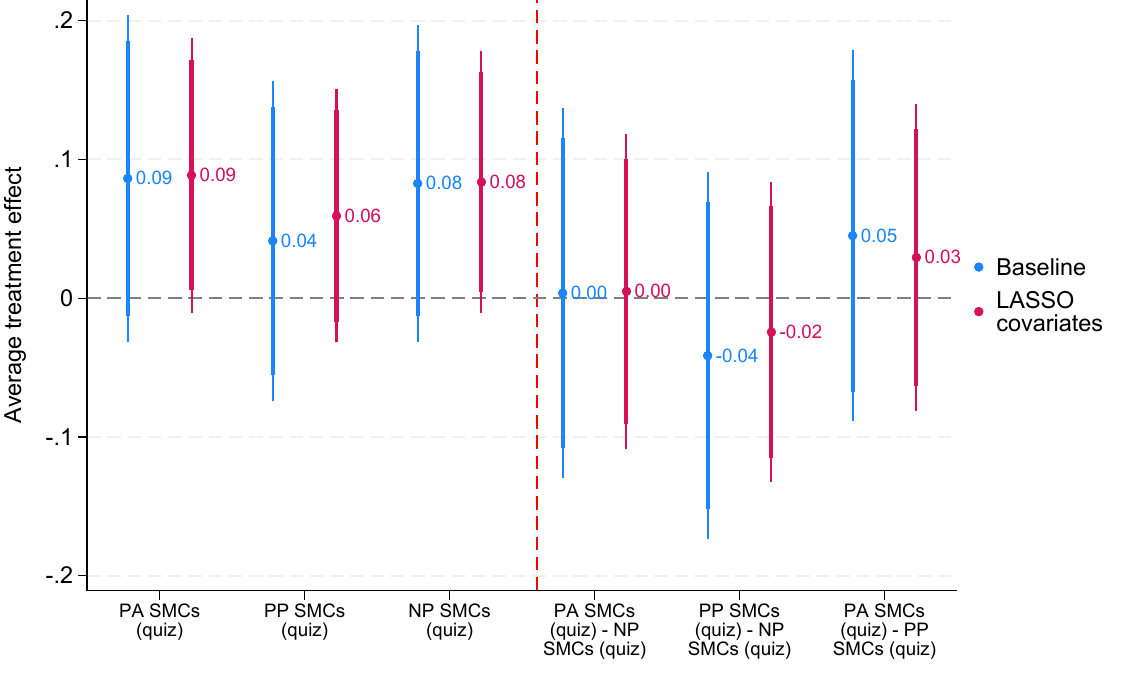}
\label{figure:midline_log_social_media_hours}
\end{center}
\end{subfigure}%
\begin{subfigure}{0.5\textwidth}
\begin{center}
\caption{Endline total weekly hours of social media (log)}
\vspace{6pt}
\includegraphics[scale=0.42]{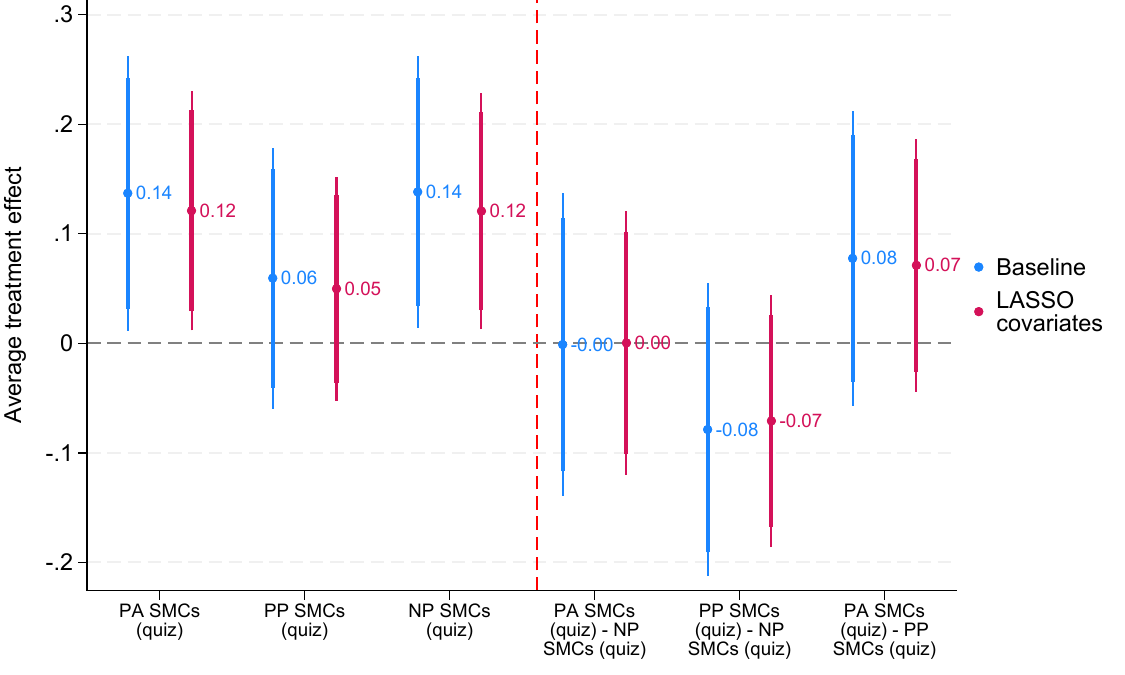}
\label{figure:endline_log_social_media_hours}
\end{center}
\end{subfigure}%

\vspace{-18pt}

\begin{center}
\caption{Effects of quiz-based incentives to consume SMCs on social media consumption}
\label{figure:first_stage}
\end{center}

\begin{tablenotes}
\vspace{-16pt}
\item {\footnotesize \textit{Notes}: Each graph plots the estimates of equation (\ref{equation:full}) reported in Tables \ref{table:FS1} and \ref{table:FS2}, with 90\% (thick lines) and 95\% (thin lines) confidence intervals. The three sets of estimates to the left of the vertical dotted line compute differences between treatment groups and the pure control group; the three sets of estimates to the right compute differences between quiz-incentivized treatment groups.}
\end{tablenotes}

\end{figure}

Figure \ref{figure:first_stage} summarizes these ``first stage'' effects among participants receiving quiz incentives to follow assigned SMCs. Each row corresponds to an outcome, while columns report midline and then endline results. Within each subfigure, the three groups of coefficients to the left of the vertical line show ATEs relative to the pure control group; the three groups of coefficients to the right show differences in ATEs between quiz-incentivized treatment conditions. 

Together, we observe substantial increases in consumption and factual recall of the content produced by assigned SMCs across the PA, PP, and NP conditions as well as a 10\% increase in social media consumption in general. Only participants assigned to follow PP SMCs perceived encountering significantly more liberal-leaning content.

\section{Political consequences of following progressive-minded SMCs}

The previous section demonstrated that biweekly quiz incentives generated substantial exposure to five assigned SMCs and greater exposure to online content more broadly during the US election campaign and its aftermath. We henceforth focus on quiz-incentivized respondents, reporting the estimates for participants receiving only recommendations or a following incentive in Appendix Tables \ref{table:H1_full}-\ref{table:H10_full}.\footnote{Another approach could be to use all treatments as instruments, but this is unlikely to capture a desired estimand. First, since treatment effects could reflect assigned SMCs or non-assigned content, it is hard define an endogenous treatment condition. Second, it is hard to interpret estimands in multi-instrument settings, particularly when imposing a linear functional form on an endogenous treatment. Third, effectively normalizing the reduced form by the first stage when pooling low- and high-intensity treatments risks introducing imprecision and obscuring effects of the high-intensity treatment when the first stages differ substantially.} Our main findings next show that quiz incentives to start following progressive-minded SMCs increased political knowledge and more progressive policy preferences, before showing that this translated into partisan preferences without affecting electoral or non-electoral political participation. The section concludes by aggregating these results and contextualizing effect sizes.

\subsection{Political engagement and policy preferences}

\subsubsection{Topical knowledge and political interest}

We begin by examining two measures of political engagement. First, our midline and endline surveys asked respondents three multiple-choice questions to gauge their knowledge of topical political news \textit{not} covered by BII-sponsored content.\footnote{The midline survey asked which state was not a battleground state (Colorado), which state Tim Walz was governor of (Minnesota), and which newspaper did not endorse a presidential candidate (Washington Post). The endline survey asked who Trump asked to lead DOGE (Elon Musk and Vivek Ramaswamy), who Special Counsel Jack Smith dropped charges against in November (Hunter Biden), and which Trump nominee for Attorney General withdrew (Matt Gaetz). Each question had four answers, as well as an option to say ``don't know.'' Participants received no incentives for providing correct answers.} Second, we asked participants how closely they followed politics on a four-point scale from ``not at all closely'' to ``very closely.'' We combine these two measures in an ICW index, standardized to the control group mean and standard deviation.

Turning first to the index, we find a substantial increase in this aspect of political engagement. Columns (1)–(2) and (7)–(8) of Table \ref{table:H1} show that quiz incentives to follow PP SMCs increased engagement by 0.36 and 0.28 standard deviations relative to the pure control group at midline and endline, respectively. As columns (3)-(4) and (9)-(10) show at midline and endline, this is driven by increasing the probability of a correct answer by 19 and 14 percentage points (or 44\% and 25\%) from around 50\% in the control group. PA SMCs boosted the overall index and topical political knowledge by about half as much, a notable increase given that PA SMCs produced only one quarter of PP SMCs' political content.

Increased political knowledge partly reflects direct exposure to assigned SMCs, but our two comparison groups suggest that greater social media use in general and exposure to political content contribute to almost half the increase in topical political knowledge. Specifically, participants assigned to follow NP SMCs did not view a larger \textit{share} of political content on YouTube---as indicated by our browsing data---but they spent more time on the platform during the campaign, which may have increased their topical knowledge. Similarly, participants in the SMS/email condition gained knowledge even though the newsletters they were sent did not include answers to our survey questions, suggesting that this treatment increased interest in current affairs.

These knowledge gains did not necessarily translate into participants \textit{perceiving} themselves as more politically attentive. In line with participants' perception that the PA and NP SMCs' content was not political, columns (5)-(6) and (11)-(12) show that neither group claimed to follow politics more closely. The same holds for receiving biweekly messages about BII topics. In contrast, participants incentivized to follow PP SMCs reported a significant increase of about 10\%. This suggests that exposure to both explicitly political and largely-apolitical SMCs can spark political engagement among young adults, possibly without them even recognizing it.

\begin{sidewaystable}
\caption{Treatment effects on political engagement}
\vspace{-12pt}
\label{table:H1}
\begin{center}
\scalebox{0.83}{
\begin{tabular}{lcccccc|cccccc}
\toprule
& \multicolumn{6}{c|}{\textbf{\textit{Midline survey outcomes}}} & \multicolumn{6}{c}{\textbf{\textit{Endline survey outcomes}}} \\
& \multicolumn{2}{c}{\textbf{Political engagement}} & \multicolumn{2}{c}{\textbf{Share topical political}} & \multicolumn{2}{c|}{\textbf{Follow politics}} & \multicolumn{2}{c}{\textbf{Political engagement}} & \multicolumn{2}{c}{\textbf{Share topical political}} & \multicolumn{2}{c}{\textbf{Follow politics}}  \\
& \multicolumn{2}{c}{\textbf{ICW index}} & \multicolumn{2}{c}{\textbf{questions correct}} & \multicolumn{2}{c|}{\textbf{closely scale}} & \multicolumn{2}{c}{\textbf{ICW index}} & \multicolumn{2}{c}{\textbf{questions correct}} & \multicolumn{2}{c}{\textbf{closely scale}} \\
& (1) & (2) & (3) & (4) & (5) & (6) & (7) & (8) & (9) & (10) & (11) & (12) \\
\midrule
\textit{Panel A: differences relative to control condition} \\ Predominantly-apolitical SMCs (quiz)&    0.187***   &    0.184***   &    0.098***   &    0.099***   &      0.076*   &      0.066*   &     0.114**   &     0.109**   &    0.074***   &    0.083***   &       0.000   &      -0.010   \\
                    &     (0.058)   &     (0.053)   &     (0.028)   &     (0.025)   &     (0.054)   &     (0.050)   &     (0.060)   &     (0.055)   &     (0.027)   &     (0.024)   &     (0.059)   &     (0.055)   \\
%p value            &       0.001   &       0.000   &       0.000   &       0.000   &       0.078   &       0.091   &       0.029   &       0.024   &       0.003   &       0.000   &       0.499   &       0.850   \\
Predominantly-political SMCs (quiz)&    0.360***   &    0.361***   &    0.188***   &    0.189***   &    0.149***   &    0.151***   &    0.277***   &    0.277***   &    0.142***   &    0.142***   &     0.098**   &     0.102**   \\
\end{tabular}
}
\end{center}
\vspace{-10pt}
\footnotesize \textit{Notes}: Each specification is estimated using OLS, and includes randomization block fixed effects. Even-numbered columns additionally include predetermined covariates selected by the \cite{belloni2014inference} LASSO procedure, and their (demeaned) interaction with each treatment condition. Estimates from the recommendation-only and incentivized follower encouragement conditions are suppressed and reported in Table \ref{table:H1_full}. Robust standard errors are in parentheses. Pre-specified one- and two-sided tests: * $p<0.1$, ** $p<0.05$, *** $p<0.01$. Two-sided tests of estimates in the opposite of a pre-specified one-sided test: $^{+}$ $p<0.1$, $^{++}$ $p<0.05$, $^{+++}$ $p<0.01$.
\end{sidewaystable}

\subsubsection{Policy positions on the BII's core issues}

We now turn to policy preferences relating to the BII's core policy issues. Table \ref{table:H2} first reports effects on policy preference indices for climate (1 item), democracy (2 items), economic (2 items), and health (3 items at midline, plus importance of vaccination at endline).\footnote{More specifically, we measure policy preferences relating to curbing climate change, commitment to democracy, agreement that violence cannot be justified, perceptions of corporate power, faith in free markets (reversed), support for family leave, belief that abortion should be guaranteed in all states, and support for increased government spending on health insurance. Appendix Table \ref{table:H3_full} reports no systematic changes in issue prioritization.} Each scale is normalized to range from the most conservative (0) to most liberal (1) position, and the ICW index aggregates all four scales. 

These attitudinal outcomes first reveal that greater social media consumption, without incentivized changes in policy or political content, made participants somewhat \textit{more conservative} at election time. The overall index in columns (1)-(2) of panel A shows that, by the midline survey the week before the election, the NP SMCs group's policy preferences had moved almost 0.1 standard deviations to the right of the control group. This change is only statistically significant at the 10\% level before introducing covariates, but we observe small rightward shifts in each policy domain. Columns (11)-(12) indicate that this difference in policy preferences dissipated by our January 2025 endline survey. 

This small conservative shift in election-time policy preferences is consistent with at least two explanations. First, because the partisan-balanced social media content consumed by our young and liberal-leaning sample was to the right of their predispositions, increased social media consumption could have updated participants' policy preferences in a conservative direction. %This balance may reflect platforms' structural bias toward promoting conservative content during the 2024 campaign \citep{brown2022echo, gauthier2025X, huszar2022algorithmic}. 
Second, the conservative content participants consumed outside of our intervention may have been more persuasive than liberal content, even when consumed in similar quantities. Indeed, the disproportionate influence of right-wing SMCs has been widely suggested.\footnote{See \textit{New York Times}, ``Republicans Built an Ecosystem of Influencers. Some Democrats Want One, Too.'', \href{https://www.nytimes.com/2024/11/28/us/politics/democratic-influencers.html}{www.nytimes.com/2024/11/28/us/politics/democratic-influencers.html}; Pew Research Center, ``How news influencers talked about Trump and Harris during the 2024 election'', \href{https://www.pewresearch.org/short-reads/2025/02/06/how-news-influencers-talked-about-trump-and-harris-during-the-2024-election/}{www.pewresearch.org/short-reads/2025/02/06/how-news-influencers-talked-about-trump-and-harris-during-the-2024-election}.} However, the effect is unlikely to be driven by the placebo SMCs group consuming a different ideological or partisan mix of social media content. Tables \ref{table:FS2} and \ref{table:FS2_nonassigned} detect no such differences, either before November or over the entire intervention period. While our AI classifications could miss political videos relying solely on memes or visuals without any meaningful transcripts, the evidence suggests the former two interpretations are more likely to explain why greater time spent on social media shifted participants to the right.  

\begin{sidewaystable}
\caption{Treatment effects on liberal policy attitudes}
\vspace{-12pt}
\label{table:H2}
\begin{center}
\scalebox{0.63}{
\begin{tabular}{lcccccccccc|cccccccccc}
\toprule
& \multicolumn{10}{c|}{\textbf{\textit{Midline survey outcomes}}} & \multicolumn{10}{c}{\textbf{\textit{Endline survey outcomes}}} \\
& \multicolumn{2}{c}{\textbf{Liberal}} & \multicolumn{2}{c}{\textbf{}} & \multicolumn{2}{c}{\textbf{}} & \multicolumn{2}{c}{\textbf{}} & \multicolumn{2}{c|}{\textbf{}} & \multicolumn{2}{c}{\textbf{Liberal}} & \multicolumn{2}{c}{\textbf{}} & \multicolumn{2}{c}{\textbf{}} & \multicolumn{2}{c}{\textbf{}} & \multicolumn{2}{c}{\textbf{}} \\
& \multicolumn{2}{c}{\textbf{policy}} & \multicolumn{2}{c}{\textbf{Liberal}} & \multicolumn{2}{c}{\textbf{Liberal}} & \multicolumn{2}{c}{\textbf{Liberal}} & \multicolumn{2}{c|}{\textbf{Liberal}} & \multicolumn{2}{c}{\textbf{policy}} & \multicolumn{2}{c}{\textbf{Liberal}} & \multicolumn{2}{c}{\textbf{Liberal}} & \multicolumn{2}{c}{\textbf{Liberal}} & \multicolumn{2}{c}{\textbf{Liberal}} \\
& \multicolumn{2}{c}{\textbf{attitudes}} & \multicolumn{2}{c}{\textbf{climate}} & \multicolumn{2}{c}{\textbf{democracy}} & \multicolumn{2}{c}{\textbf{economic}} & \multicolumn{2}{c|}{\textbf{health}} & \multicolumn{2}{c}{\textbf{attitudes}} & \multicolumn{2}{c}{\textbf{climate}} & \multicolumn{2}{c}{\textbf{democracy}} & \multicolumn{2}{c}{\textbf{economic}} & \multicolumn{2}{c}{\textbf{health}} \\
& \multicolumn{2}{c}{\textbf{ICW index}} & \multicolumn{2}{c}{\textbf{attitudes}} & \multicolumn{2}{c}{\textbf{attitudes}} & \multicolumn{2}{c}{\textbf{attitudes}} & \multicolumn{2}{c|}{\textbf{attitudes}} & \multicolumn{2}{c}{\textbf{ICW index}} & \multicolumn{2}{c}{\textbf{attitudes}} & \multicolumn{2}{c}{\textbf{attitudes}} & \multicolumn{2}{c}{\textbf{attitudes}} & \multicolumn{2}{c}{\textbf{attitudes}} \\
& (1) & (2) & (3) & (4) & (5) & (6) & (7) & (8) & (9) & (10) & (11) & (12) & (13) & (14) & (15) & (16) & (17) & (18) & (19) & (20) \\
\midrule
\textit{Panel A: differences relative to control condition} \\
\end{tabular}
}
\end{center}
\vspace{-10pt}
\footnotesize \textit{Notes}: Each specification is estimated using OLS, and includes randomization block fixed effects and adjusts for lagged outcomes and their (demeaned) interaction with each treatment condition. Even-numbered columns additionally include predetermined covariates selected by the \cite{belloni2014inference} LASSO procedure, and their (demeaned) interaction with each treatment condition. Estimates from the recommendation-only and incentivized follower encouragement conditions are suppressed and reported in Table \ref{table:H2_full}. Robust standard errors are in parentheses. Pre-specified one- and two-sided tests: * $p<0.1$, ** $p<0.05$, *** $p<0.01$. Two-sided tests of estimates in the opposite of a pre-specified one-sided test: $^{+}$ $p<0.1$, $^{++}$ $p<0.05$, $^{+++}$ $p<0.01$.
\end{sidewaystable}

In contrast, quiz incentives to follow progressive-minded SMCs led participants to adopt more liberal policy positions at election time. Panel B of Table \ref{table:H2} shows that the PA and PP SMCs groups became between 0.11 and 0.15 standard deviations more liberal \textit{relative to the NP SMCs group} at midline, reflecting small progressive differentials across all four policy domains. Holding encouragement to use social media constant, progressive-minded SMCs thus shifted policy attitudes to the left. Panel A indicates that these effects are smaller relative to the pure control group.\footnote{In both cases, Appendix Table \ref{table:H3_full} shows no significant change in issue prioritization.} The statistically significant difference between the PA and PP SMCs conditions and the placebo SMCs condition appears to reflect two countervailing forces: progressive-minded SMCs made their followers modestly more liberal, whereas more time spent on politically-balanced social media made the NP SMCs group modestly more conservative. In other words, greater social media consumption during the 2024 election campaign period pushed participants to the right, unless it was counterbalanced by consuming progressive-minded SMCs.  

To obtain a behavioral measure of policy preferences, we asked participants in the endline survey to allocate \$100 between ten non-profit organizations. Four organizations advocated for progressive policy positions on each of the BII's core topics, four advocated for conservative policy positions on the same topics, and two were non-political.\footnote{The organizations advocating progressive positions on climate, democracy, economic, and health issues were, respectively, the Environmental Defense Fund, American Civil Liberties Union, American for Financial Reform, and Planned Parenthood. The organizations advocating for conservative positions on these issues were the Heartland Institute, True the Vote, Cato Institute, and Heritage Foundation. The non-political organizations were the American Red Cross and Wikipedia. Before making their elections, participants were provided with a sentence summarizing the mission of each organization.} With a sentence summarizing each organization's mission, participants could select 0, 1, or 2 organizations to donate to, and were informed that the research team would implement 50 randomly-selected allocations. To measure liberal cause donations in each or any policy domain, we combined an indicator for donating to a liberal organization and (reversed) indicator for donating to a conservative organization; an overall ICW index combines all policy domains. 

\begin{sidewaystable}
\caption{Treatment effects on endline cause donation decisions}
\vspace{-12pt}
\label{table:cause_donations}
\begin{center}
\scalebox{0.81}{
\begin{tabular}{lcccccccccccccc}
\toprule
& \multicolumn{2}{c}{\textbf{Liberal}} & \multicolumn{2}{c}{\textbf{Liberal over}} & \multicolumn{2}{c}{\textbf{Liberal over}} & \multicolumn{2}{c}{\textbf{Liberal over}} & \multicolumn{2}{c}{\textbf{Liberal over}} & \multicolumn{2}{c}{\textbf{Donated to}} & \multicolumn{2}{c}{\textbf{Donated to}} \\
& \multicolumn{2}{c}{\textbf{causes}} & \multicolumn{2}{c}{\textbf{conservative}} & \multicolumn{2}{c}{\textbf{conservative}} & \multicolumn{2}{c}{\textbf{conservative}} & \multicolumn{2}{c}{\textbf{conservative}} & \multicolumn{2}{c}{\textbf{any liberal}} & \multicolumn{2}{c}{\textbf{any conservative}} \\
& \multicolumn{2}{c}{\textbf{ICW index}} & \multicolumn{2}{c}{\textbf{climate causes}} & \multicolumn{2}{c}{\textbf{democracy causes}} & \multicolumn{2}{c}{\textbf{economic causes}} & \multicolumn{2}{c}{\textbf{health causes}} & \multicolumn{2}{c}{\textbf{cause}} & \multicolumn{2}{c}{\textbf{cause}} \\
& (1) & (2) & (3) & (4) & (5) & (6) & (7) & (8) & (9) & (10) & (11) & (12) & (13) & (14) \\
\midrule
\textit{Panel A: differences relative to control condition} \\ Predominantly-apolitical SMCs (quiz)&    0.209***   &    0.206***   &      -0.059   &      -0.049   &    0.201***   &    0.184***   &     0.178**   &     0.175**   &     0.121**   &     0.112**   &    0.127***   &    0.124***   &      -0.024   &      -0.024   \\
                    &     (0.067)   &     (0.060)   &     (0.076)   &     (0.071)   &     (0.069)   &     (0.062)   &     (0.082)   &     (0.078)   &     (0.065)   &     (0.060)   &     (0.031)   &     (0.029)   &     (0.026)   &     (0.024)   \\
\end{tabular}
}
\end{center}
\vspace{-10pt}
\footnotesize \textit{Notes}: Each specification is estimated using OLS, and includes randomization block fixed effects, an approximate baseline outcome, and the interaction between the (demeaned) baseline outcome and each treatment condition. Even-numbered columns additionally include predetermined covariates selected by the \cite{belloni2014inference} LASSO procedure, and their (demeaned) interaction with each treatment condition. Estimates from the recommendation-only and incentivized follower encouragement conditions are suppressed and reported in Table \ref{table:cause_donations_full}. Robust standard errors are in parentheses. Pre-specified one- and two-sided tests: * $p<0.1$, ** $p<0.05$, *** $p<0.01$. Two-sided tests of estimates in the opposite of a pre-specified one-sided test: $^{+}$ $p<0.1$, $^{++}$ $p<0.05$, $^{+++}$ $p<0.01$.
\end{sidewaystable}

The results in Table \ref{table:cause_donations} show that quiz incentives to start following progressive-minded SMCs also increased donations to liberal over conservative causes. Columns (1)-(2) reveal that PA SMCs increased followers' liberal cause donation index by 0.21 standard deviations relative to the control group, while PP SMCs generated a 0.15 standard deviation increase. Relative to the control level of 61\%, columns (11)-(12) shows that the probability of donating to any liberal cause increased by 13 percentage points in the PA SMCs group and 8 percentage points in the PP SMCs group; the 17\% probability of donating to a conservative cause was largely unaffected by treatment. We observe similar increases relative to participants assigned to receive SMS/email messages or follow NP SMCs. Columns (3)-(10) show that these results are largely driven by a greater propensity to donate to organizations advocating for liberal positions on democracy and economic policy issues, with limited effects on climate causes---the topic discussed least by SMCs. The pronounced effects on cause donations at endline demonstrate enduring changes in political preferences that our more complex attitudinal scales may have struggled to pick up. 

Panel B of Tables \ref{table:H2} and \ref{table:cause_donations} finds the effects of incentives to follow PA and PP SMCs to be statistically indistinguishable. Given that participants consumed both types of SMCs at similar rates (see Table \ref{table:FS1}), this implies that PA SMCs---who produced around four times fewer political videos, focusing exclusively on the BII's four topic areas in their programmatic content---were more persuasive \textit{per political video}. Election campaigns might then maximize their impact per dollar of social media spending by collaborating with SMCs who rarely cover politics. But SMCs are not the only available option: while PA and PP SMCs did more to change policy positions, biweekly SMS/email messages also generated more liberal donations than the pure control group. 

\subsubsection{Progressive outlooks on the political system}

Beyond the specific domains covered by BII fellows, we further examined progressive outlooks more generally. In terms of policy perspectives, we created (i) a broader policy position index that added tax and racial justice policy questions to our climate, economic, and health policy preferences and (ii) a more systemic narrative index combining four questions gauging respondents' agreement on a five-point scale with the claims that addressing major social issues requires collective rather than individual action, active government intervention is required to constrain market power, the system needs to be reformed to reduce economic and racial inequities, and democracy will only serve the masses if they are informed and actively express their demands. In terms of efficacy within the political system, we created indexes of (iii) perceived individual, collective, and youth capacity to affect political change and (iv) trust in the political and social establishment (i.e. election fairness, election boards, scientists, doctors, and journalists). We combine these four items in an overall progressive outlook index.

\begin{sidewaystable}
\caption{Treatment effects on progressive outlooks}
\vspace{-12pt}
\label{table:H4}
\begin{center}
\scalebox{0.62}{
\begin{tabular}{lcccccccccc|cccccccccc}
\toprule
& \multicolumn{10}{c|}{\textbf{\textit{Midline survey outcomes}}} & \multicolumn{10}{c}{\textbf{\textit{Endline survey outcomes}}} \\
& \multicolumn{2}{c}{\textbf{Progressive}} & \multicolumn{2}{c}{\textbf{Progressive}} & \multicolumn{2}{c}{\textbf{Progressive}} & \multicolumn{2}{c}{\textbf{Sense of}} & \multicolumn{2}{c|}{\textbf{Systemic trust}} & \multicolumn{2}{c}{\textbf{Progressive}} & \multicolumn{2}{c}{\textbf{Progressive}} & \multicolumn{2}{c}{\textbf{Progressive}} & \multicolumn{2}{c}{\textbf{Sense of}} & \multicolumn{2}{c}{\textbf{Systemic trust}} \\
& \multicolumn{2}{c}{\textbf{overall ICW index}} & \multicolumn{2}{c}{\textbf{policy ICW index}} & \multicolumn{2}{c}{\textbf{worldview}} & \multicolumn{2}{c}{\textbf{efficacy}} & \multicolumn{2}{c|}{\textbf{ICW index}} & \multicolumn{2}{c}{\textbf{overall ICW index}} & \multicolumn{2}{c}{\textbf{policy ICW index}} & \multicolumn{2}{c}{\textbf{worldview}} & \multicolumn{2}{c}{\textbf{efficacy}} & \multicolumn{2}{c}{\textbf{ICW index}} \\
& (1) & (2) & (3) & (4) & (5) & (6) & (7) & (8) & (9) & (10) & (11) & (12) & (13) & (14) & (15) & (16) & (17) & (18) & (19) & (20) \\
\midrule
\textit{Panel A: differences relative to control condition} \\ Predominantly-apolitical SMCs (quiz)&       0.034   &       0.009   &       0.060   &       0.050   &       0.045   &       0.018   &       0.013   &      -0.027   &       0.045   &       0.048   &       0.037   &       0.023   &       0.034   &       0.023   &       0.003   &      -0.011   &       0.052   &       0.029   &       0.038   &       0.038   \\
                    &     (0.063)   &     (0.057)   &     (0.057)   &     (0.052)   &     (0.049)   &     (0.045)   &     (0.060)   &     (0.052)   &     (0.059)   &     (0.054)   &     (0.061)   &     (0.055)   &     (0.056)   &     (0.051)   &     (0.051)   &     (0.046)   &     (0.064)   &     (0.057)   &     (0.054)   &     (0.050)   \\
\end{tabular}
}
\end{center}
\vspace{-10pt}
\footnotesize \textit{Notes}: Each specification is estimated using OLS, and includes randomization block fixed effects and adjusts for lagged outcomes and their (demeaned) interaction with each treatment condition. Even-numbered columns additionally include predetermined covariates selected by the \cite{belloni2014inference} LASSO procedure, and their (demeaned) interaction with each treatment condition. Estimates from the recommendation-only and incentivized follower encouragement conditions are suppressed and reported in Table \ref{table:H4_full}. Robust standard errors are in parentheses. Pre-specified one- and two-sided tests: * $p<0.1$, ** $p<0.05$, *** $p<0.01$. Two-sided tests of estimates in the opposite of a pre-specified one-sided test: $^{+}$ $p<0.1$, $^{++}$ $p<0.05$, $^{+++}$ $p<0.01$.
\end{sidewaystable}

The results in Table \ref{table:H4} show that following progressive-minded SMCs largely affected policy preferences and worldviews, rather than followers' faith in the system or in their capacity to change it. As with views on issue domains covered by BII fellows, columns (3)-(4) in panel B show that quiz-based incentives to follow PA or PP SMCs led participants to express more liberal positions relative to the placebo NP SMCs group at election time. Columns (5)-(6) show a similar significant effect on midline agreement with progressive economic and political narratives, shifting their systemic understandings a little more than 0.1 standard deviations toward progressive worldviews. We observe smaller differences in the same direction at endline, with only increased support for progressive narratives due to PP SMCs remaining statistically significant at the 10\% level. The simpler SMS/email messages also achieved similar effects on progressive narratives. 

In contrast, columns (7)-(10) and (17)-(20) show no significant effects on participants' sense of efficacy or trust in established political and social systems. These negligible effects emerge in a challenging moment for generating efficacy or systemic trust. The close and polarizing 2024 presidential election campaign may have already increased levels of perceived efficacy and narrowed perspectives on how change could occur. Perhaps more importantly, the intervention coincided with a period where liberal voices---including our progressively-minded SMCs---were themselves expressing dissatisfaction with the system and their own party following Biden's debate and protracted decision to drop out, Harris becoming candidate without competition, and Trump's ultimate election victory. It may thus be premature to dismiss SMCs influence on followers' engagement with the political system in other contexts.

\subsubsection{Summary}

\begin{figure}

\begin{subfigure}{0.5\textwidth}
\begin{center}
\caption{Midline political engagement (ICW index)}
\vspace{6pt}
\includegraphics[scale=0.4]{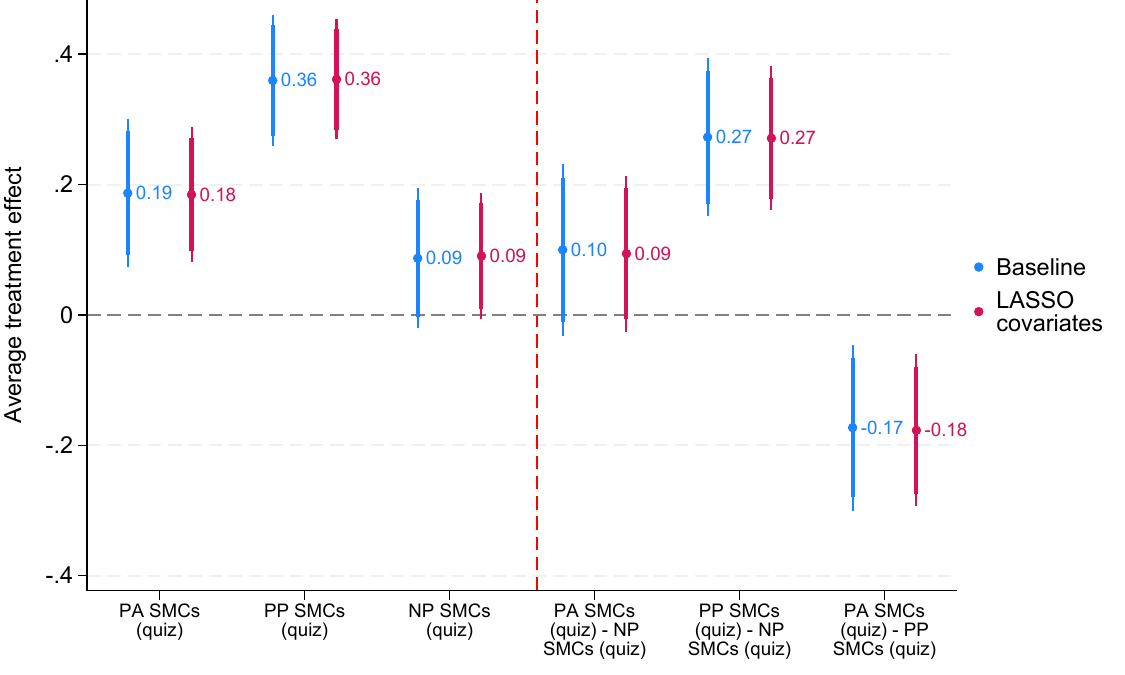}
\label{figure:midline_political_engagement}
\end{center}
\end{subfigure}%
\begin{subfigure}{0.5\textwidth}
\begin{center}
\caption{Endline political engagement (ICW index)}
\vspace{6pt}
\includegraphics[scale=0.4]{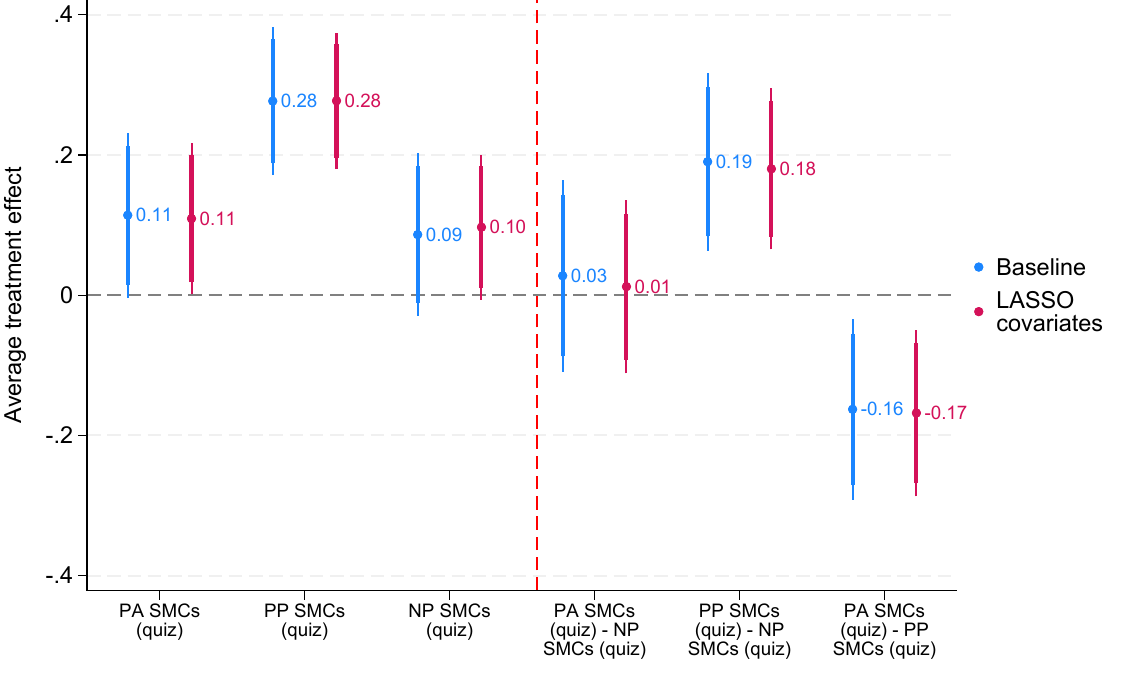}
\label{figure:endline_political_engagement}
\end{center}
\end{subfigure}%

\vspace{-4pt}

\begin{subfigure}{0.5\textwidth}
\begin{center}
\caption{Midline liberal policy preferences (ICW index)}
\vspace{6pt}
\includegraphics[scale=0.4]{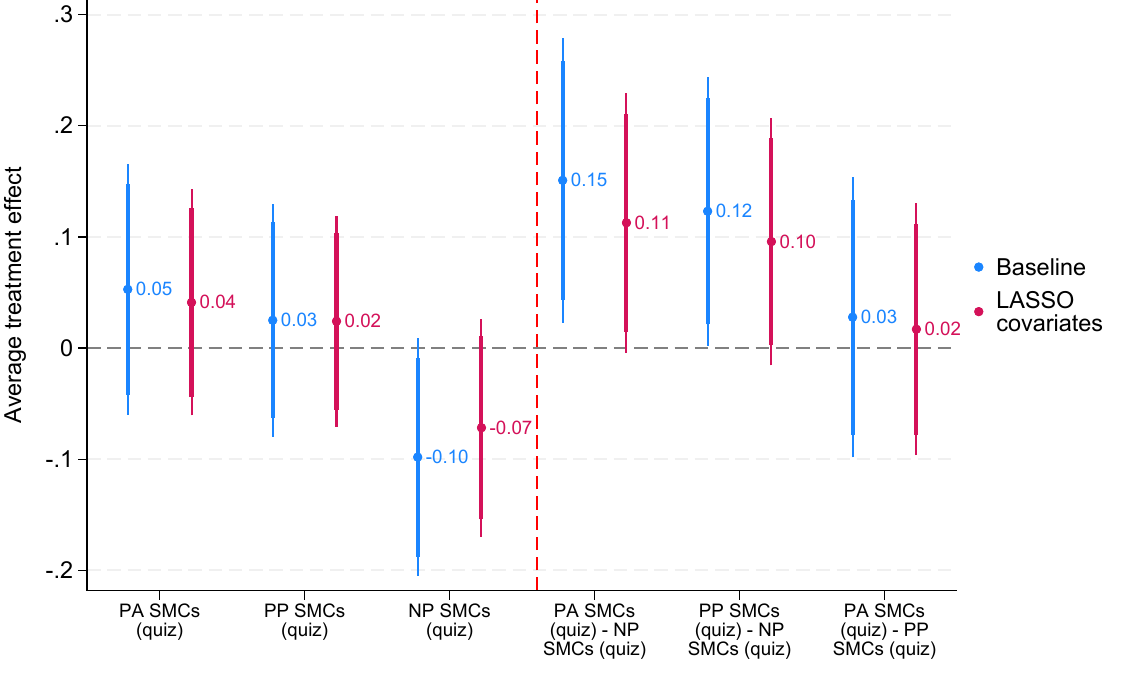}
\label{figure:midline_policy_preferences}
\end{center}
\end{subfigure}%
\begin{subfigure}{0.5\textwidth}
\begin{center}
\caption{Endline liberal policy preferences (ICW index)}
\vspace{6pt}
\includegraphics[scale=0.4]{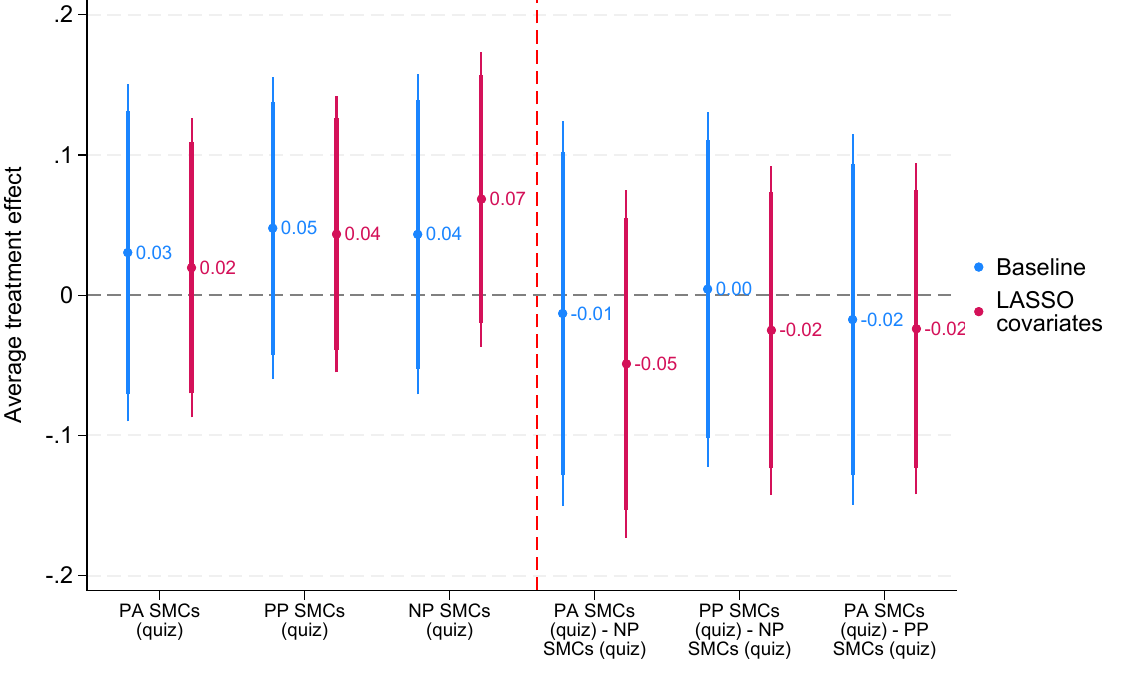}
\label{figure:endline_policy_preferences}
\end{center}
\end{subfigure}%

\vspace{-4pt}

\begin{subfigure}{0.5\textwidth}
\begin{center}
\caption{Midline progressive worldview (5-point scale)}
\vspace{6pt}
\includegraphics[scale=0.4]{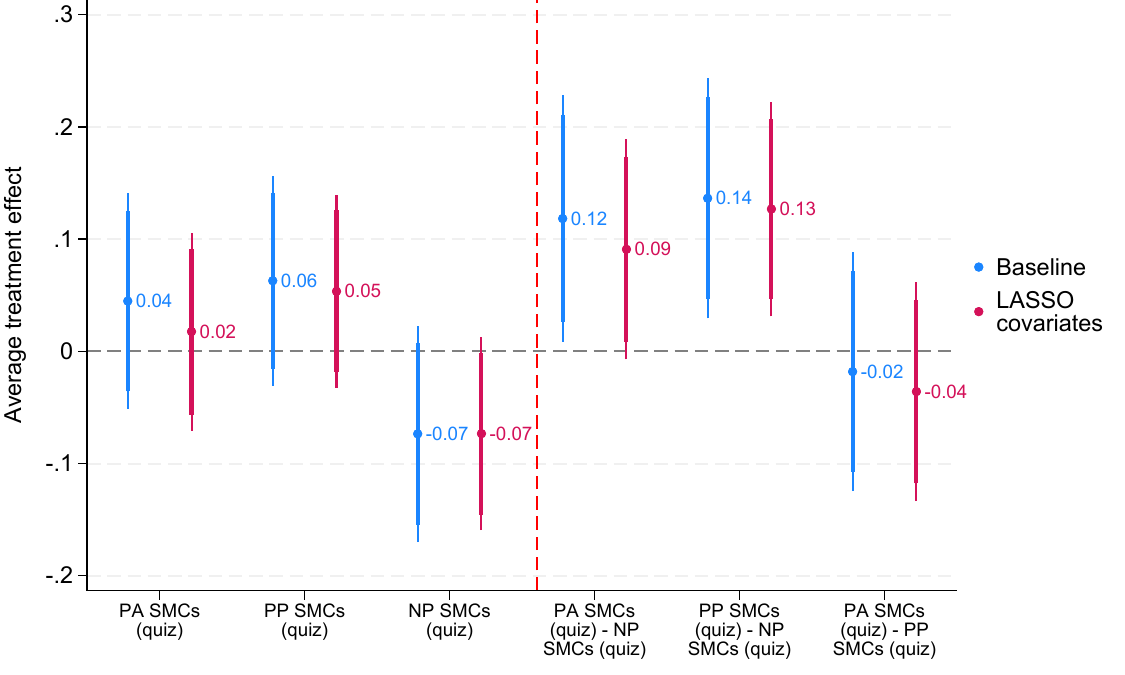}
\label{figure:midline_progressive_worldview}
\end{center}
\end{subfigure}%
\begin{subfigure}{0.5\textwidth}
\begin{center}
\caption{Endline progressive worldview (5-point scale)}
\vspace{6pt}
\includegraphics[scale=0.4]{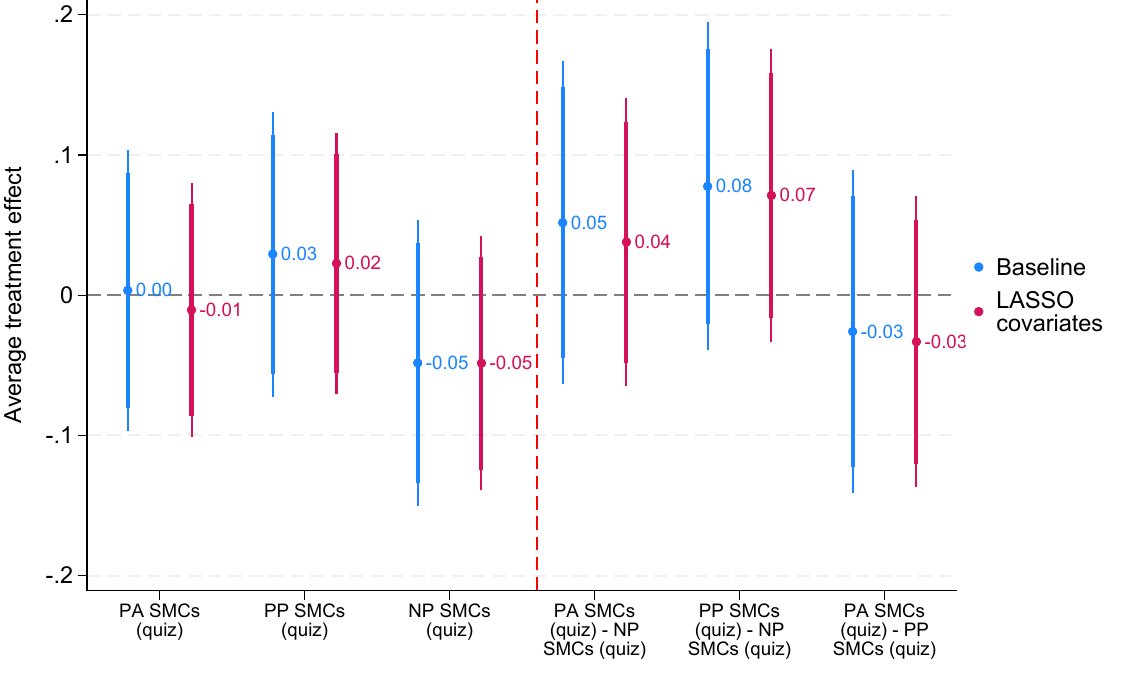}
\label{figure:endline_progressive_worldview}
\end{center}
\end{subfigure}%

\vspace{-4pt}

\begin{subfigure}{0.5\textwidth}
\begin{center}
\vspace{6pt}
\end{center}
\end{subfigure}%
\begin{subfigure}{0.5\textwidth}
\begin{center}
\caption{Endline donations to liberal causes (ICW index)}
\vspace{6pt}
\includegraphics[scale=0.4]{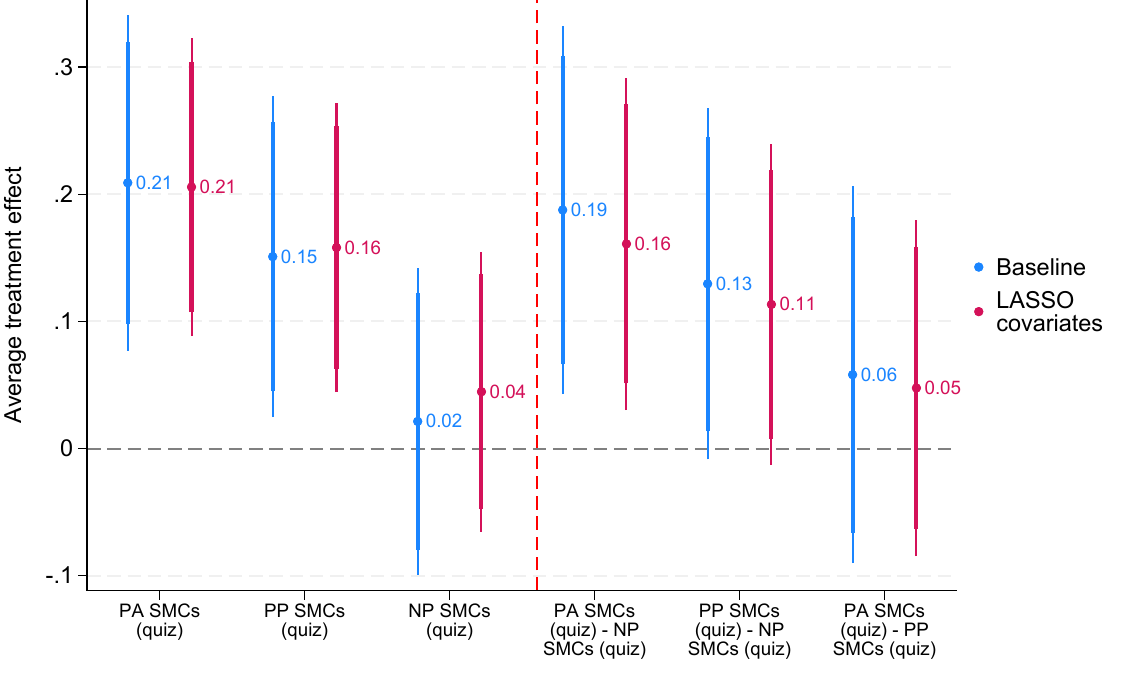}
\label{figure:endline_cause_donations}
\end{center}
\end{subfigure}%

\vspace{-18pt}

\begin{center}
\caption{Effects of quiz-based incentives to consume SMCs on political engagement and policy attitudes}
\label{figure:policy_outcomes}
\end{center}

\begin{tablenotes}
\vspace{-16pt}
\item {\footnotesize \textit{Notes}: Each graph plots the estimates of equation (\ref{equation:full}) reported in Tables \ref{table:H1}-\ref{table:H4}, with 90\% (thick lines) and 95\% (thin lines) confidence intervals. The three sets of estimates to the left of the vertical dotted line compute differences between treatment groups and the pure control group; the three sets of estimates to the right compute differences between quiz-incentivized treatment groups.}
\end{tablenotes}

\end{figure}

Figure \ref{figure:policy_outcomes} summarizes these effects on engagement and policy preferences. The first row shows that quiz-based incentives to follow all---but especially highly political---SMCs increased topical political engagement. Across attitudes and donation behaviors, the remaining rows show that incentives to start following progressive-minded SMCs translated into more liberal policy positions and worldviews, particularly around election time and relative to participants instead induced to consume NP SMCs. Moreover, PA SMCs produced similar effects to PP SMCs despite exposing participants to far less political content, suggesting that PP SMCs are more effective messengers per political video than PP SMCs. Sustained SMS/email messages also proved an effective dissemination strategy. In sum, these results show that SMCs can meaningfully influence their followers, even in the saturated and polarized environment of the 2024 presidential election campaign. 

\subsection{Partisan preferences}

We next estimate the effects of quiz incentives to start following progressive-minded SMCs on partisan evaluations. We begin by examining favorability toward the Democratic and Republican parties, before analyzing self-reported vote choices. 

\subsubsection{Favorability toward the Democratic and Republican parties}

We measured favorability toward the Democratic and Republican parties in three ways. First, we asked respondents to rate each party (as distinct from its supporters) on a feeling thermometer scale from 0 (coldest) to 100 (warmest). Second, we provided respondents with five-point Likert scales to express the extent to which they agreed that Harris and Trump had effective economic policies, would manage foreign policy well, exhibited trustworthiness and integrity, had a clear vision for America, and understood people like you. We use the mean ratings across these five attributes. Third, we created an index capturing approval of performance in office. For Democrats, this index includes an indicator for approval of President Biden and a five-point scale eliciting trust in the President at midline and endline; in the pre-election midline survey, the index also included whether the nation was heading in the right direction. For Republicans, we measured approval of Trump's transition process at endline (after the election, when the transition was almost complete). Finally, we again aggregate these measures as ICW indexes of Democrat and Republican favorability. 

\begin{sidewaystable}
\caption{Treatment effects on favorability toward the Democratic Party}
\vspace{-12pt}
\label{table:H5}
\begin{center}
\scalebox{0.73}{
\begin{tabular}{lcccccccc|cccccccc}
\toprule
& \multicolumn{8}{c|}{\textbf{\textit{Midline survey outcomes}}} & \multicolumn{8}{c}{\textbf{\textit{Endline survey outcomes}}} \\
& \multicolumn{2}{c}{\textbf{Democratic}} & \multicolumn{2}{c}{\textbf{Democratic}} & \multicolumn{2}{c}{\textbf{Harris}} & \multicolumn{2}{c|}{\textbf{Government}} & \multicolumn{2}{c}{\textbf{Democratic}} & \multicolumn{2}{c}{\textbf{Democratic}} & \multicolumn{2}{c}{\textbf{Harris}} & \multicolumn{2}{c}{\textbf{Government}} \\
& \multicolumn{2}{c}{\textbf{favorability}} & \multicolumn{2}{c}{\textbf{party}} & \multicolumn{2}{c}{\textbf{attribute}} & \multicolumn{2}{c|}{\textbf{performance}} & \multicolumn{2}{c}{\textbf{favorability}} & \multicolumn{2}{c}{\textbf{party}} & \multicolumn{2}{c}{\textbf{attribute}} & \multicolumn{2}{c}{\textbf{performance}} \\
& \multicolumn{2}{c}{\textbf{ICW index}} & \multicolumn{2}{c}{\textbf{thermometer}} & \multicolumn{2}{c}{\textbf{rating}} & \multicolumn{2}{c|}{\textbf{ICW index}} & \multicolumn{2}{c}{\textbf{ICW index}} & \multicolumn{2}{c}{\textbf{thermometer}}& \multicolumn{2}{c}{\textbf{rating}} & \multicolumn{2}{c}{\textbf{ICW index}}  \\
& (1) & (2) & (3) & (4) & (5) & (6) & (7) & (8) & (9) & (10) & (11) & (12) & (13) & (14) & (15) & (16) \\
\midrule
\textit{Panel A: differences relative to control condition} \\ Predominantly-apolitical SMCs (quiz)&       0.009   &      -0.012   &       0.133   &      -1.188   &      -0.022   &      -0.037   &       0.053   &       0.041   &      -0.002   &      -0.021   &       1.639   &       0.398   &      -0.044   &      -0.064   &       0.003   &      -0.006   \\
                    &     (0.041)   &     (0.036)   &     (1.925)   &     (1.638)   &     (0.058)   &     (0.052)   &     (0.052)   &     (0.047)   &     (0.045)   &     (0.041)   &     (1.829)   &     (1.580)   &     (0.065)   &     (0.057)   &     (0.059)   &     (0.055)   \\
\end{tabular}
}
\end{center}
\vspace{-10pt}
\footnotesize \textit{Notes}: Each specification is estimated using OLS, and includes randomization block fixed effects and adjusts for lagged outcomes and their (demeaned) interaction with each treatment condition. Even-numbered columns additionally include predetermined covariates selected by the \cite{belloni2014inference} LASSO procedure, and their (demeaned) interaction with each treatment condition. Estimates from the recommendation-only and incentivized follower encouragement conditions are suppressed and reported in Table \ref{table:H5_full}. Robust standard errors are in parentheses. Pre-specified one- and two-sided tests: * $p<0.1$, ** $p<0.05$, *** $p<0.01$. Two-sided tests of estimates in the opposite of a pre-specified one-sided test: $^{+}$ $p<0.1$, $^{++}$ $p<0.05$, $^{+++}$ $p<0.01$.
\end{sidewaystable}

In line with changes in participants' policy preferences, quiz incentives to follow progressive-minded SMCs led individuals to become more favorable toward the Democratic Party and more unfavorable toward the Republican Party. These treatment effects are most pronounced for PP SMCs: the ICW indexes in columns (1)-(2) and (9)-(10) of Table \ref{table:H5} show that the PP SMCs group became around 0.1 standard deviations more favorable toward the Democrats, particularly relative to the placebo SMCs group that became significantly less favorable to the Democrats at election time (panel B) but also compared with the control group (panel A). The analogous indexes in Table \ref{table:H6} show slightly larger decreases in favorability toward the Republican Party. These effects, especially reduced favorability toward the Republican party, are largely sustained across the midline and endline surveys and are primarily driven by changes in the feeling thermometer. We find smaller effects in the same direction for PA SMCs, who---consistent with their balanced partisan content (see Appendix Table \ref{table:FS2_nonassigned})---were more likely to influence policy positions. Normalizing by the number of political videos nevertheless suggests that PA SMCs were three times more persuasive per political video at midline. The entirely non-partisan SMS/email messages produced little effect. These findings suggest that, while both PA and PP SMCs can shape policy views, explicitly political content helps to connect SMCs' content to partisan preferences. 

\begin{sidewaystable}
\caption{Treatment effects on favorability toward the Republican Party}
\vspace{-12pt}
\label{table:H6}
\begin{center}
\scalebox{0.76}{
\begin{tabular}{lcccccc|cccccccc}
\toprule
& \multicolumn{6}{c|}{\textbf{\textit{Midline survey outcomes}}} & \multicolumn{8}{c}{\textbf{\textit{Endline survey outcomes}}} \\
& \multicolumn{2}{c}{\textbf{Republican}} & \multicolumn{2}{c}{\textbf{Republican}} & \multicolumn{2}{c|}{\textbf{Trump}} & \multicolumn{2}{c}{\textbf{Republican}} & \multicolumn{2}{c}{\textbf{Republican}} & \multicolumn{2}{c}{\textbf{Trump}} & \multicolumn{2}{c}{\textbf{Trump}} \\
& \multicolumn{2}{c}{\textbf{favorability}} & \multicolumn{2}{c}{\textbf{party}} & \multicolumn{2}{c|}{\textbf{attributes}} & \multicolumn{2}{c}{\textbf{favorability}} & \multicolumn{2}{c}{\textbf{party}} & \multicolumn{2}{c}{\textbf{attributes}} & \multicolumn{2}{c}{\textbf{transition}} \\
& \multicolumn{2}{c}{\textbf{ICW index}} & \multicolumn{2}{c}{\textbf{thermometer}} & \multicolumn{2}{c|}{\textbf{rating}} & \multicolumn{2}{c}{\textbf{ICW index}} & \multicolumn{2}{c}{\textbf{thermometer}} & \multicolumn{2}{c}{\textbf{rating}} & \multicolumn{2}{c}{\textbf{approval}} \\
& (1) & (2) & (3) & (4) & (5) & (6) & (7) & (8) & (9) & (10) & (11) & (12) & (13) & (14) \\
\midrule
\textit{Panel A: differences relative to control condition} \\ Predominantly-apolitical SMCs (quiz)&      -0.035   &      -0.038   &      -1.255   &      -1.512   &      -0.045   &      -0.033   &       0.017   &       0.019   &      -1.330   &      -1.518   &       0.037   &       0.044   &       0.034   &       0.033   \\
                    &     (0.035)   &     (0.030)   &     (1.627)   &     (1.417)   &     (0.050)   &     (0.046)   &     (0.038)   &     (0.034)   &     (1.596)   &     (1.419)   &     (0.055)   &     (0.050)   &     (0.024)   &     (0.022)   \\
\end{tabular}
}
\end{center}
\vspace{-10pt}
\footnotesize \textit{Notes}: Each specification is estimated using OLS, and includes randomization block fixed effects and adjusts for lagged outcomes and their (demeaned) interaction with each treatment condition. Even-numbered columns additionally include predetermined covariates selected by the \cite{belloni2014inference} LASSO procedure, and their (demeaned) interaction with each treatment condition. Robust standard errors are in parentheses. Estimates from the recommendation-only and incentivized follower encouragement conditions are suppressed and reported in Table \ref{table:H6_full}. Pre-specified one- and two-sided tests: * $p<0.1$, ** $p<0.05$, *** $p<0.01$. Two-sided tests of estimates in the opposite of a pre-specified one-sided test: $^{+}$ $p<0.1$, $^{++}$ $p<0.05$, $^{+++}$ $p<0.01$.
\end{sidewaystable}

\subsubsection{Voting behavior}

Having observed liberal shifts in policy preferences and partisan evaluations, we finally assess voting behavior. In the midline survey, conducted the week before the 2024 elections, we asked respondents about their voter registration status, their intention to vote in the presidential election, and who they intended to vote for. The endline survey asked participants whether they had voted and for whom they had voted, both in the presidential election and in the concurrent House and Senate races.\footnote{We code voting for Trump, voting for others, and not voting as zero. For Senate elections, we condition the sample on respondents residing in states holding elections; for our index, we impute this outcome by averaging presidential and House vote choice.} Registration and turnout are validated using Target Smart's voter file, which was successfully matched to 86\% of participants who permitted Bovitz to link the voter file to their responses for \$5 in credits.
%1,313 of the 1,522 were successfully matched
 
\begin{table}[!h]
\caption{Treatment effects on self-reported voting decisions}
\vspace{-12pt}
\label{table:H8}
\begin{center}
\scalebox{0.7}{
\begin{tabular}{lcc|cccccccc}
\toprule
& \multicolumn{2}{c|}{\textbf{\textit{Midline survey outcome}}} & \multicolumn{8}{c}{\textbf{\textit{Endline survey outcomes}}} \\
& \multicolumn{2}{c|}{\textbf{Intend to vote}} & \multicolumn{2}{c}{\textbf{Voted}} & \multicolumn{2}{c}{\textbf{Voted}} & \multicolumn{2}{c}{\textbf{Voted}} & \multicolumn{2}{c}{\textbf{Voted}} \\
& \multicolumn{2}{c|}{\textbf{Democrat}} & \multicolumn{2}{c}{\textbf{Democrat}} & \multicolumn{2}{c}{\textbf{Democrat}} & \multicolumn{2}{c}{\textbf{Democrat}} & \multicolumn{2}{c}{\textbf{Democrat}} \\
& \multicolumn{2}{c|}{\textbf{for President}} & \multicolumn{2}{c}{\textbf{ICW index}} & \multicolumn{2}{c}{\textbf{for President}} & \multicolumn{2}{c}{\textbf{for House}} & \multicolumn{2}{c}{\textbf{for Senate}} \\
& (1) & (2) & (3) & (4) & (5) & (6) & (7) & (8) & (9) & (10) \\
\midrule
\textit{Panel A: differences relative to control condition} \\ Predominantly-apolitical SMCs (quiz)&       0.015   &       0.000   &      -0.061   &-0.103$^{++}$   &      -0.016   &      -0.030   &      -0.047   &-0.045$^{+}$   &      -0.015   &       0.002   \\
                    &     (0.023)   &     (0.020)   &     (0.056)   &     (0.048)   &     (0.026)   &     (0.023)   &     (0.033)   &     (0.027)   &     (0.038)   &     (0.030)   \\
\end{tabular}
}
\end{center}
\vspace{-10pt}
\footnotesize \textit{Notes}: Each specification is estimated using OLS, and includes randomization block fixed effects and adjusts for lagged outcomes and their (demeaned) interaction with each treatment condition. Even-numbered columns additionally include predetermined covariates selected by the \cite{belloni2014inference} LASSO procedure, and their (demeaned) interaction with each treatment condition. Estimates from the recommendation-only and incentivized follower encouragement conditions are suppressed and reported in Table \ref{table:H8_full}. Robust standard errors are in parentheses. Pre-specified one- and two-sided tests: * $p<0.1$, ** $p<0.05$, *** $p<0.01$. Two-sided tests of estimates in the opposite of a pre-specified one-sided test: $^{+}$ $p<0.1$, $^{++}$ $p<0.05$, $^{+++}$ $p<0.01$.
\end{table}

We find that changes in partisan preferences are largely reflected in presidential vote choice. Columns (1)-(2) of Table \ref{table:H8} show that quiz incentives to follow PA or PP SMCs increased intentions to vote for Kamala Harris before the election by about 1.5 percentage points relative to the NP SMCs placebo group or pure control group, though these effects are not statistically significant. Turning to self-reported vote choices, participants assigned to follow PP SMCs became significantly more likely to vote for Harris in the presidential election; this effect is most pronounced relative to the NP SMCs group, where it registers at six percentage points. While this pattern is consistent with progressive-minded SMCs' content being most relevant for national debates and the presidential election, we are cautious in interpreting this estimate because we do not observe comparable effects on House or Senate vote choices.

The intervention did not affect electoral or non-electoral participation. As shown in Appendix Table \ref{table:H7_full}, we find no significant effects across treatment conditions on voter registration or voter turnout at either midline or endline, using both self-reported and validated measures. Appendix Table \ref{table:H9_full} further reports no significant effects on other self-reported forms of non-electoral participation, including attending protests. The limited effect of SMCs on citizen mobilization may partly reflect the unusually intense campaign environment, which produced the second-highest rate of turnout (as a proportion of the vote-eligible population) since 1980.

\subsubsection{Summary}

\begin{figure}

\begin{subfigure}{0.5\textwidth}
\begin{center}
\caption{Midline evaluation of the Democratic Party (ICW index)}
\vspace{6pt}
\includegraphics[scale=0.43]{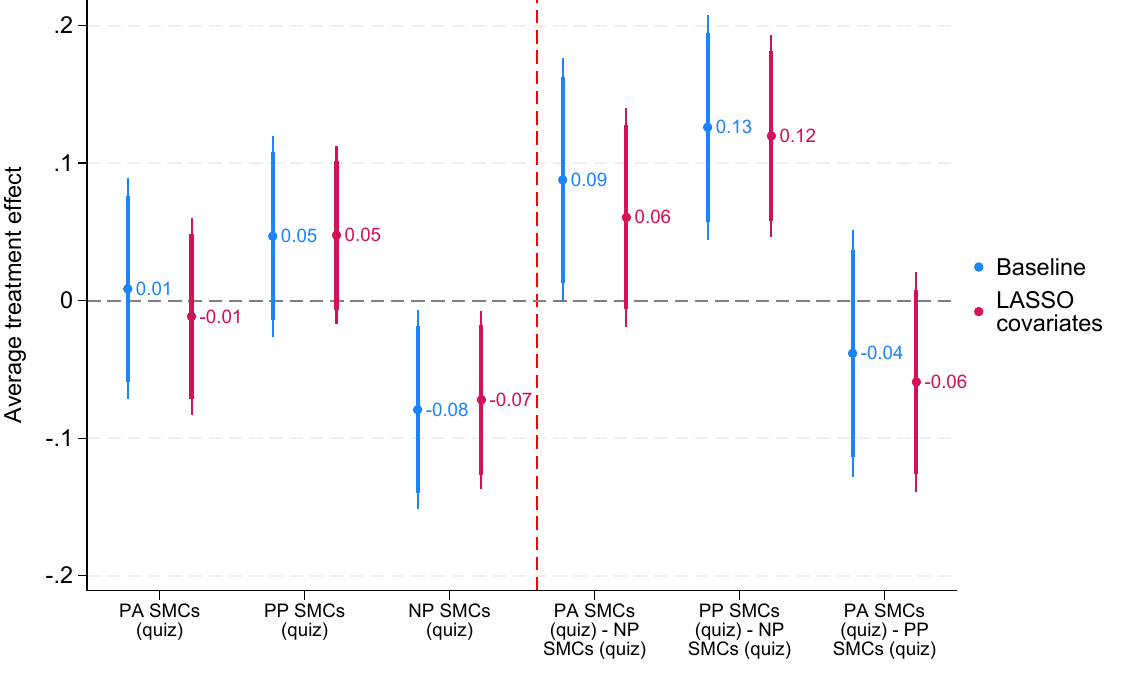}
\label{figure:midline_democratic_party}
\end{center}
\end{subfigure}%
\begin{subfigure}{0.5\textwidth}
\begin{center}
\caption{Endline evaluation of the Democratic Party (ICW index)}
\vspace{6pt}
\includegraphics[scale=0.43]{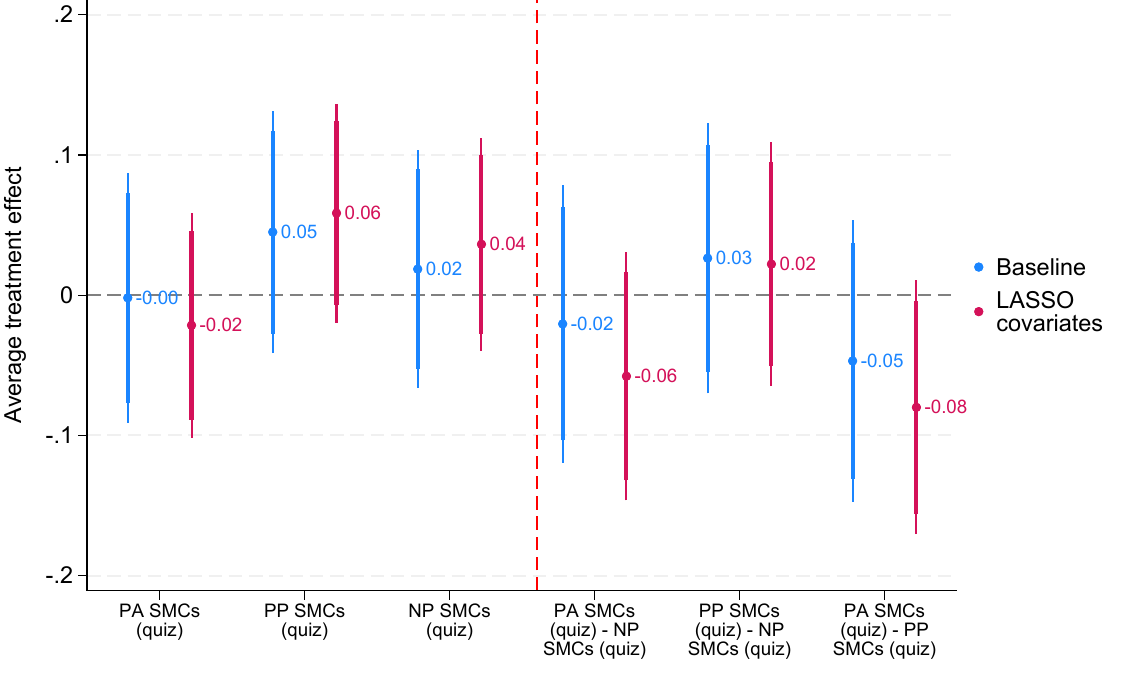}
\label{figure:endline_democratic_party}
\end{center}
\end{subfigure}%

\begin{subfigure}{0.5\textwidth}
\begin{center}
\caption{Midline evaluation of the Republican Party (ICW index)}
\vspace{6pt}
\includegraphics[scale=0.43]{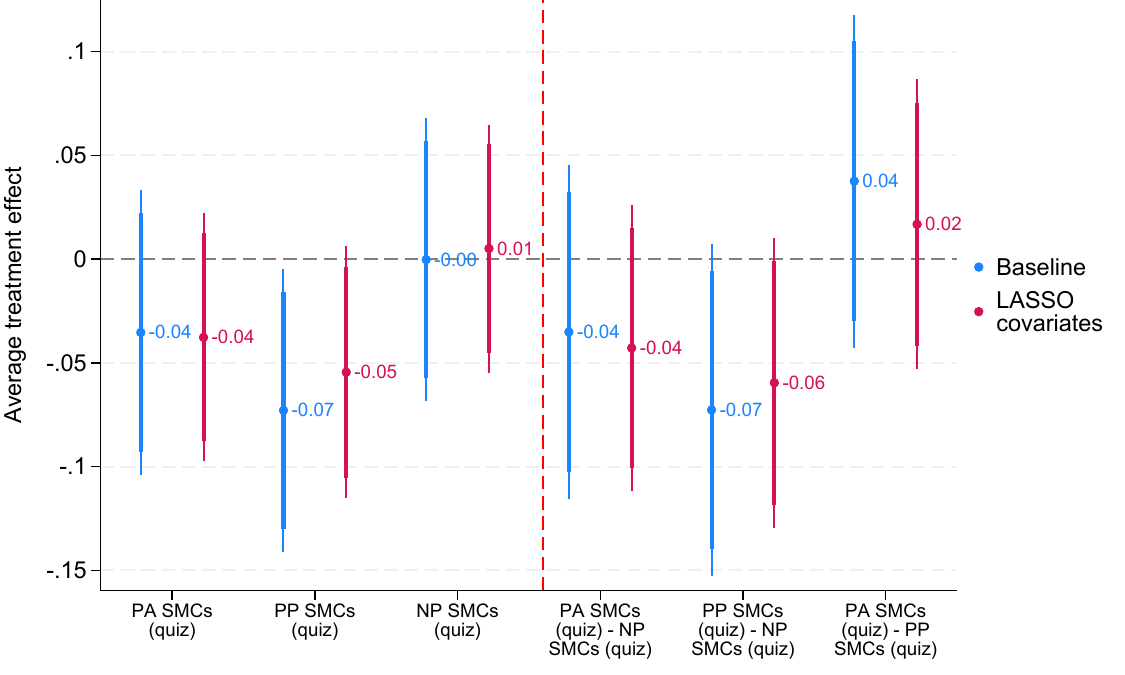}
\label{figure:midline_republican_party}
\end{center}
\end{subfigure}%
\begin{subfigure}{0.5\textwidth}
\begin{center}
\caption{Endline evaluation of the Republican Party (ICW index)}
\vspace{6pt}
\includegraphics[scale=0.43]{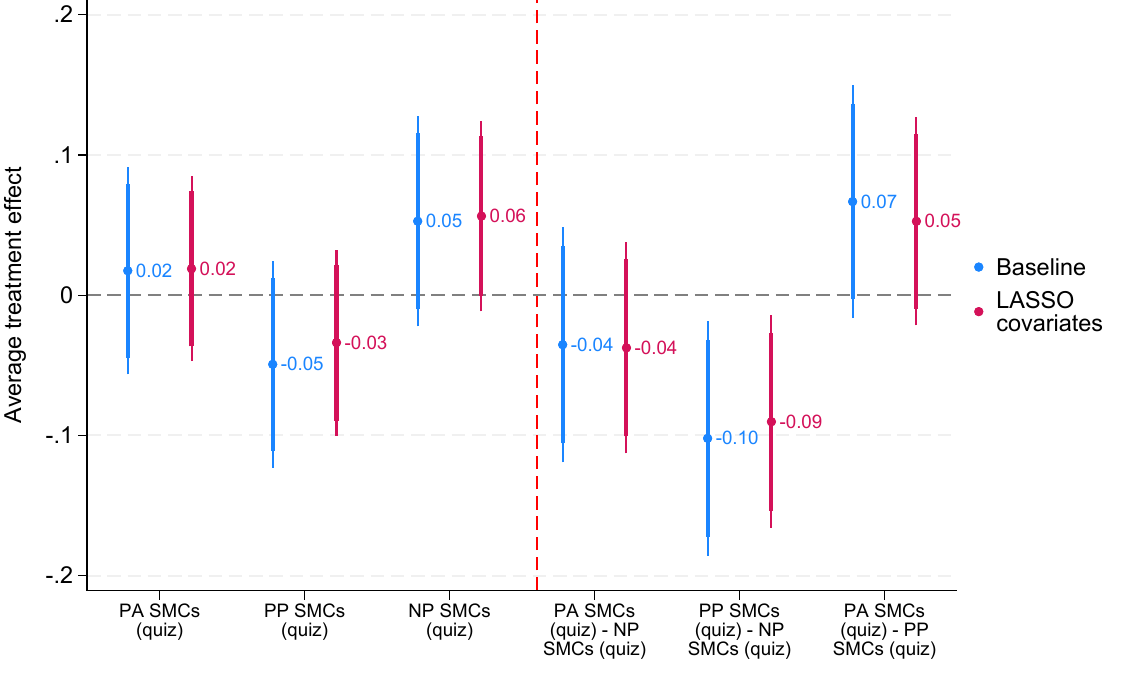}
\label{figure:endline_republican_party}
\end{center}
\end{subfigure}%

\begin{subfigure}{0.5\textwidth}
\begin{center}
\caption{Midline intention to vote for Kamala Harris}
\vspace{6pt}
\includegraphics[scale=0.43]{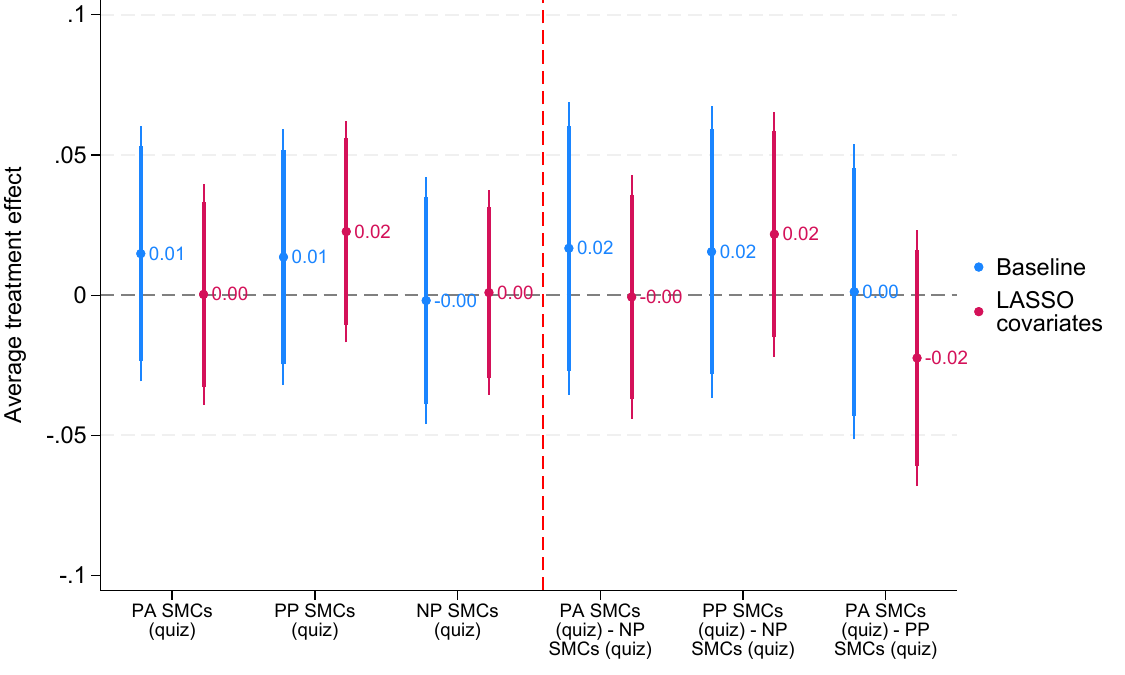}
\label{figure:midline_harris}
\end{center}
\end{subfigure}%
\begin{subfigure}{0.5\textwidth}
\begin{center}
\caption{Endline report of voting for Kamala Harris}
\vspace{6pt}
\includegraphics[scale=0.43]{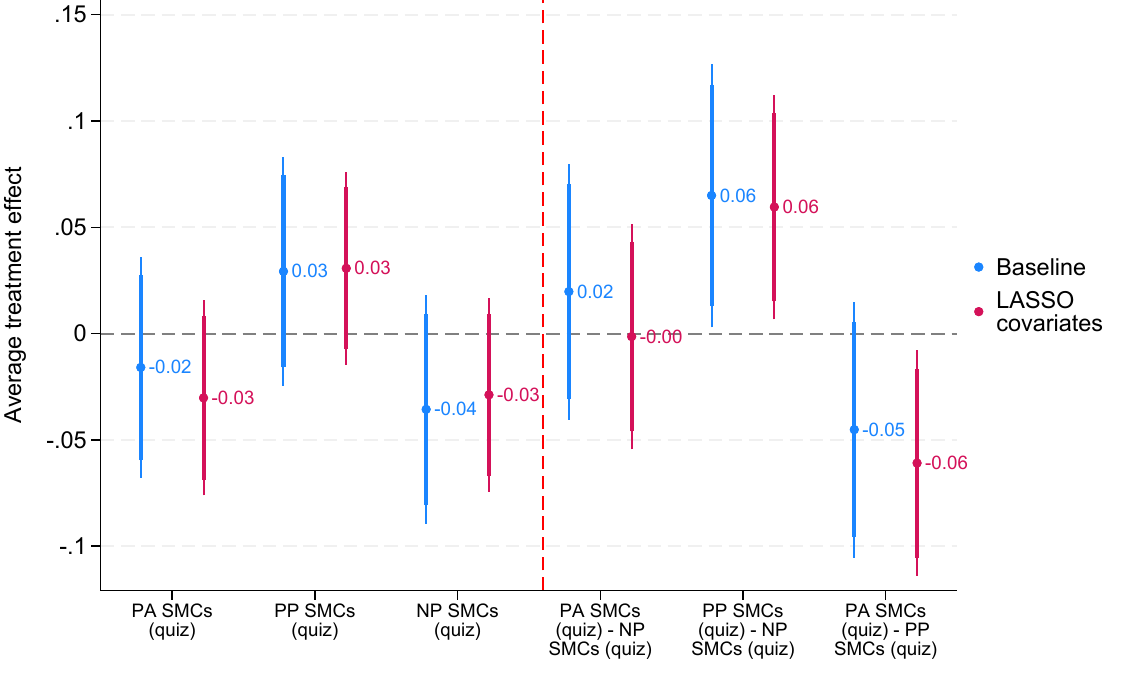}
\label{figure:endline_harris}
\end{center}
\end{subfigure}%

\vspace{-18pt}

\begin{center}
\caption{Effects of quiz-based incentives to consume SMCs on partisan preferences}
\label{figure:partisan_outcomes}
\end{center}

\begin{tablenotes}
\vspace{-16pt}
\item {\footnotesize \textit{Notes}: Each graph plots the estimates of equation (\ref{equation:full}) reported in Tables \ref{table:H5}-\ref{table:H8}, with 90\% (thick lines) and 95\% (thin lines) confidence intervals. The three sets of estimates to the left of the vertical dotted line compute differences between treatment groups and the pure control group; the three sets of estimates to the right compute differences between quiz-incentivized treatment groups.}
\end{tablenotes}

\end{figure}

Figure \ref{figure:partisan_outcomes} summarizes the partisan effects of starting to follow progressive-minded SMCs. Participants exposed to PP SMCs became more favorable toward Democrats, less favorable toward Republicans, and more likely to support Harris in the presidential election. PA SMCs shifted policy preferences in a cost-efficient way but had weaker effects on partisan preferences. These findings suggest that SMCs are powerful tools of political influence, but shaping partisan preferences may require more explicitly partisan content or more sustained, vote-oriented persuasion.

\subsection{Overall effects}

The preceding analyses found that biweekly quiz-based incentives to follow progressive-minded SMCs during the 2024 election season led participants to engage with politics more, to adopt more progressive policy stances, and to become more favorable toward the Democratic Party relative to the Republican Party. While both policy and partisan preferences shifted to the left, we do not observe systematic changes in issue salience, electoral and non-electoral participation, or institutional trust in this fraught and---for many progressive-leaning SMCs---dispiriting political context (see Appendix Tables \ref{table:H3_full}-\ref{table:H10_full}).

With a large number of outcomes and treatment conditions, it is important to probe robustness to multiple comparison corrections. We first follow prior studies \citep[e.g.][]{allcott2020welfare, broockman2025consuming, chen2019impact} in implementing \citeauthor{anderson2008multiple}'s \citeyearpar{anderson2008multiple} approach to adjusting for the false discovery rate with correlated treatment effects.\footnote{Our pre-analysis plan specified we would use the \cite{benjamini1995} approach, but we instead use \citeauthor{anderson2008multiple}'s \citeyearpar{anderson2008multiple} better-powered update of this method; we obtain similar results using the \cite{benjamini1995} approach. We conservatively corrected for 94 comparisons: between the quiz-incentivized PA and PP SMCs and both the NP SMCs and control conditions (4 comparisons, except for hypotheses FS1, FS2, and H10) for each of our 27 main outcomes indexes (13 pre-specified first stage and main outcomes at midline, 13 pre-specified first stage and main outcomes at endline, and the endline cause donation index).} Appendix Table \ref{table:sharpened} reports these ``sharpened'' $q$ values for the quiz-incentivized treatment conditions, where $q$\% of results will be false positives. Adjusting for 94 pre-specified comparisons, the most robust findings are the large increase in topical political knowledge and increased donations to liberal causes across the PA and PP SMC conditions. We obtain $q$ values of up to 0.3 for policy and partisan attitude outcomes that were statistically significant at the 10\% level without adjustment, implying that the probability that these results are false positives is at most 30\%.

\begin{table}[!h]
\caption{Treatment effects on overall indexes of political position and political participation}
\vspace{-12pt}
\label{table:overall}
\begin{center}
\scalebox{0.86}{
\begin{tabular}{lcccc|cccc}
\toprule
& \multicolumn{4}{c|}{\textbf{\textit{Midline survey outcomes}}} & \multicolumn{4}{c}{\textbf{\textit{Endline survey outcomes}}} \\
& \multicolumn{2}{c}{\textbf{Liberal}} & \multicolumn{2}{c|}{} & \multicolumn{2}{c}{\textbf{Liberal}} \\
& \multicolumn{2}{c}{\textbf{position}} & \multicolumn{2}{c|}{\textbf{Participation}} & \multicolumn{2}{c}{\textbf{position}} & \multicolumn{2}{c}{\textbf{Participation}} \\
& \multicolumn{2}{c}{\textbf{ICW index}} & \multicolumn{2}{c|}{\textbf{ICW index}} & \multicolumn{2}{c}{\textbf{ICW index}} & \multicolumn{2}{c}{\textbf{ICW index}} \\
& (1) & (2) & (3) & (4) & (5) & (6) & (7) & (8) \\
\midrule
\textit{Panel A: differences relative to control condition} \\ Predominantly-apolitical SMCs (quiz)&       0.040   &       0.026   &      -0.041   &      -0.044   &      0.080*   &       0.063   &      -0.021   &      -0.022   \\
                    &     (0.038)   &     (0.035)   &     (0.057)   &     (0.053)   &     (0.044)   &     (0.040)   &     (0.065)   &     (0.057)   \\
\end{tabular}
}
\end{center}
\vspace{-10pt}
\footnotesize \textit{Notes}: Each specification is estimated using OLS, and includes randomization block fixed effects and adjusts for lagged outcomes and their (demeaned) interaction with each treatment condition. Even-numbered columns additionally include predetermined covariates selected by the \cite{belloni2014inference} LASSO procedure, and their (demeaned) interaction with each treatment condition. Robust standard errors are in parentheses. Two-sided tests (not pre-specified): * $p<0.1$, ** $p<0.05$, *** $p<0.01$.
\end{table}

Our second approach, which was not pre-specified but has greater statistical power, aggregates outcome indexes into two conceptual groupings: political positions and participation. The former group includes policy preferences (Table \ref{table:H2}), endline cause donation decisions (Table \ref{table:cause_donations}), outlooks (Table \ref{table:H4}), partisan favorability (Tables \ref{table:H5} and \ref{table:H6}), and vote choice (Table \ref{table:H8}); the latter group includes voter registration and turnout (Appendix Table \ref{table:H7_full}) and non-electoral political participation (Appendix Table \ref{table:H9_full}).\footnote{In terms of our pre-analysis plan, the former group combines hypotheses H2, H4, H5, H6, and H8, while the latter combines hypotheses H7 and H9. Increased topical political knowledge (H1) is clear from Table \ref{table:H1}, and we exclude issue salience (H3, Appendix Table \ref{table:H3_full}) and trust outcomes (H10, Appendix Table \ref{table:H10_full}) from this analysis because neither conceptually coheres with policy and partisan positions or participation. Supporting this separation, factor analysis yields two factors with eigenvalues above 1 loading roughly along this division.} Each grouping is combined as an ICW index.

Table \ref{table:overall} reports the overall effects of the intervention on both indexes. The results confirm that quiz incentives to follow PA and PP SMCs significantly increased liberal positions by around a 0.10 standard deviations (using two-tailed tests), both relative to the pure control group and---to a somewhat greater extent, especially at midline---the NP SMCs placebo group. In contrast, we find fairly precise null effects on political participation, with our negative point estimates ruling out increases in participation above 0.1 standard deviations. To show that these results are not driven by the regression weights generated by different attrition rates across randomization blocks, Appendix Table \ref{table:overall_no_block_FEs} reports similar results when excluding randomization block fixed effects. We also obtain similar estimates when reweighting our sample to match social media users aged 18-45 in the nationally-representative 2020 American National Election Survey (see Appendix Section \ref{appendix:ANES_comparison}). 

%IV per video terms (as in recent Facebook ads paper)? or calculate the persuasion rate for each additional ad (see Spenkuch and Toniatti approach)?

While seemingly modest in magnitude, these changes in policy attitudes and partisan favorability are large by the standard of US political communication. Comparing with cable news, \cite{broockman2025consuming} show that watching CNN instead of Fox News for almost six hours per week for four weeks early in the 2020 US election campaign produced standardized effects on attitudes toward issues covered in the news and partisan evaluations that were slightly smaller than ours.\footnote{Figure OA5 of \cite{broockman2025consuming} reports an insignificant liberal shift of around 0.05 standard deviations on attitudes toward issues covered in the news, a significant decrease in evaluation of Trump of around 0.07 standard deviations, an insignificant increase in evaluation of Biden by around 0.05 standard deviations, and a significant increase in Democratic-leaning preferences of 0.07 standard deviations.} Comparing with partisan TV ads, \cite{spenkuch2018political} estimate that seeing 21 more Democrat than Republican ads in presidential campaigns from 2004-2012 increased the Democratic vote share by 0.3 percentage points (or 0.02 standard deviations), largely by mobilizing supporters and demobilizing opponents.\footnote{The 0.6 percentage point differential effect comes from Section V.B of \cite{spenkuch2018political}; the standard deviation in the Democratic vote share is from Table I. \cite{sides2022effect} report similar estimates over more presidential elections, but detect effects several times large on down-ballot elections. Other studies similarly detect short-lived effects of partisan ads \citep{gerber2011, kalla2018minimal}.} Online political advertising appears even weaker: exposure to about 140 mostly co-partisan ads on Facebook or Instagram produced negligible changes in political attitudes during the 2020 election campaign \citep{allcott2025effects}, consistent with other field experiments on sustained digital partisan ad campaigns \citep{aggarwal2023digital, coppock2022does}. More generally, \citeauthor{kalla2018minimal}'s \citeyearpar{kalla2018minimal} meta-analysis concludes that partisan campaign contact and advertising field experiments have, on average, found effects on vote choice that are indistinguishable from zero. While these studies are not perfectly comparable, our results show that exposure to roughly 60 videos from five progressive-minded SMCs---of which 20\% and 77\% of videos were political for the PA and PP groups, respectively---over five months shifted partisan preferences by magnitudes that generally exceed the most well-studied campaign strategies in the United States.\footnote{Political campaigns and slanted media content generally produces larger effects outside consolidated democracies \citep[e.g.][]{da2011campaign, enikolopov2011media, enriquez2024mass, larreguy2018leveling}.}

\section{Mechanisms}

To understand how progressive-minded SMCs influenced political positions, we examine the features of both SMCs and participants that drive persuasion. The following exploratory analyses suggest two pathways to SMC impact: a quality channel, in which parasocial relationships and trust make PA SMCs more persuasive per video than PP SMCs; and a quantity channel, in which the volume or depth of political information eventually generates persuasion. We also show that consumption of assigned SMCs persisted after the intervention concluded. On the receiver side, participants with more conservative or less precise prior beliefs were somewhat more influenced by progressive-minded SMCs, while experimenter demand is unlikely to drive the results.

\subsection{Quality and quantity sources of SMC influence}

On the messenger side, a key feature of SMCs is their ability to establish parasocial connections with followers that build credibility. To explore this channel of potential influence, our midline and endline surveys asked participants to evaluate their assigned SMCs overall on a 1-10 scale, rate their connection to them on a five-point scale, and assess whether they found their general content interesting, informative, and trustworthy on five-point agreement scales. We restrict our analyses to quiz-incentivized participants, who consumed assigned SMCs at similar rates across the PA, PP, and NP SMCs conditions; respondents in the SMS/text and control groups were not asked these questions.

Table \ref{table:smc_appraisals} first shows that assigned SMCs were viewed favorably at midline and endline, with some participants reporting strong parasocial connections. The outcome means in columns (1) and (6) indicate that the average assigned SMC received an overall rating of almost 7 out of 10, while columns (2)-(4) and (7)-(9) further show that the average respondent thought their assigned SMCs were interesting, informative, and trustworthy. Consistent with developing a parasocial connection with some assigned SMCs over five months, columns (5) and (10) show that the average respondent rated their mean connection to assigned SMCs as closer to ``somewhat connected'' than ``a little connected''; respondents reported that they were ``very connected'' or ``extremely connected'' to around a quarter of assigned SMCs. 

\begin{table}
\caption{Differences in appraisal of SMCs across quiz-incentivized treatment conditions}
\vspace{-12pt}
\label{table:smc_appraisals}
\begin{center}
\scalebox{0.62}{
\begin{tabular}{lccccc|ccccc}
\toprule
& \multicolumn{5}{c|}{\textbf{\textit{Midline survey average SMC appraisal}}} & \multicolumn{5}{c}{\textbf{\textit{Endline survey average SMC appraisal}}} \\
& \textbf{Overall} & \textbf{Finds} & \textbf{Finds} & \textbf{Finds} & \textbf{Feels} & \textbf{Overall} & \textbf{Finds} & \textbf{Finds} & \textbf{Finds} & \textbf{Feels} \\
& \textbf{rating} & \textbf{interesting} & \textbf{informative} & \textbf{trustworthy} & \textbf{connected} & \textbf{rating} & \textbf{interesting} & \textbf{informative} & \textbf{trustworthy} & \textbf{connected} \\
& (1) & (2) & (3) & (4) & (5) & (6) & (7) & (8) & (9) & (10) \\
\midrule
\textit{Panel A: differences relative to placebo condition} \\
Predominantly-apolitical SMCs (quiz)&       0.271   &       0.025   &       0.196** &       0.151*  &       0.232*  &       0.194   &       0.032   &       0.247***&       0.231***&       0.152   \\
                    &     (0.211)   &     (0.085)   &     (0.080)   &     (0.086)   &     (0.126)   &     (0.208)   &     (0.091)   &     (0.085)   &     (0.082)   &     (0.127)   \\
Predominantly-political SMCs (quiz)&      -0.081   &      -0.104   &       0.154*  &      -0.090   &       0.171   &      -0.264   &      -0.120   &       0.153   &      -0.059   &       0.096   \\
\end{tabular}
}
\end{center}
\vspace{-10pt}
\footnotesize \textit{Notes}: Each specification is estimated using OLS, and includes randomization block fixed effects. Restricting the sample to quiz-incentivized conditions, the baseline condition is the non-political SMCs (quiz) group. Robust standard errors are in parentheses. Two-sided tests (not pre-specified): * $p<0.1$, ** $p<0.05$, *** $p<0.01$. 
\end{table}

While SMCs were generally viewed as credible among participants, PA SMCs outscored PP SMCs on this dimension. The differences between these treatment conditions in panels A and B of Table \ref{table:smc_appraisals} show that PA SMCs were rated highest on average, and significantly higher than PP SMCs by endline. At both midline and endline, PA SMCs were regarded as more informative and trustworthy than NP SMCs and more trustworthy than PP SMCs.\footnote{While SMCs appeal to different audiences, some SMCs may simply have been more persuasive than others. We investigate this by restricting attention to the quiz-incentivized PA and PP SMC groups and then correlating indicators for the SMCs individuals were assigned to with their midline and endline political preferences (adjusting for baseline preferences). Joint tests then fail to reject the restriction that political preferences do not vary across assigned SMCs ($ p = 0.74$ at midline and $ p = 0.17$ at endline), suggesting that SMCs within treatment conditions produce similar effects. This likely reflects SMCs producing relatively homogeneous high-quality content and being followed by individuals with matching interests to achieve political persuasion.} These results are consistent with PA SMCs' persuasing via a quality channel: cultivating parasocial bonds and perceived authenticity, without relying on overt partisan cues (see Tables \ref{table:FS2} and \ref{table:FS2_nonassigned}).

By contrast, PP SMCs appear to derive influence through a quantity channel. While PP SMCs inspired less trust than PA SMCs, the large quantity of political content they produced became central to the political information participants received. Quiz-incentivized participants assigned to PP SMCs were 9.5 percentage points (about 50\%) more likely to say that SMCs were the first place they would turn for political information at both midline and endline. Such a change is not observed among participants encouraged to follow PA SMCs.

Heterogeneity in treatment effects across participants reinforces these channels of SMCs' persuasion. Appendix Table \ref{table:HEs_appraisal} shows that quiz-incentivized participants who rated their assigned PA and PP SMCs as more interesting, informative, trustworthy, and connected experienced significantly larger treatment effects than quiz-incentivized participants assigned to NP SMCs, especially before the election. Consumption rates were not similarly heterogeneous, suggesting that these effects are unlikely to reflect consumption differences. Since SMC ratings are necessarily measured after exposure to treatment, we use LASSO to predict these moderators using predetermined covariates. Appendix Table \ref{table:HEs_appraisal_predicted} reports substantively similar, but less precise, moderator estimates. %This further suggests that PA SMCs worked primarily through the credibility channel, whereas PP SMCs influenced participants by providing more political information. 

\subsection{Continued consumption of assigned SMCs' content}

In the face of declining interest in politics and increased media choice \citep{prior2007, toff2023avoiding}, it is important to assess whether SMCs also sustain engagement after incentives are withdrawn. Continued consumption could reflect a followers' connection to the content, habit, or limited resistance to platforms' algorithmic recommendations. Prior evidence of sustained demand is mixed: in China, encouraging exposure to news and political content activated latent demand \cite{chen2019impact}; but in the US, most Fox News viewers incentivized to watch CNN quickly returned to prior media consumption habits \citep{broockman2025consuming}. To examine whether engagement with assigned SMCs persisted after our quizzes had concluded, we use YouTube browsing data to measure participants' average weekly consumption of assigned SMCs' videos after the last quiz until they shared their browsing history (in January or February). In addition, our endline survey elicited participants' intentions to continue following assigned SMCs as well as testing their knowledge of post-quiz videos.

The results in Appendix Table \ref{table:continued} show that quiz incentives to start following all types of SMCs generated persisting engagement in the month after quizzes ceased. First, participants reported a significantly higher intention to continue following assigned SMCs of about one point on a five-point scale. Second, and more tellingly, YouTube browsing histories reveal that quiz-incentivized participants watched several assigned SMC videos per week after incentives ended. Third, these participants were about 40 percentage points more likely than the control group to correctly answer (unincentivized) endline quiz questions about content posted after incentives were withdrawn. In short, participants continued consuming both PA and PP SMCs, consistent with their favorable appraisals of assigned content. Unlike traditional political information sources, which often see their effects dissipate quickly \citep{broockman2025consuming}, SMCs---and the algorithms promoting them---can sustain political engagement among young adults. 

\subsection{Heterogeneity by participant type}

The political impact of following assigned SMCs could vary with participant characteristics. On one hand, progressive-minded SMCs may be more likely to influence initially-conservative individuals with greatest scope to update from the signal or non-Democrats with imprecise prior beliefs. Conversely, if participants engage in motivated reasoning \citep[e.g.][]{kunda1990} or conclude that counter-attitudinal SMC content lacks credibility \citep[e.g.][]{gentzkow2025ideological}, SMCs may instead shift the views of initially-liberal individuals or Democrats through positive reinforcement. To assess these possibilities, Appendix Tables \ref{table:HEs_consumption}-\ref{table:HEs_participation} report heterogeneous effects on our aggregated ICW indexes of overall SMC consumption, liberal positions, and political participation by participants' baseline characteristics. 

Our estimates more closely align with Bayesian updating. Relative to the NP SMCs placebo group, quiz incentives to follow PA SMCs produced smaller effects on liberal positions among Democrats and men. Similarly, quiz-based incentives to follow PP SMCs produced smaller effects among above-median social media users, Democrats, and men. Both conditions generated larger effects among participants with below-median prior levels of social media use, consistent with the effects being driven by individuals with priors beliefs not already informed by SMC content. These differences are imprecise because this study was not powered to detect such interaction effects. Nevertheless, our estimates suggest that progressive-minded SMCs most strongly influenced more conservative participants initially receiving less political information. As such, incentives to consume counter-attitudinal SMCs' content helps to depolarize the electorate, although it could also increase polarization when consumption is unrestricted. 

\subsection{No evidence of experimenter demand}

A natural concern is that our findings reflect experimenter demand: participants could have reported more liberal positions because they believed the intervention intended to find this. This study is more vulnerable to this potential concern because its focus on the role of SMCs was clear from the outset (e.g. all participants were informed they might be asked to follow SMCs) and treatment conditions were clearly delivered by the research team.\footnote{Unsurprisingly, 73\% of respondents listed understanding the effects of SMCs on political attitudes as one perceived motivation for this study.}

However, four pieces of evidence suggest that experimenter demand is unlikely to drive our findings. First, the main outcome groupings all contain behavioral measures---YouTube watch histories, political knowledge, donation choices, and validated turnout---that corroborate the self-reported outcomes. Second, if social desirability were driving responses, we might also expect inflated reports of electoral participation or protest attendance, yet we observe no such effects. Third, experimenter demand effects require participants to infer the purpose of the treatment. While this could plausibly explain the PP SMCs condition, participants in the PA SMCs condition did not perceive their content as political nor did they perceive a liberal shift in social media consumption, but still shifted their policy preferences to the left. Finally, we used validated turnout in the 2024 general election to identify the 28\% of participants who reported voting at endline but in fact did not. Using this proxy for the risk of providing socially desirable responses, Appendix Table \ref{table:overall_misreported_HEs} shows that treatment effects on both our overall political positions and participation outcomes are nearly identical for accurate reporters and for misreporters. This further suggests that our main findings do not hinge on patterns of overreporting.

\section{Conclusion}

This study provides the first field experimental evidence documenting the political consequences of starting to follow short-form SMCs for a sustained period. Over four and a half months during the 2024 US presidential election campaign, we encouraged participants to follow five progressive-minded SMCs whose content is native to social media, varying whether their content was predominantly apolitical or political. Both types of SMCs increased political engagement and influenced policy positions, narrative understandings of power and the economy, and partisan preferences among younger adults. For PA SMCs, policy and political persuasion appears to reflect trust and credibility, derived in part from establishing parasocial connections. PP SMCs were instead less impactful per political video but achieved influence by producing greater volume and more partisan content. Although we observed limited effects on electoral and non-electoral participation in this highly-charged political context, the effects on political preferences of this relatively small change in social media consumption exceed many traditional campaign tools. Together, these results highlight why SMCs have emerged as modern opinion leaders: exposure to their particular content shapes political attitudes and behaviors, influence can arise through either credibility or volume, and SMCs can sustain attention in today's competitive and decentralized media market. 

%Together, these findings illuminate why SMCs are becoming today's public opinion leaders. First, we show that specific content shifts political attitudes and behaviors beyond the political knowledge found by studies of using social media platforms in general. Consequently, understanding what content users are exposed to may be just as important as how much they are using social media. Second, we demonstrate that SMCs influence citizens' policy and partisan views, especially in comparison with other common communication strategies. Third, our comparison of PA and PP SMCs indicates that influence can be achieved through both quality---trust and credibility---or quantity of political content. Based on user engagement in our study and our SMCs' high follower counts, both models seem capable of attracting and sustaining attention in the highly competitive and decentralized market for engagement on social media.

Our findings have important implications. For election and issue campaigns, they suggest high returns to hiring or supporting SMCs, especially those who appear politically neutral. A key organizational question is how to optimally allocate financial resources between one-off ``pay to play'' advertising, longer-term contracts, or training stables for aligned SMCs to produce political content themselves. Our results most clearly demonstrate the effectiveness of the latter strategies, but may also underestimate the value of PA SMCs beyond this study: in the wild, they can reach larger and less politically-committed audiences. They can also set the agenda of political debates by affecting which topics get attention. 

For legislators and regulators, the persuasive power of SMCs combined with the very limited transparency provided by most technology platforms constitute a potential risk vector for illegal influence, including by foreign actors. This was recently illustrated by Romania's recent coordinated TikTok campaign in violation of the EU Digital Services Act's disclosure policies. Legislators and regulators may also note that the effects of our relatively light touch intervention (compared to the totality of what individuals consume online) suggest that platform exposure may be associated with larger societal effects, such as on trust in institutions, belief in public health, or belief in the rule of law. Further research is needed to establish such effects. 

Beyond its substantive contributions, this study provides methodological insights for designing interventions on social media. First, the algorithm we developed to match participants to interest-aligned SMCs attempts to approximate how platform recommendation systems maximize engagement \citep{aridor2024economics, guess2023social}. As such, it enables researchers to vary specific content consumed on social media without partnering with platforms. Although only financial incentives generated substantial consumption of assigned content, recommendation-only encouragement may be well-powered for more naturalistic interventions with larger samples (e.g. in partnership with social media companies). Second, our evidence suggests that researchers studying the political effects of digital content could pursue shorter interventions without sacrificing impact. We cannot pinpoint exactly how long it took for followers to connect with SMCs and alter political views, but three months of incentivized exposure caused measurable shifts in political attitudes during an intense election campaign. Third, around a third of our participants shared their YouTube histories for \$5 after completing our three surveys. As social media platforms limit access to data and scope for academic collaboration, this method suggests a cost-effective avenue for collecting rich trace data beyond desktop use. Fourth, we demonstrate the value of working with SMCs. Working with the BII identified a relatively homogeneous group of creators in terms of what content they produced when.

Several limitations of this study's design suggest directions for future research. First, the experiment took place during the 2024 US presidential campaign, a saturated media environment where some platforms altered their algorithms to downweight political content and the Democratic Party's campaign muted enthusiasm among young progressives (including SMCs). The effects we observe might be larger in other settings. Second, by encouraging participants to follow SMCs they did not already know, our design illuminates early-stage parasocial relationships. This likely understates the influence of the trusted SMCs users naturally opt to follow, although marginal effects may also decline as an SMC's messages become repetitive over time. Future work should examine how relational depth and message variation shape durability or decay, ideally by recruiting followers directly from creators' existing audiences. Third, we found limited effects on conventional participation, suggesting that SMCs may be more effective in mobilizing symbolic or issue-based activism rather than registration and turnout.\footnote{\cite{derksen2025instagram} find that restoring Instagram’s algorithmic recommendations modestly increased verified turnout, especially among younger Trump-leaning users. Such platform-level designs highlight the potential for exposure at scale but offer limited insight into mechanisms, since content is shaped by opaque algorithms and user variation. By contrast, our SMC-level design allows precise assignment of content and messenger, providing clearer evidence on which narratives resonate. We view the two approaches as complementary.} Finally, while our study centered on progressive SMCs, conservative creators may be at least as important. Figures like Joe Rogan, Ben Shapiro, and the Nelk Boys were widely credited with shifting online discourse toward Trump, often with more emotionally charged or conspiratorial rhetoric than the fact-checked, issue-based approach of BII fellows. The rightward shift we observed among placebo participants suggests conservative content may have exerted greater influence during this period, underscoring the need for systematic study of diverse content styles as actors of all types increasingly turn to SMCs or become SMCs themselves.

Our results also raise questions about the equilibrium in a market where SMC content production and audience demand are jointly determined. On the demand side, future work should map SMC consumption bundles in more representative samples beyond the 2024 election, clarifying how much content is explicitly political, how it is segmented by demographic and ideological profiles, and how it compares with other media sources. Political content can draw some audiences in while alienating others, shape perceptions of credibility, and influence demand for further political material. On the supply side, creators face incentives shaped by platform algorithms and policies, audience demand, advertiser preferences, and their own capacity to produce engaging political content, with strategic choices about politicization and timing across the electoral cycle. As SMCs become central voices in the media ecosystem, understanding how audience reactions and creator incentives interact in equilibrium is essential for assessing their broader political consequences.

Ultimately, our study speaks to a moment of profound change in if and how citizens consume political information. Growing evidence of news avoidance and political apathy signals a fundamental challenge to democracy: how can citizens be sufficiently informed in a media environment where opting out of politics is easier than ever? The rise of engaging intermediaries like SMCs suggests one pathway, but it also raises questions about the quality and diversity of public discourse as the boundary between news and entertainment blurs. Our finding that SMCs can wield significant political influence underscores the urgency of these questions in an era where new voices, compelling visuals, and short-form content distill complex ideas into fleeting moments of attention. What seem like micro-narratives are now reshaping the macro contours of political communication.

\newpage
\bibliographystyle{apsr}
\putbib[rsf.bib]
\end{bibunit}

\newpage

\appendix

\section{Online appendix}

\pagenumbering{arabic}
\renewcommand*{\thepage}{A\arabic{page}}

\setcounter{footnote}{0}
\setcounter{table}{0}
\setcounter{figure}{0}
\setcounter{equation}{0}
\renewcommand{\thetable}{\Alph{section}\arabic{table}}
\renewcommand{\thefigure}{\Alph{section}\arabic{figure}}
\renewcommand{\theequation}{\Alph{section}\arabic{equation}}

\addtocontents{toc}{\protect\setcounter{tocdepth}{3}}

\thispagestyle{empty}

\vspace{-10pt}

\tableofcontents

\newpage

\setcounter{page}{1}

\singlespacing

\begin{bibunit}[apsr]

\subsection{SMCs included in the intervention}\label{appendix:smc_descriptives}

\subsubsection{List of predominantly-apolitical SMCs}

{\small
\begin{tabular}{lcp{3cm}p{3cm}p{3cm}}
\toprule
\textbf{SMC} & \textbf{Assigned Total} & \textbf{Instagram} & \textbf{TikTok} & \textbf{YouTube} \\
\midrule
\multicolumn{5}{l}{\textit{Shorter-form SMCs:}} \\
Ambers Closet & 299 (27.7\%) &  & \href{https://www.tiktok.com/@amberscloset33}{@amberscloset33} & \href{https://tinyurl.com/smcamberscloset}{@amberscloset} \\
Anania Williams & 141 (13.1\%) &  & \href{https://www.tiktok.com/@anania00}{@anania00} &  \\
Brown Girl Green & 382 (35.4\%) & \href{https://www.instagram.com/browngirl_green/}{@browngirl\_green} & \href{https://www.tiktok.com/@browngirlgreen}{@browngirlgreen} &  \\
Cathy Pedrayes & 205 (19.0\%) & \href{https://www.instagram.com/cathypedrayes/reels/}{@cathypedrayes} & \href{https://www.tiktok.com/@cathypedrayes}{@cathypedrayes} & \href{https://www.youtube.com/@CathyPedrayesMomFriend/shorts}{@CathyPedrayesMomFriend} \\
Christine Lan & 162 (15.0\%) & \href{https://www.instagram.com/ecoamical/reels/}{@ecoamical} & \href{https://www.tiktok.com/@eco.amical}{@eco.amical} & \href{https://www.youtube.com/@eco.amical/shorts}{@eco.amical} \\
Daisy Foko & 514 (47.6\%) & \href{https://www.instagram.com/daisfoko/reels/}{@daisfoko} & \href{https://www.tiktok.com/@daisyfoko}{@daisyfoko} &  \\
\bottomrule
\end{tabular}}

\vspace{5pt}
{\noindent \footnotesize \textit{Notes}: ``Assigned Total'' indicates the number of participants to whom each creator was assigned in the quiz, recommendation, and screenshot validation conditions. The percentage in brackets indicates the share of participants assigned to Predominantly-apolitical SMCs who were recommended to follow each creator ($N$ = 1,080).}

\clearpage
\subsubsection{List of predominantly-political SMCs}

{\small
\begin{tabular}{lc p{3cm} p{3cm} p{3cm}}
\toprule
\textbf{SMC} & \textbf{Assigned Total} & \textbf{Instagram} & \textbf{TikTok} & \textbf{YouTube} \\
\midrule
\multicolumn{5}{l}{\textit{Shorter-form SMCs:}} \\
Alex & 42 (3.9\%) & \href{https://www.instagram.com/the_casalmon/}{@the\_casalmon} & \href{https://www.tiktok.com/@casalmon}{@casalmon} &  \\
andr3wsky & 90 (8.3\%) &  & \href{https://www.tiktok.com/@andr3wsky}{@andr3wsky} &  \\
Bimbo University & 235 (21.6\%) & \href{https://www.instagram.com/nikitadumptruck/?hl=en}{@nikitadumptruck} & \href{https://www.tiktok.com/@nikitadumptruck}{@nikitadumptruck} &  \\
Brittany Cunningham & 409 (37.5\%) & \href{https://www.instagram.com/mspackyetti/}{@mspackyetti} &  &  \\
Daria Rose & 269 (24.7\%) &  & \href{https://www.tiktok.com/@dariarosereal}{@dariarosereal} &  \\
Dasia Sade & 741 (68.0\%) &  &  & \href{https://www.youtube.com/@DasiaDoesIt/videos}{@DasiaDoesIt} \\
\bottomrule
\end{tabular}}

\vspace{5pt}
{\noindent \footnotesize \textit{Notes}: ``Assigned Total'' indicates the number of participants to whom each creator was assigned in the quiz, recommendation, and screenshot validation conditions. The percentage in brackets indicates the share of participants assigned to Predominantly-political SMCs who were recommended to follow each creator ($N$ = 1,090).}

\clearpage

\subsubsection{List of non-political (placebo) SMCs}

{\small
\begin{tabular}{lc p{3cm} p{3cm} p{3cm}}
\toprule
\textbf{SMC} & \textbf{Assigned Total} & \textbf{Instagram} & \textbf{TikTok} & \textbf{YouTube} \\
\midrule
\multicolumn{5}{l}{\textit{Shorter-form SMCs:}} \\
Browney Shorts & 911 (83.7\%) & \href{https://www.instagram.com/stanbrowney/reels/}{@stanbrowney} & \href{https://www.tiktok.com/@stanbrowney}{@stanbrowney} & \href{https://www.youtube.com/@BrowneyShorts/shorts}{@BrowneyShorts} \\
Bryce Xavier & 444 (40.8\%) & \href{https://www.instagram.com/brycexavier/reels/}{@brycexavier} & \href{https://www.tiktok.com/@brycexavier}{@brycexavier} &  \\
Caitlin and Tiptoe & 574 (52.7\%) & \href{https://www.instagram.com/caitlinandtiptoe/}{@caitlinandtiptoe} & \href{https://www.tiktok.com/@caitlinandtiptoe}{@caitlinandtiptoe} & \href{https://www.youtube.com/@caitlinandtiptoe/videos}{@caitlinandtiptoe} \\
Culter35 & 51 (4.7\%) & \href{https://www.instagram.com/culter35}{@culter35} &  & \href{https://www.youtube.com/channel/UC-r0FkdQhOfE367edUho7Aw}{@culter35} \\
Evelyn Gonzalez & 353 (32.4\%) & \href{https://www.instagram.com/theevelyng/reels}{@theevelyng} & \href{https://www.tiktok.com/@evelyngonzalez}{@evelyngonzalez} & \href{https://www.youtube.com/@evelyn_gonzalez/shorts}{@evelyn\_gonzalez} \\
First to Eleven & 711 (65.3\%) & \href{https://www.instagram.com/firsttoeleven/}{@firsttoeleven} & \href{https://www.tiktok.com/@firsttoeleven}{@firsttoeleven} & \href{https://www.youtube.com/@FirstToEleven/featured}{@FirstToEleven} \\
\bottomrule
\end{tabular}}

\vspace{5pt}
{\noindent \footnotesize \textit{Notes}: ``Assigned Total'' indicates the number of participants to whom each creator was assigned in the quiz, recommendation, and screenshot validation conditions. The percentage in brackets indicates the share of participants assigned to Placebo SMCs who were recommended to follow each creator ($N$ = 1,089).}

\clearpage

\subsection{Comparison of sample to nationally-representative ANES sample} \label{appendix:ANES_comparison}

\begin{table}
\centering
\caption{Summary statistics for regular social media users aged 18-45}
\label{tab:samplecompare}
\scalebox{1}{
\begin{tabular}{lcccccc}
\toprule
& \multicolumn{3}{c}{\textbf{2020 ANES sample}} & \multicolumn{3}{c}{\textbf{2024 study sample}} \\
\textbf{Variable} & \textbf{Mean} & \textbf{SD} & \textbf{Median} & \textbf{Mean} & \textbf{SD} & \textbf{Median} \\
\midrule
\textit{Panel A: Socioeconomics and demographics} & & & & & & \\
Age & 31.51 & 7.74 & 32.00 & 32.64 & 6.40 & 33 \\
Male & 0.49 & 0.50 & 0.00 & 0.44 & 0.50 & 0 \\
Female & 0.51 & 0.50 & 1.00 & 0.54 & 0.50 & 1 \\
White & 0.60 & 0.49 & 1.00 & 0.71 & 0.46 & 1 \\
Black & 0.13 & 0.33 & 0.00 & 0.23 & 0.42 & 0 \\
Hispanic & 0.18 & 0.38 & 0.00 & 0.16 & 0.37 & 0 \\
At least some college & 0.59 & 0.49 & 1.00 & 0.75 & 0.43 & 1 \\
Bachelor's degree or higher & 0.28 & 0.45 & 0.00 & 0.38 & 0.49 & 0 \\
Postgraduate degree & 0.10 & 0.30 & 0.00 & 0.13 & 0.33 & 0 \\
Heterosexual & 0.86 & 0.35 & 1.00 & 0.76 & 0.43 & 1 \\
Homosexual & 0.05 & 0.21 & 0.00 & 0.05 & 0.21 & 0 \\
Bisexual & 0.09 & 0.28 & 0.00 & 0.12 & 0.33 & 0 \\
Annual household income & 42,410 & 32,744 & 32,500 & 70,895 & 54,223 & 62,500 \\
Has children & 0.71 & 0.45 & 1.00 & 0.45 & 0.50 & 0 \\
& & & & & & \\
\textit{Panel B: Political engagement and preferences} & & & & & & \\
Follow politics (4-point scale) & 1.70 & 0.78 & 2.00 & 1.58 & 1.02 & 1 \\
Liberal-conservative (7-point scale) & 3.68 & 1.71 & 4.00 & 3.59 & 1.59 & 4 \\
Democrat & 0.52 & 0.50 & 1.00 & 0.46 & 0.50 & 0 \\
Independent & 0.17 & 0.37 & 0.00 & 0.18 & 0.38 & 0 \\
Republican & 0.32 & 0.46 & 0.00 & 0.26 & 0.44 & 0 \\
Approve of Biden & 0.43 & 0.31 & 0.50 & 0.31 & 0.46 & 0 \\
Registered to vote & 0.83 & 0.38 & 1.00 & 0.85 & 0.36 & 1 \\
& & & & & & \\
\textit{Panel C: Social media consumption} & & & & & & \\
Instagram hours/week & 8.26 & 8.90 & 3.00 & 6.05 & 9.71 & 3 \\
TikTok hours/week & 6.05 & 8.50 & 1.00 & 6.05 & 10.60 & 2 \\
YouTube hours/week & 9.88 & 8.58 & 10.00 & 12.62 & 15.42 & 8 \\
\bottomrule
\end{tabular}
}
\end{table}

\noindent To gauge the representativeness of our panel of participants, we compared it to regular social media users aged 18-45 from the nationally representative 2020 American National Election Studies (ANES). We applied several data processing steps to align the ANES variables with our survey questions. The ANES data was filtered to include only respondents aged 18-45 who were classified as regular social media users, defined as those using Instagram, TikTok, or YouTube at least a few times per week.\footnote{This filtering criterion was operationalized using Wave 3 social media usage variables (w3inst, w3tik, w3tube), where response codes 1-3 indicated usage frequency exceeding two days per week. Given that our sample definition required Wave 3 participation, we applied Wave 3 weights (w3weight) to all variables.} 

We then identified variables in the ANES that were measured similarly in our surveys. For socioeconomic and demographic covariates, we utilized pre-election profile variables for age, gender, race/ethnicity, education, sexual orientation, and household income.\footnote{Income was converted to continuous values using category midpoints and topcoded at \$150,000. Household size served as a proxy for having children, with households of three or more members coded as having children.} For political engagement and preferences, we combined pre-election and post-election variables that were measured similarly to our baseline survey.\footnote{These are a four-point scale for political attention (derived from polattrev), a seven-point ideology scale (w3lcself), party identification dummies (pid7x), and Biden approval ratings converted to a 0-1 scale. Voting behavior was captured through both self-reported turnout (w3turnout) and validated voter file matches (vote20\_match).} For social media consumption, we approximated weekly usage hours from frequency categories in the ANES, which does not directly measure time spent on platforms.\footnote{We mapped frequency responses to estimated hours per week: ``more than once a day" (21 hours, assuming 3 hours per day), ``once a day" (10 hours, assuming 1.5 hours per day), ``a few times per week" (3 hours), ``about once a week" (1.5 hours), with decreasing values for less frequent usage.} The final weighted sample included 1,331 respondents. 

Table \ref{tab:samplecompare} demonstrates that our sample exhibits fairly similar demographic and political characteristics, with some notable variations. First, the higher income levels in our sample largely reflect different topcoding thresholds---our survey topcoded at \$250,000 versus the ANES's \$150,000. Second, the stark difference in reported children (45\% in our panel 71\% in the ANES) stems from measurement approaches: we directly asked respondents whether they had any children, while ANES required using household size of three or more as an imperfect proxy for having children. Third, these surveys were conducted at different time points (ANES in 2020, ours in 2024), during which social media consumption patterns likely shifted substantially---not to mention that general sentiment toward Biden underwent considerable change between these periods. Despite these measurement and temporal differences, the comparability in demographics, political orientations, and social media usage patterns suggests our sample provides a fairly reasonable approximation of young adult social media users in the United States.

We construct post-stratification raking weights to align the 2024 Forthright sample with the 2020 American National Election Studies (ANES) on key demographic margins among adults aged 18–45. Variables are harmonized across both datasets into comparable categories: age (18–24, 25–34, 35–45), gender (male, female), race/ethnicity (White, Black, Hispanic, Asian/Other; mutually exclusive), education (high school or less, some college, bachelor's degree, postgraduate), and household income (<\$50k, ≥\$50k). The \$50k income cutoff roughly corresponds to the median household income among 18–45-year-olds in the United States during 2020–2024 and was used to differentiate lower- from higher-income respondents rather than to denote poverty. A small number of Forthright respondents had missing values on race/ethnicity (n=58, 1.2\%) or household income (n=125, 2.7\%). For race/ethnicity, we imputed the modal category; for income, we imputed the sample median household income (\$62,500) before constructing the binary indicator, assigning imputed cases to the ≥\$50k category. 

We derive population targets from the ANES by computing weighted proportions using the survey weight w3weight. These proportions are translated into population totals by multiplying by the Forthright sample size. We implement iterative proportional fitting using the survey package's rake function in R, initializing with uniform weights (svydesign(ids=~1, weights=~1)) and calibrating the Forthright sample margins to match the ANES-derived population totals across all five demographic dimensions simultaneously. To prevent extreme weights from unduly influencing variance estimates, we trim the resulting weights at the 1st and 99th percentiles and normalize them to have mean=1. Post-calibration diagnostics confirm that the weighted Forthright proportions successfully reproduce the ANES target distributions for each demographic margin. 

We show that our results do not substantively change when using these trimmed, normalized weights. Specifically, Table \ref{table:overall_reweighted_to_ANES} shows that our overall findings are substantively similar.

\subsection{Matching algorithm} \label{appendix:matching_algorithm}

Our approach to recommending participants for the main survey consists of three steps. First, we characterized the content of all SMCs in our treatment and placebo groups. Second, we collected data on SMC preferences using a `cold-start’ matching approach in a pilot study, where we sought to match participants to SMCs (within their treatment condition) based on a distance metric based on the research team’s initial assumptions about user preferences for SMCs. Third, based on the pilot data, we developed a recommendation system using machine learning techniques to refine the matching algorithm. Below, we describe each step in detail for building the matching algorithm for the main study.

\subsubsection{Characterizing the content of creators}

Recommendation algorithms generally benefit from content characterization \citep{javed2021review}. Following this best practice, we initially characterized the content of all SMCs in the apolitical, political, and placebo groups by manually coding the set of characteristics in accordance with a predefined coding scheme.

Each member of the research team independently coded the following characteristics of creators: gender, sexual orientation, ethnicity, age group, ideology, and topical expertise. We also assessed the content's visual aspects (e.g. fashionable, polished), verbal style (e.g. formal, assertive), whether the content was political, and its primary topics (e.g. news, lifestyle, entertainment, humor, family, money, science, environment, culture, education). Additionally, we coded whether the content might be perceived as polarizing by social media users. When possible, we relied on account descriptions written by the creators themselves or publicly available information online; if these sources were insufficient, we made judgments based on our own assessments.

\subsubsection{Gathering data on individual preferences}

A common challenge in developing recommendation algorithms, known as the cold start problem \citep{park2009pairwise}, is the lack of data on user-item interactions. This often requires initially launching the model by mapping users to items using heuristics that predict which types of content users with certain characteristics might prefer. In our application, we addressed this issue by utilizing a pilot containing 278 participants to enhance the recommendation system.

In our pilot survey, we measured various characteristics of our participants, including age, gender, education, sexual orientation, income, race, parental status, political ideology, and engagement with politics. We also assessed whether participants follow creators, to what extent they feel connected to them, and their openness to following new creators, including political ones. Additionally, we asked participants how much they value the visual (e.g. fashionable, polished) and verbal (e.g. formal, assertive) aspects of creators' content, as well as the importance of specific content elements, such as topic, content quality, visual and verbal style, and characteristics of creators (race, popularity, gender, sexuality, age, ideology, and topical expertise). We also gathered information on topics of interest (e.g. news, lifestyle, entertainment, humor, family, money, science, environment, culture, and education) and how participants discover new creators (e.g. searching the platform or asking friends).

Based on these characteristics, our pilot study calculated a simple distance metric that compared participant preferences (weighted by the importance of each aspect) with creator demographics and content characteristics. We evaluated the performance of this cold-start matching algorithm by asking participants about their overall satisfaction with the content of assigned creators on a 1-10 scale ($\mu=6.13$, $SD=2.47$), the extent to which they engaged with the content of assigned creators on a 1-4 scale ($\mu=0.75$, $SD=0.99$), and whether they would continue following the assigned creators after the study on a 1-5 scale ($\mu=3.25$, $SD=1.36$). The resulting metrics were standardized and aggregated into an index using Principal Components Analysis (PCA), which we then used to inform the recommendation system for the main study.

\subsubsection{Building the recommendation system}

The most common approaches to recommendation systems typically involve processing either user characteristics or content characteristics, known respectively as collaborative filtering \citep{sarwar2001item} and content-based filtering \citep{pazzani1999framework}. In either approach, the general framework is to leverage available data to generate personalized recommendations. Since we obtained characteristics on both individual preferences (in our baseline survey), creator attributes (hand-coded by the research team), and various metrics of user engagement with creators in the cold-start setup, we developed our approach using the hybrid recommendation system framework \citep{he2017neural}.

The hybrid approach to recommendation systems benefits from integrating multiple types of information, enabling it to provide more accurate recommendations \citep{burke2002hybrid}. First, it addresses the cold start problem associated with limited data by integrating both user and content attributes to generate initial recommendations with limited user interaction data. This was a key advantage in our application, as we expanded the list of creators recommended in the main study beyond the pilot, meaning we lacked data on user interactions with newly added creators. Second, the hybrid approach reduces over-specialization and aims to provide more diverse recommendations, which was particularly valuable in our application given the variety of topics and creator styles within the pool of SMCs selected for the main study.

Our hybrid recommendation system is based on a neural collaborative filtering approach \citep{he2017neural}. Our model architecture takes two sparse input vectors: one representing the participant (including relevant participant characteristics) and the other representing the creators. These inputs pass through embedding layers that convert the sparse data into dense, latent vectors. The latent vectors are then fed into a multi-layer neural network that maps them to predicted probability scores, allowing the model to learn complex participant-creator relationships from the data.

\begin{figure}[t!]
    \centering
    \caption{Neural collaborative filtering model architecture}
    \includegraphics[width=1\linewidth]{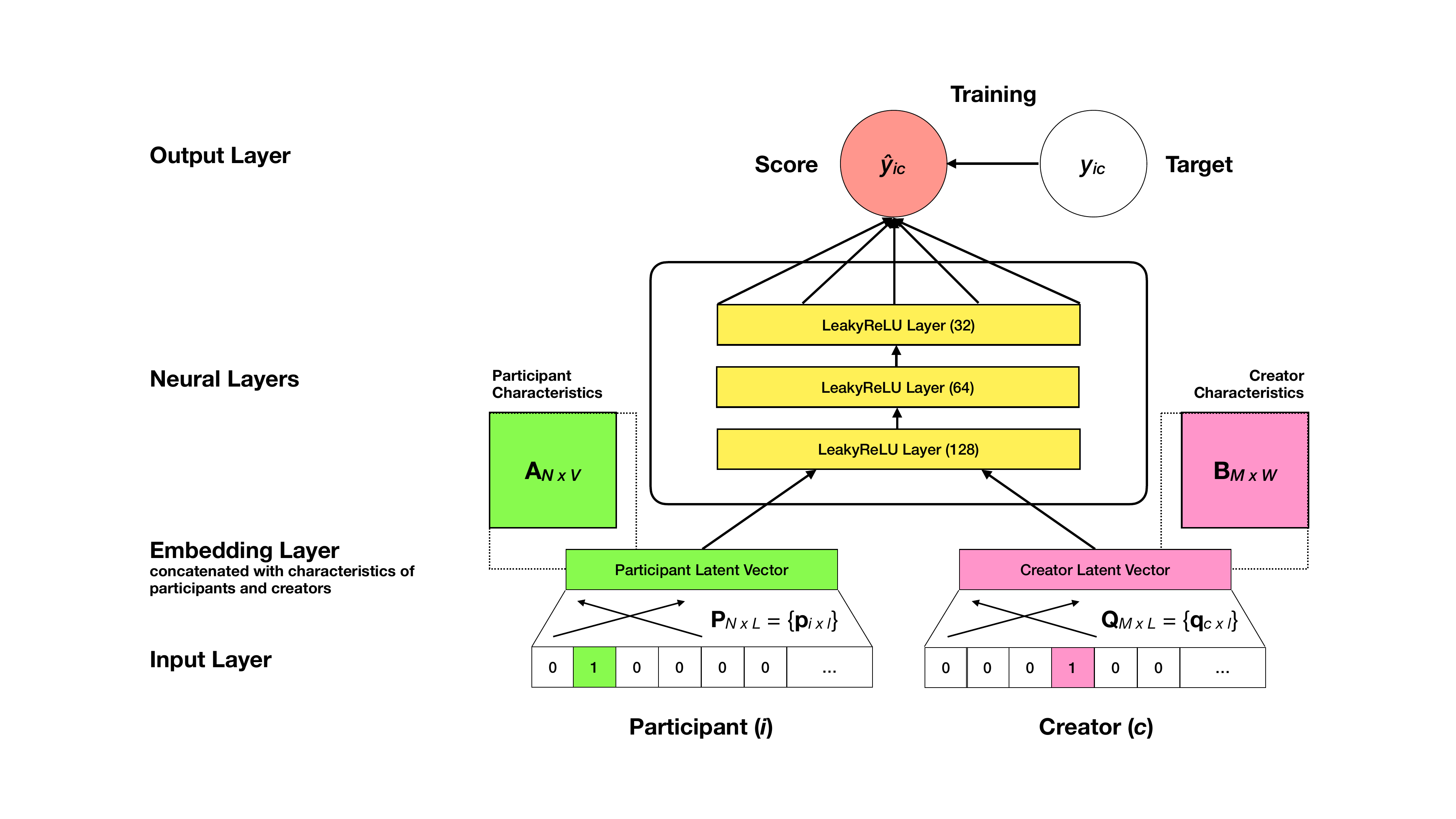}
    \label{fig:modelarch}
\end{figure}

First, two embedding layers were used to map participant and creator IDs to dense vector representations of 200 latent factors. The flattened and concatenated embedding vectors were passed through dense layers with Batch Normalization, LeakyReLU activation, and Dropout for regularization, progressively reducing the dimensions from 128 to 64 to 32. These layers captured interactions between participant-creator embeddings and participant-characteristic data, adding non-linearity to the model. The final output is a single prediction, generated using a linear activation. The model architecture is presented in Figure \ref{fig:modelarch}.

\begin{figure}[t!]
    \centering
    \caption{Neural collaborative filtering model performance evaluation}
    \includegraphics[width=1\linewidth]{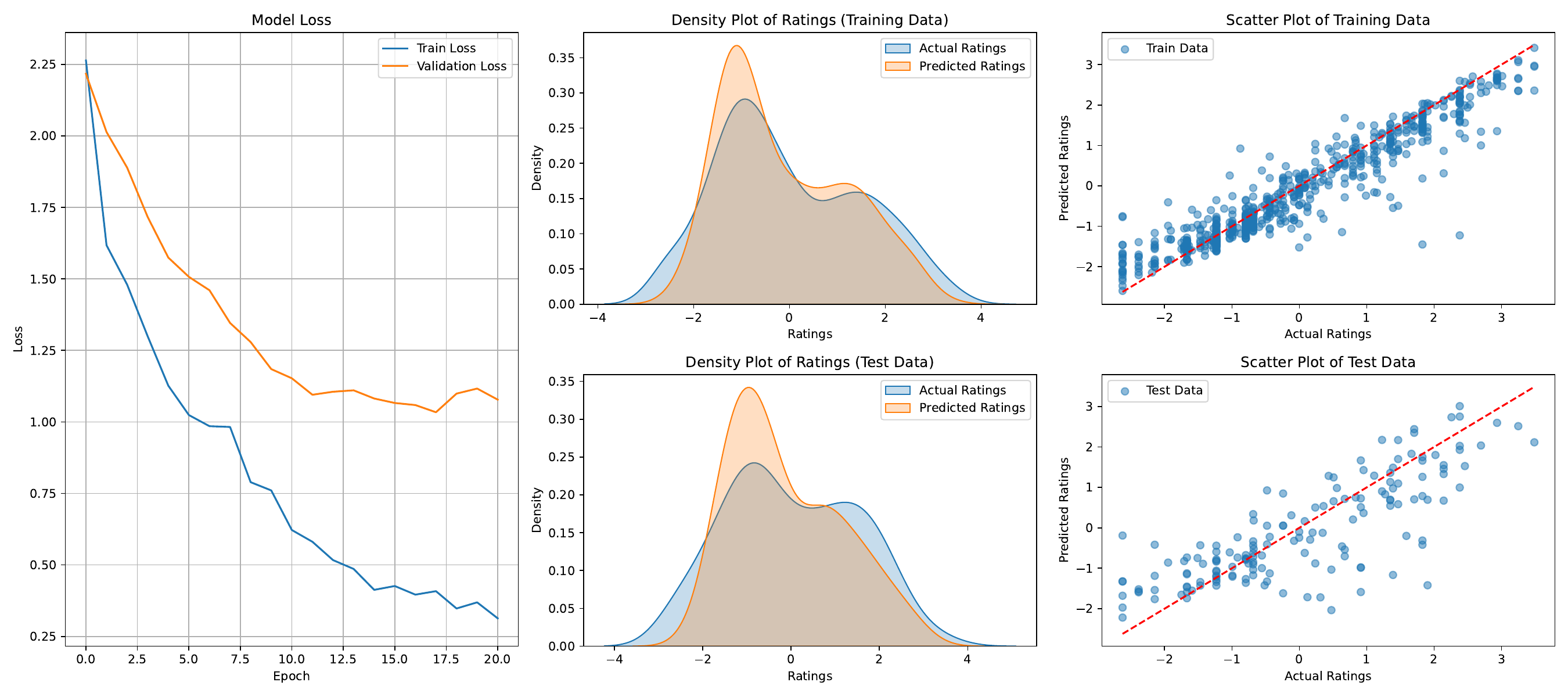}
    \label{fig:training}
\end{figure}

We trained the model using early stopping to prevent overfitting. Early stopping monitored the validation loss during training, halting the process when no further improvement was observed. In this training process, the model converged in 21 epochs, achieving stable performance on the validation set before reaching the maximum of 100 epochs. The resulting model performed well on both train and test samples, leaving no signs of overfitting. The matching algorithm performance is summarized in Figure \ref{fig:training}: the model achieved an MSE of 0.80, MAE of 0.66, RMSE of 0.89, and an $R^2$ of 0.59 on the test sample, indicating moderate predictive accuracy.

\subsubsection{Algorithm integration in the survey}

The final version of the algorithm was deployed on the Google Cloud Computing platform using Cloud Run functions and Bucket functionality. We ran the Python 3.10 function with 1GB of allocated memory to process responses from participants in real time as they took the survey on Qualtrics. Activity logs indicate that while the algorithm successfully processed all participant responses in the baseline survey, and recommendations were returned to be piped back into the Qualtrics survey, Qualtrics failed to pipe recommendations for 45 participants. These participants were dropped from the study. While the optimized version of the algorithm processed responses in an average of 0.006 seconds, the time it took for Qualtrics to pipe the responses could vary based on participants' internet connection quality, which could not be measured. Fortunately, very few participants dropped the survey due to this wait time.

\subsection{Details of content analysis} \label{appendix:content_analysis}

This section describes the two types of data underlying our content analysis: SMC-level content and respondent-level browsing histories. The SMC-level data consist of metadata and transcripts for all YouTube, Instagram, and TikTok videos, as well as Instagram posts, produced by the 60 short-form and 15 long-form SMCs that could be recommended to study participants across treatment conditions. The respondent-level data, in contrast, capture the demand side—what people actually watched. For this part of the analysis, we rely only on YouTube, since it was the only platform from which we could obtain detailed behavioral data.\footnote{While we obtained fine-grained TikTok browsing histories for 119 participants, the platform’s Research API imposed several constraints that limited the scalability of obtaining video metadata and transcripts. Data could not be retrieved for videos marked as personal or produced outside the US, and the API required time-bounded queries to be looped iteratively. Together with strict rate limits on the number of allowable requests, these restrictions made large-scale data extraction infeasible for this application. For this reason, we restrict this part of the analysis to YouTube only.}

Together, these two sources allow us to characterize both the content supplied by SMCs and the content actually consumed by participants, providing a fuller picture of the information environment during the intervention period.

\subsubsection{SMC data across platforms}

As described in Appendix \ref{appendix:smc_descriptives}, most of the SMCs included in our study maintain a presence across multiple platforms and frequently repost the same videos. Because of this high degree of cross-posting, we prioritized assembling a more complete dataset from YouTube and TikTok. By contrast, our Instagram coverage is more limited. In practice, Instagram was less central to our analysis both because it often duplicates content from the other two platforms and because obtaining systematic data from Instagram proved more difficult. Consequently, we focused Instagram data collection on SMCs who only produced content on Instagram; this limitation is mitigated by the substantial overlap in content across platforms.

We collected all YouTube, Instagram, and TikTok videos, as well as Instagram posts, produced by SMCs between August 1 and December 31, 2024. The content data was compiled in September and October 2025. To obtain transcripts of video content, specifically YouTube videos and shorts, TikTok videos, Instagram reels and videos embedded in posts, we relied on platform-generated captions (the built-in voice-to-text feature) whenever available. Such captions are often unavailable when content includes music tracks or when the audio quality is too low for reliable recognition. For videos without platform-generated captions, we used Whisper-1 from OpenAI.\footnote{OpenAI. \textit{Whisper-1}. 2023. Available at: \url{https://platform.openai.com/docs/models/whisper-1} (accessed September 29, 2025).} This approach allowed us to construct a comprehensive dataset of SMC content across platforms.

\begin{itemize}
    \item \textbf{YouTube.} In total, we collected 4,619 videos from 43 SMCs with active YouTube accounts. Two PA SMCs (Domo Wilson and Kelly Jakle) did not produce content during the study period, and one PP SMC (Quintein R. Jiles) deleted his account and later relaunched a family-oriented channel. For creators who changed their handles (e.g. Sean Pan, Unique Daily), we verified continuity and collected content from the new accounts.
    
    \item \textbf{TikTok.} The metadata for 4,726 videos from 52 SMCs on TikTok was obtained through the official TikTok Research API used for this project. The 2,606 videos lacking voice-to-text data, as well as those from accounts inaccessible through the official API, were scraped using Selenium and YouTubeDL. Since JR Mortimer (PP SMC) deleted his account after the intervention but before data scraping, we were unable to obtain data for his account.
    
    \item \textbf{Instagram.} While we obtained metadata for 4,657 content items (including posts, carousel posts, and reels) and reels from 51 SMCs in our sample, the constraints of data collection on Instagram meant that we collected and classified videos for only three SMCs who posted exclusively on Instagram---PP SMCs Elizabeth Booker Houston and Brittany Cunningham, and one Placebo SMC Katie Duke. We did not collect any data for culter35 (NP SMC), who did not produce any Instagram content during the intervention period, or for Alex @the\_casalmon (PP SMC), who deleted her Instagram account prior to data collection; both still produced content on other platforms. To obtain the Instagram data, we used Rocket API.
\end{itemize}

\begin{table}
\centering
\caption{Summary statistics for video length (duration in seconds) by treatment group, short-form creators only.} 
\begin{tabular}{lcccccccc}
\toprule
\textbf{Treatment condition} & $N$ & \textbf{Mean} & \textbf{Median} & \textbf{SD} & \textbf{Minimum} & \textbf{Maximum} & \textbf{Q1} & \textbf{Q3} \\ 
\midrule
\multicolumn{9}{l}{\textit{Panel A: YouTube}} \\ 
Predominantly-apolitical SMCs  & 648  & 136.91 & 45.00  & 278.83  & 6.00 & 2332.00  & 27.00 & 61.00  \\ 
Predominantly-political SMCs   & 536  & 959.16 & 114.50 & 1948.42 & 6.00 & 18374.00 & 60.00 & 340.50 \\ 
Placebo SMCs                   & 1000 & 376.08 & 65.50  & 673.62  & 6.00 & 8203.00  & 57.00 & 539.50 \\ 
\midrule
\multicolumn{9}{l}{\textit{Panel B: TikTok}} \\ 
Predominantly-apolitical SMCs  & 1089 & 70.10 & 47.00 & 140.51 & 0.00 & 3591.00 & 21.00 & 83.00 \\ 
Predominantly-political SMCs   & 1746 & 82.04 & 73.00 & 67.99  & 0.00 & 1221.00 & 59.00 & 90.00 \\ 
Placebo SMCs                   & 985  & 69.05 & 60.00 & 68.37  & 0.00 & 597.00  & 30.00 & 80.00 \\ 
\midrule
\multicolumn{9}{l}{\textit{Panel C: Instagram}} \\ 
Predominantly-apolitical SMCs  & 878  & 57.09  & 34.07  & 226.45 & 1.87 & 4199.85 & 17.70 & 58.43 \\ 
Predominantly-political SMCs   & 1526 & 88.34  & 65.92  & 147.55 & 1.90 & 3542.64 & 34.81 & 97.46 \\ 
Placebo SMCs                   & 933  & 49.26  & 49.72  & 36.48  & 0.60 & 408.60  & 21.27 & 66.40 \\ 
\bottomrule
\end{tabular}
\label{tab:length_summary}
\end{table}

\subsubsection{Respondent-level YouTube watch history data}

Out of 3,167 endline respondents who reached the question asking them to share their YouTube history, 2,342 consented to share their complete YouTube watch histories, and 1,149 ultimately uploaded their complete watch history data. The full browsing history is used to count the number of videos consumed from assigned and non-assigned SMCs, both during and after the intervention. To classify the videos consumed, we subset these histories to the period between August 1 and December 31 and attempted to retrieve transcripts for all videos viewed during that window. Not all videos permitted transcript extraction. We document the share of videos for which transcripts were successfully gathered and explain the reasons for missing data.

The full dataset of all videos ever watched by participants comprises approximately 22 million YouTube video viewing records, each linked to a unique respondent ID, enabling us to track viewing behavior across individuals. This rich behavioral dataset provides insight into what content users consume and when they engage with it. However, given the nature of YouTube consumption patterns, many videos appear multiple times across different viewers' histories. For our classification tasks, we therefore collapsed the data to focus on unique videos, which yielded 8,881,101 distinct videos.

The initial data processing involved several filtering steps to refine our dataset for analysis. First, we removed music videos by filtering out URLs that included ``youtube.music," reducing our corpus from approximately 8.9 million to 8.6 million videos—a reduction of 2.3\%. We did so because music videos typically lack substantive spoken content relevant to our classification objectives. Subsequently, we applied a temporal filter to focus on contemporary viewing behavior, including only videos watched by respondents after January 1, 2024. This date filtering further reduced our dataset to 4,285,611 videos. 

The transcript retrieval process presented both opportunities and challenges in our data preparation workflow:
\begin{enumerate}
    \item The API we used returned approximately 70\% of the filtered videos with the status ``success,'' resulting in 3,437,627 transcripts. The remaining videos could not be processed due to several common limitations: disabled transcripts or captions set by content creators; video unavailability caused by content being unlisted, made private, or deleted; and various platform access restrictions.
    \item After removing empty transcripts and filtering out error messages, we obtained 3,068,108 successfully retrieved transcripts suitable for classification. This additional filtering step was necessary because automated retrieval processes inevitably capture some incomplete or corrupted data that could compromise subsequent analysis.
    \item We then classified 3,068,079 transcripts as described below; 29 observations were not classified, as they were marked by GPT as ``Invalid prompt,'' meaning that they could contain content potentially harmful to humans or in violation of OpenAI policies. Within this group:
    \begin{enumerate}
        \item 3,064,925 transcripts were classified without errors and were further processed.
        \item 3,154 transcripts were not classified within the Batch API 24-hour processing window and returned with errors, so they were reclassified in a separate batch. Of these, 3,096 transcripts were successfully classified, while 58 exceeded the context window and were therefore first summarized (see below) and then classified.
    \end{enumerate}
    \item As the final step, we validated the classification output to identify missing classifications or out-of-bounds model predictions. After identifying these observations, we reclassified 17,675 transcripts in a separate batch and obtained valid, non-missing classifications, which replaced the invalid ones.
\end{enumerate}

For videos lacking available transcripts—primarily those with disabled captions—we explored alternative transcript generation methods to capture content that might otherwise be lost. We implemented automated speech recognition using Whisper-1 to extract spoken content. However, most Whisper-generated transcripts were meaningless in the context of our classification task, particularly for videos containing only music without spoken content, ambient sounds, or unclear audio. Given the poor quality and limited utility of these alternative transcripts, we excluded them from our subsequent classification analysis, focusing instead on the higher-quality automated captions retrieved through our primary method.

We acknowledge that YouTube videos can convey information through multiple modalities beyond spoken content. For instance, some videos contain no audio but display textual information, such as silent tutorials, memes with embedded text, or videos with on-screen graphics and captions. For such content, an Optical Character Recognition (OCR) approach could prove valuable in extracting visual text elements. Ultimately, we envision that a comprehensive multi-modal analysis—integrating both textual transcripts and visual content through frame-by-frame analysis—would provide richer insights into video content classification. We reserve this multi-modal approach for future research endeavors (see also \citealt{Casas2025}).

\subsubsection{Prompt design and model selection}

We classified both the SMC and participant-level data described in the sections above using GPT-5-mini, the most up-to-date and cost-effective model offered by OpenAI at the time of writing.\footnote{OpenAI. \textit{Introducing GPT-5}. 2025. Available at: \url{https://openai.com/index/introducing-gpt-5/} (accessed September 29, 2025).} We used the classification prompt below for this task.

To select the model that maximized classification performance for our task, we first hand-coded a subsample of 36 observations using the same coding scheme and then compared these hand-coded labels with the classifications produced by several models, including GPT-5-mini, GPT-o4-mini, and Gemini-2.5-Flash with and without video input. Among these, GPT-5-mini provided the best predictive performance (see Table \ref{table:model_report}) within our budget and was therefore used for the classification. Because our task involved thousands of classification requests, we used the Batch API to process them asynchronously and reduce costs.\footnote{OpenAI. \textit{Batch API Guide}. Available at: \url{https://platform.openai.com/docs/guides/batch} (accessed September 29, 2025).}

Since the GPT-5-mini model is configured for reasoning, it does not expose the temperature parameter. In standard generative models, temperature controls the randomness of the next-token sampling, while reasoning models employ deterministic decoding and thus are not influenced by sampling randomness. However, to ensure reproducibility and minimize unstable output, we set the seed parameter to a fixed random value (3784653) and requested a structured response format (\verb|json_object|). The prompt was submitted as a user message.

\begin{longtblr}[
  caption = {Comparison of model performance across LLMs and tasks},
  label   = {table:model_report}
]{
  colspec = {l *{4}{Q[c,0.14\textwidth]}},
  rowhead = 1,
  stretch = 0.5,
  rows = {font=\normalsize},
  colsep = 2pt,
  hline{1}  = {1pt},
  hline{2}  = {0.6pt},
  hline{Z}  = {1pt}
}
 & Precision & Recall & $F_1$ & Accuracy \\

\SetCell[c=5]{l}\textit{Panel A: Main format of the video} \\
GPT-5-mini  & 0.525 & 0.672 & 0.555 & 0.611 \\
GPT-o4-mini & 0.444 & 0.493 & 0.460 & 0.722 \\
Gemini-2.5-Flash (no video) & 0.827 & 0.842 & 0.829 & 0.806 \\
Gemini-2.5-Flash (video)    & 0.905 & 0.911 & 0.902 & 0.889 \\

\SetCell[c=5]{l}\textit{Panel B: Primary topic} \\
GPT-5-mini  & 0.382 & 0.431 & 0.378 & 0.611 \\
GPT-o4-mini & 0.372 & 0.476 & 0.388 & 0.583 \\
Gemini-2.5-Flash (no video) & 0.411 & 0.462 & 0.378 & 0.528 \\
Gemini-2.5-Flash (video)    & 0.300 & 0.386 & 0.285 & 0.444 \\

\SetCell[c=5]{l}\textit{Panel C: Covers politics or public affairs} \\
GPT-5-mini  & 0.940 & 0.940 & 0.940 & 0.944 \\
GPT-o4-mini & 0.906 & 0.935 & 0.913 & 0.917 \\
Gemini-2.5-Flash (no video) & 0.873 & 0.824 & 0.840 & 0.861 \\
Gemini-2.5-Flash (video)    & 0.926 & 0.846 & 0.869 & 0.889 \\

\SetCell[c=5]{l}\textit{Panel D: Discusses a US primary or general election campaign} \\
GPT-5-mini  & 0.894 & 0.857 & 0.874 & 0.917 \\
GPT-o4-mini & 0.894 & 0.857 & 0.874 & 0.917 \\
Gemini-2.5-Flash (no video) & 0.870 & 0.902 & 0.884 & 0.917 \\
Gemini-2.5-Flash (video)    & 0.900 & 0.964 & 0.926 & 0.944 \\

\SetCell[c=5]{l}\textit{Panel E: Discusses the 2024 US presidential election campaign} \\
GPT-5-mini  & 0.867 & 0.795 & 0.823 & 0.889 \\
GPT-o4-mini & 0.835 & 0.732 & 0.765 & 0.861 \\
Gemini-2.5-Flash (no video) & 0.839 & 0.839 & 0.839 & 0.889 \\
Gemini-2.5-Flash (video)    & 0.870 & 0.902 & 0.884 & 0.917 \\

\SetCell[c=5]{l}\textit{Panel F: The general politically ideological slant of the video} \\
GPT-5-mini  & 0.382 & 0.420 & 0.398 & 0.583 \\
GPT-o4-mini & 0.377 & 0.291 & 0.307 & 0.556 \\
Gemini-2.5-Flash (no video) & 0.417 & 0.433 & 0.422 & 0.667 \\
Gemini-2.5-Flash (video)    & 0.395 & 0.420 & 0.389 & 0.583 \\

\SetCell[c=5]{l}\textit{Panel G: Ideological stance of policies the video supports} \\
GPT-5-mini  & 0.163 & 0.214 & 0.185 & 0.722 \\
GPT-o4-mini & 0.326 & 0.286 & 0.292 & 0.750 \\
Gemini-2.5-Flash (no video) & 0.511 & 0.449 & 0.470 & 0.722 \\
Gemini-2.5-Flash (video)    & 0.384 & 0.417 & 0.374 & 0.722 \\

\SetCell[c=5]{l}\textit{Panel H: The overall partisan slant of the video (intensity)} \\
GPT-5-mini  & 0.402 & 0.452 & 0.395 & 0.639 \\
GPT-o4-mini & 0.482 & 0.358 & 0.385 & 0.667 \\
Gemini-2.5-Flash (no video) & 0.527 & 0.532 & 0.480 & 0.667 \\
Gemini-2.5-Flash (video)    & 0.309 & 0.439 & 0.340 & 0.583 \\

\SetCell[c=5]{l}\textit{Panel I: Criticizes or opposes US political parties} \\
GPT-5-mini  & 0.577 & 0.622 & 0.594 & 0.806 \\
GPT-o4-mini & 0.456 & 0.456 & 0.456 & 0.778 \\
Gemini-2.5-Flash (no video) & 0.621 & 0.706 & 0.654 & 0.833 \\
Gemini-2.5-Flash (video)    & 0.688 & 0.872 & 0.749 & 0.861 \\

\SetCell[c=5]{l}\textit{Panel J: Endorses or defends US political parties} \\
GPT-5-mini  & 0.486 & 0.486 & 0.486 & 0.944 \\
GPT-o4-mini & 0.486 & 0.500 & 0.493 & 0.972 \\
Gemini-2.5-Flash (no video) & 1.000 & 1.000 & 1.000 & 1.000 \\
Gemini-2.5-Flash (video)    & 0.500 & 0.648 & 0.546 & 0.944 \\

\SetCell[c=5]{l}\textit{Panel K: Endorses or defends US presidential/VP candidates} \\
GPT-5-mini  & 0.750 & 0.986 & 0.826 & 0.972 \\
GPT-o4-mini & 0.486 & 0.500 & 0.493 & 0.972 \\
Gemini-2.5-Flash (no video) & 1.000 & 1.000 & 1.000 & 1.000 \\
Gemini-2.5-Flash (video)    & 0.750 & 0.986 & 0.826 & 0.972 \\

\SetCell[c=5]{l}\textit{Panel L: Attacks or criticizes US presidential/VP candidates} \\
GPT-5-mini  & 0.554 & 0.562 & 0.558 & 0.917 \\
GPT-o4-mini & 0.602 & 0.500 & 0.529 & 0.889 \\
Gemini-2.5-Flash (no video) & 0.533 & 0.616 & 0.568 & 0.917 \\
Gemini-2.5-Flash (video)    & 0.533 & 0.616 & 0.568 & 0.917 \\
\end{longtblr}
\vspace{-8pt}
{\noindent \footnotesize\textit{Note:} Model performance metrics are based on comparisons between model predictions 
and a set of 36 human-coded videos. The human-coded data were produced by the researchers 
through full-video evaluation rather than transcript-only analysis.}
\vspace{12pt}

Since the model has a context window limit of 400,000 tokens, transcripts exceeding this threshold were first summarized with the summarization prompt (using GPT-4.1-mini with temperature set to 0 and seed fixed at 3784653, see the prompt below). The resulting summaries were then processed through the classification steps.

\subsubsection{Classification Prompt}

\begin{lstlisting}[style=prompt]
You are a helpful research assistant seeking to accurately classify the content of a [YouTube video / TikTok video / Instagram post] consumed by U.S.-based [YouTube / TikTok / Instagram] users.

Below is the information available about the [video / video / post]: [the title / the video description (if available)/ the post description (if available)], publication date, and the (usually) unpunctuated transcript of this video:

Title: {title}
Publication date: {date}
Transcript (usually unpunctuated): {transcript}  

Please carefully analyze the transcript and respond to the following classification questions.

# Instructions:
1. Base your responses strictly on the transcript, title, and publication date.
Do **not** infer or hallucinate information not explicitly stated.
2. Provide your output strictly in the JSON format shown below.
3. For single-answer questions, output a single integer (e.g., 2).
4. For multiple-answer questions, output a list of integers (e.g., [1, 3, 5]).
5. Exception: For identity_foreign, output a list of strings as specified below.
6. Answers marked with (!) are mutually exclusive and only one such option can be selected.
7. For each label, internally reason step-by-step before choosing your final answer.

# Output Format:
{{
    "transcript_quality": 3,
    "format": 5,
    "topic": [1, 3],
    "primary_topic": 3,
    "geopolitics_domain": 1,
    "domestic_politics_only": 2,
    "identity_foreign": ["Palestinian", "Turkish"],
    ...
}}

# Questions start here.

transcript_quality: Considering just the video's transcript, to what extent does it make coherent grammatical sense? Select **one** of the following options.
1. Totally incoherent
2. Barely coherent
3. Somewhat coherent
4. Mostly coherent
5. Totally coherent
0. There is no text

differentiating_speakers: Considering just the video's transcript, how confidently were you able to determine if there were multiple different speakers, and then separate their speech if there were multiple speakers? Select **one** of the following options.
1. Not at all confident
2. Barely confident
3. Somewhat confident
4. Mostly confident
5. Totally confident
0. There is no text

contextual_understanding: Overall, how confidently are you able to infer the meaning and relevant context of the video? Select **one** of the following options.
1. Not at all confident
2. Barely confident
3. Somewhat confident
4. Mostly confident
5. Totally confident

format: What is the main format of this video's content? Select **one** of the following formats.
1. How-to-do tutorial
2. Entertainment
3. Educational
4. Product/service review
5. News clip
6. Sporting event clip
7. Gaming
8. Music
9. Ads
10. Interview, discussion, or commentary
11. Memes
88. Other

purpose: What is the primary purpose of this video? Select **one or more** of the following purposes.
1. Entertain
2. Educate or inform
3. Persuade or advocate
4. Commentary or analysis
5. Satirize or ridicule
6. Call to action
7. Marketing or selling
(!) 0. None of these purposes

topic: What are the primary topics of this video? Select **one or more** of the following topics.
1. International relations and policy issues
2. Domestic public policy issues
3. Government actions or performance
4. Politicians or political parties
5. Public opinion or collective action
6. News and current affairs
7. Travel or tourism
8. Music videos and performances
9. Humor, pranks, or comedy
10. Food and culinary
11. Celebrity or entertainment gossip
12. Career, professional life, or workplace
13. Science, technology, or gadgets
14. Health, fitness, or wellness
15. Emotional or inspirational content
16. Dating and relationships
17. Gaming or esports
18. Gambling
19. Challenges or viral trends
20. Creative arts or DIY
21. Anime and Japanese culture
22. Advertising or promotion
23. History or documentary
24. Weather
25. Fashion or beauty
26. Animals or nature
27. Real-life experiences
28. Personal finance or money
29. Sports commentary or events
30. Religious
31. Paranormal or supernatural
32. Children's entertainment
33. Dramas, TV shows, or films
88. Other topic or topics not listed above

primary_topic: Which of these topics is the most prevalent topic in this video? Select **one** of the topics just determined.

political: Does the video include content related to government, agents of the state, politics, current affairs, or public policy?
1. Yes
0. No

political_topic: If the video contains content related to government, agents of the state, politics, current affairs, or public policy, what topics are covered in the video? Select **one or more** of the following options or ``none''.
1. Education
2. Health
3. Social welfare
4. Foreign policy
5. Income or wealth inequality
6. Taxes
7. Economy, jobs, or wages
8. Technology
9. Inflation or prices
10. Abortion or reproductive rights
11. Gun rights or control
12. Civil rights of racial or ethnic minorities
13. Civil rights by gender
14. Civil rights by sexuality
15. Immigration
16. Democracy
17. Climate or environment
18. Crime
19. Infrastructure
20. Energy
21. Transportation
22. Agriculture
23. Labor rights
24. Housing
25. International trade or investment
26. National security or military
27. Freedom of speech
28. Presidential activity
29. Congressional activity
30. Judicial activity
31. State or local government activity
32. Legislation 
33. Government policy implementation
34. Elections
35. Election campaign
36. Foreign affairs
37. Law enforcement
38. Moral values
39. Corruption
40. Prominent politicians
88. Other topic or topics related to government, agents of the state, politics, current affairs, or public policy not listed above
(!) 0. None: video does not contain content related to government, agents of the state, politics, current affairs, or public policy

geopolitics_domain: Does the video discuss content related to U.S. domestic or international politics or current affairs? Select **one or more** of the following options.
1. Domestic politics or current affairs
2. International politics or current affairs
(!) 0. Neither

domestic_politics_only: If the video discusses U.S. domestic politics or current affairs [option 1 in the previous question], which administrative level(s) of politics? Select **one or more** of the following options.
1. U.S. local politics or current affairs
2. U.S. state/national politics or current affairs
(!) 0. Neither or not applicable

mv_universalism: Focusing on moral values in general rather than in particular contexts, to what extent does the video promote altruism toward all Americans or people equally, versus promoting altruism toward a specific group, community, or identity? Select **one** of the following options.
1. Strongly particularistic
2. Somewhat particularistic
3. Neutral / mixed
4. Somewhat universalistic
5. Strongly universalistic

mv_humanitarianism: Focusing on moral values in general rather than in particular contexts, to what extent does the video promote compassion and care for vulnerable individuals or those in need, versus promoting emotional detachment or moral indifference? Select **one** of the following options.
1. Strongly indifferent
2. Somewhat indifferent
3. Neutral / mixed  
4. Somewhat humanitarian
5. Strongly humanitarian

mv_egalitarianism: Focusing on moral values in general rather than in particular contexts, to what extent does the video promote the idea that all individuals are morally equal, versus promoting an acceptance of social rank, inequality, or hierarchy? Select **one** of the following options.
1. Strongly hierarchical
2. Somewhat hierarchical
3. Neutral / mixed
4. Somewhat egalitarian
5. Strongly egalitarian

mv_communitarianism: Focusing on moral values in general rather than in particular contexts, to what extent does the video promote individual self-expression, autonomy, and pursuing one's own goals, versus promoting collective identity, group action, and pursuing shared community goals? Select **one** of the following options.
1. Strongly individualist
2. Somewhat individualist
3. Neutral / mixed
4. Somewhat communitarian
5. Strongly communitarian

mv_libertarianism: Focusing on moral values in general rather than in particular contexts, to what extent does the video promote personal freedom and resistance to control, versus promoting obedience to authority, discipline, or social order? Select **one** of the following options.
1. Strongly authoritarian
2. Somewhat authoritarian
3. Neutral / mixed
4. Somewhat libertarian
5. Strongly libertarian

mv_globalism: Focusing on moral values in general rather than in particular contexts, to what extent does the video promote a global or border-transcending perspective, versus promoting pride in and prioritization of one's own country or national identity? Select **one** of the following options.
1. Strongly nationalistic
2. Somewhat nationalistic
3. Neutral / mixed
4. Somewhat globalistic
5. Strongly globalistic

mv_cosmopolitanism: Focusing on moral values in general rather than in particular contexts, to what extent does the video promote one ethnic/racial/cultural group as superior or encourage isolation from other groups, versus promoting cultural relativism or embracing cultural diversity? Select **one** of the following options.
1. Strongly ethnocentric
2. Somewhat ethnocentric
3. Neutral / mixed
4. Somewhat cosmopolitan
5. Strongly cosmopolitan

mv_modernism: Focusing on moral values in general rather than in particular contexts, to what extent does the video promote respect for religious, familial, or cultural traditions, versus promoting innovation, change, or modern values? Select **one** of the following options.
1. Strongly traditional
2. Somewhat traditional
3. Neutral / mixed
4. Somewhat modern
5. Strongly modern

gov_trust: Does the video convey cynicism toward or trust in U.S. government, judicial, administrative, or democratic institutions? Select **one** of the following options.
1. Strongly cynical
2. Somewhat cynical
3. Neutral / mixed
4. Somewhat trusting
5. Strongly trusting

political_narratives: Which, if any, of the following political narratives to explain how the US political system currently works is most common in the video? Select **one or more** of the following narratives or ``none''.
1. Constitutionalist narrative: the checks and balances in the political system preserve liberty and avoid tyranny by the majority or by government
2. Representative narrative: policy decisions negotiated between elected politicians largely reflect the majority preferences of the citizens who participate in democracy
3. Elitist narrative: real political power is concentrated in the hands of powerful political organizations, wealthy elites, and corporate interests, rather than the general public or their elected representatives
4. Progressive narrative: socioeconomic, racial, or gender inequalities, and the political system that supports them, require structural reforms to increase social justice
5. Gridlock narrative: political polarization, outdated government or congressional rules or regulations, or the courts prevent opportunities for reform and effective governance
6. Traditionalism narrative: the political system fails to adequately respect traditional communities, social hierarchies, or authority roles
7. Populist narrative: a corrupt and out of touch political or corporate elite has rigged the Political system to their advantage at the expense of regular voters
8. Libertarian narrative: the political system gives too much power to the state to oppress private individuals and corporations
9. Authoritarian narrative: elected executive leaders lack the political powers required to maintain social order and security
10. Fair elections: elections are free and fair, allowing all citizens to express their preferences
11. Judicialization narrative: real legislative and executive power is concentrated in the hands of the judges, rather than elected officials
12. Base narrative: politicians are beholden to non-elite constituencies within their political party, even though they are not nationally representative
(!) 0. None: none of these political narratives is explicitly or implicitly conveyed

economic_narratives: Which, if any, of the following economic narratives to explain how the U.S. economy currently works is most common in the video? Select **one or more** of the following narratives or ``none''.
1. Free market narrative: unregulated markets, private ownership, and consumption demand with minimal government intervention advances innovation and economic well-being
2. Trickle-down narrative: tax cuts and deregulation for wealthy individuals and large corporations stimulate economic growth that eventually creates jobs, raises wages, or lower prices for lower-income individuals
3. Mobility narrative: any individual can achieve upward economic mobility by working hard to pull themselves up by their bootstraps
4. Traditionalist narrative: traditional economic and social hierarchies provide the order and stability required for the economy to prosper
5. Protectionist narrative: protectionist policies will increase domestic production and jobs, restore reciprocal trade, and establish natural resource independence
6. Interventionist narrative: government intervention in free markets can boost demand in the economy, support innovation, correct market distortions, or redistribute to ensure fairer economic distribution
7. Progressive narrative: market power enables corporations to exploit and oppress workers, especially the most disadvantaged, to benefit capitalist elites
8. Sustainability narrative: economic systems fail to prioritize well-being, social equity, or environmental sustainability alongside short-term economic growth
9. Abundance narrative: relaxing government regulations that create scarce supply and administrative burdens would allow innovations and goods production which support broad-based prosperity
(!) 0. None: none of these economic narratives is explicitly or implicitly conveyed

liberal_issues: Does the video directly or indirectly promote any of the following issue positions commonly associated with left-leaning or liberal policy agendas? Select **all** issue positions below that apply.
1. Improving the economy or jobs for everyone
2. Increasing environmental protection or renewable energy
3. Maintaining election integrity or voting rights
4. Protecting U.S. democracy and condemning political violence
5. Upholding human rights or civil liberties
6. Supporting immigration or immigrant rights
7. Increasing reproductive health and abortion rights
8. Upholding LGBTQ+ rights and inclusion
9. Supporting gender or racial justice
10. Increasing gun control or firearm regulation
11. Increasing healthcare access, quality, or affordability
12. Pursuing criminal justice reform or police accountability
13. Increasing taxes on the rich relative to taxes on the poor or middle class
14. Increasing education quality, equity, or student debt forgiveness
15. Enforcing antitrust law against market power
16. Preventing price gouging or profiteering
17. Increasing affordable housing
18. Combating poverty or expanding social safety nets
19. Holding powerful political and economic actors to account
20. Promoting democracy or human rights abroad
21. Supporting Palestinians
(!) 0. Video does not directly or indirectly promote any of these liberal policy positions

conservative_issues: Does the video directly or indirectly promote any of the following issue positions commonly associated with right-leaning or conservative policy agendas? Select **all** the positions below that apply.
1. Increasing border security or immigration restrictions
2. Increasing gun rights or 2nd Amendment protections
3. Reducing election fraud or voter ID law violations
4. Supporting religious freedom or Christian values
5. Supporting traditional family values
6. Supporting anti-abortion/pro-life policies
7. Opposing government or judicial overreach
8. Supporting fiscal conservatism or balanced budgets
9. Upholding free market capitalism or deregulation
10. Supporting freedom of speech or opposing "woke culture" or political correctness
11. Reducing America's role in global conflicts or support for other countries
12. Supporting law enforcement or raising concern about crime
13. Opposing efforts to combat or mitigate climate change
14. Restoring the working class
15. Opposing LGBTQ+ rights and inclusion
16. Protecting or investing in national security or the military
17. Increasing protectionist trade barriers
18. Increasing school choice and local curricula control
19. Reducing individual or corporate taxes
20. Questioning the credibility or relevance of medical science
21. Supporting the Israeli government or combating anti-Semitism
(!) 0. Video does not directly or indirectly promote any of these right-leaning or conservative policy positions

identity: Does the video highlight or emphasize specific aspects of any personal or group identities? Select **all** of the following identity groups that are emphasized, based on content, tone, examples, or explicit references.
1. African Americans
2. Asian Americans
3. Latinos or Hispanics
4. Indigenous communities
5. Immigrants or refugees
6. Women
7. LGBTQ+ people
8. Trans or nonbinary people
9. Muslims
10. Mormons
11. Hindus
12. Jews
13. Buddhists
14. Mainstream Christians
15. Conservative or Evangelical Christians
16. Rural Americans
17. White working class
18. Veterans or military families
19. Middle class 
20. Urban poor or working poor
21. Upper class, elite, or wealthy
22. Super rich, billionaires, or 1%
23. Youth or students
24. Elderly
25. Disabled people
26. Formerly incarcerated people
27. Parents or families
28. Union members
29. Americans (as a nationality)
30. Men
31. Community (not specified further)
32. Non-white Americans (not a specific race or ethnicity)
33. Far-left liberals
34. Far-right conservatives
35. Political independents
36. Democrat voters
37. Republican voters
(!) 0. No identity group is highlighted or emphasized

identity_foreign: Does the video highlight or emphasize the identity, culture, or perspective of specific non-American national or ethnic groups? If yes, list **all** of the groups that are highlighted or emphasized, based on content, tone, examples, or explicit references. If no identity group is highlighted or emphasized, write ``None''.

sources: Does the video explicitly or implicitly substantiate political claims by referencing sources with subject expertise or authority? Select **one or more** of the following options or ``none''.
1. A source with generally-recognized subject expertise or authority is referenced
2. A source that the video attributes expertise or authority to is referenced
3. A source with no generally-recognized or video-attributed expertise or authority is referenced
4. An unnamed source is referenced
(!) 0. None: video does not contain any political content, or no source with subject expertise or authority is explicitly or implicitly referenced

narrative: How does the video convey political content? Select **one or more** of the following options.
1. Political content is embedded within a broader narrative or story
2. Political content is stated directly as facts or reporting
3. Political content is stated directly as views or calls to action
4. Political content is conveyed through commentary on particular events
5. Political content is conveyed through interviews or discussions
(!) 0. Video does not contain any political content

comm_style: What communication styles are used in the video? Select **all** of the following options that apply, based on tone, structure, and delivery patterns evident in the video transcript.
1. Educational or explanatory
2. Persuasive or advocacy
3. Emotional or affective
4. Humorous, satirical, or sarcastic
5. Aggressive or confrontational
6. Relatable, informal, or colloquial
7. Dramatic or sensationalized
8. Neutral or objective
9. Other communication style
(!) 0. No meaningful communication in the transcript

emotion: What emotions is the video most likely to evoke in the audience? Select **all** of the following options that apply, based on tone, structure, and delivery patterns evident in the video transcript.
1. Anger or outrage
2. Fear or anxiety
3. Hope or inspiration
4. Pride (national, cultural, personal)
5. Sadness or sympathy
6. Humor or amusement
7. Disgust or contempt
8. Empathy or solidarity
9. Curiosity or intrigue
10. Gratitude or thankfulness
11. Joy or triumph
12. Seriousness or concern
(!) 0. No clear emotional appeal

toxicity: How would you characterize the tone of the video when addressing other individuals, organizations, or institutions? Select **one** of the following options.
1. Highly toxic or aggressive tone
2. Dismissive or disrespectful tone
3. Critical but civil tone
4. Neutral or professional tone
5. Friendly, warm, or respectful tone
0. No clear evaluative tone

masculinity: Does the video promote, approve, or reinforce traditional masculine norms? Select **one** of the following options.
1. Strongly opposes masculine norms
2. Some explicit opposition of masculine norms
3. Implicit challenge to masculine norms
4. Neutral / does not clearly reference masculine norms
5. Implicit reinforcement of masculine norms
6. Some explicit promotion to  masculine norms
7. Strongly promotes masculine norms

mobilization: Does the video include any of the following types of mobilization messages or calls to action? Select **all** of the following actions that the video explicitly or implicitly encourages the audience to do.
1. Vote or participate in elections
2. Participate in a protest or march
3. Attend a political campaign event or rally
4. Work or volunteer for a political campaign
5. Donate to a social or political organization
6. Contact or sign a petition sent to a politician or decision-maker
7. Attempt to persuade a family member or friend to change a political position
8. Share or amplify the video or its message
9. Obtain further political information
10. General emotional or moral call to stand up or speak out
(!) 0. No explicit or implicit mobilization message or call to action

political_party: Does the video reference any U.S. political party or any of its prominent party members? Select **at least one** of the following options or select ``none''.
1. Democratic Party
2. Republican Party
3. Green Party
4. Libertarian Party
5. Constitution Party
6. Another U.S. political party
(!) 0. None

election: Does the video reference any aspect of a U.S. primary or general election campaign for President, House, Senate, or local elections?
0. No
1. Yes

election_2024: Does the video reference the 2024 U.S. presidential election campaign? Select ``1'' only if the transcript clearly references the U.S. presidential election, primaries, voting, candidates, or election results; select ``0'' otherwise.
0. No
1. Yes

advocate_candidate: Does the video encourage viewers to vote for one or more candidates (without challenging, questioning, or being satirical or sarcastic about this view)? Select **at least one** type of candidate from the following list or select ``none''.
1. Democratic Party presidential candidate
2. Republican Party presidential candidate
3. Libertarian Party presidential candidate
4. Green Party presidential candidate
5. Independent presidential candidate
6. Any other Democrat candidate in a non-presidential race
7. Any other Republican candidate in a non-presidential race
8. Any non-Democrat or non-Republican candidate in a non-presidential race
(!) 0. None: the video did not encourage viewers to vote for any candidate

sponsored: Does the video say that any of its political content was sponsored or paid for by a particular individual or organization?
0. No
1. Yes

ideology_slant: Considering the issues covered, positions taken, and language of this video, what is the general politically ideological slant of this video (without challenging, questioning, or being satirical or sarcastic about this view)? Select **one** of the following options.
1. Strongly liberal
2. Liberal
3. Slightly liberal
4. Neutral or balanced
5. Slightly conservative
6. Conservative
7. Strongly conservative
0. Video does not contain politically ideological content

policy_slant: What is the ideological position of any public policies the video supports or advocates for (without challenging, questioning, or being satirical or sarcastic about this view)? Select **one** of the following options.
1. Strongly liberal
2. Liberal
3. Slightly liberal
4. Neutral or balanced
5. Slightly conservative
6. Conservative
7. Strongly conservative
0. Video does not support or advocate for any public policies

partisan_slant_form: What is the partisan slant of this video? Select **at least one** of the following options.
1. Explicitly supports or endorses the Democratic Party, a Democratic Party candidate or member, actions the video specifically attributes to the Democratic Party or its members, or a position the video specifically attributes to the Democratic Party in the video (without challenging, questioning, or being satirical or sarcastic about this view)
2. Implicitly supports the Democratic Party by favorably framing or editorializing about the Democratic Party, a Democratic Party candidate or member, or a position on an issue the video specifically attributes to the Democratic Party
3. Implicitly supports the Democratic Party by unfavorably framing or editorializing about the Republican Party, a Republican Party candidate or member, or a position on an issue the video specifically attributes to the Republican Party
4. Implicitly supports the Democratic Party by advocating for a position on an issue widely associated with the Democratic Party or its members
5. Implicitly supports the Democratic Party by advocating against a position on an issue widely associated with the Democratic Party or its members
(!) 6. Neutral or balanced
7. Implicitly supports the Republican Party by advocating for a position on an issue widely associated with the Republican Party or its members
8. Implicitly supports the Republican Party by advocating against a position on an issue widely associated with the Democratic Party or its members
9. Implicitly supports the Republican Party by favorably framing or editorializing about the Republican Party, a Republican Party candidate or member, or a position on an issue the video specifically attributes to the Republican Party
10. Implicitly supports the Republican Party by unfavorably framing or editorializing about the Democratic Party, a Democratic Party candidate or member, or a position on an issue the video specifically attributes to the Democratic Party
11. Explicitly supports or endorses the Republican Party, a Republican Party candidate or member, actions the video specifically attributes to the Republican Party or its members, or a position the video specifically attributes to the Republican Party in the video (without challenging, questioning, or being satirical or sarcastic about this view)
(!) 0. Video does not explicitly or implicitly reference a U.S. political party, candidate, or member, or their positions on an issue

partisan_slant_intensity: What is the overall partisan slant of this video? Select **one** of the following options, based on the selection of topics or aspects of the topics to cover, how topics are framed and scrutinized, commentary or endorsements, how topics are connected to U.S. political parties, and the tone of the language used.
1. Strongly pro-Democrat
2. Pro-Democrat
3. Slightly pro-Democrat
4. Neutral or balanced
5. Slightly pro-Republican
6. Pro-Republican
7. Strongly pro-Republican
0. Video does not contain partisan content

party_endorsement: Does the video include an explicit endorsement or defense of any U.S. political parties or their members (without challenging, questioning, or being satirical or sarcastic about this view)? Select **at least one** of the following options. 
1. Endorses or defends the Democratic Party or its members
2. Endorses or defends the Republican Party or its members
3. Endorses both parties or their members / promotes bipartisanship
4. Endorses or defends another US political party or its members
(!) 0. No U.S. political party or its members is endorsed or defended

party_critical: Does the video explicitly criticize, attack, or oppose any U.S. political parties or their members (without challenging, questioning, or being satirical or sarcastic about this view)? Select **at least one** of the following options.
1. Criticizes, attacks, or opposes the Democratic Party or its members
2. Criticizes, attacks, or opposes the Republican Party or its members
3. Criticizes both political parties or their members
4. Criticizes, attacks, or opposes another US political party or its members
(!) 0. No U.S. political party or its members is criticized, attacked, or opposed

candidate_endorsement: Does the video include an explicit endorsement or defense of any U.S. presidential or vice-presidential candidates (without challenging, questioning, or being satirical or sarcastic about this view)? Select **at least one** of the following options.
1. Endorses or defends a Democratic presidential or vice-presidential candidate
2. Endorses or defends a Republican presidential or vice-presidential candidate
3. Endorses or defends a third-party or independent presidential or vice-presidential candidate 
4. Endorses or defends multiple candidates
(!) 0. Video does not clearly endorse or defend any U.S. presidential or vice-presidential candidates

candidate_attack: Does the video explicitly criticize, attack, or oppose any U.S. presidential or vice-presidential candidates (without challenging, questioning, or being satirical or sarcastic about this view)? Select **at least one** of the following options.
1. Criticizes, attacks, or opposes a Democratic presidential or vice-presidential candidate
2. Criticizes, attacks, or opposes a Republican presidential or vice-presidential candidate
3. Criticizes, attacks, or opposes a third-party or independent presidential or vice-presidential candidate
4. Criticizes, attacks, or opposes multiple candidates
(!) 0. Video does not criticize, attack, or oppose any U.S. presidential or vice-presidential candidate
\end{lstlisting}

\subsubsection{Summarization Prompt}

\begin{lstlisting}[style=prompt]
You are a helpful research assistant. Your task is to summarize the transcript of a YouTube video consumed by U.S.-based viewers. You will be given the video's title, publication date, and its transcript (often unpunctuated).

# Input:
Title: {title}
Publication date: {date}
Transcript (may lack punctuation): {transcript}
    
# Instructions:
1. Review the entire transcript and reconstruct meaning where punctuation is missing.
2. Write a concise but thorough summary that captures the video's key points, main ideas, and overall message.
3. When relevant, incorporate examples, evidence, or explanations from the transcript that clarify or strengthen the main points.
4. Do not include filler, tangents, or repeated phrases.
5. Provide a short summary for simple transcripts, and a longer, more detailed overview for complex or lengthy ones.
6. Preserve the intent and meaning of the original text without adding interpretation or opinion.
\end{lstlisting}

\subsection{Example of SMS/email treatment message \label{appendix:sms_example}}

The SMS/email treatment group received biweekly messages containing core content share with BII fellows. Below is an example of the message sent on October 25 by our survey partners to participants, with the text at the topic being common across all nine messages and the five bullet points varying by message.

\begin{quote}
Thank you again for participating in the social media influencers study! Here's our bi-weekly newsletter summarizing popular recent content on social media.

\begin{itemize}
\item After an election, it can take longer than a single night to count all of the votes accurately and to complete all the verification steps. This work is called a canvass and certification process.
\item There are many reasons for the current affordable housing shortages in the USA. In good news, the current administration proposed a plan to cap rent hikes by 5\% per year!
\item In the US, about 17 million people earn less than \$15/hour. Raising the wage floor to \$15 would give them a pay boost and help increase it for millions of other workers too.
\item Exciting news, by 2028 the Postal Service will make the majority of its fleet electric! That's a big factor when it comes to reducing carbon emissions by 40\% by 2030.
\item We can make progress on climate change when we acknowledge we need big solutions — corporate innovation, government regulation, and community organizing. Let’s do it together.
\end{itemize}

\end{quote}

\subsection{Description of main outcomes \label{appendix:outcomes}}

We next compile the survey questions underpinning our main outcomes. Full survey instruments are available upon request. 

\begin{longtblr}[
  caption = {Description of main outcomes},
  label   = {outcomes_table},
]{
  colspec = {
    Q[l,0.17\textwidth]
    Q[l,0.4\textwidth]
    Q[l,0.4\textwidth]
  },
  rowhead = 1,
  stretch = 0.5,
  rows = {font=\small},
  colsep = 2pt,
  hline{1}  = {1pt},
  hline{2}  = {0.6pt},
  hline{Z}  = {1pt}
}
%--------------------------------------------------------------------------
%--------------------------------------------------------------------------
\textbf{Outcome} & \textbf{Midline Survey} & \textbf{Endline Survey} \\ 
%--------------------------------------------------------------------------
\SetCell[c=3]{l} \textit{H1: increased political engagement in general} \\ \hline

Knowledge of current affairs &
Which is not a key battleground state that is frequently discussed in 2024 due to its close polling and potential to swing the election?
\begin{itemize}[nosep]
    \item Colorado \textit{(correct)}
    \item Georgia
    \item Pennsylvania
    \item Wisconsin
    \item Don't know
\end{itemize}

\vspace{3pt}
Democratic Vice-Presidential candidate Tim Walz has been serving as the Governor of which U.S. state since 2019?
\begin{itemize}[nosep]
    \item Minnesota \textit{(correct)}
    \item Michigan
    \item Ohio
    \item Wisconsin
    \item Don't know
\end{itemize}

\vspace{3pt}
Which of the following newspapers has said it will not endorse a candidate for president in this election?
\begin{itemize}[nosep]
    \item Washington Post \textit{(correct)}
    \item Boston Globe
    \item New York Post
    \item New York Times
    \item Don't know
\end{itemize} &
Donald Trump asked Elon Musk and Vivek Ramaswamy to lead what organization? 
\begin{itemize}[nosep]
    \item Department of Government Efficiency \textit{(correct)}
    \item Department of the Treasury
    \item Republican National Committee
    \item National Republican Congressional Committee 
    \item Don't know 
\end{itemize}

\vspace{3pt}
In November, Special Counsel Jack Smith filed a motion to drop all federal charges against who? 
\begin{itemize}[nosep]
    \item Donald Trump \textit{(correct)}
    \item Hunter Biden
    \item Steve Bannon
    \item Bob Menendez
    \item Don't know
\end{itemize}

\vspace{3pt}
Which Trump cabinet nominee withdrew from consideration for the position of Attorney General?
\begin{itemize}[nosep]
    \item Matt Gaetz \textit{(correct)}
    \item Robert F. Kennedy 
    \item William Barr 
    \item JD Vance 
    \item Don't know
\end{itemize} \\

Interest in politics &
Thinking back over the past three months, how closely did you follow US politics?
\begin{enumerate}[leftmargin=*, nosep]
    \item[0.] Not at all closely
    \item[1.] Somewhat closely
    \item[2.] Rather closely
    \item[3.] Very closely
\end{enumerate} &
Thinking back over the past 2 months, from the elections in November until today, how closely did you follow U.S. politics?
\begin{enumerate}[leftmargin=*, nosep]
    \item[0.] Not at all closely
    \item[1.] Somewhat closely
    \item[2.] Rather closely
    \item[3.] Very closely
\end{enumerate} \\ \hline
%--------------------------------------------------------------------------
\SetCell[c=3]{l} \textit{H2: increased progressive policy attitudes} \\ \hline

Progressive climate preferences &
To what extent do you agree or disagree with the following statements?
\begin{itemize}[nosep]
    \item \textit{Government should do more to curb climate change, even at the expense of economic growth}
\end{itemize}

\begin{enumerate}[leftmargin=*, nosep]
    \item Strongly disagree
    \item Somewhat disagree
    \item Neither agree nor disagree
    \item Somewhat agree
    \item Strongly agree
\end{enumerate} &
To what extent do you agree or disagree with the following statements?
\begin{itemize}[nosep]
    \item \textit{Government should do more to curb climate change, even at the expense of economic growth}
\end{itemize}

\begin{enumerate}[leftmargin=*, nosep]
    \item Strongly disagree
    \item Somewhat disagree
    \item Neither agree nor disagree
    \item Somewhat agree
    \item Strongly agree
\end{enumerate} \\

Progressive democracy preferences &
Is democracy preferable to other kinds of government?
\begin{enumerate}[leftmargin=*, nosep]
    \item[2.] Democracy may have its problems, but it is the best system of government
    \item[1.] In some circumstances, a non-democratic government can be preferable
    \item[0.] It doesn’t matter what kind of government we have
\end{enumerate}

\vspace{3pt}
To what extent do you agree or disagree with the following statements? 
\begin{itemize}[nosep]
    \item \textit{It is never justified, under any circumstances, for Americans to take violent action against the government}
\end{itemize}

\begin{enumerate}[leftmargin=*, nosep]
    \item Strongly disagree
    \item Somewhat disagree
    \item Neither agree nor disagree
    \item Somewhat agree
    \item Strongly agree
\end{enumerate} &
Is democracy preferable to other kinds of government?
\begin{enumerate}[leftmargin=*, nosep]
    \item[2.] Democracy may have its problems, but it is the best system of government
    \item[1.] In some circumstances, a non-democratic government can be preferable
    \item[0.] It doesn’t matter what kind of government we have
\end{enumerate}

\vspace{3pt}
To what extent do you agree or disagree with the following statements? 
\begin{itemize}[nosep]
    \item \textit{It is never justified, under any circumstances, for Americans to take violent action against the government}
\end{itemize}

\begin{enumerate}[leftmargin=*, nosep]
    \item Strongly disagree
    \item Somewhat disagree
    \item Neither agree nor disagree
    \item Somewhat agree
    \item Strongly agree
\end{enumerate} \\

Progressive economic justice preferences &
To what extent do you agree or disagree with the following statements?
\begin{itemize}[nosep]
    \item \textit{Government regulators should do more to prevent corporations using their market power to raise prices or reduce the size or quality of their products}
    \item \textit{Free markets should be entrusted to determine the cost of housing and home insurance}
\end{itemize}
\begin{enumerate}[leftmargin=*, nosep]
    \item Strongly disagree
    \item Somewhat disagree
    \item Neither agree nor disagree
    \item Somewhat agree
    \item Strongly agree
\end{enumerate} &
To what extent do you agree or disagree with the following statements?
\begin{itemize}[nosep]
    \item \textit{Government regulators should do more to prevent corporations using their market power to raise prices or reduce the size or quality of their products}
    \item \textit{Free markets should be entrusted to determine the cost of housing and home insurance}
\end{itemize}
\begin{enumerate}[leftmargin=*, nosep]
    \item Strongly disagree
    \item Somewhat disagree
    \item Neither agree nor disagree
    \item Somewhat agree
    \item Strongly agree
\end{enumerate} \\

Progressive public health preferences &
To what extent do you agree or disagree with the following statements?
\begin{itemize}[nosep]
    \item \textit{The government should intervene to reduce childcare costs and require employers to provide more family leave}
\end{itemize}
\begin{enumerate}[leftmargin=*, nosep]
    \item Strongly disagree
    \item Somewhat disagree
    \item Neither agree nor disagree
    \item Somewhat agree
    \item Strongly agree
\end{enumerate}

\vspace{3pt}
Do you favor an increase, decrease, or no change in government spending to help people pay for health insurance when they can’t pay for it all themselves?
\begin{enumerate}[leftmargin=*, nosep]
    \item Decrease
    \item No change
    \item Increase
    \item[88.] Don’t know
\end{enumerate}

\vspace{3pt}
Should access to abortion be guaranteed in all U.S. states?
\begin{enumerate}[leftmargin=*, nosep]
    \item[0.] No
    \item[1.] Yes
    \item[88.] Don't know
    \item[99.]Prefer not to say
\end{enumerate} &
To what extent do you agree or disagree with the following statements?
\begin{itemize}[nosep]
    \item \textit{The government should intervene to reduce childcare costs and require employers to provide more family leave}
\end{itemize}
\begin{enumerate}[leftmargin=*, nosep]
    \item Strongly disagree
    \item Somewhat disagree
    \item Neither agree nor disagree
    \item Somewhat agree
    \item Strongly agree
\end{enumerate}

\vspace{3pt}
Do you favor an increase, decrease, or no change in government spending to help people pay for health insurance when they can’t pay for it all themselves?
\begin{enumerate}[leftmargin=*, nosep]
    \item Decrease
    \item No change
    \item Increase
    \item[88.] Don’t know
\end{enumerate}

\vspace{3pt}
Should access to abortion be guaranteed in all U.S. states?
\begin{enumerate}[leftmargin=*, nosep]
    \item[0.] No
    \item[1.] Yes
    \item[88.] Don't know
    \item[99.]Prefer not to say
\end{enumerate}

\vspace{3pt}
To what extent do you agree or disagree that it is important for most people to receive annual vaccinations against the flu and COVID-19? 
\begin{enumerate}[leftmargin=*, nosep]
    \item Strongly disagree
    \item Somewhat disagree
    \item Neither agree nor disagree
    \item Somewhat agree
    \item Strongly agree
    \item[88.] Don't know
\end{enumerate} \\ \hline
%--------------------------------------------------------------------------
\SetCell[c=3]{l} \textit{H3: increased concern for progressive policy issues} \\ \hline

Salience of climate issues

Salience of democracy issues

Salience of economic justice issues

Salience of public health issues & 
Which of the following policy issues are most important to you? Pick up to 5
issues.
\begin{enumerate}[leftmargin=*, nosep]
    \item Health
    \item Foreign policy
    \item Income inequality
    \item Taxes
    \item Jobs and wage growth
    \item Inflation and prices
    \item Abortion and reproductive rights
    \item Gun rights
    \item Civil rights of racial or ethnic, gender, or sexual minorities
    \item Immigration
    \item Democracy
    \item Climate and the environment
    \item Crime
    \item Education
    \item[0.] None of the above
\end{enumerate} & 
Which of the following policy issues are most important to you? Pick up to 5
issues.
\begin{enumerate}[leftmargin=*, nosep]
    \item Health
    \item Foreign policy
    \item Income inequality
    \item Taxes
    \item Jobs and wage growth
    \item Inflation and prices
    \item Abortion and reproductive rights
    \item Gun rights
    \item Civil rights of racial or ethnic, gender, or sexual minorities
    \item Immigration
    \item Democracy
    \item Climate and the environment
    \item Crime
    \item Education
    \item[0.] None of the above
\end{enumerate}\\ \hline
%--------------------------------------------------------------------------
\SetCell[c=3]{l} \textit{H4: increased progressive political perspective} \\ \hline

Progressive policy preferences in general &
{To what extent do you agree or disagree with the following statements? 
\begin{itemize}[nosep]
    \item \textit{Government should do more to curb climate change, even at the expense of economic growth}
    \item \textit{The government should intervene to reduce childcare costs and require employers to provide more family leave}
    \item \textit{Government regulators should do more to prevent corporations using their market power to raise prices or reduce the size or quality of their products}
    \item \textit{Free markets should be entrusted to determine the cost of housing and home insurance}
    \item \textit{Cutting taxes is an effective way to increase economic growth}
\end{itemize}

\begin{enumerate}[leftmargin=*, nosep]
    \item Strongly disagree
    \item Somewhat disagree
    \item Neither agree nor disagree
    \item Somewhat agree
    \item Strongly agree
\end{enumerate}

\vspace{3pt}
How much do you support the implementation of policies aimed at reducing racial inequalities?
\begin{enumerate}[leftmargin=*, nosep]
    \item Do not support at all
    \item Support a little
    \item Moderately support
    \item Support a lot
    \item Fully support
\end{enumerate}

\vspace{3pt}
Do you favor an increase, decrease, or no change in government spending to help people pay for health insurance when they can’t pay for it all themselves?
\begin{enumerate}[leftmargin=*, nosep]
    \item Decrease
    \item No change
    \item Increase
    \item[88.] Don’t know
\end{enumerate}

\vspace{3pt}
Should access to abortion be guaranteed in all U.S. states?
\begin{enumerate}[leftmargin=*, nosep]
    \item[0.] No
    \item[1.] Yes
    \item[88.] Don't know
    \item[99.] Prefer not to say
\end{enumerate}} &
{To what extent do you agree or disagree with the following statements? 
\begin{itemize}[nosep]
    \item \textit{Government should do more to curb climate change, even at the expense of economic growth}
    \item \textit{The government should intervene to reduce childcare costs and require employers to provide more family leave}
    \item \textit{Government regulators should do more to prevent corporations using their market power to raise prices or reduce the size or quality of their products}
    \item \textit{Free markets should be entrusted to determine the cost of housing and home insurance}
    \item \textit{Cutting taxes is an effective way to increase economic growth}
\end{itemize}

\begin{enumerate}[leftmargin=*, nosep]
    \item Strongly disagree
    \item Somewhat disagree
    \item Neither agree nor disagree
    \item Somewhat agree
    \item Strongly agree
\end{enumerate}

\vspace{3pt}
How much do you support the implementation of policies aimed at reducing racial inequalities?
\begin{enumerate}[leftmargin=*, nosep]
    \item Do not support at all
    \item Support a little
    \item Moderately support
    \item Support a lot
    \item Fully support
\end{enumerate}

\vspace{3pt}
Do you favor an increase, decrease, or no change in government spending to help people pay for health insurance when they can’t pay for it all themselves?
\begin{enumerate}[leftmargin=*, nosep]
    \item Decrease
    \item No change
    \item Increase
    \item[88.] Don’t know
\end{enumerate}

\vspace{3pt}
Should access to abortion be guaranteed in all U.S. states?
\begin{enumerate}[leftmargin=*, nosep]
    \item[0.] No
    \item[1.] Yes
    \item[88.] Don't know
    \item[99.] Prefer not to say
\end{enumerate}

\vspace{3pt}
To what extent do you agree or disagree that it is important for most people to receive annual vaccinations against the flu and COVID-19? 
\begin{enumerate}[leftmargin=*, nosep]
    \item Strongly disagree
    \item Somewhat disagree
    \item Neither agree nor disagree
    \item Somewhat agree
    \item Strongly agree
    \item[88.] Don't know
\end{enumerate}} \\

Progressive worldview &
To what extent do you agree or disagree with the following statements?
\begin{itemize}[nosep]
    \item \textit{Addressing major challenges in society (like inflation or climate change) requires systemic changes, rather than individual actions by citizens or corporations}
    \item \textit{Active government intervention is needed to constrain the power of free markets and special interests}
    \item \textit{The current political system needs to be transformed to reduce long-running economic and racial inequalities in U.S. society}
    \item \textit{Democracy will only serve the interests of the masses if many citizens are informed and actively express their demands}
\end{itemize}

\begin{enumerate}[leftmargin=*, nosep]
    \item Strongly disagree
    \item Somewhat disagree
    \item Neither agree nor disagree
    \item Somewhat agree
    \item Strongly agree
\end{enumerate} &
To what extent do you agree or disagree with the following statements?
\begin{itemize}[nosep]
    \item \textit{Addressing major challenges in society (like inflation or climate change) requires systemic changes, rather than individual actions by citizens or corporations}
    \item \textit{Active government intervention is needed to constrain the power of free markets and special interests}
    \item \textit{The current political system needs to be transformed to reduce long-running economic and racial inequalities in U.S. society}
    \item \textit{Democracy will only serve the interests of the masses if many citizens are informed and actively express their demands}
\end{itemize}

\begin{enumerate}[leftmargin=*, nosep]
    \item Strongly disagree
    \item Somewhat disagree
    \item Neither agree nor disagree
    \item Somewhat agree
    \item Strongly agree
\end{enumerate} \\

Individual and group political efficacy &
To what extent do you agree or disagree with the following statements?
\begin{itemize}[nosep]
    \item \textit{I, as an individual, can influence political outcomes through my actions}
    \item \textit{Groups of like-minded can collectively bring about meaningful political change}
\end{itemize}
\begin{enumerate}[leftmargin=*, nosep]
    \item Strongly disagree
    \item Somewhat disagree
    \item Neither agree nor disagree
    \item Somewhat agree
    \item Strongly agree
\end{enumerate}

\vspace{3pt}
To what extent do you agree or disagree with the following statements about younger generations (e.g. Millennials, Gen Z) and their role in politics?
\begin{itemize}[nosep]
    \item \textit{Younger generations have the power to make a difference in politics}
\end{itemize}
\begin{enumerate}[leftmargin=*, nosep]
    \item Strongly disagree
    \item Somewhat disagree
    \item Neither agree nor disagree
    \item Somewhat agree
    \item Strongly agree
\end{enumerate} &
To what extent do you agree or disagree with the following statements?
\begin{itemize}[nosep]
    \item \textit{I, as an individual, can influence political outcomes through my actions}
    \item \textit{Groups of like-minded can collectively bring about meaningful political change}
\end{itemize}
\begin{enumerate}[leftmargin=*, nosep]
    \item Strongly disagree
    \item Somewhat disagree
    \item Neither agree nor disagree
    \item Somewhat agree
    \item Strongly agree
\end{enumerate}

\vspace{3pt}
To what extent do you agree or disagree with the following statements about younger generations (e.g. Millennials, Gen Z) and their role in politics?
\begin{itemize}[nosep]
    \item \textit{Younger generations have the power to make a difference in politics}
\end{itemize}

\begin{enumerate}[leftmargin=*, nosep]
    \item Strongly disagree
    \item Somewhat disagree
    \item Neither agree nor disagree
    \item Somewhat agree
    \item Strongly agree
\end{enumerate} \\

Trust in the system of electoral administration and experts &
To what extent do you agree or disagree with the following statements?
\begin{itemize}[nosep]
    \item \textit{Votes in U.S. elections are counted accurately and fairly}
\end{itemize}

\begin{enumerate}[leftmargin=*, nosep]
    \item Strongly disagree
    \item Somewhat disagree
    \item Neither agree nor disagree
    \item Somewhat agree
    \item Strongly agree
\end{enumerate}

\vspace{3pt}
Now please tell us the degree to which you personally trust or distrust the following institutions: \textit{State and local election boards}.
\begin{enumerate}[leftmargin=*, nosep]
    \item Completely distrust
    \item Somewhat distrust
    \item Neither distrust nor trust
    \item Somewhat trust
    \item Completely trust
\end{enumerate}

\vspace{3pt}
Next is a list of groups in society. Please tell us the degree to which you personally trust or distrust members of these groups: \textit{Scientists}, \textit{Doctors}, \textit{Journalists}.
\begin{enumerate}[leftmargin=*, nosep]
    \item Completely distrust
    \item Somewhat distrust
    \item Neither distrust nor trust
    \item Somewhat trust
    \item Completely trust
\end{enumerate} &
To what extent do you agree or disagree with the following statements?
\begin{itemize}[nosep]
    \item \textit{Votes in U.S. elections are counted accurately and fairly}
\end{itemize}

\begin{enumerate}[leftmargin=*, nosep]
    \item Strongly disagree
    \item Somewhat disagree
    \item Neither agree nor disagree
    \item Somewhat agree
    \item Strongly agree
\end{enumerate}

\vspace{3pt}
Now please tell us the degree to which you personally trust or distrust the following institutions: \textit{State and local election boards}.
\begin{enumerate}[leftmargin=*, nosep]
    \item Completely distrust
    \item Somewhat distrust
    \item Neither distrust nor trust
    \item Somewhat trust
    \item Completely trust
\end{enumerate}

\vspace{3pt}
Next is a list of groups in society. Please tell us the degree to which you personally trust or distrust members of these groups: \textit{Scientists}, \textit{Doctors}, \textit{Journalists}.
\begin{enumerate}[leftmargin=*, nosep]
    \item Completely distrust
    \item Somewhat distrust
    \item Neither distrust nor trust
    \item Somewhat trust
    \item Completely trust
\end{enumerate} \\ \hline
%--------------------------------------------------------------------------
\SetCell[c=3]{l} \textit{H5: increased favorability toward Democrats} \\ \hline

Feeling toward the Democratic party &
Next, we'd like you to rate how you feel toward political parties on a scale from 0 to 100, which we call a “feeling thermometer.” On this feeling thermometer, ratings between 0 and 49 degrees mean that you feel unfavorable and cold (with 0 being the most unfavorable/coldest). Ratings between 51 and 100 degrees mean that you feel favorable and warm (with 100 being the most favorable/warmest). A rating of 50 means you have no feelings one way or the other. Now, please tell me how warmly you feel toward
\begin{itemize}[nosep]
    \item The Democratic Party
\end{itemize} &
Next, we'd like you to rate how you feel toward political parties on a scale from 0 to 100, which we call a “feeling thermometer.” On this feeling thermometer, ratings between 0 and 49 degrees mean that you feel unfavorable and cold (with 0 being the most unfavorable/coldest). Ratings between 51 and 100 degrees mean that you feel favorable and warm (with 100 being the most favorable/warmest). A rating of 50 means you have no feelings one way or the other. Now, please tell me how warmly you feel toward
\begin{itemize}[nosep]
    \item The Democratic Party
\end{itemize} \\

Appraisal of Kamala Harris as presidential candidate &
In your opinion, how do you think the following statements apply to  Kamala Harris?
\begin{itemize}[nosep]
    \item \textit{Harris has effective economic policies that would benefit the country}
    \item \textit{Harris would manage international relations and foreign policy well}
    \item \textit{Harris exhibits trustworthiness and integrity in her actions and statements}
    \item \textit{Harris has a clear and beneficial vision for the future of America}
    \item \textit{Harris understands and looks out for people like you}
\end{itemize}

\begin{enumerate}[leftmargin=*, nosep]
    \item Strongly disagree
    \item Somewhat disagree
    \item Neither agree nor disagree
    \item Somewhat agree
    \item Strongly agree
\end{enumerate} &
In your opinion, how do you think the following statements apply to  Kamala Harris?
\begin{itemize}[nosep]
    \item \textit{Harris has effective economic policies that would benefit the country}
    \item \textit{Harris would manage international relations and foreign policy well}
    \item \textit{Harris exhibits trustworthiness and integrity in her actions and statements}
    \item \textit{Harris has a clear and beneficial vision for the future of America}
    \item \textit{Harris understands and looks out for people like you}
\end{itemize}

\begin{enumerate}[leftmargin=*, nosep]
    \item Strongly disagree
    \item Somewhat disagree
    \item Neither agree nor disagree
    \item Somewhat agree
    \item Strongly agree
\end{enumerate} \\

Appraisal of incumbent government performance &
Do you approve or disapprove of the way Joe Biden is handling his job as president?
\begin{enumerate}[leftmargin=*, nosep]
    \item[1.] Approve
    \item[0.] Disapprove
    \item[88.] Don't know
\end{enumerate}

\vspace{3pt}
Now please tell us the degree to which you personally trust or distrust the following institutions: \textit{President}
\begin{enumerate}[leftmargin=*, nosep]
    \item Completely distrust
    \item Somewhat distrust
    \item Neither distrust nor trust
    \item Somewhat trust
    \item Completely trust
\end{enumerate}

\vspace{3pt}
All in all, do you think that things in the nation are …?
\begin{enumerate}[leftmargin=*, nosep]
    \item[2.] Generally headed in the right direction
    \item[0.] On the wrong track
    \item[1.] Not sure what direction the country is headed in
\end{enumerate} &
Do you approve or disapprove of the way Joe Biden is handling his job as president?
\begin{enumerate}[leftmargin=*, nosep]
    \item[1.] Approve
    \item[0.] Disapprove
    \item[88.] Don't know
\end{enumerate}

\vspace{3pt}
Now please tell us the degree to which you personally trust or distrust the following institutions: \textit{President}
\begin{enumerate}[leftmargin=*, nosep]
    \item Completely distrust
    \item Somewhat distrust
    \item Neither distrust nor trust
    \item Somewhat trust
    \item Completely trust
\end{enumerate} \\ \hline
%--------------------------------------------------------------------------
\SetCell[c=3]{l} \textit{H6: decreased favorability toward Republicans} \\ \hline

Feeling toward the Republican party &
Next, we'd like you to rate how you feel toward political parties on a scale from 0 to 100, which we call a “feeling thermometer.” On this feeling thermometer, ratings between 0 and 49 degrees mean that you feel unfavorable and cold (with 0 being the most unfavorable/coldest). Ratings between 51 and 100 degrees mean that you feel favorable and warm (with 100 being the most favorable/warmest). A rating of 50 means you have no feelings one way or the other. Now, please tell me how warmly you feel toward
\begin{itemize}[nosep]
    \item The Republican Party
\end{itemize} &
Next, we'd like you to rate how you feel toward political parties on a scale from 0 to 100, which we call a “feeling thermometer.” On this feeling thermometer, ratings between 0 and 49 degrees mean that you feel unfavorable and cold (with 0 being the most unfavorable/coldest). Ratings between 51 and 100 degrees mean that you feel favorable and warm (with 100 being the most favorable/warmest). A rating of 50 means you have no feelings one way or the other. Now, please tell me how warmly you feel toward
\begin{itemize}[nosep]
    \item The Republican Party
\end{itemize} \\

Appraisal of Donald Trump as presidential candidate &
In your opinion, how do you think the following statements apply to Donald Trump?
\begin{itemize}[nosep]
    \item \textit{Trump has effective economic policies that would benefit the country}
    \item \textit{Trump would manage international relations and foreign policy well}
    \item \textit{Trump exhibits trustworthiness and integrity in her actions and statements}
    \item \textit{Trump has a clear and beneficial vision for the future of America}
    \item \textit{Trump understands and looks out for people like you}
\end{itemize}

\begin{enumerate}[leftmargin=*, nosep]
    \item Strongly disagree
    \item Somewhat disagree
    \item Neither agree nor disagree
    \item Somewhat agree
    \item Strongly agree
\end{enumerate} &
In your opinion, how do you think the following statements apply to Donald Trump?
\begin{itemize}[nosep]
    \item \textit{Trump has effective economic policies that would benefit the country}
    \item \textit{Trump would manage international relations and foreign policy well}
    \item \textit{Trump exhibits trustworthiness and integrity in her actions and statements}
    \item \textit{Trump has a clear and beneficial vision for the future of America}
    \item \textit{Trump understands and looks out for people like you}
\end{itemize}

\begin{enumerate}[leftmargin=*, nosep]
    \item Strongly disagree
    \item Somewhat disagree
    \item Neither agree nor disagree
    \item Somewhat agree
    \item Strongly agree
\end{enumerate} \\

Appraisal of Trump’s transition & &
Do you approve or disapprove of the way Donald Trump is handling his presidential transition? 
\begin{enumerate}[leftmargin=*, nosep]
    \item[1.] Approve
    \item[0.] Disapprove
    \item[88.] Don't know
\end{enumerate} \\ \hline
%--------------------------------------------------------------------------
\SetCell[c=3]{l} \textit{H7: increased electoral participation} \\ \hline

Probability of being registered to vote &
Many people are not, or not yet, registered to vote. Are you registered to vote for the upcoming elections?
\begin{enumerate}[leftmargin=*, nosep]
    \item[1.] Yes
    \item[0.] No
    \item[88.] Don’t know
    \item[99.] Prefer not to say
\end{enumerate} & \\

Probability of turning out to vote &
How likely is it that you will vote in the 2024 election for President?
\begin{enumerate}[leftmargin=*, nosep]
    \item[5.] I have already voted early or by mail
    \item[4.] Definitely will be voting
    \item[3.] Probably will be voting
    \item[2.] 50-50
    \item[1.] Probably won’t be voting
    \item[0.] Definitely won’t be voting
    \item[99.] Prefer not to say
\end{enumerate}
&
Did you vote in the 2024 election for President?
\begin{enumerate}[leftmargin=*, nosep]
    \item[1.] Yes
    \item[0.] No
    \item[99.] Prefer not to say
\end{enumerate}

\vspace{3pt}
Did you vote for a member of Congress in your district? 
\begin{enumerate}[leftmargin=*, nosep]
    \item[1.] Yes
    \item[0.] No
    \item[99.] Prefer not to say
\end{enumerate}

\vspace{3pt}
Did you vote for a member of the Senate in your state? 
\begin{enumerate}[leftmargin=*, nosep]
    \item[1.] Yes
    \item[0.] No
    \item[99.] Prefer not to say
\end{enumerate}
\\ \hline
%--------------------------------------------------------------------------
\SetCell[c=3]{l} \textit{H8: increased probability of voting Democrat} \\ \hline

Probability of voting for Kamala Harris for president &
{If the 2024 election for president were held today between former President Donald Trump (Republican) and Vice-President Kamala Harris (Democrat), who would you vote for or haven’t you decided?
\begin{enumerate}[leftmargin=*, nosep]
    \item Kamala Harris
    \item Donald Trump
    \item Someone else
    \item[88.] I haven’t decided yet
    \item[0.] I would not vote
    \item[99.] Prefer not to say
\end{enumerate}

\vspace{3pt}
Did you vote for former President Donald Trump (Republican), Vice-President Kamala Harris (Democrat), or someone else?
\begin{enumerate}[leftmargin=*, nosep]
    \item Kamala Harris
    \item Donald Trump
    \item Someone else
    \item[99.] Prefer not to say
\end{enumerate}}
&
Did you vote for former President Donald Trump (Republican), Vice-President Kamala Harris (Democrat), or someone else?
\begin{enumerate}[leftmargin=*, nosep]
    \item Kamala Harris
    \item Donald Trump
    \item Someone else
    \item[99.] Prefer not to say
\end{enumerate} \\

Probability of voting Democrat for House & &
Which party did you vote for in your district's election for the House of Representatives? 
\begin{enumerate}[leftmargin=*, nosep]
    \item Democratic Party
    \item Republican Party
    \item Another party or candidate
    \item[99.] Prefer not to say
\end{enumerate} \\

Probability of voting Democrat for Senate & &
Which party did you vote for in your state's election for the Senate? 
\begin{enumerate}[leftmargin=*, nosep]
    \item Democratic Party
    \item Republican Party
    \item Another party or candidate
    \item[99.] Prefer not to say
\end{enumerate} \\ \hline
%--------------------------------------------------------------------------
\SetCell[c=3]{l} \textit{H9: increased non-electoral participation} \\ \hline

Probability of non-electoral political action
&
Which, if any, of the following activities have you participated in over the last 6 months? Please select all that apply.
\begin{enumerate}[leftmargin=*, nosep]
    \item Participated in a march or protest
    \item Attended a political campaign event or rally
    \item Worked or volunteered for a political campaign
    \item Donated to a social or political organization
    \item Contacted or signed a petition sent to a politician
    \item Attempted to persuade a family member or friend to change a political position
    \item[0.] None of the above
    \item[99.] Prefer not to say
\end{enumerate}
&
Which, if any, of the following activities have you participated in between the elections in November and today? Please select all that apply.
\begin{enumerate}[leftmargin=*, nosep]
    \item Participated in a march or protest
    \item Attended a political campaign event or rally
    \item Worked or volunteered for a political campaign
    \item Donated to a social or political organization
    \item Contacted or signed a petition sent to a politician
    \item Attempted to persuade a family member or friend to change a political position
    \item[0.] None of the above
    \item[99.] Prefer not to say
\end{enumerate} \\

Probability of participating in a protest for a liberal cause
&
Which, if any, of the following activities have you participated in over the last 6 months? Please select all that apply.
\begin{itemize}
    \item Participated in a march or protest
\end{itemize}

What type of cause (or types of causes) were you protesting about?
\begin{enumerate}[leftmargin=*, nosep]
    \item Conservative
    \item Non-political
    \item Liberal
    \item Other
\end{enumerate}

&
Which, if any, of the following activities have you participated in between the elections in November and today? Please select all that apply. 
\begin{itemize}
    \item Participated in a march or protest
\end{itemize}

What type of cause (or types of causes) were you protesting about?
\begin{enumerate}[leftmargin=*, nosep]
    \item Conservative
    \item Non-political
    \item Liberal
    \item Other
\end{enumerate} \\

Probability of hypothetically resisting Trump administration
& &
If the incoming Trump administration implemented policies you strongly opposed, how do you think you would react? Please select all that apply. 
\begin{enumerate}[leftmargin=*, nosep]
    \item[0.] Do nothing
    \item[1.] Attend a protest
    \item[2.] Donate to an opposing political party or organization
    \item[3.] Contact one of your elected representatives
    \item[4.] Volunteer to work with an organization resisting these policies
    \item[5.] Sign a petition
    \item[6.] Criticize the policy or Trump administration online
    \item[7.] Criticize the policy or Trump administration in discussions with friends and family
    \item[8.] Do something else
    \item[88.] Don't know
\end{enumerate} \\

Probability of donating to a progressive cause & &
Next, we would like to know which organizations you would donate \$100 to. We will conduct a lottery to make a \$100 donation on behalf of 50 randomly selected participants. Please select up to 2 organizations to donate to. 
\begin{enumerate}[leftmargin=*, nosep]
    \item American Civil Liberties Union (ACLU) - Defends civil liberties, focusing on voting rights, access to abortions, racial justice, LGBTQ+ rights, and immigrant protections 
    \item Americans for Financial Reform (AFRED) - Advocates for stronger regulation of Wall Street to protect consumers and promote a fair and just financial system
    \item Environmental Defense Fund (EDF) - Combats climate change and supports renewable energy and environmental justice
    \item Planned Parenthood - Offers reproductive healthcare, including contraception, abortion services, and sexual education
    \item American Red Cross - Provides disaster relief, blood donations, and emergency assistance
    \item Wikipedia - Maintains a free online encyclopedia written by a community of volunteers
    \item Cato Institute - Advocates for libertarian ideas of smaller government, free markets, and reduced regulation
    \item Heartland Institute - Promotes free-market policies, questions climate change, and opposes many environmental regulations
    \item Heritage Foundation - Supports traditional values and free markets, wrote "Project 2025", and opposes Obamacare
    \item True the Vote - Advocates for stricter voting laws and focuses on preventing election fraud by training election monitors 
    \item[0.] Prefer not to donate to any of these causes
\end{enumerate} \\ \hline
%--------------------------------------------------------------------------
\SetCell[c=3]{l} \textit{H10: predominantly apolitical SMCs are more likely to increase trust} \\ \hline

Institutional trust in government
& 
Now please tell us the degree to which you personally trust or distrust the following institutions: \textit{President}, \textit{Congress}, \textit{Supreme Court}, \textit{State and local election boards}.
\begin{enumerate}[leftmargin=*, nosep]
    \item Completely distrust
    \item Somewhat distrust
    \item Neither distrust nor trust
    \item Somewhat trust
    \item Completely trust
\end{enumerate}
&
Now please tell us the degree to which you personally trust or distrust the following institutions: \textit{President}, \textit{Congress}, \textit{Supreme Court}, \textit{State and local election boards}.
\begin{enumerate}[leftmargin=*, nosep]
    \item Completely distrust
    \item Somewhat distrust
    \item Neither distrust nor trust
    \item Somewhat trust
    \item Completely trust
\end{enumerate}
\\

Interpersonal trust 
&
To what extent do you agree or disagree with the following statements?
\begin{itemize}[nosep]
    \item \textit{Most people can be trusted}
\end{itemize}
\begin{enumerate}[leftmargin=*, nosep]
    \item Strongly disagree
    \item Somewhat disagree
    \item Neither agree nor disagree
    \item Somewhat agree
    \item Strongly agree
\end{enumerate}
&
To what extent do you agree or disagree with the following statements?
\begin{itemize}[nosep]
    \item \textit{Most people can be trusted}
\end{itemize}
\begin{enumerate}[leftmargin=*, nosep]
    \item Strongly disagree
    \item Somewhat disagree
    \item Neither agree nor disagree
    \item Somewhat agree
    \item Strongly agree
\end{enumerate}
\end{longtblr}

\subsection{Correspondence with the pre-analysis plan \label{appendix:deviations}}

This study was preregistered in the Social Science Registry (\href{https://www.socialscienceregistry.org/trials/13994}{www.socialscienceregistry.org/trials/13994}); our pre-analysis plan (PAP) is provided in Appendix Section \ref{appendix:PAP}. Beyond our estimation strategy, the PAP provided a detailed plan for coding, aggregating, and testing variables relating to our ten primary hypotheses and three ``first stage'' hypotheses. Analysis followed the PAP exactly, except for several minor deviations we justify below.

We first explain how our results correspond to the estimation strategy detailed in section 4.2 of the PAP. Our tables report two types of pre-specified results: panel A reports estimates from the first and most general regression equation in the PAP (i.e. equation (1) in the PAP); and panel B focuses on comparisons between the quiz-incentivized conditions. The PAP had also proposed estimating equations that pool across encouragements (i.e. equations (2)-(4) in the PAP), but we elected to save space by not reporting these because Table \ref{table:FS1} demonstrates substantially greater compliance in quiz-incentivized groups. As anticipated in the PAP, this led us to focus on the quiz-incentivized conditions because these are the most powerful forms of treatment. Pooling across treatment encouragements would thus mask important differences in compliance. Nevertheless, pooled estimates can approximately be recovered by averaging across the encouragement-specific ATEs in panel A of our tables for each treatment condition; full tables for the pooled results are available upon request. 

Finally, we deviated from the PAP in three ways: \begin{enumerate}
    \item \textit{First stage outcome coding}: First, given the clear first stage effects, we save space in Tables \ref{table:FS1}, \ref{table:FS2}, and \ref{table:FS3} by not reporting ICW indexes aggregating the variables specified in hypotheses FS1-FS3. Second, we added the natural logarithm of the number of TikTok and YouTube videos consumed by assigned SMCs to Table \ref{table:FS1} (in addition to the pre-specified level) to reflect the right skew in the observed data. Third, for hypothesis FS2, we replaced our pre-specified measure of the quantity of news consumed on social media (given the difficulty of classifying news for respondents) with the total number of hours spent per week on Instagram, TikTok, and YouTube and added behavioral measures of (log) total social media consumption on TikTok and YouTube to Table \ref{table:FS2} in order to illuminate \textit{total} social media consumption. %Fourth, for hypothesis FS2, we also added behavioral measure of political and liberal content consumption based on GPT-4o-mini's classification of participants' YouTube watch histories.
    \item \textit{Cause donation outcome coding}: The sole change to our primary outcomes was to analyze the cause donation lottery endline outcome separately, instead of as part of hypothesis H9 concerning non-electoral political participation. We did so because this is an artificial form of non-electoral participation: donating is part of the survey, not something people organically opted to do. Consequently, it better serves as a behavioral measure of policy preferences than a mode of political participation. In addition to our pre-specified indicator for donating to any of the four liberal causes in columns (11)-(12), the cause donation results in Table \ref{table:cause_donations} include several further outcomes that were not pre-specified: the analogous indicator for donating to any of the four conservative causes in columns (13)-(14); topic-specific ICW indexes combining an indicator for donating to the liberal cause and an (reversed) indicator for donating to the conservative cause; and an overall ICW index that combine all the topic-specific indicators in columns (1)-(2). 
    \item \textit{Multiple testing}: Our PAP proposed applying the \cite{benjamini1995} correction for multiple hypothesis testing to our primary outcomes and treatment groups. We instead follow similar recent interventions \citep[e.g.][]{allcott2020welfare, broockman2025consuming, chen2019impact} in applying \citeauthor{anderson2008multiple}'s \citeyearpar{anderson2008multiple} slightly sharper approach to applying the \cite{benjamini1995} correction. Our adjustments for the false discovery rate focuses on 94 tests of the effect of the quiz-incentivized treatment conditions: up to 4 tests (PA and PP SMCs conditions relative to the pure control and NP SMCs condition, without LASSO covariates) for each of the 27 primary outcome indices (13 pre-specified midline hypotheses, 13 pre-specified endline hypotheses, and the cause donation lottery index just described). The sharpened $q$ values are reported in Table \ref{table:sharpened}, and are only slightly less conservative than results using the \cite{benjamini1995} correction.
\end{enumerate}

\subsection{Research ethics \label{appendix:ethics}}

Beyond receiving IRB approval, our study reflected careful attention to the ethics of field experimentation and associated data collection consistent with the American Political Science Association's \textit{Principles and Guidance for Human Subjects Research}. Below, we outline the considerations and measures taken to ensure that our field experiment is ethical and aligned with established social scientific principles. 

First, participants in this study were encouraged, but not forced, to follow SMCs without deception. All participants provided informed consent before the study began, understanding that their participation---including the possibility of being asked to follow new SMCs---was voluntary and that they could withdraw from the study at any time without penalty. Our encouragements to follow new accounts mirror the recommendations participants encounter organically in their daily lives on platforms such as Instagram, TikTok, and YouTube or via friends. The additional provision of financial incentives to consume media content for some participants is a common approach in the literature to ensure high levels of treatment exposure \citep[e.g.][]{bowles2023sustaining, broockman2025consuming, chen2019impact}.

Second, the study's SMC recommendations were tailored to match participants' interests and preferences, as determined through a baseline survey and the machine learning matching algorithm described in Appendix Section \ref{appendix:matching_algorithm}. This not only increased the likelihood that participants would enjoy recommended content, but also reduced any perception of manipulation or undue pressure. Since treated participants could choose not to consume recommended content (or SMS/email messages in that case of that treatment), our expectation was that participants would not be harmed---and would likely benefit---by participating in the study.

Third, participants received appropriate compensation for their time completing surveys, with payment levels designed to be fair and non-coercive. Participants were paid \$3.43 in survey platform credits for completing the baseline survey and \$5.40 for completing the midline and endline surveys; additional incentives were provided for participants who are slow to respond to surveys.

Fourth, quota sampling ensured that our participant pool was diverse and broadly representative of American adults aged 18–45 in terms of gender, age, education, and partisan identification. Efforts were also made to exclude individuals who are not aligned with the study's focus, namely those who infrequently use Instagram, TikTok, or YouTube. This approach ensured that participants could engage meaningfully with the study while minimizing unnecessary burden on individuals outside the intended scope.

Fifth, all collected data were anonymized and securely stored to safeguard participants' privacy. We also collected some personally identifiable information---such as social media account details and corresponding platform-level engagement data---to verify participants' social media user status and engagement beyond survey responses, which participants were informed about before starting the study. The endline survey offered participants the option to receive \$5 per platform upon sharing their TikTok and YouTube video histories; step-by-step instructions were provided on how to do this and avoid sharing additional personal information.

Finally, the small scale of the randomized intervention and the geographic dispersion of study participants across the US ensured that the intervention could not plausibly have influenced 2024 election outcomes.

\subsection{Identification checks}

To validate the experimental design, we examine differential attrition and covariate balance. First, the tests of the null hypothesis at the foot of Table \ref{table:attrition} largely confirm---most clearly at midline---that treatment conditions experienced indistinguishable (and generally low) levels of attrition from the baseline sample. The endline survey shows slightly higher survey completion rates among participants who received incentives to consume PP SMCs. Turning to our behavioral data, quiz-incentivized participants were somewhat more willing to provide TikTok and YouTube browsing history data, although differences across quiz-incentivized conditions were statistically indistinguishable. Second, Table \ref{table:baseline_balance} shows that our treatment conditions are mean-balanced across 50 predetermined covariates, while Tables \ref{table:midline_balance}-\ref{table:endline_balance} show that this continues to hold in the midline and endline survey samples.

\begin{table}
\caption{Attrition}
\vspace{-12pt}
\label{table:attrition}
\begin{center}
\scalebox{0.88}{
\begin{tabular}{lcc|cc|ccc}
\toprule
& \multicolumn{2}{c|}{\textbf{\textit{Midline survey}}} & \multicolumn{2}{c|}{\textbf{\textit{Endline survey}}} & \multicolumn{3}{c}{\textbf{\textit{Provided behavioral data}}} \\
& \textbf{Started} & \textbf{Completed all} & \textbf{Started} & \textbf{Completed all} & \textbf{TikTok} & \textbf{YouTube} & \textbf{Matched} \\
& \textbf{survey} & \textbf{questions} & \textbf{survey} & \textbf{questions} & \textbf{history} & \textbf{history} & \textbf{voter file} \\
& (1) & (2) & (3) & (4) & (5) & (6) & (7) \\
\midrule
PA SMCs (quiz)      &      -0.027   &      -0.030   &       0.009   &       0.026   &       0.029** &       0.054*  &       0.048   \\
                    &     (0.027)   &     (0.027)   &     (0.026)   &     (0.026)   &     (0.013)   &     (0.028)   &     (0.031)   \\
PA SMCs (follow)    &      -0.011   &      -0.003   &       0.034   &       0.051** &       0.009   &       0.052*  &       0.006   \\
                    &     (0.025)   &     (0.025)   &     (0.025)   &     (0.026)   &     (0.011)   &     (0.028)   &     (0.030)   \\
\end{tabular}
}
\end{center}
\vspace{-10pt}
\footnotesize \textit{Notes}: Each specification is estimated using OLS, and includes randomization block fixed effects. Sample sizes differ between we exclude respondents who failed both attention checks with the corresponding survey. Two-sided tests: * $p<0.1$, ** $p<0.05$, *** $p<0.01$. 
\end{table}

\begin{sidewaystable}
\caption{Baseline balance tests}
\vspace{-12pt}
\label{table:baseline_balance}
\begin{center}
\scalebox{0.55}{
\begin{tabular}{lcccccccccccccccccc}
\toprule
& \textbf{Days using} & \textbf{Days using} & \textbf{Days using} & \textbf{Hours using} & \textbf{Hours using} & \textbf{Hours using} & \textbf{Regularly} & \textbf{Regularly} & \textbf{Regularly} & \textbf{Weekly videos} & \textbf{Encounter} & \textbf{Encounter} & \textbf{Encounter} & \textbf{Encounter} & \textbf{Encounter} & \textbf{Connectedness} & \textbf{Social media} \\
& \textbf{Instagram} & \textbf{TikTok} & \textbf{YouTube} & \textbf{Instagram} & \textbf{TikTok} & \textbf{YouTube} & \textbf{use social} & \textbf{use social} & \textbf{use social} & \textbf{watched from} & \textbf{SMCs on} & \textbf{SMCs through} & \textbf{SMCs by} & \textbf{SMCs on} & \textbf{SMCs in} & \textbf{to} & \textbf{accounts} \\
& \textbf{in average} & \textbf{in average} & \textbf{in average} & \textbf{per} & \textbf{per} & \textbf{per} & \textbf{media on} & \textbf{media on} & \textbf{media on} & \textbf{followed} & \textbf{social media} & \textbf{friends} & \textbf{searching} & \textbf{other} & \textbf{other} & \textbf{SMCs} & \textbf{watch} \\
& \textbf{week} & \textbf{week} & \textbf{week} & \textbf{week} & \textbf{week} & \textbf{week} & \textbf{phone} & \textbf{tablet} & \textbf{computer} & \textbf{accounts} & \textbf{platforms} & \textbf{or family} & \textbf{social media} & \textbf{forums} & \textbf{ways} & \textbf{followed} & \textbf{most of} \\
& (1) & (2) & (3) & (4) & (5) & (6) & (7) & (8) & (9) & (10) & (11) & (12) & (13) & (14) & (15) & (16) & (17) \\
\midrule
PA SMCs (quiz)      &      -0.285   &       0.036   &      -0.007   &      -0.043   &      -0.068   &       0.816   &       0.007   &      -0.014   &       0.005   &       0.362   &       0.021   &      -0.004   &       0.012   &       0.063** &       0.019   &       0.002   &      -0.703   \\
\end{tabular}
}
\end{center}
\vspace{-10pt}
\footnotesize \textit{Notes}: Each specification is estimated using OLS, and includes randomization block fixed effects. Two-sided tests: * $p<0.1$, ** $p<0.05$, *** $p<0.01$. 
\end{sidewaystable}

\begin{sidewaystable}
\caption*{Table \ref{table:baseline_balance}: Baseline balance tests (continued)}
\vspace{-12pt}
\begin{center}
\scalebox{0.62}{
\begin{tabular}{lcccccccccccccccccc}
\toprule
& \textbf{Preference} & \textbf{Openness} & & & & & & & & & & & & & \textbf{American} & \textbf{At least a} & \textbf{At least} \\
& \textbf{for SMCs to} & \textbf{to} & \textbf{Prefers} & \textbf{Prefers} & \textbf{Prefers} & \textbf{Prefers} & & & & & & & & & \textbf{Indian, Alaskan,} & \textbf{high school} & \textbf{some} \\
& \textbf{talk about} & \textbf{following} & \textbf{fashionable} & \textbf{polished} & \textbf{formal} & \textbf{assertive} & & & & & & & & & \textbf{Pacific} & \textbf{diploma (or} & \textbf{college} \\
& \textbf{politics} & \textbf{new SMCs} & \textbf{SMCs} & \textbf{SMCs} & \textbf{SMCs} & \textbf{SMCs} & \textbf{Age} & \textbf{Male} & \textbf{Female} & \textbf{Non-binary} & \textbf{White} & \textbf{Black} & \textbf{Hispanic} & \textbf{Asian} & \textbf{Islander} & \textbf{equivalent)} & \textbf{credit} \\
& (1) & (2) & (3) & (4) & (5) & (6) & (7) & (8) & (9) & (10) & (11) & (12) & (13) & (14) & (15) & (16) & (17) \\
\midrule
PA SMCs (quiz)      &       0.052   &      -0.040   &       0.121** &      -0.022   &       0.044   &       0.016   &       0.372   &      -0.043   &       0.035   &       0.008   &      -0.029   &      -0.001   &      -0.022   &       0.002   &      -0.021** &      -0.008   &      -0.017   \\
                    &     (0.062)   &     (0.071)   &     (0.059)   &     (0.052)   &     (0.058)   &     (0.074)   &     (0.371)   &     (0.032)   &     (0.032)   &     (0.008)   &     (0.029)   &     (0.027)   &     (0.023)   &     (0.015)   &     (0.011)   &     (0.011)   &     (0.028)   \\
PA SMCs (follow)    &       0.116*  &       0.057   &       0.054   &      -0.024   &      -0.019   &       0.016   &      -0.503   &       0.005   &      -0.012   &       0.008   &      -0.039   &       0.015   &      -0.036   &       0.016   &      -0.021*  &       0.005   &      -0.050*  \\
\end{tabular}
}
\end{center}
\vspace{-10pt}
\footnotesize \textit{Notes}: Each specification is estimated using OLS, and includes randomization block fixed effects. Two-sided tests: * $p<0.1$, ** $p<0.05$, *** $p<0.01$. 
\end{sidewaystable}

\begin{sidewaystable}
\caption*{Table \ref{table:baseline_balance}: Baseline balance tests (continued)}
\vspace{-12pt}
\begin{center}
\scalebox{0.58}{
\begin{tabular}{lccccccccccccccccc}
\toprule
& \textbf{At least a} & \textbf{Completed} & & & & & \textbf{Has at} & & \textbf{Think of} & \textbf{Think} & \textbf{Think of} & & \textbf{Approve} & & & \textbf{Intend} & \textbf{Intend} \\
& \textbf{complete} & \textbf{masters,} & & & & \textbf{Annual} & \textbf{least} & & \textbf{self as} & \textbf{of self} & \textbf{self as} & \textbf{Liberal-} & \textbf{of} & & \textbf{Intend} & \textbf{to} & \textbf{to} \\
& \textbf{bachelor's} & \textbf{professional,} & & & & \textbf{household} & \textbf{one} & \textbf{Follow} & \textbf{Democrat or} & \textbf{as} & \textbf{Republican or} & \textbf{conservative} & \textbf{President} & \textbf{Registered} & \textbf{to} & \textbf{vote} & \textbf{vote} \\
& \textbf{degree} & \textbf{or doctoral} & \textbf{Heterosexual} & \textbf{Homosexual} & \textbf{Bisexual} & \textbf{income} & \textbf{child} & \textbf{politics} & \textbf{strong Democrat} & \textbf{independent} & \textbf{strong Republican} & \textbf{scale} & \textbf{Biden} & \textbf{to vote} & \textbf{vote} & \textbf{Harris} & \textbf{Trump} \\
& (1) & (2) & (3) & (4) & (5) & (6) & (7) & (8) & (9) & (10) & (11) & (12) & (13) & (14) & (15) & (16) & (17) \\
\midrule
PA SMCs (quiz)      &      -0.029   &      -0.009   &      -0.070***&       0.002   &       0.046** &   -2180.221   &       0.012   &       0.002   &       0.017   &      -0.032   &       0.016   &      -0.017   &      -0.023   &      -0.001   &      -0.012   &      -0.024   &       0.018   \\
                    &     (0.031)   &     (0.022)   &     (0.026)   &     (0.014)   &     (0.022)   &  (3539.723)   &     (0.031)   &     (0.063)   &     (0.029)   &     (0.024)   &     (0.024)   &     (0.036)   &     (0.027)   &     (0.024)   &     (0.027)   &     (0.028)   &     (0.027)   \\
\end{tabular}
}
\end{center}
\vspace{-10pt}
\footnotesize \textit{Notes}: Each specification is estimated using OLS, and includes randomization block fixed effects. Two-sided tests: * $p<0.1$, ** $p<0.05$, *** $p<0.01$. 
\end{sidewaystable}

\begin{sidewaystable}
\caption{Baseline balance tests among midline sample}
\vspace{-12pt}
\label{table:midline_balance}
\begin{center}
\scalebox{0.55}{
\begin{tabular}{lcccccccccccccccccc}
\toprule
& \textbf{Days using} & \textbf{Days using} & \textbf{Days using} & \textbf{Hours using} & \textbf{Hours using} & \textbf{Hours using} & \textbf{Regularly} & \textbf{Regularly} & \textbf{Regularly} & \textbf{Weekly videos} & \textbf{Encounter} & \textbf{Encounter} & \textbf{Encounter} & \textbf{Encounter} & \textbf{Encounter} & \textbf{Connectedness} & \textbf{Social media} \\
& \textbf{Instagram} & \textbf{TikTok} & \textbf{YouTube} & \textbf{Instagram} & \textbf{TikTok} & \textbf{YouTube} & \textbf{use social} & \textbf{use social} & \textbf{use social} & \textbf{watched from} & \textbf{SMCs on} & \textbf{SMCs through} & \textbf{SMCs by} & \textbf{SMCs on} & \textbf{SMCs in} & \textbf{to} & \textbf{accounts} \\
& \textbf{in average} & \textbf{in average} & \textbf{in average} & \textbf{per} & \textbf{per} & \textbf{per} & \textbf{media on} & \textbf{media on} & \textbf{media on} & \textbf{followed} & \textbf{social media} & \textbf{friends} & \textbf{searching} & \textbf{other} & \textbf{other} & \textbf{SMCs} & \textbf{watch} \\
& \textbf{week} & \textbf{week} & \textbf{week} & \textbf{week} & \textbf{week} & \textbf{week} & \textbf{phone} & \textbf{tablet} & \textbf{computer} & \textbf{accounts} & \textbf{platforms} & \textbf{or family} & \textbf{social media} & \textbf{forums} & \textbf{ways} & \textbf{followed} & \textbf{most of} \\
& (1) & (2) & (3) & (4) & (5) & (6) & (7) & (8) & (9) & (10) & (11) & (12) & (13) & (14) & (15) & (16) & (17) \\
\midrule
PA SMCs (quiz)      &      -0.364*  &       0.022   &      -0.089   &      -0.321   &       0.252   &       0.235   &      -0.008   &      -0.030   &       0.009   &       0.544   &       0.043   &      -0.004   &      -0.013   &       0.081** &       0.021   &       0.069   &      -0.147   \\
\end{tabular}
}
\end{center}
\vspace{-10pt}
\footnotesize \textit{Notes}: Each specification is estimated using OLS, and includes randomization block fixed effects. Two-sided tests: * $p<0.1$, ** $p<0.05$, *** $p<0.01$. 
\end{sidewaystable}

\begin{sidewaystable}
\caption*{Table \ref{table:midline_balance}: Baseline balance tests among midline sample (continued)}
\vspace{-12pt}
\begin{center}
\scalebox{0.62}{
\begin{tabular}{lcccccccccccccccccc}
\toprule
& \textbf{Preference} & \textbf{Openness} & & & & & & & & & & & & & \textbf{American} & \textbf{At least a} & \textbf{At least} \\
& \textbf{for SMCs to} & \textbf{to} & \textbf{Prefers} & \textbf{Prefers} & \textbf{Prefers} & \textbf{Prefers} & & & & & & & & & \textbf{Indian, Alaskan,} & \textbf{high school} & \textbf{some} \\
& \textbf{talk about} & \textbf{following} & \textbf{fashionable} & \textbf{polished} & \textbf{formal} & \textbf{assertive} & & & & & & & & & \textbf{Pacific} & \textbf{diploma (or} & \textbf{college} \\
& \textbf{politics} & \textbf{new SMCs} & \textbf{SMCs} & \textbf{SMCs} & \textbf{SMCs} & \textbf{SMCs} & \textbf{Age} & \textbf{Male} & \textbf{Female} & \textbf{Non-binary} & \textbf{White} & \textbf{Black} & \textbf{Hispanic} & \textbf{Asian} & \textbf{Islander} & \textbf{equivalent)} & \textbf{credit} \\
& (1) & (2) & (3) & (4) & (5) & (6) & (7) & (8) & (9) & (10) & (11) & (12) & (13) & (14) & (15) & (16) & (17) \\
\midrule
PA SMCs (quiz)      &       0.038   &      -0.009   &       0.162** &       0.042   &       0.067   &       0.005   &       0.281   &      -0.029   &       0.018   &       0.010   &      -0.023   &       0.002   &      -0.023   &      -0.007   &      -0.021   &      -0.014   &      -0.010   \\
                    &     (0.073)   &     (0.082)   &     (0.069)   &     (0.061)   &     (0.069)   &     (0.089)   &     (0.425)   &     (0.037)   &     (0.037)   &     (0.010)   &     (0.034)   &     (0.032)   &     (0.027)   &     (0.017)   &     (0.014)   &     (0.014)   &     (0.033)   \\
PA SMCs (follow)    &       0.065   &       0.075   &       0.066   &       0.001   &      -0.017   &      -0.014   &      -0.625   &      -0.011   &      -0.004   &       0.015   &      -0.043   &       0.016   &      -0.034   &       0.015   &      -0.025*  &       0.011   &      -0.054   \\
\end{tabular}
}
\end{center}
\vspace{-10pt}
\footnotesize \textit{Notes}: Each specification is estimated using OLS, and includes randomization block fixed effects. Two-sided tests: * $p<0.1$, ** $p<0.05$, *** $p<0.01$. 
\end{sidewaystable}

\begin{sidewaystable}
\caption*{Table \ref{table:midline_balance}: Baseline balance tests among midline sample (continued)}
\vspace{-12pt}
\begin{center}
\scalebox{0.58}{
\begin{tabular}{lccccccccccccccccc}
\toprule
& \textbf{At least a} & \textbf{Completed} & & & & & \textbf{Has at} & & \textbf{Think of} & \textbf{Think} & \textbf{Think of} & & \textbf{Approve} & & & \textbf{Intend} & \textbf{Intend} \\
& \textbf{complete} & \textbf{masters,} & & & & \textbf{Annual} & \textbf{least} & & \textbf{self as} & \textbf{of self} & \textbf{self as} & \textbf{Liberal-} & \textbf{of} & & \textbf{Intend} & \textbf{to} & \textbf{to} \\
& \textbf{bachelor's} & \textbf{professional,} & & & & \textbf{household} & \textbf{one} & \textbf{Follow} & \textbf{Democrat or} & \textbf{as} & \textbf{Republican or} & \textbf{conservative} & \textbf{President} & \textbf{Registered} & \textbf{to} & \textbf{vote} & \textbf{vote} \\
& \textbf{degree} & \textbf{or doctoral} & \textbf{Heterosexual} & \textbf{Homosexual} & \textbf{Bisexual} & \textbf{income} & \textbf{child} & \textbf{politics} & \textbf{strong Democrat} & \textbf{independent} & \textbf{strong Republican} & \textbf{scale} & \textbf{Biden} & \textbf{to vote} & \textbf{vote} & \textbf{Harris} & \textbf{Trump} \\
& (1) & (2) & (3) & (4) & (5) & (6) & (7) & (8) & (9) & (10) & (11) & (12) & (13) & (14) & (15) & (16) & (17) \\
\midrule
PA SMCs (quiz)      &      -0.005   &       0.023   &      -0.069** &       0.003   &       0.050*  &     -52.136   &      -0.013   &       0.032   &       0.053   &      -0.037   &      -0.015   &      -0.037   &       0.021   &       0.014   &       0.032   &       0.006   &      -0.025   \\
                    &     (0.036)   &     (0.026)   &     (0.030)   &     (0.016)   &     (0.026)   &  (3880.494)   &     (0.036)   &     (0.075)   &     (0.034)   &     (0.028)   &     (0.027)   &     (0.042)   &     (0.033)   &     (0.027)   &     (0.032)   &     (0.032)   &     (0.031)   \\
\end{tabular}
}
\end{center}
\vspace{-10pt}
\footnotesize \textit{Notes}: Each specification is estimated using OLS, and includes randomization block fixed effects. Two-sided tests: * $p<0.1$, ** $p<0.05$, *** $p<0.01$. 
\end{sidewaystable}

\begin{sidewaystable}
\caption{Baseline balance tests among endline sample}
\vspace{-12pt}
\label{table:endline_balance}
\begin{center}
\scalebox{0.55}{
\begin{tabular}{lcccccccccccccccccc}
\toprule
& \textbf{Days using} & \textbf{Days using} & \textbf{Days using} & \textbf{Hours using} & \textbf{Hours using} & \textbf{Hours using} & \textbf{Regularly} & \textbf{Regularly} & \textbf{Regularly} & \textbf{Weekly videos} & \textbf{Encounter} & \textbf{Encounter} & \textbf{Encounter} & \textbf{Encounter} & \textbf{Encounter} & \textbf{Connectedness} & \textbf{Social media} \\
& \textbf{Instagram} & \textbf{TikTok} & \textbf{YouTube} & \textbf{Instagram} & \textbf{TikTok} & \textbf{YouTube} & \textbf{use social} & \textbf{use social} & \textbf{use social} & \textbf{watched from} & \textbf{SMCs on} & \textbf{SMCs through} & \textbf{SMCs by} & \textbf{SMCs on} & \textbf{SMCs in} & \textbf{to} & \textbf{accounts} \\
& \textbf{in average} & \textbf{in average} & \textbf{in average} & \textbf{per} & \textbf{per} & \textbf{per} & \textbf{media on} & \textbf{media on} & \textbf{media on} & \textbf{followed} & \textbf{social media} & \textbf{friends} & \textbf{searching} & \textbf{other} & \textbf{other} & \textbf{SMCs} & \textbf{watch} \\
& \textbf{week} & \textbf{week} & \textbf{week} & \textbf{week} & \textbf{week} & \textbf{week} & \textbf{phone} & \textbf{tablet} & \textbf{computer} & \textbf{accounts} & \textbf{platforms} & \textbf{or family} & \textbf{social media} & \textbf{forums} & \textbf{ways} & \textbf{followed} & \textbf{most of} \\
& (1) & (2) & (3) & (4) & (5) & (6) & (7) & (8) & (9) & (10) & (11) & (12) & (13) & (14) & (15) & (16) & (17) \\
\midrule
PA SMCs (quiz)      &      -0.367*  &       0.097   &      -0.027   &       0.222   &       0.311   &       0.818   &      -0.005   &      -0.024   &       0.014   &       0.469   &       0.021   &      -0.006   &      -0.007   &       0.075** &       0.025   &       0.086   &      -0.128   \\
\end{tabular}
}
\end{center}
\vspace{-10pt}
\footnotesize \textit{Notes}: Each specification is estimated using OLS, and includes randomization block fixed effects. Two-sided tests: * $p<0.1$, ** $p<0.05$, *** $p<0.01$. 
\end{sidewaystable}

\begin{sidewaystable}
\caption*{Table \ref{table:endline_balance}: Baseline balance tests among endline sample (continued)}
\vspace{-12pt}
\begin{center}
\scalebox{0.62}{
\begin{tabular}{lcccccccccccccccccc}
\toprule
& \textbf{Preference} & \textbf{Openness} & & & & & & & & & & & & & \textbf{American} & \textbf{At least a} & \textbf{At least} \\
& \textbf{for SMCs to} & \textbf{to} & \textbf{Prefers} & \textbf{Prefers} & \textbf{Prefers} & \textbf{Prefers} & & & & & & & & & \textbf{Indian, Alaskan,} & \textbf{high school} & \textbf{some} \\
& \textbf{talk about} & \textbf{following} & \textbf{fashionable} & \textbf{polished} & \textbf{formal} & \textbf{assertive} & & & & & & & & & \textbf{Pacific} & \textbf{diploma (or} & \textbf{college} \\
& \textbf{politics} & \textbf{new SMCs} & \textbf{SMCs} & \textbf{SMCs} & \textbf{SMCs} & \textbf{SMCs} & \textbf{Age} & \textbf{Male} & \textbf{Female} & \textbf{Non-binary} & \textbf{White} & \textbf{Black} & \textbf{Hispanic} & \textbf{Asian} & \textbf{Islander} & \textbf{equivalent)} & \textbf{credit} \\
& (1) & (2) & (3) & (4) & (5) & (6) & (7) & (8) & (9) & (10) & (11) & (12) & (13) & (14) & (15) & (16) & (17) \\
\midrule
PA SMCs (quiz)      &       0.055   &      -0.029   &       0.165** &       0.023   &       0.085   &      -0.005   &       0.161   &      -0.034   &       0.024   &       0.010   &      -0.020   &       0.021   &      -0.026   &      -0.010   &      -0.026*  &      -0.019   &      -0.023   \\
                    &     (0.072)   &     (0.081)   &     (0.068)   &     (0.061)   &     (0.069)   &     (0.088)   &     (0.417)   &     (0.037)   &     (0.037)   &     (0.009)   &     (0.033)   &     (0.032)   &     (0.026)   &     (0.017)   &     (0.014)   &     (0.013)   &     (0.033)   \\
PA SMCs (follow)    &       0.110   &       0.060   &       0.067   &      -0.006   &      -0.007   &      -0.002   &      -1.122** &       0.007   &      -0.022   &       0.014   &      -0.040   &       0.016   &      -0.017   &       0.016   &      -0.030** &       0.006   &      -0.054*  \\
\end{tabular}
}
\end{center}
\vspace{-10pt}
\footnotesize \textit{Notes}: Each specification is estimated using OLS, and includes randomization block fixed effects. Two-sided tests: * $p<0.1$, ** $p<0.05$, *** $p<0.01$. 
\end{sidewaystable}

\begin{sidewaystable}
\caption*{Table \ref{table:endline_balance}: Baseline balance tests among endline sample (continued)}
\vspace{-12pt}
\begin{center}
\scalebox{0.58}{
\begin{tabular}{lccccccccccccccccc}
\toprule
& \textbf{At least a} & \textbf{Completed} & & & & & \textbf{Has at} & & \textbf{Think of} & \textbf{Think} & \textbf{Think of} & & \textbf{Approve} & & & \textbf{Intend} & \textbf{Intend} \\
& \textbf{complete} & \textbf{masters,} & & & & \textbf{Annual} & \textbf{least} & & \textbf{self as} & \textbf{of self} & \textbf{self as} & \textbf{Liberal-} & \textbf{of} & & \textbf{Intend} & \textbf{to} & \textbf{to} \\
& \textbf{bachelor's} & \textbf{professional,} & & & & \textbf{household} & \textbf{one} & \textbf{Follow} & \textbf{Democrat or} & \textbf{as} & \textbf{Republican or} & \textbf{conservative} & \textbf{President} & \textbf{Registered} & \textbf{to} & \textbf{vote} & \textbf{vote} \\
& \textbf{degree} & \textbf{or doctoral} & \textbf{Heterosexual} & \textbf{Homosexual} & \textbf{Bisexual} & \textbf{income} & \textbf{child} & \textbf{politics} & \textbf{strong Democrat} & \textbf{independent} & \textbf{strong Republican} & \textbf{scale} & \textbf{Biden} & \textbf{to vote} & \textbf{vote} & \textbf{Harris} & \textbf{Trump} \\
& (1) & (2) & (3) & (4) & (5) & (6) & (7) & (8) & (9) & (10) & (11) & (12) & (13) & (14) & (15) & (16) & (17) \\
\midrule
PA SMCs (quiz)      &      -0.000   &       0.023   &      -0.079***&       0.004   &       0.050** &   -1179.323   &      -0.003   &       0.044   &       0.027   &      -0.029   &      -0.004   &      -0.029   &       0.016   &       0.008   &       0.017   &      -0.022   &       0.006   \\
                    &     (0.037)   &     (0.026)   &     (0.030)   &     (0.016)   &     (0.025)   &  (4008.687)   &     (0.036)   &     (0.074)   &     (0.034)   &     (0.028)   &     (0.027)   &     (0.042)   &     (0.033)   &     (0.027)   &     (0.032)   &     (0.033)   &     (0.031)   \\
\end{tabular}
}
\end{center}
\vspace{-10pt}
\footnotesize \textit{Notes}: Each specification is estimated using OLS, and includes randomization block fixed effects. Two-sided tests: * $p<0.1$, ** $p<0.05$, *** $p<0.01$. 
\end{sidewaystable}

\subsubsection{Additional results}

\begin{figure}

\begin{subfigure}{0.5\textwidth}
\begin{center}
\caption{Compliance with predominantly-apolitical SMCs (quiz) treatment}
\vspace{6pt}
\includegraphics[scale=0.42]{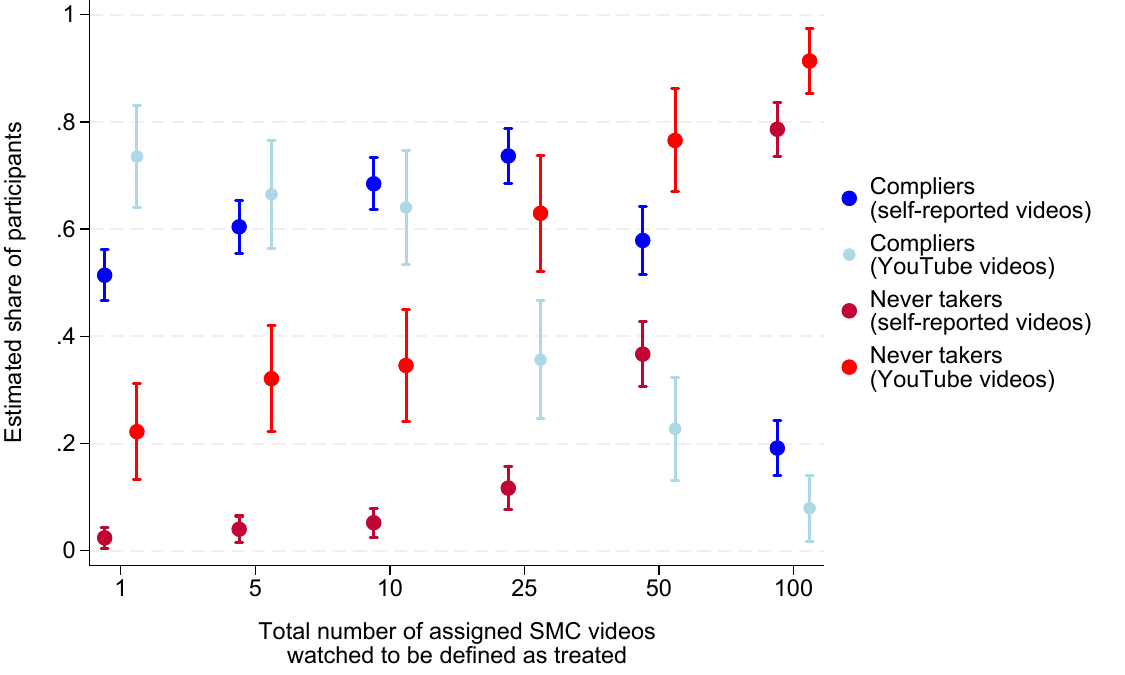}
\label{figure:complier_rate_apolitical}
\end{center}
\end{subfigure}%
\begin{subfigure}{0.5\textwidth}
\begin{center}
\caption{Compliance with predominantly-political SMCs (quiz) treatment}
\vspace{6pt}
\includegraphics[scale=0.42]{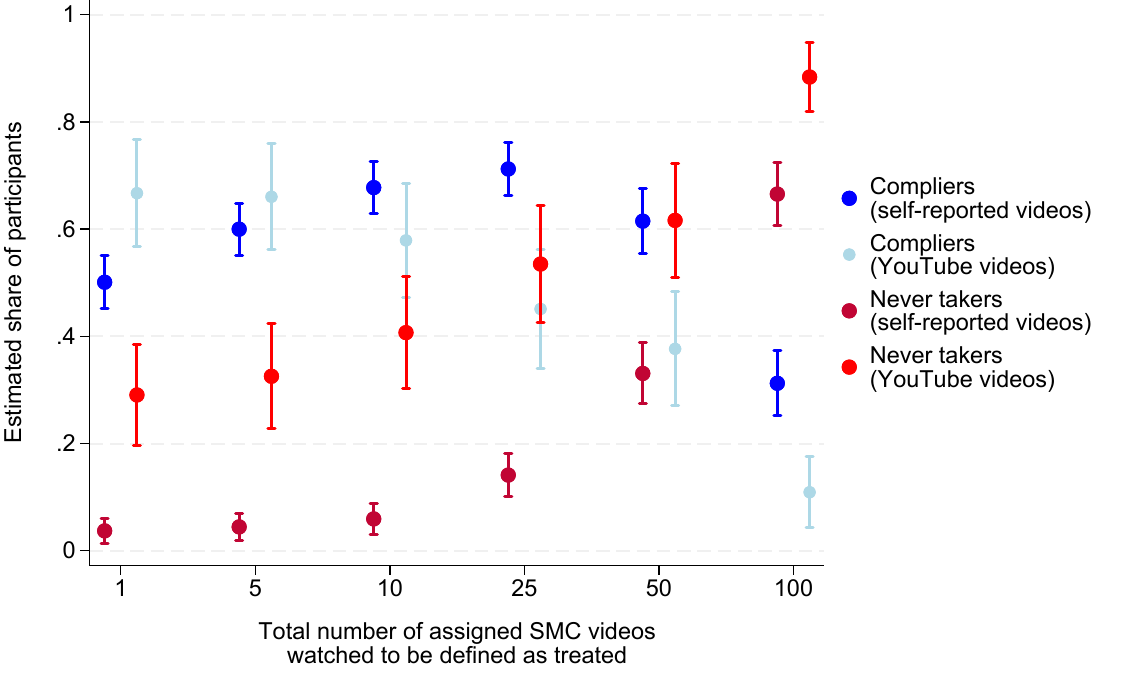}
\label{figure:complier_rate_political}
\end{center}
\end{subfigure}%

\vspace{-4pt}

\begin{subfigure}{0.5\textwidth}
\begin{center}
\caption{Compliance with placebo SMCs (quiz) treatment}
\vspace{6pt}
\includegraphics[scale=0.42]{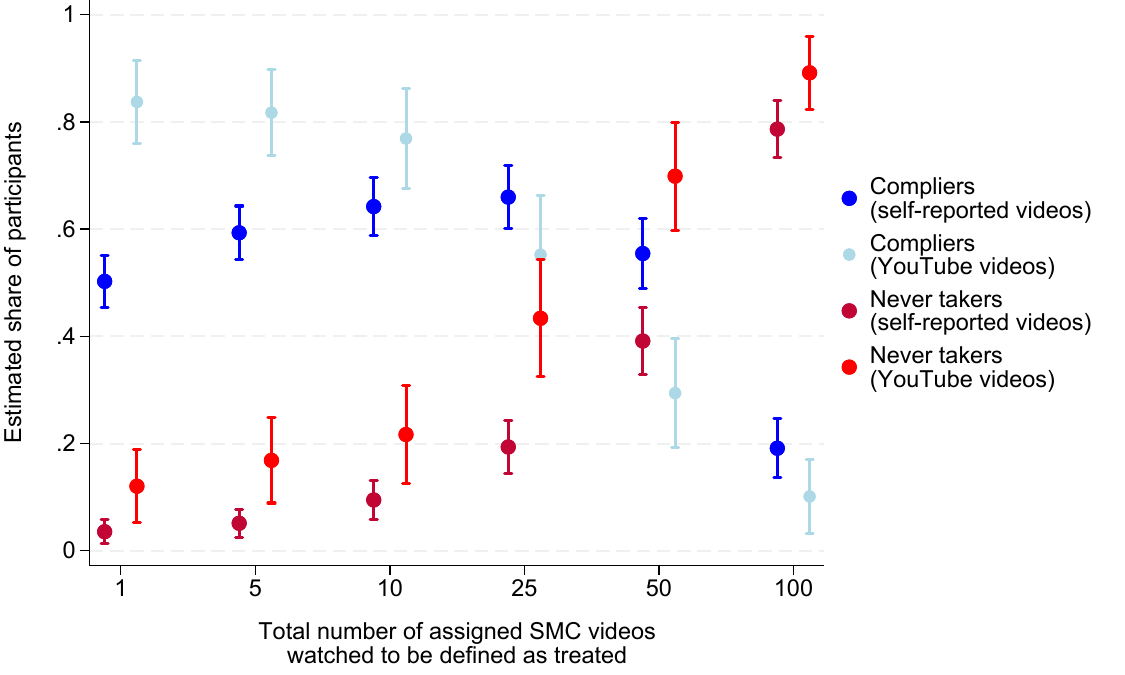}
\label{figure:complier_rate_placebo}
\end{center}
\end{subfigure}%

\vspace{-18pt}

\begin{center}
\caption{Compliance with quiz-incentivized treatment conditions as a function of number of videos watched}
\label{figure:compliance_rates}
\end{center}

\begin{tablenotes}
\vspace{-16pt}
\item {\footnotesize \textit{Note}: All estimates implement the method describing in \cite{marbach2020profiling} in Stata, varying binary definitions of the exposure treatment by the number of videos watched and whether this is self-reported and based on YouTube watch histories.}
\end{tablenotes}

\end{figure}

The remaining figures and tables report additional results of interest. These include levels of compliers (Figure \ref{figure:compliance_rates}) and types of compliance (Tables \ref{table:complier_types_self_reports} and \ref{table:complier_types_youtube}), effects on consumption of longer-form SMCs (Table \ref{table:FS1_long_form}) and non-assigned SMCs (Table \ref{table:FS1_other}), effects on non-assigned content more generally (Table \ref{table:FS2_nonassigned}), effects on offline activity (Table \ref{table:FS3}), full estimates from our main analyses for all encouragement conditions (Tables \ref{table:H1_full}-\ref{table:H10_full}), sharpened $q$ values generated by the \cite{anderson2008multiple} procedure for multiple testing (Table \ref{table:sharpened}), various analyses of potential mechanisms, and effects on continued consumption of assigned SMCs (Table \ref{table:continued}). 

\begin{sidewaystable}
\caption{Treatments effects on consumption of recommended long-form SMCs}
\vspace{-12pt}
\label{table:FS1_long_form}
\begin{center}
\scalebox{0.82}{
\begin{tabular}{lcccc|cccc|cc}
\toprule
& \multicolumn{4}{c|}{\textbf{\textit{Midline survey outcomes}}} & \multicolumn{4}{c|}{\textbf{\textit{Endline survey outcomes}}} & \multicolumn{2}{c}{\textbf{\textit{Browsing history watch counts during intervention}}} \\
& \multicolumn{2}{c}{\textbf{Longer-form SMC}} & \multicolumn{2}{c|}{\textbf{Longer-form SMC}} & \multicolumn{2}{c}{\textbf{Longer-form SMC}} & \multicolumn{2}{c|}{\textbf{Longer-form SMC}} & \multicolumn{1}{c}{\textbf{YouTube videos}} & \multicolumn{1}{c}{\textbf{YouTube videos}} \\
& \multicolumn{2}{c}{\textbf{consumption}} & \multicolumn{2}{c|}{\textbf{consumption}} & \multicolumn{2}{c}{\textbf{consumption}} & \multicolumn{2}{c|}{\textbf{consumption}} & \multicolumn{1}{c}{\textbf{by assigned}} & \multicolumn{1}{c}{\textbf{by assigned}} \\
& \multicolumn{2}{c}{\textbf{frequency}} & \multicolumn{2}{c|}{\textbf{at all}}& \multicolumn{2}{c}{\textbf{frequency index}} & \multicolumn{2}{c|}{\textbf{at all}} & \multicolumn{1}{c}{\textbf{SMCs}} & \multicolumn{1}{c}{\textbf{SMCs (log)}} \\
& (1) & (2) & (3) & (4) & (5) & (6) & (7) & (8) & (9) & (10) \\
\midrule
\textit{Panel A: differences relative to control condition} \\ Predominantly-apolitical SMCs (quiz)&    0.277***   &    0.261***   &    0.163***   &    0.154***   &     0.187**   &     0.154**   &    0.087***   &    0.080***   &       0.106   &       0.030   \\
                    &     (0.093)   &     (0.079)   &     (0.036)   &     (0.031)   &     (0.084)   &     (0.071)   &     (0.034)   &     (0.029)   &     (0.144)   &     (0.043)   \\
\end{tabular}
}
\end{center}
\vspace{-10pt}
\footnotesize \textit{Notes}: Each specification is estimated using OLS, and includes randomization block fixed effects (with the exception of the smaller samples for YouTube behavioral outcomes). Except for the YouTube behavioral outcome, even-numbered columns additionally include predetermined covariates selected by the \cite{belloni2014inference} LASSO procedure, and their (demeaned) interaction with each treatment condition. Robust standard errors are in parentheses. Two-sided tests (not pre-specified): * $p<0.1$, ** $p<0.05$, *** $p<0.01$.
\end{sidewaystable}

\begin{sidewaystable}
\begin{center}
\caption{Baseline characteristics of compliers, never-takers, and always-takers, where compliance is defined by consuming at least five videos from assigned SMCs (self-reported)}
\label{table:complier_types_self_reports}
\scalebox{0.67}{
\begin{tabular}{lccccccccc}
\toprule
& \multicolumn{3}{c}{\textbf{Predominantly-apolitical SMCs}} & \multicolumn{3}{c}{\textbf{Predominantly-political SMCs}} & \multicolumn{3}{c}{\textbf{Placebo SMCs}} \\
& \textbf{Compliers} & \textbf{Never-takers} & \textbf{Always-takers} & \textbf{Compliers} & \textbf{Never-takers} & \textbf{Always-takers} & \textbf{Compliers} & \textbf{Never-takers} & \textbf{Always-takers} \\
\midrule
Days using Instagram in average week&4.01&4.80&3.40&4.04&4.80&3.50&4.01&4.80&3.31\\
Days using TikTok in average week&2.92&4.04&3.00&2.78&4.04&2.17&2.76&4.04&2.08\\
Days using YouTube in average week&5.49&5.59&6.10&5.63&5.59&5.33&5.69&5.59&6.08\\
Hours using Instagram per week&5.33&7.35&4.30&4.76&7.35**&2.58&4.40&7.35&3.46\\
Hours using TikTok per week&4.75&7.23&3.30&4.27&7.23&2.92&4.29&7.23&2.62\\
Hours using YouTube per week&12.01&12.30&11.20&12.35&12.30**&8.83&11.08&12.30&13.46\\
\end{tabular}
}
\end{center}
\vspace{-4pt}
\footnotesize \textit{Notes}: Compliance shares are computed according to \cite{marbach2020profiling}. Two-sided tests of equality with each covariate complier mean: $^{*}$ $p<0.1$, $^{**}$ $p<0.05$, $^{***}$ $p<0.01$.
\end{sidewaystable}

\begin{sidewaystable}
\begin{center}
\caption{Baseline characteristics of compliers, never-takers, and always-takers, where compliance is defined by consuming at least five videos from assigned SMCs (YouTube browsing history)}
\label{table:complier_types_youtube}
\scalebox{0.67}{
\begin{tabular}{lccccccccc}
\toprule
& \multicolumn{3}{c}{\textbf{Predominantly-apolitical SMCs}} & \multicolumn{3}{c}{\textbf{Predominantly-political SMCs}} & \multicolumn{3}{c}{\textbf{Placebo SMCs}} \\
& \textbf{Compliers} & \textbf{Never-takers} & \textbf{Always-takers} & \textbf{Compliers} & \textbf{Never-takers} & \textbf{Always-takers} & \textbf{Compliers} & \textbf{Never-takers} & \textbf{Always-takers} \\
\midrule
Days using Instagram in average week&4.05&3.50&4.81&3.82&3.50&5.21&3.99&3.50&5.43\\
Days using TikTok in average week&3.29&3.50&4.00&3.24&3.50&3.75&3.72&3.50&2.93\\
Days using YouTube in average week&6.60&7.00***&4.58&6.80&7.00***&4.57&6.17&7.00&5.57\\
Hours using Instagram per week&7.54&5.00**&4.50&6.12&5.00&6.36&5.37&5.00&6.21\\
Hours using TikTok per week&6.18&5.00&5.15&5.53&5.00&4.73&5.72&5.00&3.93\\
Hours using YouTube per week&14.37&16.50&10.96&16.39&16.50***&8.21&13.85&16.50&12.00\\
\end{tabular}
}
\end{center}
\vspace{-4pt}
\footnotesize \textit{Notes}: Compliance shares are computed according to \cite{marbach2020profiling}. Two-sided tests of equality with each covariate complier mean: $^{*}$ $p<0.1$, $^{**}$ $p<0.05$, $^{***}$ $p<0.01$.
\end{sidewaystable}

\begin{sidewaystable}
\caption{Treatments effects on consumption of non-assigned SMCs}
\vspace{-12pt}
\label{table:FS1_other}
\begin{center}
\scalebox{0.65}{
\begin{tabular}{lcccccccc|cccccccc}
\toprule
& \multicolumn{8}{c|}{\textbf{\textit{Midline survey outcomes}}} & \multicolumn{8}{c}{\textbf{\textit{Endline survey outcomes}}} \\
& \multicolumn{2}{c}{\textbf{Average other SMC}} & \multicolumn{2}{c}{\textbf{Average other SMC}} & \multicolumn{2}{c}{\textbf{Average famous SMC}} & \multicolumn{2}{c|}{\textbf{Average famous SMC}} & \multicolumn{2}{c}{\textbf{Average other SMC}} & \multicolumn{2}{c}{\textbf{Average other SMC}} & \multicolumn{2}{c}{\textbf{Average political SMC}} & \multicolumn{2}{c}{\textbf{Average political SMC}} \\
& \multicolumn{2}{c}{\textbf{consumption}} & \multicolumn{2}{c}{\textbf{consumption}} & \multicolumn{2}{c}{\textbf{consumption}} & \multicolumn{2}{c|}{\textbf{consumption}} & \multicolumn{2}{c}{\textbf{consumption}} & \multicolumn{2}{c}{\textbf{consumption}} & \multicolumn{2}{c}{\textbf{consumption}} & \multicolumn{2}{c}{\textbf{consumption}} \\
& \multicolumn{2}{c}{\textbf{frequency index}} & \multicolumn{2}{c}{\textbf{at all}}& \multicolumn{2}{c}{\textbf{frequency index}} & \multicolumn{2}{c|}{\textbf{at all}}& \multicolumn{2}{c}{\textbf{frequency index}} & \multicolumn{2}{c}{\textbf{at all}} & \multicolumn{2}{c}{\textbf{frequency index}} & \multicolumn{2}{c}{\textbf{at all}} \\
& (1) & (2) & (3) & (4) & (5) & (6) & (7) & (8) & (9) & (10) & (11) & (12) & (13) & (14) & (15) & (16) \\
\midrule
\textit{Panel A: differences relative to control condition} \\ Predominantly-apolitical SMCs (quiz)&     0.204**   &    0.187***   &    0.090***   &    0.085***   &       0.060   &       0.049   &       0.032   &       0.031   &       0.076   &       0.055   &     0.059**   &     0.053**   &      -0.021   &      -0.045   &       0.003   &      -0.006   \\
\end{tabular}
}
\end{center}
\vspace{-10pt}
\footnotesize \textit{Notes}: Each specification is estimated using OLS, and includes randomization block fixed effects. In even columns, we use the \cite{belloni2014inference} LASSO procedure to further select predetermined covariates and their (demeaned) interaction with each treatment condition. Robust standard errors are in parentheses. Two-sided tests (not pre-specified): * $p<0.1$, ** $p<0.05$, *** $p<0.01$.
\end{sidewaystable}

\begin{sidewaystable}
\caption{Treatment effects on non-assigned YouTube content (including recommendation and follower incentive conditions)}
\vspace{-12pt}
\label{table:FS2_nonassigned}
\begin{center}
\scalebox{0.79}{
\begin{tabular}{lccccc|ccccc}
\toprule
& \multicolumn{5}{c|}{\textbf{YouTube videos watched before November 2024}} & \multicolumn{5}{c}{\textbf{YouTube videos watched throughout the intervention}} \\
& \multicolumn{1}{c}{\textbf{Political}} & \multicolumn{1}{c}{\textbf{Liberal}} & \multicolumn{1}{c}{\textbf{Liberal}} & \multicolumn{1}{c}{\textbf{Democrat}} & \multicolumn{1}{c|}{} & \multicolumn{1}{c}{\textbf{Political}} & \multicolumn{1}{c}{\textbf{Liberal}} & \multicolumn{1}{c}{\textbf{Liberal}} & \multicolumn{1}{c}{\textbf{Democrat}} \\
& \multicolumn{1}{c}{\textbf{content}} & \multicolumn{1}{c}{\textbf{values}} & \multicolumn{1}{c}{\textbf{lean}} & \multicolumn{1}{c}{\textbf{support}} & \multicolumn{1}{c|}{\textbf{Mobilization}} & \multicolumn{1}{c}{\textbf{content}} & \multicolumn{1}{c}{\textbf{values}} & \multicolumn{1}{c}{\textbf{lean}} & \multicolumn{1}{c}{\textbf{support}} & \multicolumn{1}{c}{\textbf{Mobilization}} \\
& \multicolumn{1}{c}{\textbf{index}} & \multicolumn{1}{c}{\textbf{index}} & \multicolumn{1}{c}{\textbf{index}} & \multicolumn{1}{c}{\textbf{index}} & \multicolumn{1}{c|}{\textbf{index}} & \multicolumn{1}{c}{\textbf{index}} & \multicolumn{1}{c}{\textbf{index}} & \multicolumn{1}{c}{\textbf{index}} & \multicolumn{1}{c}{\textbf{index}} & \multicolumn{1}{c}{\textbf{index}} \\
& (1) & (2) & (3) & (4) & (5) & (6) & (7) & (8) & (9) & (10) \\
\midrule
\textit{Panel A: differences relative to control condition} \\ Predominantly-apolitical SMCs (quiz)&     -0.153*   &       0.002   &       0.001   &       0.001   &      -0.122   &    -0.197**   &       0.018   &      -0.054   &       0.073   &    -0.201**   \\
                    &     (0.085)   &     (0.113)   &     (0.101)   &     (0.081)   &     (0.106)   &     (0.083)   &     (0.118)   &     (0.089)   &     (0.085)   &     (0.102)   \\
\end{tabular}
}
\end{center}
\vspace{-10pt}
\footnotesize \textit{Notes}: Each specification is estimated using OLS. Robust standard errors are in parentheses. Pre-specified one- and two-sided tests: * $p<0.1$, ** $p<0.05$, *** $p<0.01$. Two-sided tests of estimates in the opposite of a pre-specified one-sided test: $^{+}$ $p<0.1$, $^{++}$ $p<0.05$, $^{+++}$ $p<0.01$.
\end{sidewaystable}

\begin{table}
\caption{Treatment effects on off-social media activity}
\vspace{-12pt}
\label{table:FS3}
\begin{center}
\scalebox{0.82}{
\begin{tabular}{lcccc|cccc}
\toprule
& \multicolumn{4}{c|}{\textbf{\textit{Midline survey outcomes}}} & \multicolumn{4}{c}{\textbf{\textit{Endline survey outcomes}}} \\
& \multicolumn{2}{c}{\textbf{Traditional media}} & \multicolumn{2}{c|}{\textbf{Offline activity}} & \multicolumn{2}{c}{\textbf{Traditional media}} & \multicolumn{2}{c}{\textbf{Offline activity}}  \\
& \multicolumn{2}{c}{\textbf{consumption}} & \multicolumn{2}{c|}{\textbf{count}} & \multicolumn{2}{c}{\textbf{consumption}} & \multicolumn{2}{c}{\textbf{count}} \\
& (1) & (2) & (3) & (4) & (5) & (6) & (7) & (8) \\
\midrule
\textit{Panel A: differences relative to control condition} \\ Predominantly-apolitical SMCs (quiz)&      -0.068   &      -0.247   &       0.037   &       0.007   &       0.029   &      -0.219   &       0.052   &       0.072   \\
                    &     (1.232)   &     (0.942)   &     (0.152)   &     (0.126)   &     (1.199)   &     (1.016)   &     (0.154)   &     (0.133)   \\
%p value            &       0.478   &       0.397   &       0.809   &       0.953   &       0.981   &       0.415   &       0.737   &       0.587   \\
Predominantly-apolitical SMCs (follow)&       0.625   &       0.715   &      -0.074   &       0.001   &    -2.016**   &    -1.634**   &      -0.108   &      -0.060   \\
\bottomrule
\end{tabular}
}
\end{center}
\vspace{-10pt}
\footnotesize \textit{Notes}: Each specification is estimated using OLS, and includes randomization block fixed effects. Even-numbered columns additionally include predetermined covariates selected by the \cite{belloni2014inference} LASSO procedure, and their (demeaned) interaction with each treatment condition. Robust standard errors are in parentheses. Pre-specified one- and two-sided tests: * $p<0.1$, ** $p<0.05$, *** $p<0.01$. Two-sided tests of estimates in the opposite of a pre-specified one-sided test: $^{+}$ $p<0.1$, $^{++}$ $p<0.05$, $^{+++}$ $p<0.01$.
\end{table}

\begin{table}
\caption{Treatment effects on political engagement in general (including recommendation and follower incentive conditions)}
\vspace{-12pt}
\label{table:H1_full}
\begin{center}
\scalebox{0.58}{

}
\end{center}
\vspace{-10pt}
\footnotesize \textit{Notes}: Each specification is estimated using OLS, and includes randomization block fixed effects. Even-numbered columns additionally include predetermined covariates selected by the \cite{belloni2014inference} LASSO procedure, and their (demeaned) interaction with each treatment condition. Robust standard errors are in parentheses. Pre-specified one- and two-sided tests: * $p<0.1$, ** $p<0.05$, *** $p<0.01$. Two-sided tests of estimates in the opposite of a pre-specified one-sided test: $^{+}$ $p<0.1$, $^{++}$ $p<0.05$, $^{+++}$ $p<0.01$.
\end{table}

\begin{sidewaystable}
\caption{Treatment effects on liberal policy attitudes (including recommendation and follower incentive conditions)}
\vspace{-12pt}
\label{table:H2_full}
\begin{center}
\scalebox{0.53}{

}
\end{center}
\vspace{-10pt}
\footnotesize \textit{Notes}: Each specification is estimated using OLS, and includes randomization block fixed effects and adjusts for lagged outcomes and their (demeaned) interaction with each treatment condition. Even-numbered columns additionally include predetermined covariates selected by the \cite{belloni2014inference} LASSO procedure, and their (demeaned) interaction with each treatment condition. Robust standard errors are in parentheses. Pre-specified one- and two-sided tests: * $p<0.1$, ** $p<0.05$, *** $p<0.01$. Two-sided tests of estimates in the opposite of a pre-specified one-sided test: $^{+}$ $p<0.1$, $^{++}$ $p<0.05$, $^{+++}$ $p<0.01$.
\end{sidewaystable}

\begin{sidewaystable}
\caption{Treatment effects on progressive issue salience}
\vspace{-12pt}
\label{table:H3_full}
\begin{center}
\scalebox{0.53}{

}
\end{center}
\vspace{-10pt}
\footnotesize \textit{Notes}: Each specification is estimated using OLS, and includes randomization block fixed effects and adjusts for lagged outcomes and their (demeaned) interaction with each treatment condition. Even-numbered columns additionally include predetermined covariates selected by the \cite{belloni2014inference} LASSO procedure, and their (demeaned) interaction with each treatment condition. Robust standard errors are in parentheses. Pre-specified one- and two-sided tests: * $p<0.1$, ** $p<0.05$, *** $p<0.01$. Two-sided tests of estimates in the opposite of a pre-specified one-sided test: $^{+}$ $p<0.1$, $^{++}$ $p<0.05$, $^{+++}$ $p<0.01$.
\end{sidewaystable}

\begin{table}
\caption{Treatment effects on endline cause donation decisions (including recommendation and follower incentive conditions)}
\vspace{-12pt}
\label{table:cause_donations_full}
\begin{center}
\scalebox{0.56}{

}
\end{center}
\vspace{-10pt}
\footnotesize \textit{Notes}: Each specification is estimated using OLS, and includes randomization block fixed effects, an approximate baseline outcome, and the interaction between the (demeaned) baseline outcome and each treatment condition. Even-numbered columns additionally include predetermined covariates selected by the \cite{belloni2014inference} LASSO procedure, and their (demeaned) interaction with each treatment condition. Robust standard errors are in parentheses. Pre-specified one- and two-sided tests: * $p<0.1$, ** $p<0.05$, *** $p<0.01$. Two-sided tests of estimates in the opposite of a pre-specified one-sided test: $^{+}$ $p<0.1$, $^{++}$ $p<0.05$, $^{+++}$ $p<0.01$.
\end{table}

\begin{sidewaystable}
\caption{Treatment effects on progressive outlooks (including recommendation and follower incentive conditions)}
\vspace{-12pt}
\label{table:H4_full}
\begin{center}
\scalebox{0.53}{

}
\end{center}
\vspace{-10pt}
\footnotesize \textit{Notes}: Each specification is estimated using OLS, and includes randomization block fixed effects and adjusts for lagged outcomes and their (demeaned) interaction with each treatment condition. Even-numbered columns additionally include predetermined covariates selected by the \cite{belloni2014inference} LASSO procedure, and their (demeaned) interaction with each treatment condition. Robust standard errors are in parentheses. Pre-specified one- and two-sided tests: * $p<0.1$, ** $p<0.05$, *** $p<0.01$. Two-sided tests of estimates in the opposite of a pre-specified one-sided test: $^{+}$ $p<0.1$, $^{++}$ $p<0.05$, $^{+++}$ $p<0.01$.
\end{sidewaystable}

\begin{sidewaystable}
\caption{Treatment effects on favorability toward the Democratic Party (including recommendation and follower incentive conditions)}
\vspace{-12pt}
\label{table:H5_full}
\begin{center}
\scalebox{0.53}{

}
\end{center}
\vspace{-10pt}
\footnotesize \textit{Notes}: Each specification is estimated using OLS, and includes randomization block fixed effects and adjusts for lagged outcomes and their (demeaned) interaction with each treatment condition. Even-numbered columns additionally include predetermined covariates selected by the \cite{belloni2014inference} LASSO procedure, and their (demeaned) interaction with each treatment condition. Robust standard errors are in parentheses. Pre-specified one- and two-sided tests: * $p<0.1$, ** $p<0.05$, *** $p<0.01$. Two-sided tests of estimates in the opposite of a pre-specified one-sided test: $^{+}$ $p<0.1$, $^{++}$ $p<0.05$, $^{+++}$ $p<0.01$.
\end{sidewaystable}

\begin{table}
\caption{Treatment effects on favorability toward the Republican Party (including recommendation and follower incentive conditions)}
\vspace{-12pt}
\label{table:H6_full}
\begin{center}
\scalebox{0.53}{

}
\end{center}
\vspace{-10pt}
\footnotesize \textit{Notes}: Each specification is estimated using OLS, and includes randomization block fixed effects and adjusts for lagged outcomes and their (demeaned) interaction with each treatment condition. Even-numbered columns additionally include predetermined covariates selected by the \cite{belloni2014inference} LASSO procedure, and their (demeaned) interaction with each treatment condition. Robust standard errors are in parentheses. Pre-specified one- and two-sided tests: * $p<0.1$, ** $p<0.05$, *** $p<0.01$. Two-sided tests of estimates in the opposite of a pre-specified one-sided test: $^{+}$ $p<0.1$, $^{++}$ $p<0.05$, $^{+++}$ $p<0.01$.
\end{table}

\begin{table}
\caption{Treatment effects on electoral participation (including recommendation and follower incentive conditions)}
\vspace{-12pt}
\label{table:H7_full}
\begin{center}
\scalebox{0.67}{

}
\end{center}
\vspace{-10pt}
\footnotesize \textit{Notes}: Each specification is estimated using OLS, and includes randomization block fixed effects (with the exception of the smaller samples for voter file outcomes) and adjusts for lagged outcomes and their (demeaned) interaction with each treatment condition. Except for the voter file data, even-numbered columns additionally include predetermined covariates selected by the \cite{belloni2014inference} LASSO procedure, and their (demeaned) interaction with each treatment condition. Robust standard errors are in parentheses. Pre-specified one- and two-sided tests: * $p<0.1$, ** $p<0.05$, *** $p<0.01$. Two-sided tests of estimates in the opposite of a pre-specified one-sided test: $^{+}$ $p<0.1$, $^{++}$ $p<0.05$, $^{+++}$ $p<0.01$.
\end{table}

\begin{table}
\caption{Treatment effects on self-reported voting decisions (including recommendation and follower incentive conditions)}
\vspace{-12pt}
\label{table:H8_full}
\begin{center}
\scalebox{0.62}{

}
\end{center}
\vspace{-10pt}
\footnotesize \textit{Notes}: Each specification is estimated using OLS, and includes randomization block fixed effects and adjusts for lagged outcomes and their (demeaned) interaction with each treatment condition. Even-numbered columns additionally include predetermined covariates selected by the \cite{belloni2014inference} LASSO procedure, and their (demeaned) interaction with each treatment condition. Robust standard errors are in parentheses. Pre-specified one- and two-sided tests: * $p<0.1$, ** $p<0.05$, *** $p<0.01$. Two-sided tests of estimates in the opposite of a pre-specified one-sided test: $^{+}$ $p<0.1$, $^{++}$ $p<0.05$, $^{+++}$ $p<0.01$.
\end{table}

\begin{table}
\caption{Treatment effects on non-electoral participation (including recommendation and follower incentive conditions)}
\vspace{-12pt}
\label{table:H9_full}
\begin{center}
\scalebox{0.53}{

}
\end{center}
\vspace{-10pt}
\footnotesize \textit{Notes}: Each specification is estimated using OLS, and includes randomization block fixed effects and adjusts for lagged outcomes and their (demeaned) interaction with each treatment condition. Even-numbered columns additionally include predetermined covariates selected by the \cite{belloni2014inference} LASSO procedure, and their (demeaned) interaction with each treatment condition. Robust standard errors are in parentheses. Pre-specified one- and two-sided tests: * $p<0.1$, ** $p<0.05$, *** $p<0.01$. Two-sided tests of estimates in the opposite of a pre-specified one-sided test: $^{+}$ $p<0.1$, $^{++}$ $p<0.05$, $^{+++}$ $p<0.01$.
\end{table}

\begin{table}
\caption{Treatment effects on institutional and interpersonal trust (including recommendation and follower incentive conditions)}
\vspace{-12pt}
\label{table:H10_full}
\begin{center}
\scalebox{0.63}{

}
\end{center}
\vspace{-10pt}
\footnotesize \textit{Notes}: Each specification is estimated using OLS, and includes randomization block fixed effects and adjusts for lagged outcomes and their (demeaned) interaction with each treatment condition. Even-numbered columns additionally include predetermined covariates selected by the \cite{belloni2014inference} LASSO procedure, and their (demeaned) interaction with each treatment condition. Robust standard errors are in parentheses. pre-specified two-sided tests: * $p<0.1$, ** $p<0.05$, *** $p<0.01$.
\end{table}

\begin{table}
\caption{Sharpened $q$ values for multiple testing of pre-specified hypotheses}
\vspace{-12pt}
\label{table:sharpened}
\begin{center}
\scalebox{0.96}{

}
\end{center}
\vspace{-10pt}
\footnotesize \textit{Notes}: The $q$ value is computed using the \cite{anderson2008multiple} approach to correcting for the false discovery rate. The coefficients $\tau_A$, $\tau_P$, and $\tau_O$ denote the estimated ATE of the quiz-incentivized PA, PP, and NP SMC conditions, respectively, from specifications without LASSO-select covariates. The hypotheses draw from our pre-analysis plan.
\end{table}

\begin{table}
\caption{Treatment effects on overall political position and political participation, reporting all treatment conditions}
\vspace{-12pt}
\label{table:overall_all_treatments}
\begin{center}
\scalebox{0.86}{

}
\end{center}
\vspace{-10pt}
\footnotesize \textit{Notes}: Each specification is estimated using OLS, and adjusts for lagged outcomes and their (demeaned) interaction with each treatment condition. Even-numbered columns additionally include predetermined covariates selected by the \cite{belloni2014inference} LASSO procedure, and their (demeaned) interaction with each treatment condition. Robust standard errors are in parentheses. Two-sided tests (not pre-specified): * $p<0.1$, ** $p<0.05$, *** $p<0.01$.
\end{table}

\begin{table}
\caption{Treatment effects on overall political position and political participation, excluding randomization block fixed effects}
\vspace{-12pt}
\label{table:overall_no_block_FEs}
\begin{center}
\scalebox{0.86}{
\begin{tabular}{lcccc|cccc}
\toprule
& \multicolumn{4}{c|}{\textbf{\textit{Midline survey outcomes}}} & \multicolumn{4}{c}{\textbf{\textit{Endline survey outcomes}}} \\
& \multicolumn{2}{c}{\textbf{Liberal}} & \multicolumn{2}{c|}{} & \multicolumn{2}{c}{\textbf{Liberal}} \\
& \multicolumn{2}{c}{\textbf{position}} & \multicolumn{2}{c|}{\textbf{Participation}} & \multicolumn{2}{c}{\textbf{position}} & \multicolumn{2}{c}{\textbf{Participation}} \\
& \multicolumn{2}{c}{\textbf{ICW index}} & \multicolumn{2}{c|}{\textbf{ICW index}} & \multicolumn{2}{c}{\textbf{ICW index}} & \multicolumn{2}{c}{\textbf{ICW index}} \\
& (1) & (2) & (3) & (4) & (5) & (6) & (7) & (8) \\
\midrule
 
& \multicolumn{2}{c}{[-3.14,2.15]} & \multicolumn{2}{c|}{[-2.17,5.15]} & \multicolumn{2}{c}{[-2.87,2.12]} & \multicolumn{2}{c}{[-1.36,4.20]} \\
\bottomrule
\vspace{-6pt} \\ 
\end{tabular}
}
\end{center}
\vspace{-10pt}
\footnotesize \textit{Notes}: Each specification is estimated using OLS, and adjusts for lagged outcomes and their (demeaned) interaction with each treatment condition. Even-numbered columns additionally include predetermined covariates selected by the \cite{belloni2014inference} LASSO procedure, and their (demeaned) interaction with each treatment condition. Robust standard errors are in parentheses. Two-sided tests (not pre-specified): * $p<0.1$, ** $p<0.05$, *** $p<0.01$.
\end{table}

\begin{table}
\caption{Treatment effects on overall political position and political participation, reweighted to match the nationally-representative 2020 ANES sample}
\vspace{-12pt}
\label{table:overall_reweighted_to_ANES}
\begin{center}
\scalebox{0.86}{

}
\end{center}
\vspace{-10pt}
\footnotesize \textit{Notes}: Each specification is estimated using OLS, and adjusts for lagged outcomes and their (demeaned) interaction with each treatment condition. Even-numbered columns additionally include predetermined covariates selected by the \cite{belloni2014inference} LASSO procedure, and their (demeaned) interaction with each treatment condition. Robust standard errors are in parentheses. Two-sided tests (not pre-specified): * $p<0.1$, ** $p<0.05$, *** $p<0.01$.
\end{table}

\begin{table}
\caption{Heterogeneous treatment effects of quiz incentives to follow SMCs on overall outcomes, by appraisal of SMCs}
\vspace{-12pt}
\label{table:HEs_appraisal}
\begin{center}
\scalebox{0.61}{

}
\end{center}
\vspace{-10pt}
\footnotesize \textit{Notes}: Each specification is estimated using OLS, and includes randomization block fixed effects and adjusts for lagged outcomes and their (demeaned) interaction with each treatment condition. Robust standard errors are in parentheses. Two-sided tests (not pre-specified): * $p<0.1$, ** $p<0.05$, *** $p<0.01$. 
\end{table}

\begin{table}
\caption{Heterogeneous treatment effects of quiz incentives to follow SMCs on overall outcomes, by machine-learned predicted appraisal of SMCs}
\vspace{-12pt}
\label{table:HEs_appraisal_predicted}
\begin{center}
\scalebox{0.61}{

}
\end{center}
\vspace{-10pt}
\footnotesize \textit{Notes}: Each specification is estimated using OLS, and includes randomization block fixed effects and adjusts for lagged outcomes and their (demeaned) interaction with each treatment condition. Robust standard errors are in parentheses. Two-sided tests (not pre-specified): * $p<0.1$, ** $p<0.05$, *** $p<0.01$. 
\end{table}

\begin{sidewaystable}
\caption{Heterogeneous treatment effects on midline and endline consumption of assigned SMCs}
\vspace{-12pt}
\label{table:HEs_consumption}
\begin{center}
\scalebox{0.64}{

}
\end{center}
\vspace{-10pt}
\footnotesize \textit{Notes}: Each specification is estimated using OLS, and includes randomization block fixed effects and adjusts for lagged outcomes and their (demeaned) interaction with each treatment condition. Robust standard errors are in parentheses. Two-sided tests (not pre-specified): * $p<0.1$, ** $p<0.05$, *** $p<0.01$.
\end{sidewaystable}

\begin{sidewaystable}
\caption{Heterogeneous treatment effects on midline and endline political preferences ICW indexes}
\vspace{-12pt}
\label{table:HEs_preferences}
\begin{center}
\scalebox{0.64}{

}
\end{center}
\vspace{-10pt}
\footnotesize \textit{Notes}: Each specification is estimated using OLS, and includes randomization block fixed effects and adjusts for lagged outcomes and their (demeaned) interaction with each treatment condition. Robust standard errors are in parentheses. Two-sided tests (not pre-specified): * $p<0.1$, ** $p<0.05$, *** $p<0.01$.
\end{sidewaystable}

\begin{sidewaystable}
\caption{Heterogeneous treatment effects on midline and endline political participation ICW indexes}
\vspace{-12pt}
\label{table:HEs_participation}
\begin{center}
\scalebox{0.64}{

}
\end{center}
\vspace{-10pt}
\footnotesize \textit{Notes}: Each specification is estimated using OLS, and includes randomization block fixed effects and adjusts for lagged outcomes and their (demeaned) interaction with each treatment condition. Robust standard errors are in parentheses. Two-sided tests (not pre-specified): * $p<0.1$, ** $p<0.05$, *** $p<0.01$.
\end{sidewaystable}

\begin{table}
\caption{Treatment effects on consumption of assigned SMCs and overall political position and political participation among respondents matched to the voter file, by (demeaned) misreported validated turnout}
\vspace{-12pt}
\label{table:overall_misreported_HEs}
\begin{center}
\scalebox{0.6}{

}
\end{center}
\vspace{-10pt}
\footnotesize \textit{Notes}: Each specification is estimated using OLS, and adjusts for lagged outcomes and their (demeaned) interaction with each treatment condition. Odd columns report average effects for the sample of participants matched to Target Smart's voter file. Even-numbered columns additionally include an interaction between treatment conditions and our indicator for misreporting 2024 turnout. Robust standard errors are in parentheses. Two-sided tests (not pre-specified): * $p<0.1$, ** $p<0.05$, *** $p<0.01$.
\end{table}

\begin{table}
\caption{Treatment effects on continued consumption after quiz incentives were withdrawn}
\vspace{-12pt}
\label{table:continued}
\begin{center}
\scalebox{0.82}{

}
\end{center}
\vspace{-10pt}
\footnotesize \textit{Notes}: Each specification is estimated using OLS, and includes randomization block fixed effects (with the exception of the smaller samples for YouTube behavioral outcomes). Except for the YouTube behavioral outcomes, even-numbered columns additionally include predetermined covariates selected by the \cite{belloni2014inference} LASSO procedure, and their (demeaned) interaction with each treatment condition. Robust standard errors are in parentheses. Two-sided tests (not pre-specified): * $p<0.1$, ** $p<0.05$, *** $p<0.01$.
\end{table}

\clearpage

%pap


\section*{Supplementary material}

# 3 Research design

During the 2024 US election campaign and its immediate aftermath, we investigate the consequences of following SMCs producing progressive political content among adults aged 18-45. Drawing information from the Better Internet Initiative (BII), our RCT evaluates the extent to which five months of exposure to SMCs—who produce either predominantly apolitical or predominantly political content—influences public opinion and political behaviors. Our field experiment promises to illuminate the effects of sustained exposure to progressive-slanted SMC content, and distinguish this from receiving similar content not provided by an SMC or consuming non-political content from SMCs.

## 3.1 The BII fellows program

The BII fellowship program helps online content creators produce videos to educate their audiences about politically salient progressive issues. The BII has expanded its annual cohorts from 35 SMCs in 2021 to over 100 SMCs in 2024, with content being posted most regularly on YouTube and Tik-Tok as well as Instagram.[5] Applicants are selected on the basis of their reach, audience rapport, and compatibility with the BII's values. Selected fellows differ considerably in the extent to which their normal content is political, with some focusing on political or news-heavy content and others focusing on predominantly apolitical content. They also differ in their followings, ranging from tens of thousands to tens of millions; collectively, BII fellows have more than 150 million subscribers.

The BII and the average SMC collaboratively produce around 10 videos annually, blending them alongside or within the creators' regular content. BII-sponsored content is fact-checked and non-partisan, but progressive in nature, and covers four themes: democracy, climate change, public health, and social and economic justice. The BII draws from partner organizations that specialize in these areas to source topical issues to SMCs for their content. Quality control is maintained by the BII's internal project manager, who closely monitors and provides feedback on all sponsored content to ensure its quality, accuracy, and informativeness. In addition to supporting content production in the SMC's own voice, fellows are paid by the BII for the content they post (at below-market rates). All content posted by BII fellows is educational, factual, and non-partisan, in compliance with section 501(c)(3) of the Internal Revenue Code.

For the purposes of this project, we selected 32 BII fellows whose normal content is either predominantly political or apolitical and strikes a good balance for audience engagement—substantial enough to be meaningful, but not so extensive as to be onerous. We further selected BII fellows for whom the bulk of their sponsored content was expected to be broadcast during the period of the proposed intervention. This set of BII fellows is augmented by 18 SMCs producing predominantly political content.

---

[5]Most BII fellows have cross-platform presence. For instance, many SIIs post their TikTok videos on their Instagram feeds.

7

We divided these SMCs along two dimensions. The first group of 25 apolitical BII fellows are those who rarely posts news or political content except as part of the BII program. The second group of political SMCs comprises 7 BII fellows and 18 others producing predominantly political content with similarly-sized followers and at a similar cadence.[6] Second, we distinguished 20 lower-intensity and 5 higher-intensity content producers within each group. The lower-intensity SMCs generally produce no more than 20 minutes of content a week, whereas higher-intensity SMCs often produce long-form content such as 30- or 60-minute shows or podcasts or produce many shorter videos each week.

### 3.2 Sample

Bovitz's Forthright pool of ∼300,000 US panelists, supplemented by small samples from Esearch (2.6% of the sample) and SurveySavvy (19.8% of the sample), provided a 4,716-person panel for this study (having targeted 4,550 baseline completes and obtained 4,566 baseline completes).[7] These panelists were appealing because they usually take non-political consumer surveys and their comparatively high pay for completing online surveys generates relatively low attrition rates (only about 40% within panel studies). We recruited American citizens aged 18-45 with the goal of them taking baseline, midline, and endline surveys. Quota sampling was used to ensure balance over gender and a final sample that is not too skewed in terms of age, education, or partisan identification. To increase the likelihood that treatment leads participants to consume novel SMC content, the 14.3% of individuals who use neither Instagram, TikTok, nor YouTube more than two days a week. We also screened out the 1.2% of participants who did not provide the name of their social media handle and the 12.0% of participants who failed at least one of two attention checks. Participants were, or will be, paid in redeemable credits within their platforms for completing the 15-minute baseline, midline, and endline surveys.

### 3.3 Treatment conditions

The core treatment condition encouraged participants to follow five assigned SMCs of a particular type they did not already follow for almost five months from mid-August until late December 2024. Subject to these SMCs being on the platform(s) regularly used by the participant, we used machine learning (based on feedback from a pilot study) to identify the SMCs predicted to be of greatest interest to each participant.[8] The account details of the SMCs an individual was assigned to follow were communicated at the end of the baseline survey, with additional reminders sent by SMS and email (both immediately and around a month into the intervention). The choice of five SMCs aimed

---

[6] The non-BII fellows produce some content that may not be compliant with section 501(c)(3) of the Internal Revenue Code.

[7] We exclude 28 ineligible participants, who were disqualified for one of the following reasons: out of study age range; consumed social media irregularly and were not screened out due to a technical error; SMC recommendations (which are generated for all participants regardless of treatment status) did not pipe within Qualtrics; multiple responses from the same participant.

[8] In the pilot, this involved first characterizing the regular content of all BII fellows (and applicants, for the placebo group) and then using baseline survey questions to minimize the distance between participants' preferences over multiple dimensions of SMC types and our characterizations. We then used feedback from our pilot study to train a neural network to predict the SMCs of the greatest participants with particular preferences.

8

to ensure that participants would encounter BII-sponsored content at least once a week during the election campaign and its aftermath. This intensity of treatment is moderate in the context of the potentially large amount of time participants spend on social media, but it may be compounded by Instagram, TikTok, and YouTube's algorithms feeding participants more similar content over time. Like most interventions on social media, the treatment should be considered as the bundle of targeted content together with the content that leads individuals to consume.

This core treatment comes in two variants. The first "political SMCs" condition encourages participants to follow five SMCs who all specialize in political content, such as day-to-day politics, economic and social issues, or topical news. Since the BII selects aligned fellows, their content is likely to complement BII-sponsored content; the non-BII fellows were selected because they produce similarly progressively-slanted political content. The second "apolitical SMCs" version instead encourages participants to follow five BII fellows who usually produce apolitical content, such as fashion and lifestyle, popular culture, or more niche areas. Whereas the former treatment condition assesses the appeal and effects of predominantly political SMCs, the latter assesses whether political content is more effectively delivered alongside content appealing to other interests (e.g. Baum and Jamison 2006; Kim 2022).

In addition to encouraging participants to follow progressively-minded SMCs, a third treatment sends the same information distilled into biweekly SMS and (where possible) email messages. Although it is impossible to do so fully, the purpose of this treatment condition is to help distinguish an SMC's delivery of content within their platform from the impact of the content itself.

Finally, there are two types of control group: a placebo control group was similarly encouraged to follow five entirely apolitical SMCs for five months (partly drawn from applicants for the BII program who were not selected); a pure control group received no content at all. The placebo group is included to hold constant the encouragement to use social media more regularly.

### 3.4 Encouragements to follow assigned SMCs

Although an SMC's BII-sponsored and other content is designed to be engaging, following five SMCs who regularly produce content is time-consuming and may involve substituting away from other content or activities. To maintain engagement with this content, we provide three types of incentives for project participants to follow the five SMCs they were assigned to follow based on their treatment condition and matched interest.

First, the most powerful but least natural encouragement rewards participants with $20 for correctly answering 4 of 5 questions in a biweekly quiz on Qualtrics about the content of their assigned SMCs' posts (with one question per SMC). Similar financial incentives have achieved high and sustained compliance rates when encouraging study participants to watch CNN instead of Fox News (Broockman and Kalla forthcoming) as well as use VPNs to access foreign news in China (Chen and Yang 2019) and consume fact-checking podcasts in South Africa (Bowles et al. 2023). Quiz ques-

tions will focus on the apolitical components of posts featuring BII messaging to avoid informing participants through the quiz itself. Furthermore, to reduce attrition in this expensive category of participants, they will be informed that $5 of each financial incentive won will be withheld until completion of the following midline or endline survey. Five biweekly quizzes were held before the midline, and a further four quizzes will be held after the midline survey.

Second, we attempt to generate exposure to SMC content algorithmically by asking participants to subscribe to the channel of each of their assigned SMCs. Toward the end of the baseline survey, and again within the midline survey, participants will be offered $1 per SMC for uploading a screenshot verifying that they now follow or have subscribed to each assigned SMC. This lighter-touch but more naturalistic encouragement, which approximates a positive online encounter with content produced by an account someone does not normally follow, draws inspiration from Levy's (2021) successful (unincentivized) nudge to subscribe to news sites on Facebook. We expect this engagement will be logged by Instagram, TikTok, or YouTube, whose algorithms may then show the participant more content from this account in future.

Third, we will simply recommend the five SMCs to participants, who will be informed that these recommendations follow from a matching algorithm. This version of the encouragement most closely mimics the algorithmic and word-of-mouth means through which individuals come to naturally follow SMCs producing content of interest to them. Unlike the previous two encouragements, it encourages participants to opt in to following an SMC without providing incentives to do so, and thus better captures the organic nature of following SMCs.

To equalize the maximum potential payment for participants across treatment conditions, treated and control participants who do not receive the quiz encouragement will be entered into lotteries. These lotteries will reward participants who complete the endline survey with entrance into a lottery with a prize that would sum up to the maximum $180 incentive available to those in the quiz condition (i.e. $170 for those in the algorithmic feedback condition and $180 for those in any other condition).

## 3.5 Randomization

The treatment and incentive conditions were cross-randomized to generate the intended distribution of participants shown in brackets in Table 1; the actual numbers assigned to each condition on our baseline survey are shown in each cell. Treatments were randomized within blocks of 13 participants with the same baseline political partisanship and who consumed similar levels of social media (more or less than 10 hours per week). These blocks were generated sequentially within Qualtrics, allowing for treatment to be assigned at the end of the baseline survey.

10

**Table 1:** Distribution of panel survey participants by treatment and encouragement condition (targeted numbers in brackets)

| | | *Treatment type:* | | | | |
| --- | --- | --- | --- | --- | --- | --- |
| | | Political BII fellows | Apolitical BII fellows | Placebo SMCs | BII messaging via SMS and email | Pure control |
| *Encouragement type:* | Biweekly $20 quiz | 355 [350] | 364 [350] | 362 [350] | | |
| | Matched interest | 364 [350] | 366 [350] | 365 [350] | | |
| | Algorithmic feedback | 361 [350] | 360 [350] | 362 [350] | | |
| | None | | | | 726 [700] | 731 [700] |

## 3.6 Data collection

We will conduct a pre-election midline survey (starting in late October 2024 and concluding around a week after this pre-analysis plan was uploaded) and a post-intervention endline survey (in January 2025) to measure outcomes and mediators, while the baseline survey (in August 2024) was used to measure pre-treatment outcomes, covariates not already available from Forthright, and potential moderators. The primary groups of self-reported outcomes are: (i) measures of engagement with assigned SMCs (e.g. videos watched per week, accurate recall of key content); (ii) broader media consumption outcomes that capture substitution away from other SMCs, TV, and offline activities; (iii) political interest and knowledge; (iv) attitudes relating to the BII themes of democracy and election integrity, public health, climate change, and social and economic justice, as well as their issue salience; (v) broader attitudes relating to progressive worldviews; (vi) evaluations of and affect toward political parties and presidential candidates; (vii) electoral participation; (viii) voting for the Democratic party; and (ix) non-electoral political participation and support for progressive causes. As described in detail below, groups of outcomes will be aggregated using inverse-covariance weighted indexes to increase precision and mitigate multiple comparison issues.

Several behavioral outcomes will supplement these self-reported outcomes. First, to the extent possible, we will match participants (who are willing to share identifying information) to voter files to measure turnout in 2024 elections and, where relevant, political disposition based on participation in prior primary elections. Second, the endline survey will ask participants to allocate $100 between 10 non-profit organizations whose causes that are aligned with BII messaging, neutral, or opposed to BII messaging; participants will be informed that the research team will make donations on behalf of 50 randomly-selected participants. Third, at the end of the endline survey, participants will be asked to share video history data from their TikTok and YouTube accounts; participants will be offered $5 to download and share their video watch histories on these platforms. Among those willing to do so, this will enrich the characterization of the treatment.

### 3.7 Missing data

Finally, before turning to our main analyses, we outline our protocol for addressing several forms of missing data in our outcome variables and covariates:

- Unobserved data: observations for which data are unobserved, due to survey attrition, refusal to answer questions, or unavailable administrative or audit records, will be treated as missing data and dropped from our analyses.

- Don't know responses: don't know survey responses will be assigned the median value on survey scales (except likelihood scales) and assigned zeros for binary or binarized variables (e.g. yes/no, correct/incorrect, etc. questions) and the lower end of the likelihood scales.

- Unreached survey branches: where a question is not relevant to a respondent (e.g. because they did not vote), we assign zeros for binary or binarized variables and the lower end of likelihood scales.

For our analysis, we aggregate some outcomes into scales. Since this requires values for each item included in the scale, we must also address aggregation across items suffering unobserved data. We will drop participants when over half of the index items are unobserved.

## 4 Analysis

### 4.1 Estimation samples

Our final sample varies by the outcome, depending on whether an outcome is measured in the midline survey, online survey, or administratively (in the case of validated turnout). For survey responses, we will further drop respondents who fail *both* the attention checks in a given survey. We will then analyze the remaining data in three ways:

- *Full sample of participants*: all participants who do not fail both attention checks.

- *Participants engaged with the study*: all participants who do not fail both attention checks *and* did not drop out of the study after treatment assignment (provided this does not generate meaningful covariate imbalances).

- *Sample reweighting*: reweighting the samples above to match population marginal distributions over key covariates (gender, partisanship, age, and race/ethnicity).

The former two approaches provide the best-powered estimates for sample treatment effects, while (marginal distribution) reweighting will help us to generalize the results to the more representative group of Millennial and Gen Z Americans.

### 4.2 Estimation

We will estimate the average intent to treat (ITT) effects of the various treatment conditions in a given survey wave using OLS regressions of the following form:

$$Y_i = \alpha_b + \beta_0 Y_i^{pre} + \boldsymbol{\beta}_1[(Y_i^{pre} - \overline{Y}^{pre}) \times \mathbf{T}_i] + \boldsymbol{\gamma}\mathbf{X}_i + \boldsymbol{\tau}\mathbf{T}_i + \varepsilon_i, \tag{1}$$

where $Y_i$ is an outcome (usually an index of outcomes) for respondent $i$, $\alpha_b$ is a randomization block fixed effect, $Y_i^{pre}$ is the closest pre-treatment measure of the outcome (wherever possible) and $\overline{Y}^{pre}$ is its sample mean, $\mathbf{X}_i$ is a vector of pre-treatment covariates (and their demeaned interaction with treatment conditions) selected by the Belloni, Chernozhukov and Hansen (2014) double-LASSO selection procedure, and $\mathbf{T}_i$ is the vector of treatment assignments. The demeaned covariate-by-treatment interaction terms are included to increase precision (Lin 2013). In the event of significant attrition, we will estimate specifications both with and without randomization block fixed effects (because differences in attrition across blocks would effectively alter observation weights). Robust standard errors will be computed throughout, reflecting the individual-level randomization.

In addition to analyzing each treatment $\times$ encouragement cell in Table 1 separately, better-powered analyses will pool across encouragement types using the following specifications:

$$Y_i = \alpha_b + \beta_0 Y_i^{pre} + \boldsymbol{\beta}_1[(Y_i^{pre} - \overline{Y}^{pre}) \times \mathbf{T}_i] + \boldsymbol{\gamma}\mathbf{X}_i + \tau_P P_i + \tau_A A_i + \tau_O O_i + \tau_M M_i + \varepsilon_i, \tag{2}$$

$$Y_i = \alpha_b + \beta_0 Y_i^{pre} + \boldsymbol{\beta}_1[(Y_i^{pre} - \overline{Y}^{pre}) \times \mathbf{T}_i] + \boldsymbol{\gamma}\mathbf{X}_i + \tau_P P_i + \tau_A A_i + \tau_M M_i + \varepsilon_i, \tag{3}$$

$$Y_i = \alpha_b + \beta_0 Y_i^{pre} + \boldsymbol{\beta}_1[(Y_i^{pre} - \overline{Y}^{pre}) \times \mathbf{T}_i] + \boldsymbol{\gamma}\mathbf{X}_i + \tau_{PA} 1[P_i = 1 \cup A_i = 1] + \tau_M M_i + \varepsilon_i, \tag{4}$$

where $P_i$ is an indicator for the pooled political SMC condition, $A_i$ is an indicator for the pooled apolitical SMC condition, $O_i$ is an indicator for the pooled placebo SMC condition, and $M_i$ is an indicator for the SMS and email messaging condition. The second of these specifications will pool the placebo and pure control groups to increase statistical power in the event that the midline and endline participants are balanced in terms of pre-treatment covariates and there are no significant differences in outcomes between the two groups, while the last specification further pools across the apolitical and political SMCs conditions. Finally, to focus on the best-powered tests, we will restrict attention to participants receiving incentivized quizzes, treating the placebo condition as the control group.

Given the limited risk of spillovers between disparate panelists, $\boldsymbol{\tau}$ identifies the set of average treatment effects—and differences between treatment effects—if the randomization survives attrition. We will assess this by testing for differential attrition across treatment conditions and examining covariate balance at every stage of analysis. While encouragement incentives are designed to be parallel across most treatment conditions, this may not apply to the pure control group. If differential attrition emerges, we will focus on subgroups where this is not a problem or, in a worst-case

13

scenario, use bounding techniques.

Since the encouragements may fail to generate exposure to assigned SMCs, ITT estimates may not capture the relevant estimand. To adjust for non-compliance, we will use instrumental variable methods to rescale the ITT by the number of videos that a participant reports watching from their five assigned SMCs. Since binarizing a multivalued treatment biases estimates when the instrument induces a first stage at multiple treatment intensities which affect the outcome (Marshall 2016), we will use 2SLS to estimate the complier-weighted average causal response for this effectively continuous endogenous treatment variable.

### 4.3 Hypotheses

We enumerate our main hypotheses in Table 2. The "first stage" hypotheses (denoted by FS) pertain to the uptake of the intervention, anticipating that our encouragements to follow SMCs (of all types) will increase engagement with assigned SMCs relative to the SMS/email and pure control conditions (FS1); substitute away from activity outside social media (FS3); and that the political SMCs treatment will expose participants to more progressive political content (FS2). For each overarching hypothesis, we prespecify a corresponding index of outcomes; the items within each family of outcomes will also be examined separately. The variable names in the table correspond with the survey items in the full midline and endline questionnaires included in the Appendix; endline outcomes are largely identical to midline outcomes, but with a couple of additional outcomes.

Our primary hypotheses (denoted by H) pertain to our main outcomes, positing that all treatment conditions except the placebo condition—but especially the SMC treatment conditions—will increase political engagement (H1), progressive policy attitudes (H2), the salience of progressive policy issues (H3), progressive political perspectives (H4), favorability toward the Democratic party (H5) at the expense of the Republican party (H6), electoral participation (H7), electoral support for the Democratic party (H8), progressive non-electoral political behaviors (H9), and institutional and interpersonal trust (H10). We class H1-H4 as the immediate outcomes most likely to be impacted by assignment to SMCs (or SMS/email messages), while H5-H10 are downstream outcomes only likely to change if immediate outcomes are affected. We expect SMC delivery to be more effective than the SMS/email delivery, but do not have a clear expectation as to whether the apolitical or political SMCs will be more impactful, and thus do not register a directional hypothesis for this important comparison. While political SMCs communicate more political information, they may turn respondents off or lack credibility due to their predominantly political content.

Our procedure for statistical inference is as follows. For the hypotheses where we register a directional expectation, as defined in the second column of Table 2, we will conduct one-tailed $t$ tests. For hypotheses where no direction is prespecified (e.g. the difference between the apolitical and political conditions for most outcomes), as well as estimates in the opposite direction to our hypothesis, we will conduct two-tailed tests. We will use the Benjamini and Hochberg (1995) method

14

Table 2: Hypotheses, tests, and outcome measurement

| Hypothesis | Coefficient test | Midline survey outcome measures | Endline survey outcome measures | Administrative data measures |
|---|---|---|---|---|
| **FIRST STAGE HYPOTHESES** | | | | |
| **FS1: increased engagement with the content of assigned SMCs** | | | | |
| FS1a: increased consumption of assigned SMCs' content (intensive margin) | $\{\tau_A; \tau_Y; \tau_0\} > \tau_M = 0$ | icw/mean(consump_freq_1_1, consump_freq_1_2, consump_freq_2_1, consump_freq_2_3, mean(consump_videos_1, consump_videos_2, consump_videos_3, consump_videos_4, consump_videos_5)) | icw/mean(consump_freq_1_1, consump_freq_1_2, consump_freq_2_1, consump_freq_2_3, mean(consump_videos_1, consump_videos_2, consump_videos_3, consump_videos_4, consump_videos_5)) | number of videos watched per assigned TikTok or YouTube SMC |
| FS1b: increased consumption of assigned SMGs' content (extensive margin) | $\{\tau_A; \tau_Y; \tau_0\} > \tau_M = 0$ | mean(consump_freq_1_1>0, consump_freq_1_2>0, consump_freq_2_2>0, consump_freq_2_3>0, consump_videos_1>0, consump_videos_2>0, consump_videos_3>0, consump_videos_4>0, consump_videos_5>0) | mean(consump_freq_1_1>0, consump_freq_1_2>0, consump_freq_2_2>0, consump_freq_2_3>0, consump_videos_1>0, consump_videos_2>0, consump_videos_3>0, consump_videos_4>0, consump_videos_5>0) | share of assigned TikTok accounts followed; share of assigned TikTok and YouTube accounts watched at least one video from |
| FS1c: increased knowledge of assigned SMGC content | $\{\tau_A; \tau_Y; \tau_0\} > \tau_M = 0$ | mean(quiz_1=1, quiz_x=1) | mean(quiz_1=1, quiz_x=1) | |
| **FS2: increased exposure to progressive and political content on social media in general** | $\tau_Y > \tau_0 = 0, \tau_Y > \{\tau_A; \tau_M\}$ | | | |
| FS2a: increased exposure to news content on social media | $\tau_Y > \tau_0 = 0, \tau_Y > \{\tau_A; \tau_M\}$ | sm_content_1 * (media_facebook + media_instagram + media_tiktok + media_youtube) | sm_content_1 * (media_facebook + media_instagram + media_tiktok + media_youtube) | |
| FS2b: increased share of news and political content on social media | $\tau_Y > \tau_0 = 0, \tau_Y > \{\tau_A; \tau_M\}$ | mean(sm_content_1, sm_content_2) | mean(sm_content_1, sm_content_2) | |
| FS2c: increased share of liberal political content on social media | $\tau_Y > \tau_0 = 0, \tau_Y > \{\tau_A; \tau_M\}$ | sm_content_2 * sm_liberal_lean (recoding don't know as balanced) | sm_content_2 * sm_liberal_lean (recoding dont know as balanced) | |
| **FS3: decreased activity outside social media** | | | | |
| FS3a: decreased consumption of traditional media | $\{\tau_A; \tau_Y; \tau_0\} < \tau_M = 0$ | media_tv + media_radio | media_tv + media_radio | |
| FS3b: decreased engagement in offline activities | $\{\tau_A; \tau_Y; \tau_0\} < \tau_M = 0$ | count(activities) | count(activities) | |
| **IMMEDIATE OUTCOME HYPOTHESES** | | | | |
| **H1: increased political engagement in general** | | | | |
| H1a: increased knowledge of current affairs | $\tau_Y > \{\tau_A; \tau_M\} > \tau_0 = 0$ | mean(pol_know_1=1, pol_know_3=1) | mean(pol_know_1=1, pol_know_2=1, pol_know_3=1) | |
| H1b: increased interest in politics | $\{\tau_A; \tau_Y\} > \tau_M > \tau_0 = 0$ | follow_politics | follow_politics | |
| **H2: increased progressive policy attitudes** | | | | |
| H2a: increased progressive climate preferences | $\{\tau_A; \tau_Y\} > \tau_M > \tau_0 = 0$ | policy_environment | policy_environment | |
| H2b: increased progressive democracy preferences | $\{\tau_A; \tau_Y\} > \tau_M > \tau_0 = 0$ | mean(democracy, agree_violence_unjustified) | mean(democracy, agree_violence, violence_unjustified) | |
| H2c: increased progressive economic justice preferences | $\{\tau_A; \tau_Y\} > \tau_M > \tau_0 = 0$ | mean(policy_corporate_power, policy_free_markets) | mean(policy_corporate_power, policy_free_markets) | |
| H2d: increased progressive public health preferences | $\{\tau_A; \tau_Y\} > \tau_M > \tau_0 = 0$ | mean(policy_family, health_insurance, abortion) | mean(policy_family, health, insurance, abortion, vax_importance) | |
| **H3: increased concern for progressive policy issues** | | | | |
| H3a: increased salience of climate issues | $\{\tau_A; \tau_Y\} > \tau_M > \tau_0 = 0$ | mean(salient_issues=12, policy_salience_3) | mean(salient_issues=12, policy_salience_3) | |
| H3b: increased salience of democracy issues | $\{\tau_A; \tau_Y\} > \tau_M > \tau_0 = 0$ | mean(salient_issues=11, policy_salience_2) | mean(salient_issues=11, policy_salience_2) | |
| H3c: increased salience of economic justice issues | $\{\tau_A; \tau_Y\} > \tau_M > \tau_0 = 0$ | mean(salient_issues=(5 or 6), policy_salience_1) | mean(salient_issues=(5 or 6), policy_salience_1) | |
| H3d: increased salience of public health issues | $\{\tau_A; \tau_Y\} > \tau_M > \tau_0 = 0$ | mean(salient_issues=(1 or 7), policy_salience_4) | mean(salient_issues=(1 or 7), policy_salience_4) | |
| **H4: increased progressive political perspective** | | | | |
| H4a: increased progressive policy preferences in general | $\{\tau_A; \tau_Y\} > \tau_M > \tau_0 = 0$ | icw(policy_environment, policy_family, policy_corporate_power, racial_inequality, health_insurance, abortion, policy_free_markets, policy_cut_taxes) | icw(policy_environment, policy_family, policy_racial_inequality, abortion, health_insurance, policy_cut_taxes, vax_importance) | |
| H4b: increased progressive worldview | $\{\tau_A; \tau_Y\} > \tau_M > \tau_0 = 0$ | mean(prog_worldview_systemic, prog_worldview_government, prog_worldview_inequalities, prog_worldview_participation) | mean(prog_worldview_systemic, prog_worldview_government, prog_worldview_inequalities, prog_worldview_participation) | |
| H4c: increased individual and group political efficacy | $\{\tau_A; \tau_Y\} > \tau_M > \tau_0 = 0$ | mean(ind_efficacy, young_generations, power)   coll_efficacy | mean(ind_efficacy, young_generations, power)   coll_efficacy | |
| H4d: increased trust in the system of electoral administration and experts | $\{\tau_A; \tau_Y\} > \tau_M > \tau_0 = 0$ | icw(agree_elections_fair, trust_election_boards, trust_scientists, trust_doctors, trust_journalists) | icw(agree_elections_fair, trust_election_boards, trust_scientists, trust_doctors, trust_journalists) | |

*Notes:* Coefficients refer to the pooled estimating equations above, but similar predictions hold for each encouragement group within a treatment category. "icw" denotes an inverse covariance weighting function with variables as its arguments; "mean" denotes a mean function with variables as its arguments; "count" denotes a function counting the number of items selected from a variable; "abs" denotes the absolute value function. Emboldened variable names denote variables measured in only the midline or only the endline survey.

Table 2: Hypotheses, tests, and outcome measurement (continued)

| Hypothesis | Coefficient test | Midline survey outcome measures | Endline survey outcome measures | Administrative data measures |
|---|---|---|---|---|
| **DOWNSTREAM OUTCOME HYPOTHESES** | | | | |
| *H5: increased favorability toward Democrats* | | | | |
| H5a: increased feeling toward the Democratic party | $\{\tau_A, \tau_P\} > \tau_M > \tau_0 = 0$ | icw(following items) party_affection_1 | icw(following items) party_affection_1 | |
| H5b: increased appraisal of Kamala Harris as presidential candidate | $\{\tau_A, \tau_P\} > \tau_M > \tau_0 = 0$ | mean(attributes_harris_economics, attributes_harris_foreign, attributes_harris_trust, attributes_harris_vision, attributes_harris_understands) | mean(attributes_harris_economics, attributes_harris_foreign, at- tributes_harris_trust, attributes_harris_vision, attributes_harris_understands) | |
| H5c: increased appraisal of incumbent government performance | $\{\tau_A, \tau_P\} > \tau_M > \tau_0 = 0$ | icw(aviation_direction, biden_approval, trust_president) | icw(biden_approval, trust_president) | |
| *H6: decreased favorability toward Republicans* | | | | |
| H6a: decreased feeling toward the Republican party | $\{\tau_A, \tau_P\} < \tau_M < \tau_0 = 0$ | icw(following items) party_affection_2 | icw(following items) party_affection_2 | |
| H6b: decreased appraisal of Donald Trump as presidential candidate | $\{\tau_A, \tau_P\} < \tau_M < \tau_0 = 0$ | mean(attributes_trump_economics, attributes_trump_foreign, attributes_trump_trust, attributes_trump_vision, attributes_trump_understands) | mean(attributes_trump_economics, at- tributes_trump_foreign, attributes_trump_trust, attributes_trump_vision, attributes_trump_understands) | |
| H6c: decreased appraisal of Trump's transition | $\{\tau_A, \tau_P\} > \tau_M > \tau_0 = 0$ | | trump_approval | |
| *H7: increased electoral participation* | | | | |
| H7a: increased probability of being registered to vote | $\{\tau_A, \tau_P\} > \tau_M > \tau_0 = 1$ | registered_to_vote=1 | | validated registration data (if possible) |
| H7b: increased probability of turning out to vote | $\{\tau_A, \tau_P\} > \tau_M > \tau_0 = 0$ | turnout≥3 | mean(turnout_pres=1, turnout_house=1, turnout_senate=1) | validated turnout data (if possible) |
| *H8: increased probability of voting Democrat* | | | | |
| H8a: increase probability of voting for Kamala Harris for president | $\{\tau_A, \tau_P\} > \tau_M > \tau_0 = 0$ | vote_choice=1 or vote_choice_early=1 | vote_choice_pres=1 | |
| H8b: increase probability of voting Democrat for House | $\{\tau_A, \tau_P\} > \tau_M > \tau_0 = 0$ | | vote_choice_house=1 | |
| H8c: increase probability of voting Democrat for Senate | $\{\tau_A, \tau_P\} > \tau_M > \tau_0 = 0$ | | vote_choice_senate=1 | |
| *H9: increased non-electoral participation* | | | | |
| H9a: increased probability of non-electoral political action | $\{\tau_A, \tau_P\} > \tau_M > \tau_0 = 0$ | icw(following items) count(non_election_part) | icw(following items) count(non_election_part) | |
| H9b: increased probability of participating in a protest for a liberal cause | $\{\tau_A, \tau_P\} > \tau_M > \tau_0 = 0$ | (non_election_part=1 and protest_issue≥3) | (non_election_part=1 and protest_issue≥3) | |
| H9c: increased probability of hypothetically resisting Trump administration | $\{\tau_A, \tau_P\} > \tau_M > \tau_0 = 0$ | | trump_policy_react=1 (1 or 2 or 3) | |
| H9d: increased probability of donating to a progressive cause | $\{\tau_A, \tau_P\} > \tau_M > \tau_0 = 0$ | | donation=1 (1 or 2 or 3 or 4) | |
| *H10: predominantly apolitical SMCs are more likely to increase trust* | | | | |
| H10a: predominantly apolitical SMCs increased institutional trust in government relative to predominantly political SMCs | $\tau_A > \tau_M > \tau_P$ | icw(following items) mean(trust_president, trust_congress, trust_supreme_court, trust_election_boards) agree_trust_people | icw(following items) mean(trust_president, trust_congress, trust_supreme_court, trust_election_boards) agree_trust_people | |
| H10b: predominantly apolitical SMCs increased interpersonal trust relative to predominantly political SMCs | $\tau_A > \tau_M > \tau_P$ | | | |

*Notes:* Coefficients refer to the pooled estimating equations above, but similar predictions hold for each encouragement group within a treatment category. "icw" denotes an inverse covariance weighting function with variables as its arguments; "mean" denotes a mean function with variables as its arguments; "count" denotes a function counting the number of items selected from a variable; "abs" denotes the absolute value function. Emboldened variable names denote variables measured in only the midline or only the endline survey.

to adjust $p$-values for multiple comparisons across our primary hypothesis indices and treatment groups.

We will also explore the effects on secondary outcomes of potential interest. These include continuing to follow assigned SMCs, perceptions of government responsiveness, reactions to the election and its aftermath, affective and policy political polarization, health care uptake, life satisfaction and mental health, attitudes toward social media platform regulations, and perceived social media value. We will also consider a variety of potential mediators driving any effects of SMCs on primary and secondary outcomes, including the type and political persuasiveness of content encountered on social media, quality of assigned SMC content and the connection participants feel towards assigned SMCs, trust in SMCs, and reliance on social media for political information.

We will further examine heterogeneity in treatment effects across theoretically motivated, pre-specified covariates. These include prior beliefs and attitudes, baseline levels of partisanship, political interest, age, social media use, and initial levels of trust in SMCs. Since the moderating role of these variables are generally theoretically ambiguous, we treat these analyses as exploratory.



\newpage
\bibliographystyle{apsr}
\putbib[rsf.bib]
\end{bibunit}

\end{document}